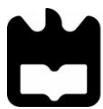

# Portuguese SKA White Book

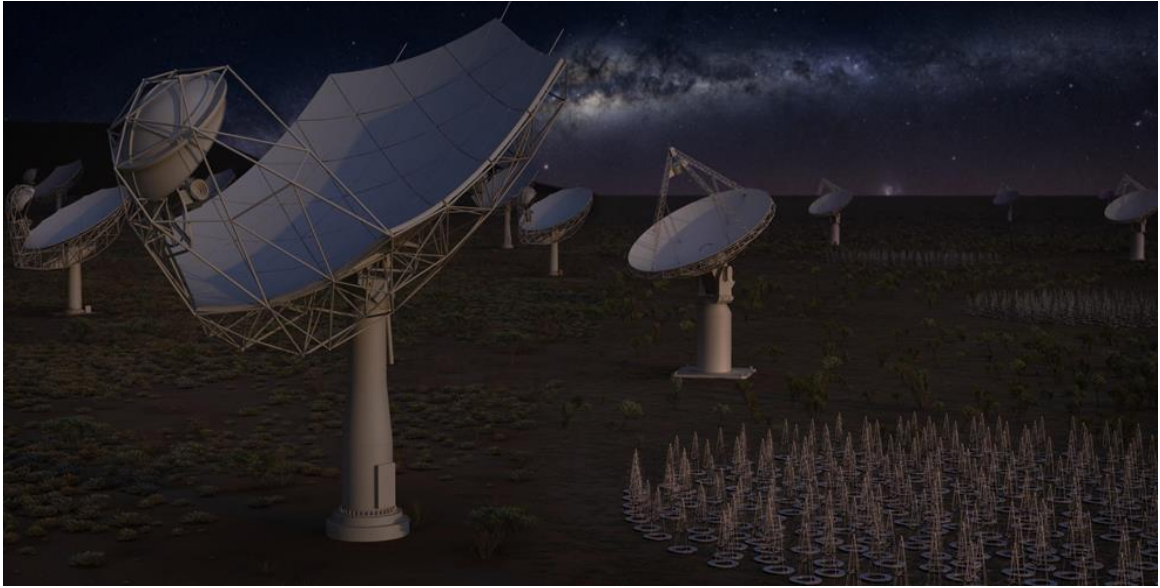

cradle of life  neutron stars
HI  cosmology  planetary missions
pulsars  virtualization
big data
galaxy evolution
atlantic connections
multi-messenger
computational astrophysics
transients  black holes
fast radio bursts
power engineering
VLBI  gravitational waves
exoplanets
heliospheric physics

**universidade de aveiro**
theoria poiesis praxis

**Title**
Portuguese SKA White Book

**Cover Image credit:** Square Kilometre Array Organisation

**Editorial Board**
Domingos Barbosa (Instituto de Telecomunicações)
Sonia Antón (Universidade of Aveiro, Instituto de Telecomunicações)
João Paulo Barraca (Instituto de Telecomunicações, Universidade of Aveiro)
Miguel Bergano (Instituto de Telecomunicações)
Alexandre Correia (Universidade de Coimbra)
Dalmiro Maia (Faculdade de Ciências da Universidade do Porto)
Valério Ribeiro (Instituto de Telecomunicações, Universidade de Aveiro)

**Publisher**
UA Editora – Universidade de Aveiro

**ISBN**
978-972-789-637-0

Agradecemos o apoio financeiro da Infraestrutura de Investigação E-Ciência Sustentável com o Square Kilometre Array (ENGAGE SKA), referência POCI-01-0145-FEDER-022217, financiada pelo Programa Operacional Competitividade e Internacionalização (COMPETE 2020) e pela Fundação para a Ciência e Tecnologia (FCT), Portugal, o apoio do Instituto de Telecomunicações, da Faculdade de Ciências da Universidade do Porto, da Universidade de Aveiro, da Universidade de Coimbra e do TICE.PT. Este trabalho foi tambem apoiado pela FCT e MCTES através de fundos nacionais e quando aplicável cofinanciado pelo FEDER, no âmbito do Acordo de Parceria PT2020 no âmbito do projeto UID/EEA/50008/2019 e projectos UIDB/50008/2020-UIDP/50008/2020.

We acknowledge financial support from Enabling Green E-science for the Square Kilometre Array Research Infrastructure (ENGAGE SKA), grant POCI-01-0145-FEDER-022217, funded by Programa Operacional Competitividade e Internacionalização (COMPETE 2020) and FCT, Portugal, and support from Instituto de Telecomunicações, Faculty of Sciences of University of Porto, University of Coimbra, University of Aveiro and TICE.PT. This work was also funded by the Portuguese Science Foundation (FCT) and Ministério da Ciência, Tecnologia e Ensino Superior (MCTES) through national funds and when applicable co-funded EU funds under the project UIDB/50008/2020-UIDP/50008/2020 and UID/EEA/50008/2019.

*"..."*

- *"SKA: will be one of the great physics machines of 21st Century and, when complete, one of the world's engineering marvels."*
- *"...Broader science range than any other science facility on Earth."*

Philip Diamond, SKA Director General,

At Portugal's SKA Meeting, Agência para o Investimento e Comércio Externo de Portugal (AICEP Portugal Global), Lisboa 2015

2020     Portuguese SKA White Book     Page 4 of 210

# Table of Contents

# Prefácio

Este livro resulta das contribuições apresentadas nos dias nacionais do Square Kilometre Array (SKA), realizados nos dias 6 e 7 de fevereiro de 2018, na presença do Vice-Director Geral do SKA Alistair McPherson e do Director para a Ciência do Projeto SKA Robert Braun. Esta iniciativa com impulso da Fundação para a Ciência e Tecnologia (FCT) foi realizada para promover o SKA - o maior radiotelescópio do mundo - entre as comunidades científicas e empresariais portuguesas, e contou com a contribuição de decisores políticos portugueses. A reunião teve muito sucesso ao fornecer uma visão geral detalhada do status, visão e objetivos da SKA e descreve a maioria das contribuições portuguesas para a ciência, a tecnologia e as aspirações relacionadas à indústria.

A coordenação coube à equipa editorial deste documento, composta por investigadores membros da infraestrutura Engage SKA Portugal e integrando entidades como o Instituto de Telecomunicações, a Universidade de Aveiro, a Faculdade de Ciências da Universidade do Porto e a Universidade de Coimbra. Naturalmente, com a evolução da ciência e a construção do SKA aproximando-se do seu inicio, este livro deve ser atualizado regularmente para refletir os resultados dos Precursores do SKA, a cada vez maior participação e integração portuguesa nas equipas científicas e de construção do SKA.

O encontro segue uma história de eventos nacionais com a presença da Organização SKA, nomeadamente desde os Encontros Nacionais Ciência 2017 e Ciência 2018 após convite do Ministério da Ciência, Tecnologia e Ensino Superior, na AICEP em 2015 com a presença do agora Presidente da AICEP, Luis Castro Henriques, a Agência de Inovação (ANI, José Caldeira) e a FCT (Ricardo Migueis); a atenção prestada nos eventos nacionais da União Internacional para a Radio Ciência (URSI) patrocinados pela ANACOM. Essas atividades foram enriquecidas ao longo dos anos por variados eventos de radioastronomia e astrofísica, tecnologia e engenharia relacionados, muitos deles patrocinados pela FCT, Radionet e outros projetos do 7º Programa-Quadro (FP7) e Horizonte 2020. O SKA contou com apresentações convidadas em eventos emblemáticos nacionais como as reuniões anuais da URSI e as Jornadas da Fundação para a Computação Cientifica Nacional (FCCN). Um marco importante foi um evento de apresentação do SKA no Museu das Comunicações com o apoio da URSI-Portugal e ANACOM e da 8ª Parceria África-Europa (Ciência, Sociedade da Informação, Espaço) patrocinada pela Comissão Europeia e com o apoio da UMIC e da FCT (Departamento da Sociedade da Informação) em Lisboa, 2012.

A reunião destes dias SKA de 2018 contou com a participação de Manuel Caldeira Cabral, Ministro da Economia, Paulo Ferrão, Presidente da FCT, Pedro Vieira, em nome da Comissão de Coordenação da da Região Centro, Daan du Toit, Diretor Geral Adjunto do Departamento de Ciência e Tecnologia do Governo da África do Sul e Presidente do Comitê de Estratégia e Negócios da SKA (STRACOM), o deputado João Marques, o Presidente da Câmara da Covilhã, Vitor Pereira, o Diretor da Parkurbis Jorge Patrão, José Pedro Borrego do regulador nacional do espectro ANACOM, João Nuno Ferreira da FCCN, Carlos Salema, Presidente de Instituto de Telecomunicações, Vasco Lagarto, Coordenador do Cluster Nacional de Competitividade TICE.PT, Susana Catita, em nome do Ministério da Ciência, Tecnologia e Ensino Superior, e contou com a presença da AICEP, do Atlantic International Research (AIR) Center, e muitos outros ilustres convidados e espectadores. Uma palavra especial é devida a Paulo Moniz pela sua hospitalidade na Universidade da Beira Interior (UBI), a Paulo Freire, do Instituto Max Planck para a Radioastronomia em Bona, Alemanha e que nos apoiou e participou online, e a Orfeu Bertolami da Faculdade de Ciências da Universidade do Porto pela sua paciência e entusiasmo e que inventariou e sumariou os resultados principais e as áreas de grande impacto do SKA apresentados nesta longa sessão. Estiveram presentes investigadores das instituições do ENGAGE SKA (Instituto de Telecomunicações, Universidade de Aveiro, Universidade de Coimbra, Faculdade de Ciências da Universidade do Porto, Instituto Politécnico de Beja), da Universidade de Lisboa e Instituto de Astrofísica e Ciências do Espaço, da Faculdade de Engenharia do Porto e representantes ao mais alto nível de empresas como a Critical Software, a Altice Labs, a Visabeira Global, o Grupo ProEF, a Voltalia Portugal, o Grupo DST, GMV, Deimos, PMEs como a Ubiwhere, Spinworks e a Innovpoint e naturalmente o cluster de competividade TICE.PT.

A discussão e as apresentações foram realizadas em dois locais diferentes: o primeiro dia foi vocacionado para a indústria e o mapeamento das grandes competências em TIC e Sistemas de Energia Sustentável e para mapear alinhamentos com as Estratégias Regionais de Especialização Inteligente (RIS3). Este primeiro dia começou com uma recepção na Universidade da Beira Interior, continuou com uma apresentação detalhada nos Paços do Concelho da Covilhã e culminou com uma visita ao Altice Data Center na Covilhã enquanto montra das infraestruturas de TIC portuguesas; o segundo dia, em Lisboa, explorou o cenário científico do SKA, com contribuições nas áreas de investigação consideradas pelo SKA como áreas chave, e que decorrem nas diferentes unidades de investigação nacionais, assim como foram explanadas as capacidades e demais aspirações da indústria.

Este evento foi organizado em conjunto pela FCT, TICE.PT, Instituto de Telecomunicações, Universidade de Aveiro, Faculdade de Ciências da Universidade do Porto, Universidade da Beira Interior, Parkurbis, Câmara Municipal da Covilhã e ENGAGE SKA. Gostaríamos também de agradecer o apoio do Altice Data Center. Detalhes do evento podem ser consultados no site:
http://www.engageska-portugal.pt/en/news/2018/02/21/successful-ska-day-national-event/

Agradecemos calorosamente a Filipa Coelho, Cristiana Leandro e Mário Amaral pela amável colaboração da FCT. Para além destas organizações, agradecemos o apoio e presença de investigadores das várias instituições e representantes de empresas.

Finalmente, agradecemos calorosamente à Organização SKA e, em particular ao seu Vice-Director Geral Alistair McPherson e ao seu Director de Ciência Robert Braun pela presença, paciência e apoio incondicional.

Domingos Barbosa, em nome da equipa editorial.

# Foreword

This book stems from the contributions presented at the Portuguese SKA Days, held on the 6th and 7th February 2018 with the presence of the SKA Deputy Director General Alistair McPherson and the SKA Science Director Robert Braun. This initiative was held to promote the Square Kilometer Array (SKA) - the world's largest radio telescope - among the Portuguese scientific and business communities with support from the Portuguese Science and Technology Foundation (FCT) with the contribution of Portuguese policy makers and researchers. The meeting was very successful in providing a detailed overview of the SKA status, vision and goals and describes most of the Portuguese contributions to science, technology and the related industry aspirations.

The coordination was handled by the editors of this document, who are researchers of the ENGAGE SKA research infrastructure and affiliated to the Instituto de Telecomunicações, the University of Aveiro, the Faculty of Sciences of University of Porto, and the University of Coimbra. Naturally, with the evolving science and with the SKA construction approaching, this book should be updated regularly to reflect the SKA pathfinder results, the Portuguese integration in SKA teams and contributions towards construction.

The meeting follows a history of national events with the participation of the SKA Organization, namely at the national Science and Technology Summits Ciência 2017 and Ciência 2018 following the invitations from the Ministry of Science, Technology and Higher Education, at AICEP in 2015 with presence of the, now, AICEP Chief Executive Officer (CEO) Luis Castro Henriques, the Innovation Agency (ANI, José Caldeira) and the FCT (Ricardo Migueis), the attention provided at the national node of the International Union of Radio Science (URSI) events sponsored by ANACOM. These activities were enriched by many radio astronomy and related technology and engineering events, many of them sponsored by FCT, Radionet and other FP7 and H2020 projects. SKA also had larger presentations at national flagship events such as the national URSI Annual meetings and FCCN Journeys. A major milestone was a side event at the Museum of Communications with support from URSI-Portugal, ANACOM and the Africa-Europe 8th Partnership (Science, ICT, Space) sponsored by the European Commission in Lisbon and by UMIC and FCT (Information Society Department) in 2012.

The SKA days in 2018 meeting had the participation of Manuel Caldeira Cabral, Minister of Economy, Paulo Ferrão, President of FCT, Pedro Vieira, on behalf of CCDR of Centro Region, Daan du Toit, Deputy Director-General of the Department of Science and Technology of Government of South Africa and Chair of the SKA Strategy and Business Committee (STRACOM), MP João Marques, the Mayor of Covilhã, Vitor Pereira, Jorge Patrão, Director Parkurbis, José Pedro Borrego from the national Spectrum regulator ANACOM (it sponsors Portuguese representation to the URSI Commission J for Radioastronomy), João Nuno Ferreira from FCCN, Carlos Salema, the President of Instituto de Telecomunicações, Vasco Lagarto, Coordinator of the national ICT Competitivity Cluster TICE.PT, Susana Catita on behalf of the Ministry of Science, Technology and Higher Education, AICEP presence, the AIR Centre, and many other distinguished guests and attendants. A special thanks is due to Paulo Moniz for his hospitality at Universidade of Beira Interior, to Paulo Freire from Max Planck für

Radioastronomie in Bonn, Germany, for his support and for joining remotely and to Orfeu Bertolami from the Faculty of Sciences of University of Porto for his patience and enthusiasm in following this long session and for summarizing the main results and impact areas of the SKA. The meeting had the presence of researchers from the ENGAGE SKA institutions (Instituto de Telecomunicações, University of Aveiro, University of Coimbra, Faculty of Sciences of the University of Porto, Instituto Politécnico de Beja), from the University of Lisbon and Institute of Astrophysics and Space Sciences, Faculty of Engineering of the University of Porto and other high level representatives of industry companies like Critical Software, Altice Labs, Visabeira Global, ProEF, Voltalia Portugal, DST Group and SMEs like Ubiwhere, Spinworks and Innovpoint and naturally the cluster TICE.PT.

The discussion and presentations were held over two days: The first day was geared towards industry and the mapping of strong competences on ICT and Sustainable Energy Systems and to map alignments with the regional Smart Specialization Strategies (RIS3). This first day started with a reception at UBI and continued with a detailed presentation at the Covilhã Town Hall. The day culminated with a visit to the Altice Data Centre in Covilhã as a showcase of Portuguese ICT infrastructures; The second day explored the SKA scientific landscape, the Portuguese contributions, the main research units' interests and the related industry experiences, capabilities and aspirations.

This event was jointly organized by the FCT, TICE.PT, Instituto de Telecomunicações, University of Aveiro, Faculty of Sciences of the University of Porto, University of Beira Interior, Parkurbis, Covilhã Town Hall and ENGAGE SKA. We acknowledge the support from Altice Data Centre. Details of the event may be consulted at the site of the event:
http://www.engageska-portugal.pt/en/news/2018/02/21/successful-ska-day-national-event/

We warmly thank Filipa Coelho, Cristiana Leandro and Mário Amaral for the kind support at FCT. We also thank all researchers and industry representatives for their presence in this meeting.

Finally, we warmly thank the SKA Organization and in particular the Deputy Director-General Alistair McPherson and the Science Director Robert Braun for their patience and support.

Domingos Barbosa, on behalf of the Editorial team.

# A importância da Radioastronomia

A Radioastronomia é o ramo da astronomia com maior impacto social e tecnológico actualmente: apesar de os seus primórdios recuarem aos anos 30 do Século XX, esta é uma ciência que se consolida no Pós-2ª Guerra Mundial. É um ramo que sempre teve uma ligação ciência–engenharia importante e uma preponderância tecnológica, num registo cultural de Ciência aberta e cidadã. As suas técnicas e descobertas foram galardoadas com o Prémio Nobel por várias vezes, tendo contribuído para a Humanidade com algumas das tecnologias com maior impacto social das últimas décadas tais como os Sistemas de referência para navegação espacial, o WiFi e a computação distribuída.

A História da Ciência prova que as observações rádio expandiram enormemente o horizonte da astronomia. A radioastronomia foi responsável pela descoberta de novos fenómenos cósmicos como pulsares, quasares e radiogaláxias. É também parcialmente responsável pela proposta de existência da matéria escura, uma componente importante do nosso Universo. A procura de vida inteligente no Universo com radiotelescópios levou ao desenvolvimento da computação distribuída através da internet: o projecto SETI@home[1] tornou-se o terceiro maior utilizador mundial de computação voluntária em larga escala para investigação cientifica através da internet. Esta atividade é apoiada pela iniciativa Breakthrough Listen[2], iniciativa científica sem precedentes com financiamento de 100 M€.

Globalmente, a radioastronomia oferece-nos uma melhor compreensão da nossa história cósmica, das várias componentes e interações físicas no Universo e abre-nos uma janela na procura de vida inteligente no Universo. A radioastronomia continuará a evoluir nas próximas décadas, aperfeiçoando novas técnicas com impacto sobretudo nas tecnologias da Informação e Comunicação (Big Data) e produzindo novas descobertas. Abaixo, a lista de algumas descobertas principais da radioastronomia:

- ❖ Prémio Nobel da Física em 1974 - Síntese de Abertura (Interferometria) e descobertas dos Pulsares (M. Ryle & A. Hewish, Cambridge, 1967)
- ❖ Prémio Nobel da Física em 1978 - Descoberta do Fundo Cósmico em Microondas (A. Penzias & R. Wilson, Bell Labs, 1965)
- ❖ Prémio Nobel da Física em 1993 – Teste da Relatividade Geral com pulsares; descoberta de nova janela para estudo das ondas gravitacionais (R. Hulse & J. Taylor, 1974)
- ❖ Prémio Nobel da Física em 2006 - Fundo Cósmico em micro-ondas: primeiras medidas das anisotropias cósmicas com o satélite COBE da NASA (J. Mather & G. Smoot, 1993).
- ❖ Primeira Imagem de um buraco negro (2019) - colaboração Event Horizon Telescope.

**Exemplos de Impacto Social e tecnológico da Radioastronomia:**

- ❖ WiFi: redes sem fios.
- ❖ Desenvolvimento de sistemas de referência para navegação espacial, usados pelo GPS, Galileo, BeiDou, Glonass, etc.
- ❖ Geodesia/sismologia: monitorização das placas tectónicas com alta precisão
- ❖ Amplificadores de baixo ruído (radar, telecomunicações e deteção remota)
- ❖ Espaço: rastreio de sondas espaciais
- ❖ Computação distribuída voluntária: plataforma avançada para aplicações distribuídas através da internet em áreas como a matemática, linguística, medicina, biologia, ciências ambientais e astrofísica.

---

[1] https://setiathome.berkeley.edu/
[2] https://breakthroughinitiatives.org/initiative/1

# The Square Kilometre Array: A Summary

*"The encyclopaedia of the Universe is written in very small (21-cm) typescript, to read it one requires a very sensitive telescope."*

Peter Wilkinson, 1990

The Square Kilometre Array (SKA) will be the world's most advanced radio observatory, many times more sensitive and hundreds of times faster at mapping the sky than today's best radio telescopes. The SKA is to be designed and built in two phases in the Southern Hemisphere, designated SKA1 and SKA2.

The SKA1 observatory comprises a global headquarters in Jodrell Bank, near Manchester, United Kingdom and two telescopes on two continents: a 197 dish interferometer in South Africa (SKA1-Mid) operating Mid frequencies between 450 and 15000 MHz, already including the 64 MeerKAT SKA precursor antennas; and a low-frequency antenna interferometer in Western Australia (SKA1-Low), operating from 50 to 350 MHz, utilizing ~131000 dipole antennas. SKA Phase 2 (SKA2, starting in +2024) will spread stations to SKA African Partners - Botswana, Ghana, Kenya, Zambia, Madagascar, Mauritius, Mozambique, Namibia - and Australia/New Zealand.

SKA1 will be a single observatory operating the two low and mid frequency telescopes with the ability to offer advanced capabilities in order to address very broad science cases. These science cases will range from fundamental questions of cosmology, astronomy and astrophysics to providing Space Weather services or even future Deep Space Network support to planetary space missions. When complete the SKA will be the most powerful radio observatory in the world. SKA1 will consist of a central core of ~200 km diameter, with 3 spiral arms of cables connecting nodes of antennas spreading over sparse territories. Eventually, SKA 2 will spread its arms connecting station nodes across several countries up to 3000km apart. Since the SKA will continuously scan the sky, it will present a strong need for high quality of service of its ICT infrastructure to achieve the required high operational availability. Therefore, the SKA presents the unique opportunity for a combination of low power computing, efficient data storage, local data services, the inclusion of newer Smart Grid power management systems, and provision of local Renewable Energies, a major milestone in scientific organizations.

|  | **SKA1_LOW (Australia)** | **SKA1_MID (South Africa)** |
|---|---|---|
| **Sensors type** | 131 000 dipoles | 197 Dishes (including 64 MeerKAT) |
| **Frequency range** | 50-350 MHz | 0.45-15 GHz |
| **Collecting Area** | 0.4 $Km^2$ | 32 000$m^2$ |
| **Max baseline** | 65 Km (between stations) | 150 Km |
| **Raw Data Output** | 7.2 Terabit/sec (estimate) | 8.8 Terabit/sec (estimate) |

**Table 1- SKA Phase 1 Telescopes Broad Characteristics**

In 2011, the SKA Organization (SKAO) was established as a UK Company Limited by Guarantee, a necessary intermediate step to enable the design and governance structure to proceed while political negotiations were being developed to establish an Inter-Governmental Organization (IGO). Since the Preparatory Phase Proposal for the SKA (PrepSKA) project, funded by the 7th EU Framework program, the IGO was widely considered as the best option to guarantee a long-term government commitment and funding stability for a global project, similar to other large research infrastructures such as ESO, ESA, CERN or ITER.

Broadly speaking, the SKA may be described as a giant network of sensors, digitisers and computers, not unlike other sensor fields or civic recording networks such as Internet of Things (IoT) networks. The sensors here are a vast collection of high frequency parabolic dish antennas with radio receivers at their foci and sparse and dense aperture arrays. The processed signals are relayed to central supercomputers over optical fibres that will extend over thousands of kilometres in its Phase 2, offering a high performance communications infrastructure in low density populated and remote areas.

The installation of the SKA in desert areas of the Karoo (South Africa) and Western Australia requires exquisitely designed mechanical and electronic components with extremely low failure rates, and with high availability similar to aerospace engineering. These components will be supported by an advanced high-availability Information and Communication Technology infrastructure.

Following the COST Workshop on the societal benefits of the SKA, and its follow-up analysis summary [5]: "**Clearly, the SKA represents more than a telescope; it is also a model for the future of global communication and information technology (ICT) and the benefits to society from ICT innovations will precede the astronomical discoveries**."

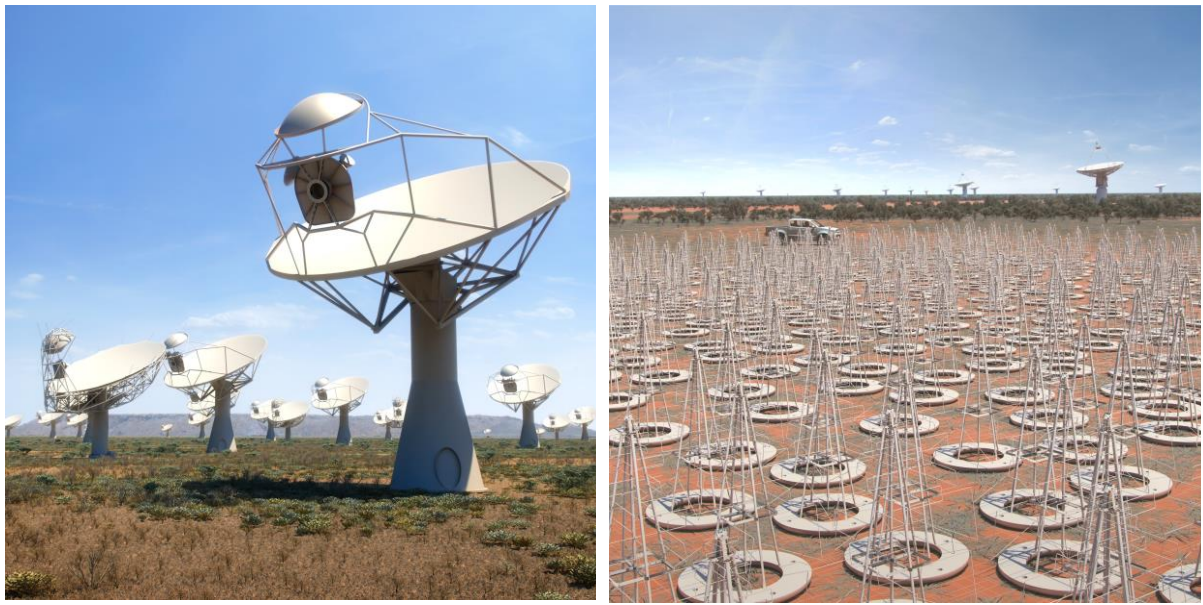

**Figure 1.** Artist impression of the SKA1 MID antennas in South Africa (left) and SKA1 LOW dipole antenna station in Australia (right).

Since 2016, the SKA has been recognized by the European Commission as a Landmark Project of the European Strategy Forum of Research Infrastructures (ESFRI) and therefore it entered the strict club of major Research Infrastructures, the only ESFRI truly global Research Infrastructure. The SKA Organization includes twelve formal members (Australia, Canada, China, India, Italy, New Zealand, South Africa, Sweden, the Netherlands, France, Spain and the United Kingdom). As of 12th of March 2019, the Convention for the establishment of the SKA Observatory IGO was signed by a group of Founding Signatories that includes Australia, Italy, China, Portugal, South Africa, The Netherlands, and the United Kingdom to be followed by several other countries such as Sweden and India. The SKA

Observatory, furthermore, involves more than 67 organizations in 20 countries, and counts with world-wide ICT industrial partners.

The ratification process for establishing the Convention will proceed throughout 2019 and early 2020 and the SKA Observatory will become the new XXIst Century big science Organization. The reason for the establishment of an IGO are listed below and are common to other well established International Conventions:

- Government commitment: Long-term political stability, funding stability
- An agreed degree of independence in structure
- Privileges and Immunities for members: functional support for project.
- Freedom to manage through procurement process, employment

### SKA: pushing the INNOVATION envelope

The SKA is rightly perceived as an iconic, and a revolutionary new radio telescope, an example for global cooperation in many 'frontier' domains of the XXIst century. The SKA project evolved to maturity though a long period of design and prototyping which will allow the SKA to achieve the enormous potential, in terms of increasing our understanding of the Universe, in exploring technologies for communication and innovation, and incorporating viable green energy supply. The design follows System Engineering rigorous principles. This design process considered the SKA geographically distributed nature and the technologies and products developed and selected for Construction were devised to minimize Capital Costs (CAPEX) and Operational Costs (OPEX) while providing high availability and reliability, good maintainability and a good upgradability path. The software developments and applications follows principles aligned with the best practices advised by the Software Engineering Institute at Carnegie Mellon University and source on the DevOps practices. These practices combine software development (Dev) and information-technology operations (Ops) aiming to shorten the systems development life cycle and provide continuous delivery with high software quality. Throughout this long design process and with the utilization of pathfinders and precursors for science and technology anticipation, we can point the following (non-exhaustive) SKA key major drivers for innovation from ICT and sensor technology to green energy systems that also generate contributions to human capital development and employment:

- Low-cost antennae: smart manufacturing and SATCOM applicability.
- Low-cost, high-performance receiver systems
- Non-cryogenic cooling
- New classes of Phased array systems, similar to Synthetic Aperture Radar systems. Impact on space Earth Observations, Space Weather monitoring
- 1Pb/s data transport (> 5x internet traffic of the Earth circa 2018): data optical transmission technologies; SKA as a market driver.
- Low-power signal processing
- Distribution of time at pico-second level over thousands of kms
- Exascale computing and Data mining analytics using Artificial Intelligence and Machine learning algorithms
- Incorporation of ICT mass market technologies: distributed databases and cloud computing, Internet of Things (IoT) smart technologies on sensor monitoring.
- Incorporation of green energy developments: sustainable energy systems and smartgrid management technologies.

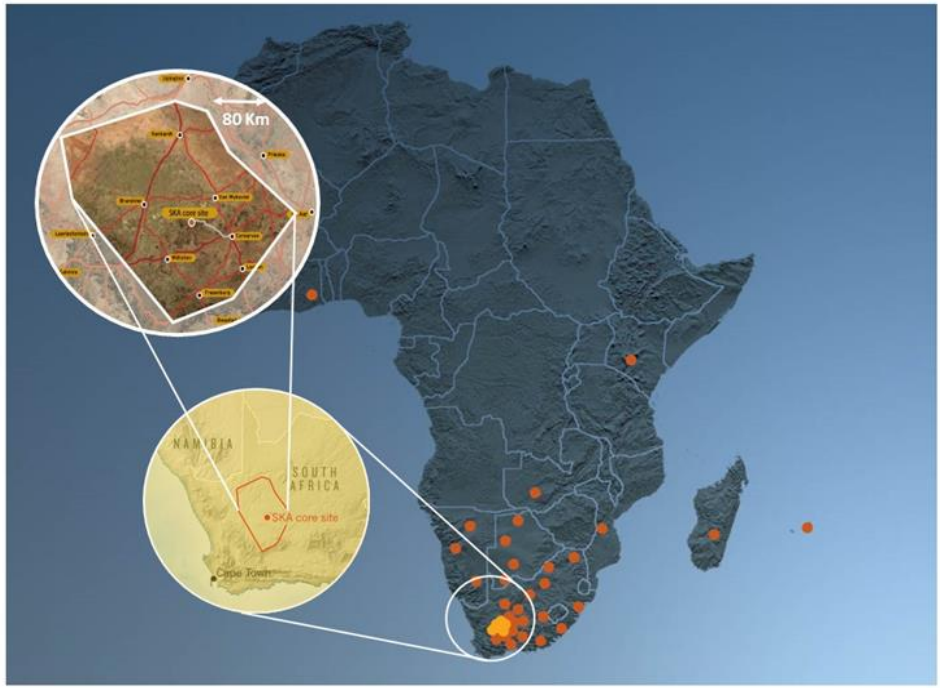

**Figure 2.** The SKA1 Radio Quiet Zone, in the Karoo, South Africa. Also shown are the planned SKA2 locations in the SKA Africa Partner Countries.

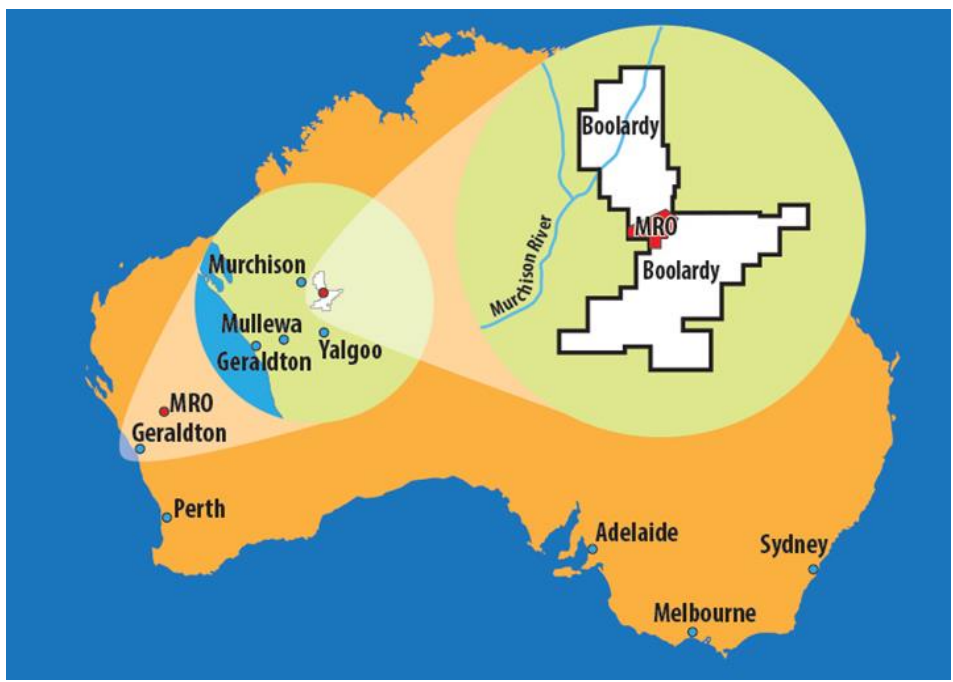

**Figure 3.** The SKA1 Radio Quiet Zone, in the Karoo, South Africa. Also shown are the planned SKA2 locations in the SKA Africa Partner Countries.

**History of the SKA**:

The first suggestions for a large radiotelescope were brought up for discussion in the early 1990's. Since then, the SKA has been developed and its scientific case refined over 25 years. A detailed historical background is found in Ekers 2012, and the summaries of the EU COST sponsored "SKA Economic Benefits" Conference, in Rome 2010. Major steps towards the SKA can be summarized below (from Ekers 2010):

- 1988 Independent suggestions for a Large Radio Telescope
- 1990 10th anniversary of VLA – the visions merge
- 1993 URSI General Assembly Kyoto resolution, Commission J Resolution for Large Telescope Group:
    - Strong scientific case
    - Internationally accessible
    - Innovative technology
    - International collaboration
- 1994 IAU forms a Large Telescope Working Group
- 1996 OECD Global Science Forum activities start
- 1998 "SKA" name adopted (1kT, SKAI, …
- 2000 International MOU signed
    - 18 members: Europe (6), United States (6), Canada (2) Australia (2), China (1), India (1), Other large members (2)
- 2001 logo competition!
- 2002 SKA activities in South Africa
- 2005 EC funding starts
- 2009 Agencies SKA Group formed (ASG)
- 2012 SKA Organization established and Site Selection concluded.
- 2019 SKA Observatory Convention signed, Rome

**Science by Design**

Since 2014 the design of SKA1 has been undertaken by a growing world-wide community of experts in the field, with the various scientific aspects being tackled by eleven design consortia following a System Engineering approach. These produced detailed Element designs, associated documentation and prototyping that have being assembled for System Critical Design Review between late 2019 and early 2020 (see Figure 4). In parallel, several SKA Science Working Groups (SWG) have been created to enable a bottom-up participation of the international scientific community. The SWGs do inform the SKA1 design, forecast SKA science by taking into account the science dynamics and the evolving engineering requirements, and absorb developments and emerging trends from other fields and enhance synergies with major observatories and experiments.

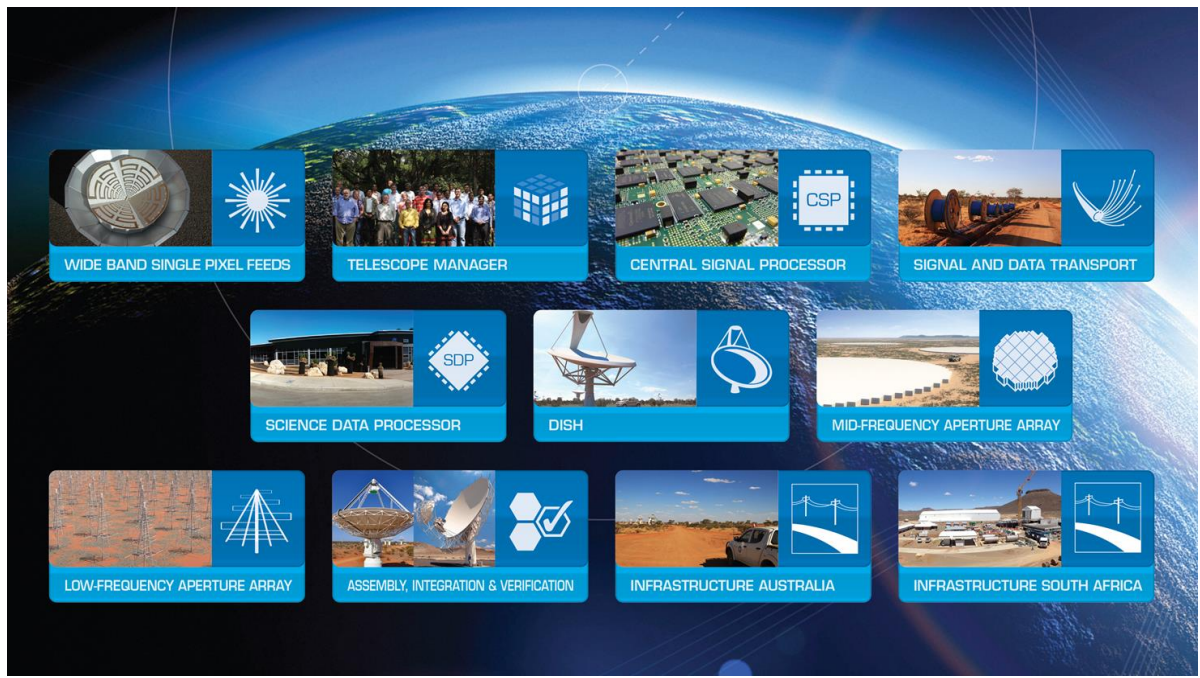

**Figure 4.** Element Design Consortia of THE SKA. Early on, SKA1 adopted a System Engineering approach in order to enable better management and informed design and minimize design, budget and construction risks.

In short, this enhances the prioritization of science topics and the emergence of candidates to large Key Science Projects (KSPs). Current SWGs represent a wide range of scientific areas:

- Cosmology
- Gravitational Waves
- Cradle of Life
- Epoch of Reionization
- Extragalactic Continuum (galaxies/AGN, galaxy clusters)
- Extragalactic Spectral Line
- HI galaxy science
- Magnetism
- Our Galaxy
- Pulsars
- Solar, Heliospheric & Ionospheric Physics
- Transients
- Transients
- Technique focused Working Group:
    - VLBI
- Topical Focus Group:
    - High Energy Cosmic Particles

Currently, as part of an inclusive, bottom-up international collaboration following an Open Science policy, anyone willing to contribute to the development of a particular science topic or domain with SKA1 is strongly encouraged to participate. Each SWG has a Chairperson or SKA Project Scientist/Science Director that can be further contacted for discussion and memberships and other scientific information. Synergy or Focus groups exploring important emergent topics that may be relevant to other large projects or science domains are also in constitution or can be proposed. In fact, the open nature of the SKA and its strong bottom-up approach sourcing in the wide scientific community opens great opportunities for the young researchers willing to contribute and build a footprint in radioastronomy.

# 1 THE OPERATIONAL MODEL

The SKA will have a uniquely distributed setup, with one observatory operating two telescopes on three continents for a global scientific community. The observatory's operations will be guided by principles which maximise scientific impact and availability while minimising radio frequency interference and ensure the data is accessible to the wide range of people globally. The SKA will be[3]:

- **Distributed** One observatory operating two telescopes on three continents for a global scientific community.
- **Accessible** Common software and user interface. Preprogramed algorithms. Training at the SKA Regional Centres.
- **Open** Open Access to non-proprietary data.
- **Citizen-ready** Access to SKA public data for citizen science projects.

Its Operational Principles can be summarized as:

- **24-hour operation** to maximise scientific impact and provide access to as much of the southern sky as possible.
- **Service Observing** (no visiting observers at the telescope) to ensure efficient operations and minimise radio frequency interference.
- **Flexible scheduling** to ensure dynamic response to observing conditions and to provide for Target-of-Opportunity and triggered events.
- **3-pronged Commensal** observing to enhance scientific productivity via sub-arrays, commensal data and processing.
- **Ability to form sub-arrays** configured and operated independently of each other.
- **Operational availability** of at least **95%**.
- **Observatory interface to users**, including data access and user support, to be provided through a network of SKA Regional Centres.

The Access Principles can be summarized as:

- **Common time allocation** process based on scientific merit and technical feasibility.
- **Access proportional** to national share in the project.
- **Up to 5% Open Time** available.
- **Key Science Projects to take up 50-75%** of observing time, with conventional PI-led projects taking up the remainder.
- **All data to be made openly** available following a proprietary period.

---

[3] SKA Operational Brochure - 1.2 August 2018. Retrieved from SKA site.

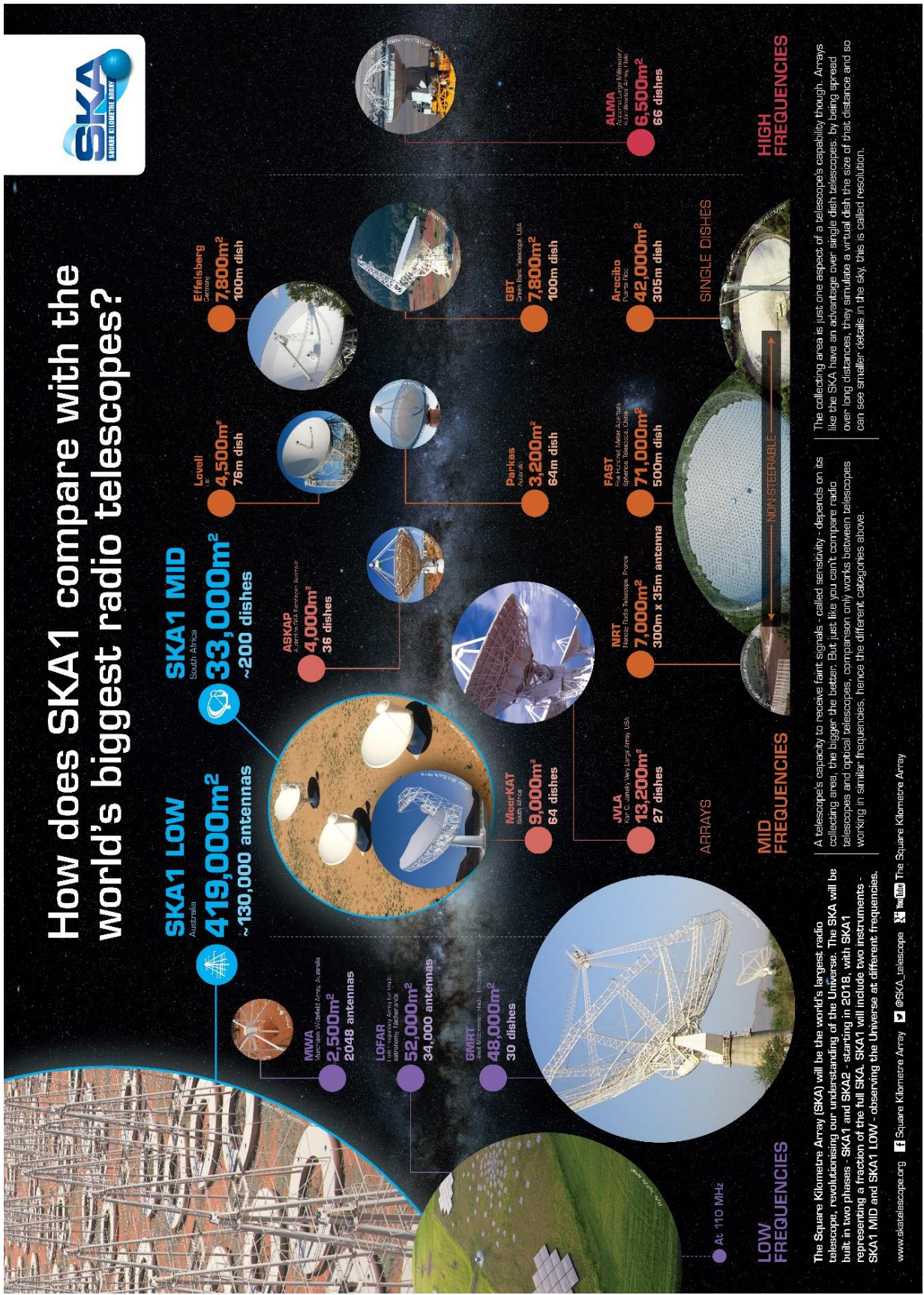

**Figure 5.** Comparison of SKA and other operating radio telescopes. Image courtesy: SKA Organisation

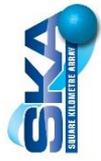
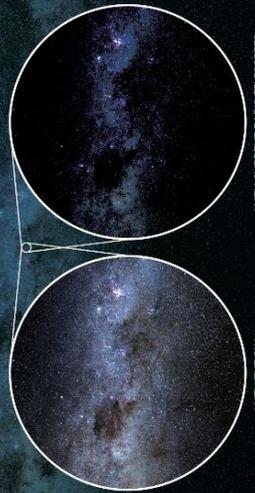
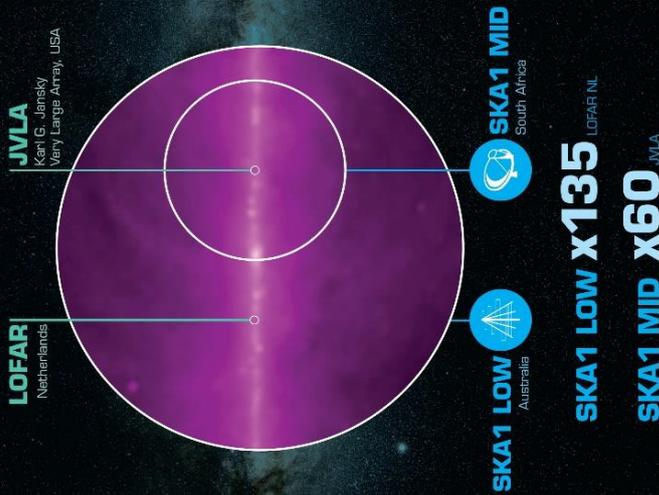
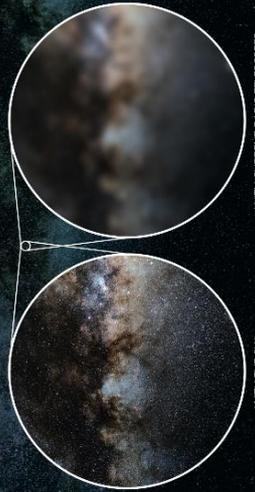

**Figure 6.** Major benefits of the SKA in angular resolution, survey speed and sensitivity. Image courtesy: SKA Organisation.

# 2 TRANSFORMATIONAL SCIENCE FROM EXA-SCALE DATA: TOWARDS REGIONAL CENTRES

Here, we have to consider that the primary activity of the SKA Observatory will be to deliver high quality primary data products. Science exploration and data provision or curation are outside its scope and have to rely on a world distributed digital infrastructure with agreed policies and sustainable funding. To tackle the dataflow/workflow deluge expected from the SKA, a network of e-infrastructures is being devised, federating high performance computing (HPC) facilities, and Data storage centres, with high speed networks designed for research workflows. This network will form the science gateway for the global SKA community to access their proprietary and public SKA data products. At the moment, several major Regional Centres are being considered, designed and prototyped in Europe, Canada, Africa, India, China and Australia. In Europe, the design is being driven through the Advanced European Network of E-infrastructures for Astronomy with the SKA (AENEAS) project, funded by EC H2020 framework program. AENEAS is designing its EU Science Data Centre (ESDC) network consulting with the major EU e-infrastructures like GEANT[4], PRACE[5], EGI[6] and RDA[7]. ESDC facilities will provide the necessary additional massive archiving capability to deal with the >600 PB/year, and the computational power to help researchers with the scientific exploration. Typical to distributed large Data Science facilities, data will be made available through a network of digital infrastructures capable of:

- **computational capacity** - post/re-processing and scientific data analysis;
- **long-term storage** - archiving of standard SKA and derived data products;
- **local user support** - scientific exploration.

Basic data products will be produced and stored in Cape Town, South Africa, for SKA1-mid and Perth, Australia, for SKA1-low. From there, these data products will be delivered to a global alliance of SKA Regional Centres for further processing and archiving and access by the user community. Development of new data models will be required as current software packages for radio astronomy data reduction do not have the capabilities to handle the SKA's large bandwidths and large Field of View datasets. The characteristics of an ESDC are currently estimated to provide the resources noted below by 2022/2023 (ie, AENEAS):

- A constantly growing amount of data beginning from tens, subsequently reaching about (estimated) ~700 Petabytes/year to Europe within a decade; and reaching >1 Exabyte after 2028.The output data rates from the Science Data Processors (SDPs, located in Australia and South Africa) to ESDC are estimated to be 55 Gbps (low) and 70 Gbps (mid) by 2023;

- Between 35-70 PetaFLOPS will be required for data processing within 2024 outside of the Science Data Processors and within the ESDC to provide high level data products;

- Data curation and stewardship will also need to be provided by the sites, leveraging a data service to the wider international communities, including support to future Large Legacy Data under OpenScience / FAIR principles.

- Portugal participates in the effort of designing a node for the SKA EU Regional Centre Network, with ENGAGE SKA as a member of AENEAS. The Portuguese contribution to the ESDC is in line with current participation in other large scale science organizations therefore capabilities must be developed to store and curate up to 5-10Petabytes/year, some of which will be at the forefront of cutting edge science and used by international science groups.

---

[4] GEANT - the Pan-European data network for the research and education community; https://www.geant.org/
[5] PRACE - Partnership for Advanced Computing in Europe; https://prace-ri.eu/
[6] EGI - European Grid Initiative; https://www.egi.eu/
[7] RDA - Research Data Alliance; https://www.rd-alliance.org/

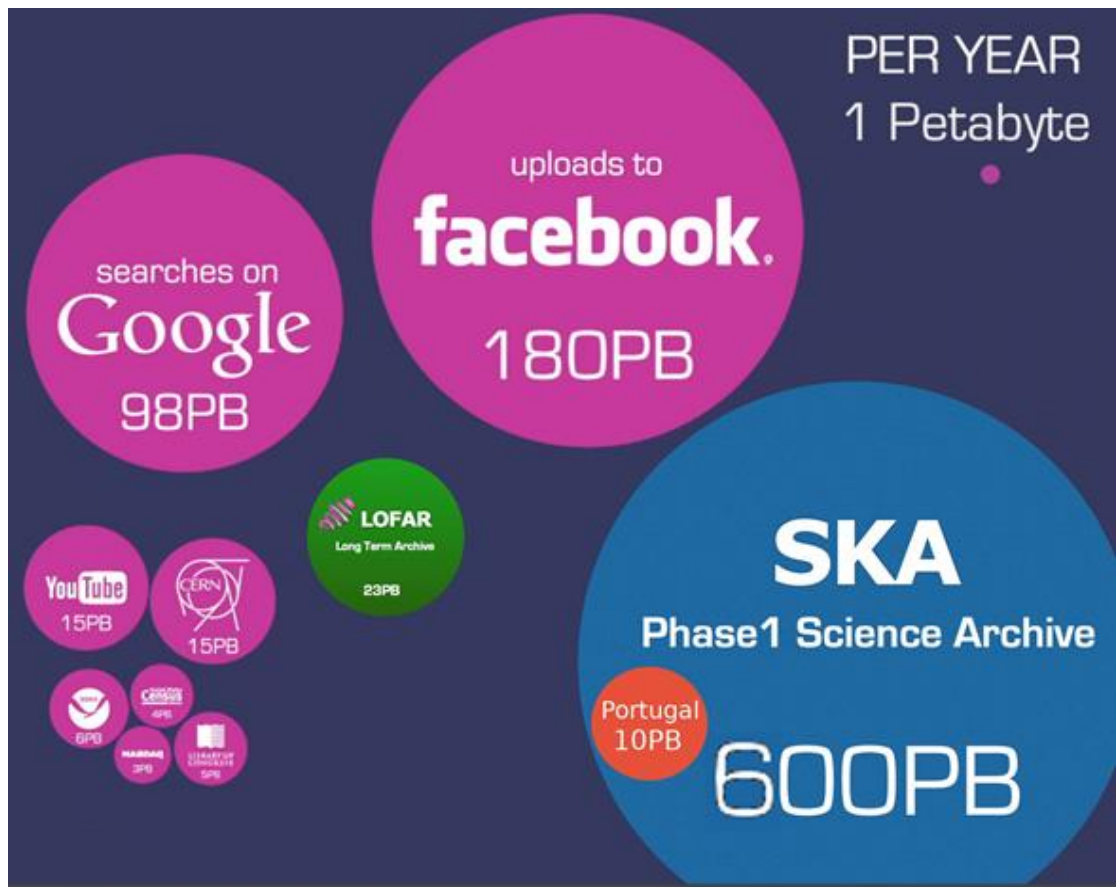

**Figure 7.** Raw estimation of SKA Phase 1 Science Archive and the potential contribution from Portugal.

To test capabilities and foster the broader scientific community science interests, the SKA Organization has organized Science and Algorithm Data Challenges, exercises on the of forefront data driven science for complexity analysis. Data Infrastructure Challenges, in the form of Proof of Concept (PoC) will also be required to test data transfer and storage performance from South Africa and Australia to Europe. The key aim of the series of Data Challenges is to prepare the wider scientific community for the kind of data products they will receive from SKA observations, and to gather valuable feedback which will inform the development of automated data reduction procedures. These procedures will rely on a large amount of automation, machine learning and artificial intelligence algorithms, for pattern recognition and faster and automatic science extraction. In particular, the way science is done is changing to a data and processing intensive approach, where small clusters are no longer viable for cutting edge science. It is anticipated the following (minimum) list of services, to be provided by the Portuguese ESDC node:

- a collaborative environment for computing, storage and software resources for the analysis and visualisation of large science data products (Peta->Exascale) that includes Machine Learning and Artificial Intelligence algorithms. This environment may evolve along interoperability principles outlined by the European Open Science Cloud

- HPC as a Service (HaaS) for on-demand numerical services;

- a Helpdesk and local user support;

- access to view project logs and track observing progress;

- accounting metrics (e.g. CPU hours, data storage, etc) at the level of individual users through groups to large projects and nations;

## ICT Technologies/Methods to use

One of main challenges for SKA related research will be the computer-assisted integrative study of large and increasingly complex combinations of data in order to understand astrophysical processes. This includes data reconfiguration to be correlated with other large datasets (data lake concept) under agreed interfaces run through the International Virtual Observatory Alliance web APIs. Deep learning techniques for feature detection, which are already being used in other scientific, medical and industrial domains - and in many of those cases surpassing the accuracy of the domain specialists - will be tweaked and applied to the specific case of data treatment. The service to be provided through the ESDCs including Portugal will include Data Products, Standard products, calibrated multi-dimensional sky images (commonly data cubes); time-series data, catalogued data for discrete sources (global sky model) and pulsar candidates. The ESDC will not hold products from individual science teams however, further Processing and Science Extraction will be performed in the ESDC. The computationally dominant steps within both of these processes are the gridding and Fourier transform steps besides on-demand numerical simulation for model sharing. The virtualisation will be based on widely adopted versions of community-supported software that will mature up to deployment, built on OpenStack, including containerization middleware like Singularity or Docker, container orchestration systems like Kubernetes (K8s) and analytic engines like SPARC. Close collaboration and convergence with large contributors (like CERN) in the open source community could lead to shorter time to market deployment and reduced long-term maintenance effort. Some of the intermediate steps will require compute intensive tasks that will greatly benefit from parallelization in a HPC environment to handle in a timely manner the massive data flow.

The implementation and developments of smart cyber-infrastructures using these technologies will source on a mix of cutting-edge industry standards like DevOps, a common practice framework that combines software development (Dev) and information-technology operations (Ops) and Agile collaborative work methodologies. DevOps aims to shorten the systems development life cycle and provide continuous delivery with high software quality by managing end-to-end engineering processes. This intends to automate the processes between software development and IT teams to enable the build, test, and release of software faster and more reliably. According to CSIRO and the Software Engineering Institute, DevOps is as "a set of practices intended to reduce the time between committing a change to a system and the change being placed into normal production, while ensuring high quality"[11]. Puppet, Chef, TeamCity OpenStack, AWS are some of the popular DevOps tools to be used.

Agile is a software development process that helps to manage complex projects like critical mission software projects and large science collaborations that emphasizes on iterative, incremental, and evolutionary development in the Software Development Life Cycle (SDLC). The Agile methodology was adopted for the construction of the SKA and makes use of such tools like JIRA, Bugzilla, Kanboard, Scrum. Inherent to the process, Agile produces better applications suites with the desired requirements. It can easily adapt according to the changes made on time, during the project life. In short, Agile offers a shorter development cycle and improved defect detection while DevOps supports Agile's release cycle.

## Positive External Impacts and Synergies

The technologies considered by SKA for deployment and development do present a wide interest to industry including in particular the telco, automotive, medical and aerospace domains. It is clear SKA will become an interesting platform of industrial relevance for advanced formation and training of skilled human resources in Information Technologies pushing forward qualified employment. The availability of an open science cloud and HPC infrastructure and the associated high performance storage data centres as part of integrated e-infrastructures, with low barriers for adoption, will allow the collaborative and synergistic integration of other major scenarios.

Key examples are the scenarios based on Smart Cities. In these scenarios the dimension of a typical Portuguese city does not require the continuous availability of highly efficient infrastructure, and the cost of deploying a sporadically used infrastructure prohibits its effective use. Here, the use of Neural Networks and Deep Learning/Artificial Intelligence techniques presents major advantages to detection and classification tasks, in particular those involved with image recognition, object classification or data analytics and forecast. This is already being deployed in pilots related to the Engage SKA infrastructure, and the Telecommunications Institute under the context of Smart Cities and Highway traffic.

The Smart Agriculture is rapidly moving towards what is nowadays called Smart Internet of Things (IoT) that is based on machine learning techniques and is a hot topic connected to Blockchain technologies. Connected to precision agriculture is the theme of water management, a critical service demanding strict security in terms of data safety. In terms of requirements, IoT and Blockchain technologies are highly demanding in terms of networking and computing resources.

# 3   PORTUGAL AND SKA: SCIENCE AND INDUSTRY RETOUR

SKA has seen growing participation from the Portuguese Scientific and industrial communities since early preparations. Portuguese participation in Science Working Groups (SWGs) is experiencing a constant growth and now SKA SWGs count with a robust presence with more Portuguese researcher contributing than any other comparable large projects in the area of Astronomy and Space Sciences subscribed by the FCT.

The SKA Science Book "Advancing Astrophysics with the Square Kilometre Array" [13] is a milestone in international scientific participation with about 135 chapters written by 1,213 contributors from 31 nationalities, adding up to some 2000 pages covering many areas of astrophysics, from cosmology to the search for life in the Universe. The book counts with articles featuring about 21 Portuguese participants in its science chapter contributions. Additionally, we also cite the participation of Portuguese scientific diaspora enabling a larger footprint of Portuguese scientific contribution, including in leadership topics. The SWGs with Portuguese presence represent a wide range of scientific areas that are synergistic with other large programs subscribed by Portugal. They are listed below and a summary of Portuguese member's activity is presented in several chapters throughout this book:

- Cosmology (FCUP, I. Telecomunicações, UP)
- Epoch of Reionization (U. Lisbon)
- Cradle of Life (U. Coimbra)
- Extragalactic Continuum, Galactic HI (U. Lisboa – I. Astrophysics & Space Science, U. Aveiro, I. Telecomunicações)
- HI Galaxy Science (FEUP)
- Our Galaxy (U. Évora)
- Solar, Heliospheric & Ionospheric Physics (FCUP)
- Transients (U. Aveiro, I. Telecomunicações)
- VLBI (Technique focused Working Group) (U. Aveiro, I. Telecomunicações)

A milestone was reached in 2015, when the ENGAGE SKA consortium of research institutions, universities and industry was selected to feature in Portugal's National Roadmap of Research Infrastructures of Strategic Relevance highlighting the impact of SKA science and technology and its opportunities for Portugal. The ENGAGE SKA is led by the Telecommunications Institute, with participation of the University of Aveiro, the Faculty of Sciences of the University of Porto, the University of Coimbra, the University of Évora, the Polytechnic Institute of Beja and more recently with Associação RAEGE-Açores. ENGAGE SKA is assisted by an industrial consortium with competences in the areas of System Engineering, information technologies, communications and electronics, energy and infrastructure, space and defence. The industries include Critical Software, Altice, Voltalia Portugal, Visabeira Global, ProEF, DST Group, SpinWorks, LC Technologies and one Science and Technology Park – Parkurbis. The industrialization of the Portuguese SKA program has been actively supported by the national Competitivity Cluster TICE.PT for ICT to promote socio-economic benefits from the scientific activities of ENGAGE SKA associates and contribute to qualified employment. ENGAGE SKA goals can be summarized below:

- Continue to support National and International Agencies in the Pre-Construction Phase of the SKA.
- Foster Science Presence and Human Capacitation, participation and leadership in SKA Working Groups
- Maximize the non-astronomy industry opportunities from SKA, including Big Data / ICT trends, Sustainable Energy systems etc.
- Foster regional research/industry connections to broader Global linkages – establish long-term tech Demonstrators and pilots in Portugal
- Provide an industry-led catalyst for informing national strategy regarding Big Data, Energy and Environmental sensors
- Develop SKA industry consortia; develop example for future industry consortia-building value chain (Research - Innovation - Market)
- Develop a long term vision of activities aligned with the United Nations Sustainable Development Goals and maximize value for society.

Portuguese researchers and industry have also been developing relevant infrastructures for radio astronomy and the SKA, including: the FCUP radio telescope for solar physics, the IT radiotelescope in Pampilhosa da Serra and mid-frequency aperture arrays testing in the Alentejo, southern Portugal as part of the SKA Advanced Instrumentation Programme. Aperture Arrays based on EMBRACE (Electronic MultiBeam Radio Astronomy ConcEpt)[8] [12] prototype telescope for the phase 2 of the Square Kilometre Array (SKA) project were publically demonstrated for the first time in the presence of the SKADS lead team for the President of the Republic Aníbal Cavaco Silva and his Presidency Delegation in 2010[9], at the inauguration of the "Centro de Acolhimento de Microempresas, Moura" as part of the then Presidency's "Roteiro da Juventude". More recently, a High Performance Computing facility, one of the most performant in the country was installed in Évora as part of the ENGAGE SKA computing efforts. This last computing facility will become one of the four poles of the National Network of Advanced Computing currently in commissioning through FCCN/FCT, as stated in the national digital Agenda INCoDe 2030[10]. The following chapters on science and industry linkages and synergies will detail and highlight show cases and contributions.

Participation in scientific preparations and design phases took advantage of the growth of science community in radiastronomical projects supported by several programs of FCT and more recently by COMPETE and EU programs (FP6, FP7 and H2020). In particular, Instituto Superior Técnico (CENTRA) has participated in the SKA Design Studies (SKADS) funded by the European Commission 6th Framework program (FP6) with a group initiated (now at Western Cape University) on the Epoch of Reionization and the 21-cm cosmology. The GEM project team (now at IT) was seeking to establish a footprint on SKA science driven design and explore capacity building with industry. This ENGAGE/IT based team made a step forward with their participation in The Preparatory Phase Proposal for the SKA (PrepSKA, FP7) focusing already in ICT domains and contributing to the delivery of the Developments plans (technical and policy) for Pre-Construction. Around the same time, by 2008, Portugal had a representative at the SKA Science and Engineering Committee (SSEC), as part of the European Delegation to the SSEC, in coordination with the European SKA Forum (ESKAF).

---

[8] https://en.wikipedia.org/wiki/EMBRACE_(telescope)
[9] http://anibalcavacosilva.arquivo.presidencia.pt/?idc=25&idt=40
[10] https://www.incode2030.gov.pt/sites/default/files/out_acp_pt.pdf

As a corollary, Portugal, through the Foundation for Science and Technology (FCT), is a founding member of the SKA Observatory, signed on the 12th of March 2019 by the Minister of Science, Technology and Higher Education Manuel Heitor in Rome, Italy (**Figure 8**).

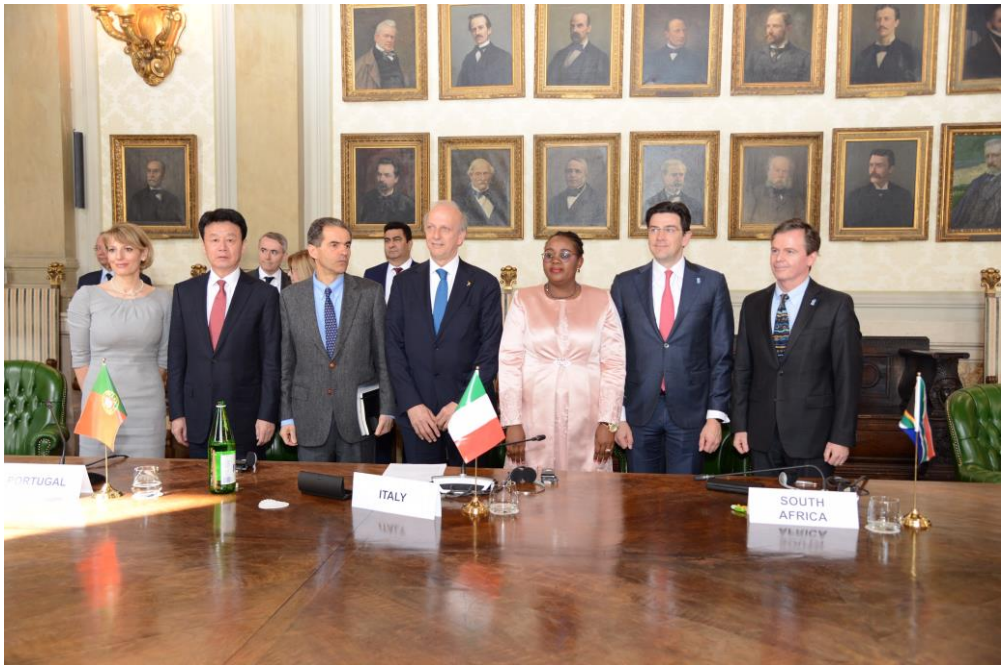

**Figure 8.** Signature Ceremony 12th of March 2019 in Rome: countries involved in the Square Kilometre Array (SKA) Project have come together in Rome for the signature of the international treaty establishing the intergovernmental organisation that will oversee the delivery of the world's largest radio telescope. Seven countries signed the treaty, including Australia, China, Italy, The Netherlands, Portugal, South Africa and the United Kingdom.

During the SKA's preconstruction phase, Portugal has been participating as an observer country of the SKA Organisation for many years. As well as contributing to the SKA's Telescope Manager, Signal and Data Transport, Science Data Processor and Mid Frequency Aperture Array consortia, Portuguese groups and contractors like Critical Software, Martifer Solar/Voltalia and Visabeira Global have also provided important consultancy on Big Data technologies and sustainable energy systems, important seeds to foster the participation of Portuguese companies in SKA1 construction. For instance, Portugal had an active role leading the LINFRA Workpackage of the Telescope Manager Element Consortium led by NCRA, India, up through the Concept Design Review (CDR) and prototyping phases. These efforts continued towards the current SKA Bridging period, where ENGAGE SKA (IT and FCUP) team is a contributor to several software SAFe Agile teams (SYSTEM, BUTTONS and in preparation for CREAM). These activities benefit from the consulting and collaboration from industrial partners working in the areas of smartcities and the space and telco sector. The Bridging Period is further supported by the ENGAGE compute Cluster at IT using a private cloud architecture that became a critical infrastructure for the SKA software development and repositories. ENGAGE SKA participated in the Advanced European Network of E-infrastructures for Astronomy with the SKA (AENEAS) project (EU H2020) investigating the feasibility plans for an EU SRC network and the technical requirements for a Portuguese node. In parallel, the project Rede Atlântica de Estações GeoEspaciais (RAEGE) started as collaboration between the Azores Autonomous Government and the Geographical Institute of Spain by installing a 13.6 metre radiotelescope in St. Maria island. This facility established applied radioastronomy for geodesy studies and applications and is becoming an asset for training and education in radioastronomy.

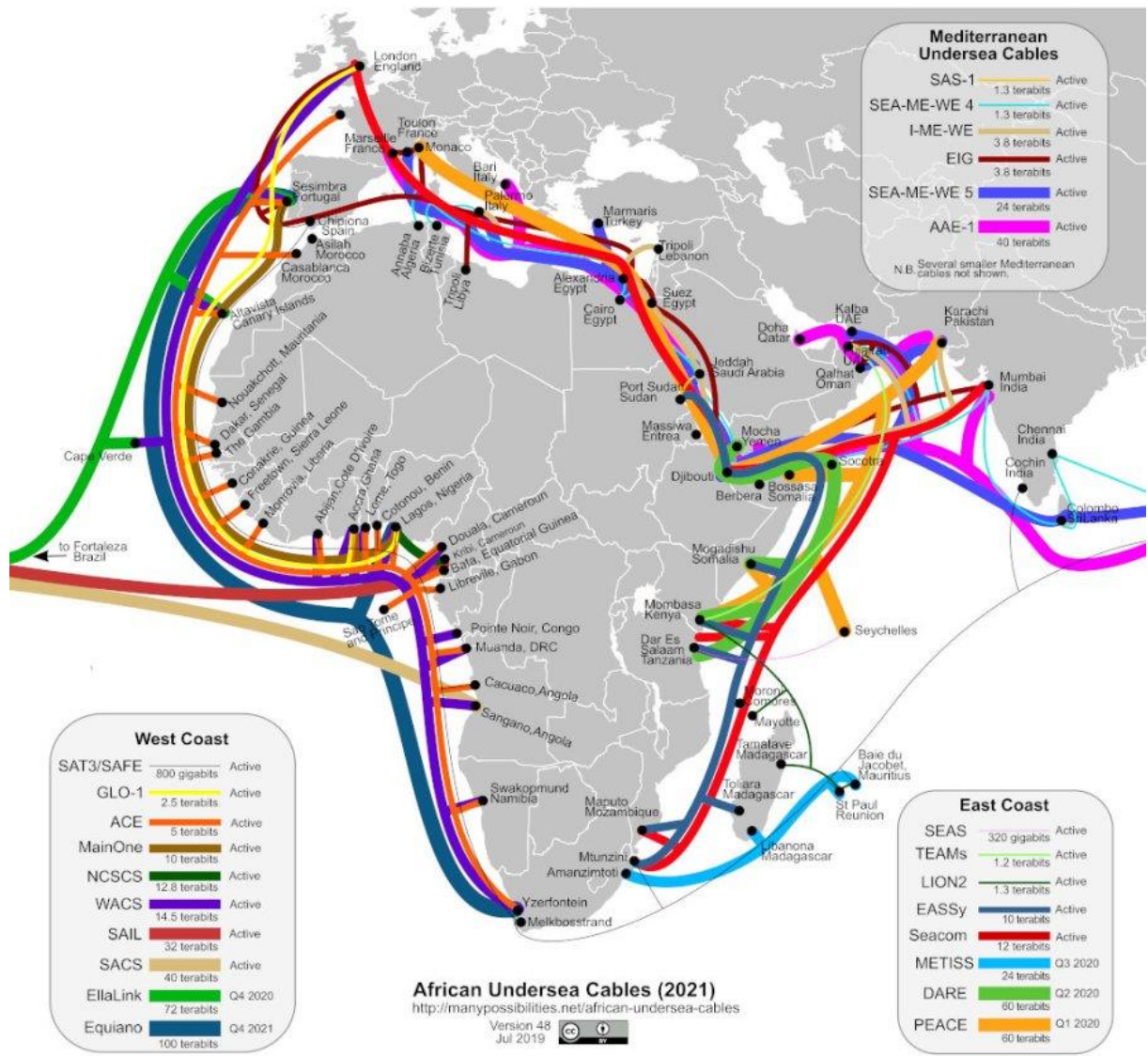

**Figure 9.** Submarine optical cable network in the Atlantic-Indian Ocean areas. Portugal is a major hub for submarine cables in the Atlantic area. This network will be of great interest for the data transmission between Africa, Australia and Europe SKA regional centres.

From early on, radioastronomy investigations with societal benefits were pursued: INESC-ID (Lisbon) was part of the Pulsar Plane project[11] (funded by the EU FP7) led by NLR in The Netherlands, addressing technologies and concepts that have the potential to bring step changes in the second half of this century and beyond. Pulsar Plane researched how new navigation systems based on the signals received from pulsars could be of interest for the aerospace sector. Pulsars are fast rotating neutron stars that emit electromagnetic radiation, which is received on Earth as a series of very stable fast periodic pulses with periods in between milliseconds and seconds (with precision better than an atomic clock) and therefore can act as celestial lighthouses. The project investigated the feasibility of an aircraft navigation system inside the Earth's atmosphere using signals from millisecond radio pulsars received by phase array antennae in aircraft.

The first world prototype of a radioastronomical system fed by solar power, an example of a green phased array, was first trialled in Moura, Portugal through the Biostirling Consortium[12] (EU FP7) (see industry linkages chapter) in collaboration with ASTRON in The Netherlands, Max Planck für

---

[11] https://cordis.europa.eu/project/id/335063/reporting
[12] https://cordis.europa.eu/project/id/309028/reporting/es

2020                           Portuguese SKA White Book                           Page 30 of 210

Radioastronomie in Germany and the Instituto de Astrofísica de Andalucía in Granada, Spain and a consortium of energy industries. The system had its full first light in 25th April 2018, Freedom Day in Portugal!

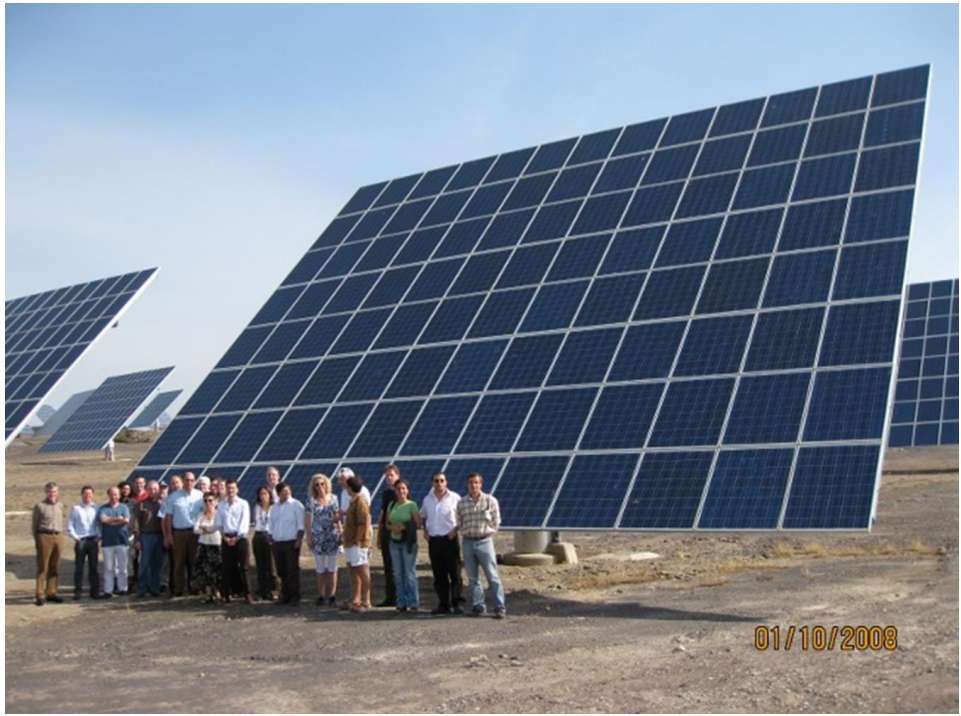

**Figure 10.** Photovoltaic Power Plant of Amareleja, Moura, Portugal: Visit from SKADS (FP6) conference held in Lisbon (2008). From early on, Portugal pioneered discussions on Green energy options for Large Scale projects.

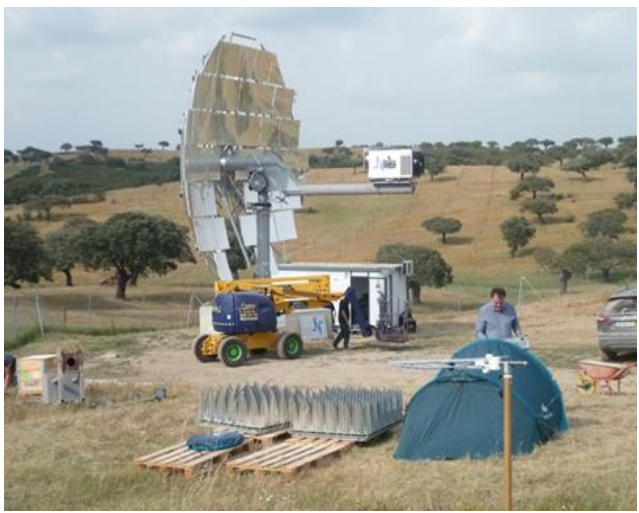
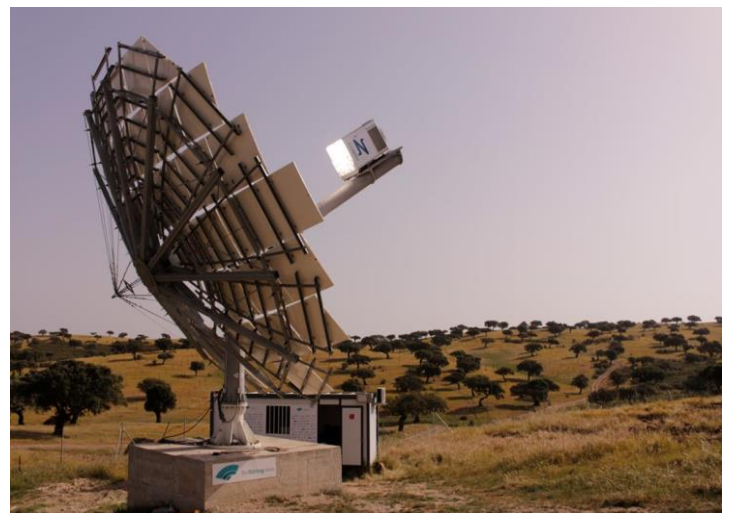

**Figure 11**. Biostirling project (FP7) in action in Herdade da Contenda, Moura. Two world first times: first time in the world a solar concentrator with a hybrid Stirling engine in its focus was trialled; first time a radioastronomical system was fed with solar power. Left: solar concentrator and modules of SKA Mid Frequency Aperture Array. Right: Dish-Stirling concentrator unit in routine Sun operation; we can see the bright focus spot, at ~700ºC.

# 4 VALUE FOR SOCIETY: INVESTING IN PEOPLE, PROSPERITY AND PEACE

The International Astronomical Union through its strategic plan "Astronomy for Development" actively promotes the use of astronomy as a tool for development by mobilizing the human and financial resources to connect science with economic growth and cultural change in society. This became particularly relevant since 2012, when an international panel of astronomers spread the SKA project between Southern Africa and Australia, maximizing its geographical impact. Besides the major socio-economic benefits that that ultimately lead to wealth and skilled jobs, SKA will strongly contribute to education and create long terms activities that will be felt over time.

Undoubtedly, the SKA will play a major role on the African scientific renaissance. Although the participation of the African Partner countries – Botswana, Ghana, Kenya, Madagascar, Mauritius, Mozambique, Namibia and Zambia – is not expected until the mid-2020s, efforts are required - right now - to transfer knowledge, technology and develop the necessary skills in radio astronomy and associated fields. Hosting part of a large scientific infrastructure like the SKA, requires major developments in cyberinfrastructures, big data, and renewable energies, which represents a major asset for improving Quality of Life (QoL).

To foster the collaboration with emergent African activities and contribute to capacitation efforts in key areas, the African European Radio Astronomy Platform (AERAP) formed in 2011 as a stakeholder forum convened to define priorities for radio astronomy cooperation between Africa and Europe. AERAP was a response to the call by the European Parliament, through the adoption of its Written Declaration 45 "on Science Capacity Building in Africa: promoting European-African radio astronomy partnerships". This call was repeated by the Heads of State of the African Union, through their decision "Assembly/AU/Dec.407 CXVIII", for radio astronomy to be a priority focus area for Africa—EU cooperation. Africa-EU collaboration in radio astronomy was rightly considered to open new innovation opportunities in ICT sectors for companies and research institutes across both continents. Similarly, AERAP promoted collaboration with the new major regions, including Australian, USA and Chinese partners.

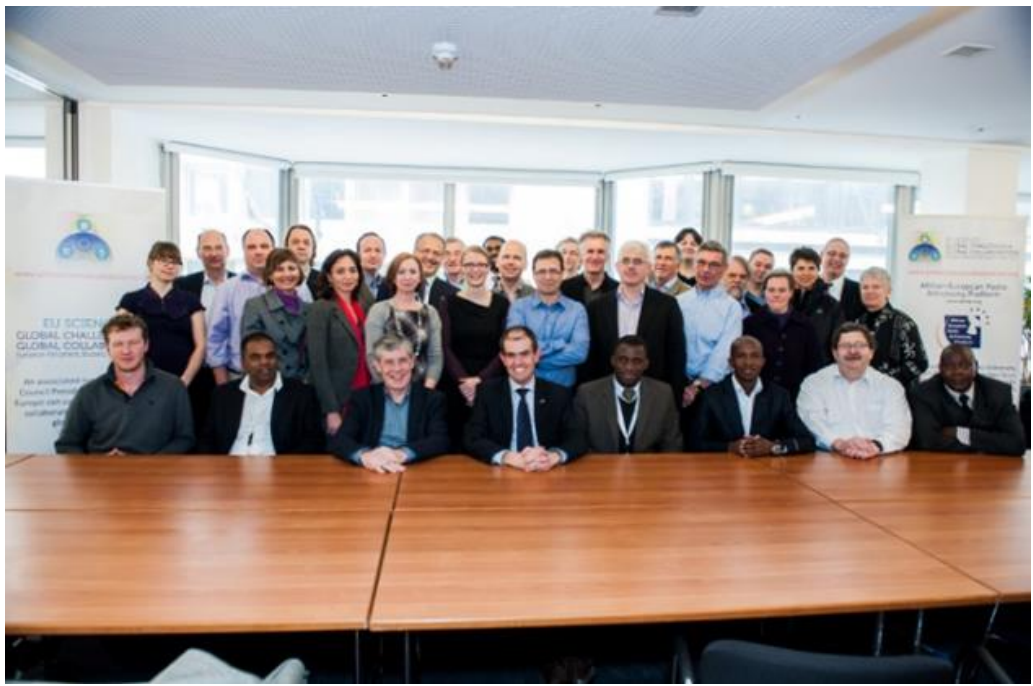

**Figure 12**. AERAP: EU participants of the Workshop to discuss the Draft AERAP Framework Program for Cooperation, South Africa Embassy, Brussels 6-7 March 2013.

Prosperity is also a key objective of AERAP, reflected in its quest for leveraging the investment in radio astronomy related technology domains such as ICT, renewable energy and advanced manufacturing for economic growth and industrial competitiveness, for the mutual benefit of Africa and Europe. Above all, these community efforts also support peace by promoting international scientific cooperation, by bringing people together, fostering international understanding, friendship and solidarity between Africa and Europe. From early on, Portuguese institutions and researchers from Instituto de Telecomunicações, Faculty of Sciences of the University of Porto and University of Aveiro in collaboration with TICE.PT fostered AERAP actions and were founding stakeholders of AERAP.

AERAP focus on research in big data and data capacity derived from the radio astronomy research and its impact on the development of astronomy science. Thus, AERAP contributes to the increased human capacity development and to bringing about win-win opportunities for Europe – Africa collaboration to develop innovative and practical applications resulting from radio astronomy collaboration. In its "Framework Programme for Cooperation" of 2013 (see Figure 9), AERAP agreed that ICT is the backbone of modern radio astronomy, enabling radio astronomers to reveal some of the most extreme events in the universe.

The existing and planned African radio astronomy facilities like KAT7, MeerKAT, the African VLBI Network or the SKA enable scientists to address a wide range of fundamental questions in physics, astrophysics, cosmology and astrobiology. They will be able to probe previously unexplored territories in the distant universe. The processing requirements for radio astronomy drive key aspects of the big data technologies in particular the analysis of "streaming data", "data analytics" as well as "exploration and visualisation" of very large-scale, but comparatively homogeneous data volumes. In the wider context, these technologies translate into areas which have broad commercial and social application. In addition to producing ground-breaking science, radio astronomy can also have a relevant and significant socio-economic impact in Africa (for example, about 40% of the components and equipment's for MeerKAT came from local industries). The operations and maintenance of radio astronomy facilities requires highly qualified engineers, scientists and other technical staff. The skills required to participate in the global radio astronomy arena are in many ways generic and applicable to other industries where high-technology skills are necessary, such as the telecommunications industry, medical devices and technologies, agriculture and the like. In parallel, as a related example of capacitation, South Africa has invested massively in the last decade on student training – both in science and engineering - in preparation for a number of astronomical projects happening on the ground. First with the creation of the National Astrophysics and Space Programme (NASSP), honours and MSc, and then the SKA Bursary program, which has given opportunities to many students from other African countries to pursue studies in astronomy, astrophysics, cosmology and engineering. Young students from the SKA African Partner countries, i.e. Botswana, Ghana, Kenya, Madagascar, Mauritius, Mozambique, Namibia and Zambia were joined by students from many other African nations. This has led to a significant step forward in human resource capacitation with hundreds of bursaries, many of them returning to their home countries as seeds to new science communities and contributing to build a local qualified labour force.

The human capacity programs have been inspirational and led to the spread of young science communities required to foster the consolidation of the Africa VLBI Network (AVN) and the local establishment of scientific groups across Southern Africa in astronomy and space in particular in the SKA African Partner countries. This tidal movement also sparked a number of initiatives to foster capacitation and economic development: The Development in Africa with radio Astronomy (DARA) program started with funding from the Newton Fund (United Kingdom) and the National Research Foundation (South Africa) and established itself as a major milestone on Advanced Training and socio-economic mapping of the AVN and SKA opportunities with Big Data and exploration of synergies and commerce with local space programs, in particular on Earth Observations and the "New space" economy thus contributing to the leveraging of the concept of co-location of astronomy, ICT and space infrastructures. This triggered a number of follow-up projects targeting Big Data (DARA Big Data) and the creation of global partnerships with Africa, Latin America and South East Asia countries like DRAGN[13]. This collaboration is building a global network of expertise in the mobilization of radio astronomy for economic development, with support from ICT and space business sectors. This global endeavour will [9] **"establish and build 'south-south' connections that will help the sharing of experiences and lessons learned in how to translate the high tech skills of radio astronomy into local job creation and entrepreneurship."**

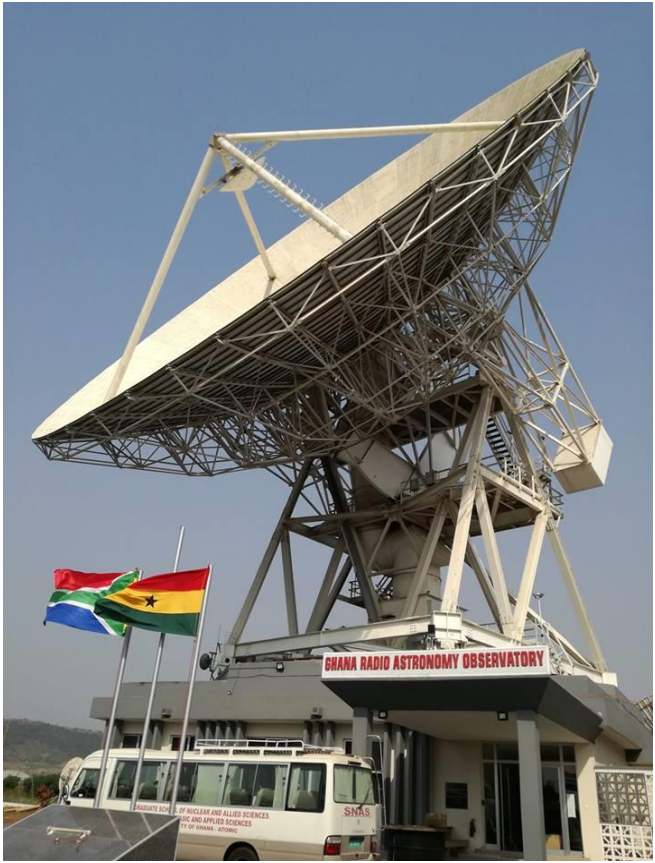

**Figure 13**. Africa VLBI Network: Ghana Radio Astronomy Observatory new radiotelescope, after refurbishments of a 32-metre SATCOM antenna similar to the Intelsat 32-metre available in S. Miguel, Azores. Photo taken during DARA sponsored Ghana Dish Conversion Workshop, Accra, January 2019.

Similarly, these partnerships paved ground for new global initiatives like the Iniciativa VLBI IBEROAMERICANA (IVIA)[14], an initiative involving collaborative work toward converting telecommunication antennas into radio telescopes, along with their commissioning for VLBI observations. IVIA will use the pool of resources of participating institutions, spread over between Europe and South and Central America to reach a critical mass in order to maximize resources and infrastructure utilization in each participating country.

Portuguese researchers have been involved and invited to DARA associated actions and contribute for North-South / South-North collaborations like IVIA. Aligned with this global trend, in Portugal the DevelOpment of PaloP knowLEdge in Radioastronomy (DOPPLER) project started as a partnership between various Portuguese and Mozambican institutions, namely the Institute of Telecommunications and University of Aveiro, the Faculty of Sciences of the University of Porto, the University of Coimbra, the University Eduardo Mondlane (Mozambique) and the Non-Governmental Organisation Osuwela (Mozambique). The partnership has the support from National Portuguese ICT Cluster, TICE.PT and is funded by FCT and by the Aga Khan Development Network (AKDN).

DOPPLER aims to promote a sustainable development agenda through human capacity development and scientific excellence. DOPPLER in collaboration with ENGAGE SKA, will address areas of strategic relevance crossing the areas of Earth Observations, Big Data and Radio Astronomy. Furthermore, DOPPLER includes initiatives for mutual capacity building in particular through capacity building around areas of biodiversity, food security and land management. DOPPLERs work plan include, pedagogical workshops, promote synergies with industry, promote Mozambique-Portugal mobility, scientific and technological demonstrations in schools, universities and industries. In fact, these capacitation efforts much benefited from a collaboration with DARA partners and related investments on the education infrastructure in Mozambique thus highlighting the global character of these partnerships (see chapter in Excellence and Society, Education and Dissemination section).

---

[13] https://dragn.info/

[14] http://www.ivia-net.org/

The common aim of this cross-fertilization of initiatives and projects is the creation of an international collaboration that can positively impact human development in the regions involved. This will be followed-up by social sciences studies to measure the societal impacts and benefits of radioastronomy at large.

# 5    SHARED SKY: ART AND DESIGN

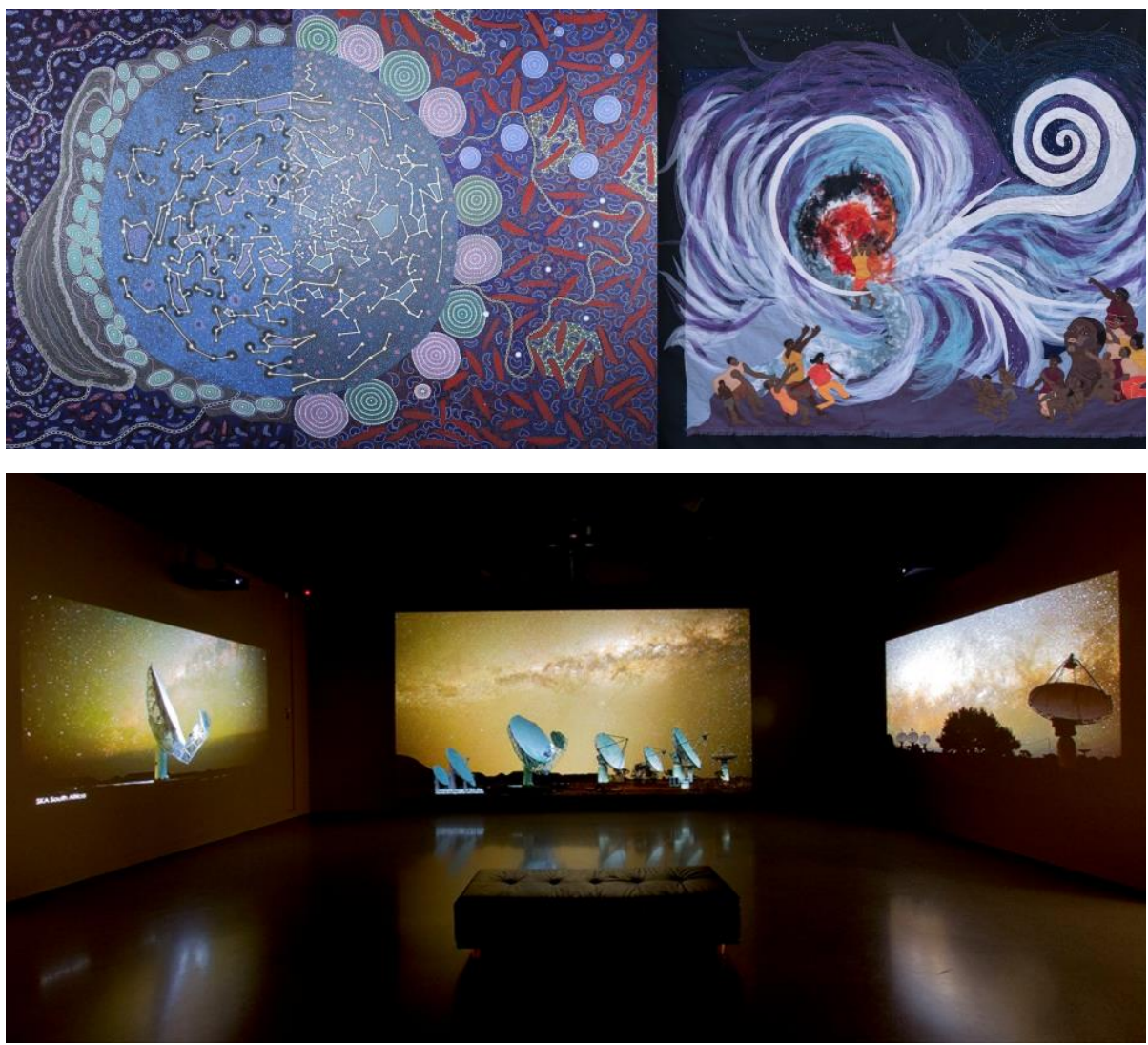

**Figure 14.** @ Shared Sky exhibition. Left: Collaborative painting of Yamaji Aboriginal artists in Australia; Right; Collaborative tapestry of artists' descendants of /Xam speaking San people of the central Karoo, South Africa. Below: video installation with the "SKA" skylines (MeerKAT and ASKAP precursors)!

In Western Australia, the onset of the SKA related activities led to a close partnership between the Commonwealth Scientific and Industrial Research Organisation (CSIRO) and other major Federal and Provincial Australian partners with the Wajarri Yamatji Aboriginal people, the custodians of the land where SKA will be deployed. Their collaboration and partnership engaged local artists to reproduce the visions of the night sky. Corollary to this magnificent collaboration, the SKA's Indigenous Art/Astronomy Exhibition Shared Sky started as an iconic way to celebrate the concept of "One Sky" shared by all Humans. This collective effort is showcasing art from descendants of some of the most ancient human populations in Australia and South Africa that stargaze at the same latitude sky as part of their life and culture for thousands of years. This collaborative exhibition joins the visions and artworks from the Wajarri Yamatji Aboriginal people from Western Australia and descendants of /Xam speaking San people from central Karoo desert region in South Africa. Paintings and kilts reflect a visual language that stretches back to a time of great antiquity.

The Shared Sky exhibition pairs the exquisite artworks, video installations with story-telling traditions from the elders and installations featuring the landscape of the Karoo and Western Australia with the new SKA precursor skyline and the wonderfully designed SKA antennas. Above all, (sic) : " It reflects the richness of the artist's ancestor's understanding of the world developed across countless generations observing the movements of the night sky. Shared Sky explores how this sophisticated understanding of celestial mechanics resonates in the work of living artists that are sharing their insights with scientists working to unlock the secrets of the Universe".

But Shared Sky is also an act of cultural dignity, a sense of pride from San South African and Wajarri Yamatji Australian people, fostering the dialogue between cultures, promoting and enriching the interactions between Art, Science and Innovation. This exhibition was inaugurated in Perth, Australia and since then it has been around the globe, including the European Commission HQ in Brussels in 2018. It is expected to be available in Portugal by 2021, an excellent opportunity to celebrate Portugal's membership to the SKA and foster friendship with the SKA nations and people.

**REFERENCES:**

[1] "The History of the SKA - born global", R. Ekers, PoS(RTS2012)007, 2012.

[2] Portuguese Roadmap of Research Infrastructures 2014-2020, FCT 2014, Available at: https://www.fct.pt/apoios/equipamento/roteiro/2013/docs/Portuguese_Roadmap_of_Research_Infrastructures.pdf

[3] Strategy report on research infrastructures; Available at: http://roadmap2018.esfri.eu/media/1066/esfri-roadmap-2018.pdf

[4] SKA Operational Brochure - 1.2 August 2018. Retrieved from the SKA site.

[5] Non-astronomy benefits of the Square Kilometre Array (SKA) radio telescope, Ver 1.6 Compiled & edited by Phil Crosby & Jo Bowler, SKA Program Development Office

[6] Shared Sky: The SKA's Indigenous Astronomy/Art Exhibition, https://www.skatelescope.org/shared-sky/

[7] African European Radio Astronomy Platform, https://www.aerap.org/

[8] Development in Africa with radio astronomy; https://www.dara-project.org/

[9] Development through Radio Astronomy – Global network; https://dragn.info/

[10] DevelOpment of PaloP knowLEdge in Radioastronomy (DOPPLER); http://doppler.av.it.pt/

[11] Bass, Len; Weber, Ingo; Zhu, Liming (2015). DevOps: A Software Architect's Perspective. ISBN 978-0134049847.

[12] Torchinsky, S. A.; Olofsson, A. O. H.; Censier, B.; Karastergiou, A.; Serylak, M.; Picard, P.; Renaud, P.; Taffoureau, C. (2016), "Characterization of a dense aperture array for radio astronomy", Astronomy & Astrophysics, 589: A77, arXiv:1602.07976, Bibcode:2016A&A...589A..77T, doi:10.1051/0004-6361/201526706

[13] SKA Science Book "Advancing Astrophysics with the Square Kilometre Array, editor: SKA Organisation, 2015, http://pos.sissa.it/cgi-bin/reader/conf.cgi?confid=215

# Scientific Challenges

# Cosmology with SKA[1]

Orfeu Bertolami and Cláudio Gomes

Departamento de Física e Astronomia, Faculdade de Ciências da Universidade do Porto, Rua do Campo Alegre s/n, 4169-007 Porto, Portugal

**ABSTRACT**

We review some of the major contributions that the Square Kilometre Array (SKA) will provide for Cosmology. We discuss the SKA measurements of the equation of state parameter for dark energy from Baryonic Acoustic Oscillations (BAO), of the dark matter power spectrum and modifications of the Poisson equation or the slip relation from weak lensing. We also comment on measurements of the cosmic magnetism and its role on the dynamics of the Universe.

**Keywords:** SKA, Cosmology, dark energy, dark matter, gravity, modified gravity, cosmic magnetism

## 1 INTRODUCTION

The Square Kilometre Array (SKA) is a worldwide collaboration with radio telescopes hosted in South Africa and Australia at a first stage, and in other eight African countries later on, namely, Botswana, Ghana, Kenya, Madagascar, Mauritius, Mozambique, Namibia and Zambia. This will be largest radio telescope array in the world and it will end up being 50 times more sensitive and 100 times faster than current best radio telescopes. Its life span is expected to be of at least 50 years, and although it uses the most recent antenna technology and signal processing and computing, it can be continuously upgraded as computing power increases. SKA will cover frequencies in the range from 70 MHz to 10 GHz.

The knowledge of the Universe will be dramatically increased with SKA, since it will provide insight on the evolution of galaxies and cosmology, on strong gravity through pulsars and black holes, on the origin and evolution of cosmic magnetism, on cosmic history of the Universe at dark ages and reionisation epochs, and on putative cradle of life [1].

In what concerns Cosmology, radio observations from SKA are expected to constrain the equation of state parameter for dark energy, hence discriminating between several models of cosmic acceleration, to observe a dark matter power spectrum, to test modified theories of gravity using weak lensing observations, and to get and insight on the cosmic magnetism [1-3].

## 2 COSMOLOGY

### Dark Universe

One of the science key goals of SKA concerns the understanding of the dark components of the Universe. Several missions have been designed to study the energy content of the Universe. The most recent, Planck mission [4], showed that the Universe is filled with 4.9% of ordinary matter, 26.8% of dark matter and 68.3% of dark energy. Dark matter is responsible for the flattening of the galactic rotation curves, and gravitational lensing which cannot be explained only by regular matter. On its turn, dark energy is the smooth energy substratum behind the current acceleration of the Universe. These two dark components dominate the Universe. Nevertheless, aside from some observational signatures and dynamical effects, their inner nature still remains unknown.

SKA will be able to provide some answers about the components of the Universe through the 21cm radiation of the neutral hydrogen at high redshifts. The cosmic evolution of such well-defined signal encodes information about the medium where it propagates, most particularly about the dark matter and dark energy effects. For instance, by measuring the BAO signature in the radio of the galaxy power spectrum, SKA will allow for precise measurements of the equation of state of dark energy, $w = p/\rho$, of the form:

$$w = w_0 + w_a(1 - a), \quad (1)$$

where $p$ and $\rho$ are the pressure and energy density of dark energy, $w_0$ and $w_a$ are some real parameters and $a$ is the scale factor. In Fig. 1 one can see that assuming a Gaussian distribution around the fiducial model $w = -1$, which corresponds to the cosmological constant as dark energy, the predicted SKA results combined with Planck [4] and Baryon Oscillation Spectroscopic Survey (BOSS, [5]) priors constrain considerably the allowed region of parameters. The expected combined precision is compared with the one from Euclid mission [6]. Furthermore, by studying weak gravitational lensing the measurement of the dark matter power spectrum will be possible, and bounds on the mass and the number of neutrinos families can be obtained.

---

[1] Based on a talk delivered by one of us (O.B.) at the SKA.PT Days at February 7th, 2018 in Lisbon, Portugal

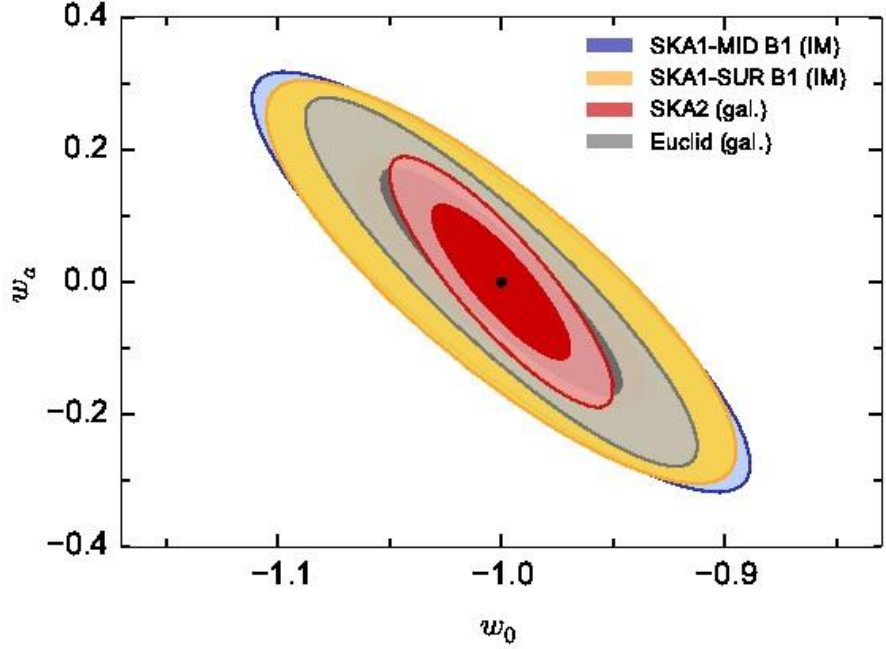

**Figure 1**. Predicted SKA constraints at various stages on the equation of state parameter of dark energy (with BOSS and Planck priors, [3]).

## Testing gravity to the limit

As stated, testing the nature of dark matter or dark energy is a key issue in Cosmology. However, the effect of these putative entities can be manifestations of a yet unknown gravity theory beyond General Relativity [7]. Thus, testing gravity at the strong limit provides a deeper understanding on the nature of this fundamental interaction. SKA, in particular, will be able to detect a vast number of pulsars, and binary systems of a pulsar orbiting a black hole, hence providing rich information about strong gravity. Thousands of millisecond pulsars might be detected, forming a "pulsar timing array" which can be a prime arena for detection of gravitational waves [8]. On the other hand, General Relativity and alternative theories of gravity will be tested through gravitational lensing observations. In general, there are three model dependent parameters which characterise the growth of structures in any modified gravity theory [6]. The first is the modified gravitational constant, $G_N \tilde{\mu}(a,k)$, where $\tilde{\mu}(a,k)$ is a model dependent correction in the Poisson equation, which can be expressed in the Fourier space as:

$$-2k^2 \Phi = 8\pi G_N a^2 \rho D \tilde{\mu}(a,k), \quad (2)$$

where $k$ is the wavenumber and $D$ is the gauge-invariant density contrast. The second is the anisotropic stress or slip relation, $\tilde{\gamma}$, which relates the two gauge invariant Bardeen potentials, $\Phi$ and $\Psi$, which come from time and spacial scalar perturbations of the metric, respectively:

$$\Psi = \tilde{\gamma}(a,k)\Phi, \quad (3)$$

and the third one which is the growth rate, $f(a,k)$, (or its index $\gamma$):

$$f(a,k) = \left( \frac{a^2 8\pi G_N \rho}{3H^2} \right)^\gamma, \quad (4)$$

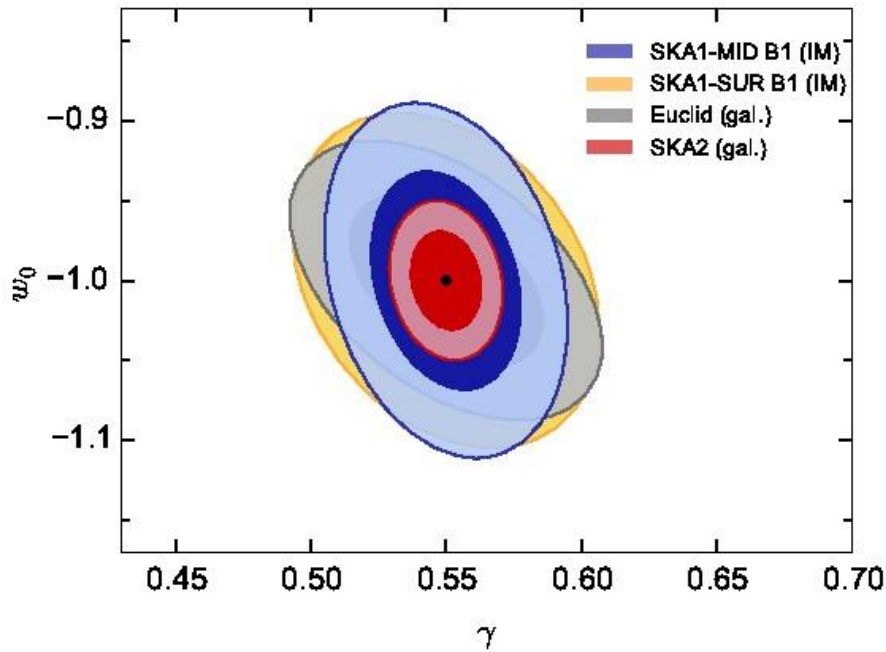

**Figure 2.** SKA predictions for the constraints on the growth of structures index (including BOSS and Planck priors, [3]).

where $H = \dot{a}/a$ is the Hubble expansion rate. Therefore, this may discriminate between models of gravity. In Fig. 2, we can see the improvement of the measurement of the equation of state parameter for dark energy over the index of structure growth. In General Relativity, $\tilde{\mu} = \tilde{\gamma} = 1$ and $\gamma \approx 0.545$. Thus, observational deviations from these values are evidences of modified gravity models, which can allow for predicting dependencies on the scale factor and on the wavenumber on the previous parameters. A particular example of such modified gravity theories is the one which admits an extension of the well-known $f(R)$ theories with a non-minimal coupling between matter and curvature [9], whose cosmological perturbations give specific relationships for those parameters [10].

Furthermore, it would also be interesting to investigate clusters rich in radio galaxies in order to assess whether there is evidence for interacting dark matter-dark energy models [11-13] or modified gravity theories [14] through deviations from the virial theorem, which, at cosmic scales, is given by the so-called Layzer-Irvine equation.

**Cosmic magnetism**

Another relevant science objective of SKA is related to the cosmic magnetism, which is ubiquitous in the Universe, since interstellar gas, planets, stars and galaxies all exhibit magnetic fields. However, the shape and strength of such magnetic field in galaxies or even its origin is not known. The Universe itself can be magnetic. Therefore, radio observations can measure the Faraday rotation, the polarised synchrotron emission and the Zeeman effect, which gives a detailed information on such fields, in particular whether they are primordial [15,16] or are generated later on via a dynamo mechanism. Another relevant issue is the impact of magnetic fields on the evolution of the Universe.

## 3 CONCLUSIONS

In this brief contribution we have presented the main contributions that SKA is expected to provide for Cosmology. As the world largest radio telescope array, SKA will be able to access the parameters that characterise the dark components of the Universe with unprecedented detail: one expects to get a power spectrum for dark matter from weak lensing observations and to considerably constrain the equation of state of dark energy (cf. Eq. 1). The acquired knowledge will allow for ruling out many models of dark matter and dark energy, including those where these two components interact with each other [11-13] or come from a unique field or a fluid model [17-19]. Moreover, SKA data will provide essential information on putative alternative models of gravity beyond General Relativity.

As mentioned, SKA will allow to look at the strong regime of Einstein's gravity and of any other gravitational models. On the other hand, by studying weak lensing distribution of matter and gravitational distortions, SKA will provide means to test the three parameters that characterise alternative theories of gravity: the growth of structure factor (or index), the modified gravitational constant and the anisotropic stress.

It is also expected that SKA will be an excellent tool to study cosmic magnetism and its characterisation over several redshifts, as well as the role of magnetic fields on the evolution of the Universe.

To conclude we can say that SKA will, most likely, revolutionise the way we perceive the Universe, as it will provide a huge amount of data on a wide range of astrophysical and cosmological issues and will allow for the understanding of many key features of the Cosmos.

## ACKNOWLEDGMENTS

C. G. acknowledges the support from Fundação para a Ciência e a Tecnologia (FCT) under the grant SFRH/BD/102820/2014.

## REFERENCES

[14] Carilli, C. and Rawlings, S. "Science with the Square Kilometer Array: Motivation, Key Science Projects, Standards and Assumptions," New Astronomy Reviews 48, 979–984 (2004).

[15] Bull, P, Ferreira, P. G., Patel, P., Santos, M. G. "Late-time cosmology with 21cm intensity mapping experiments," ApJ 803, 21 (2015).

[16] Raccanelli, A., et. al. "Measuring redshift-space distortions with future SKA surveys," PoS (AASKA14) 031 (2014).

[17] Ade, P. A. R. et al. [Planck Collaboration], "Planck 2015 results. XIII. Cosmological parameters," Astron. Astrophys. 594 A20 (2016).

[18] Samushia, L. et al. "The Clustering of Galaxies in the SDSS-III Baryon Oscillation Spectroscopic Survey (BOSS): measuring growth rate and geometry with anisotropic clustering," MNRAS, 439, 3504-3519 (2014).

[19] Amendola, L. et al. "Cosmology and fundamental physics with the Euclid satellite," Living Rev. Relativity 16, 6 (2013).

[20] **Bertolami, O.** "What if ... General Relativity is not the theory?" [arXiv:1112.2048 [gr-qc]] (2011).

[21] Kramer, M. et al. "Strong-Field Tests of Gravity Using Pulsars and Black Holes," New Astron. Rev. 48, 993-1002 (2004).

[22] **Bertolami, O.**, Boehmer, C. G., Harko, T. and Lobo, F. S. N. "Extra force in f(R) modified theories of gravity," Phys. Rev. D 75, 104016 (2007).

[23] **Bertolami**, O., Frazão, P. and Páramos, J., "Cosmological perturbations in theories with non-minimal coupling between curvature and matter," JCAP 05, 029 (2013).

[24] **coupling between curvature and matter," JCAP 05, 029 (2013).Bertolami, O.**, Pedro, F. G. and Delliou, M. L. "Dark Energy-Dark Matter Interaction and the Violation of the Equivalence Principle from the Abell Cluster A586," Phys. Lett. B 654, 165-169 (2007).

[25] **Bertolami, O.**, Pedro, F. G. and Delliou, M. L. "The Abell Cluster A586 and the Equivalence Principle," Gen. Rel. Grav. 41, 2839-2846 (2009).

[26] **Bertolami, O.**, Pedro, F. G. and Delliou, M. L. "Testing the interaction of dark energy to dark matter through the analysis of virial relaxation of clusters Abell Clusters A586 and A1689 using realistic density profiles," Gen. Rel. Grav. 44, 1073-1088 (2012).

[27] **Bertolami, O.** and **Gomes, C.** "The Layzer-Irvine equation in theories with non-minimal coupling between matter and curvature," JCAP 09, 010 (2014).

[28] Turner, M. S. and Widrow, L. M. "Inflation-produced, large-scale magnetic fields," Phys. Rev. D 37, 2743 (1988).

[29] **Bertolami, O.** and Mota, D. F. "Primordial magnetic fields via spontaneous breaking of Lorentz invariance," Phys. Lett. B 455, 96-103 (1999).

[30] Kamenshchik, A. Y., Moschella, U. and Pasquier, V. "An alternative to quintessence," Phys. Lett. B 511, 265-268 (2001).

[31] Bilić, N., Tupper, G. B. and Viollier, R. D. "Unification of Dark Matter and Dark Energy: the Inhomogeneous Chaplygin Gas," Phys. Lett. B 535, 17-21 (2002).

[32] Bento, M. C., **Bertolami, O.** and Sen, A. A. "Generalized Chaplygin Gas, Accelerated Expansion and Dark Energy-Matter Unification," Phys. Rev. D 66, 043507 (2002).

# Searching for new cosmology boundaries with SKA[1]

P. Moniz

Departamento de Física e Centro de Matemática e Aplicações (CMA - UBI), Faculdade de Ciências, Universidade da Beira Interior, 6200-001, Covilhã, Portugal

**ABSTRACT**

SKA's eventual retrieved data as well as its subsequent analysis may assist to either set guidelines or test fundamental proposals where cosmology is paramount. This brief note conveys a concrete route of exploration where SKA 'sails' could bring closer to an advantageous edge of significant discoveries. Within this purpose, our research group may contribute within a unique assembly of team work features.

**Keywords:** SKA, Cosmology, Quantum Gravity

E-mail: pmoniz@ubi.pt (ORCID: 0000-0001-7170-8952)

## 1      INTRODUCTION

Let us begin setting up the fundamental query and then comment how will SKA assist, within this research group potential.

XXIst century cosmology is a well-established quantitative area. It is really fascinating as it has thus entered a golden epoch, where future improvements (in both quantity and quality) will lead to an even clearer perspective of where we are, and why and how we come to be here: new technology has been and will be used to extract gradually more precise charts of the universe; SKA's technology, devices and contribution will be among some to emerge and provide decisive responses.

The current paradigm in cosmology is fundamentally based in the inflationary 'big bang' scenario. It has been under scrutiny and so far has successfully passed all the major tests. And it has allowed us to address some of the inconsistencies of the standard cosmological model, rendering the inflationary setting consistent with observations. So far…

Notwithstanding the merits, an apparent weakness emerges for that picture. For the paradigm to be realistic it has to be generically possible. In other simple words, we need to know the probability for the inflationary scenario to occur. The problem is that, addressing this conundrum lies beyond the scope of the inflationary paradigm itself. In fact, this is the issue of the initial conditions of the universe.

Classically speaking we have no guide as to how to choose one set of initial conditions rather than another. Additional arguments are therefore required. One option is to invoke quantum cosmological (QC) ingredients, e.g., the universe began in some sort of transition from a quantum regime, so that the initial classical parameters are determined in a probabilistic way. However, such task seems then to transfer our quest into determining (a) the most probable state (wave function) of the universe, $\Psi$, and (b) its distinctive predictive signatures. In this context, QC is basically the application of quantum mechanics to models with time reparametrization invariance (e.g., general relativity); QC can be considered as a kind of toy model attempt to obtain the relevant information for a full quantum theory of gravity. And this theory is one of the supreme challenges for fundamental science in the twenty-first century.

Given this framework, what predictions or (falsifiable) tests for the universe can be made using quantum cosmology? With regard to such question, addressing it will provide a picture of how the classically observed universe could come into being from induced (!) initial conditions and settings. An 'observational' perspective

---

[1] Based on a talk delivered at the SKA.PT Days on February 6th, 2018 in Covilhã, Portugal

for QC is required, it ought to be explored, in particular regarding the subsequent structure formation in a suitable but QC consistent inflationary stage.

"Observational" QC is therefore an intense and dramatic challenge. Let us be a bit more concrete. Assuming we achieve a satisfactory QC framework that would set the emergence for both the dynamical and the initial conditions, *what* type of predictions can be made? Would they be testable and if so *how*? In particular, can we establish that the most probable (classical) emergent spacetime will have a satisfactory inflationary phase, with suitable density fluctuations and gravitational waves? From such a route, originating from a more fundamental level, can we also thereby determine the probability distribution for the values of the constants of nature, depending on the different choices of the QC vacuum state (possibly related to different compactifications in string theory)?

Let us be even a bit more specific regarding that challenge: QC has been faced with difficulties, and still is facing problems. Our knowledge of the very early universe will surely improve if we can resolve the following points: How can we select the initial cosmological state? For this, two fundamental laws are needed: (a) a basic dynamical law, usually taken as the Wheeler-DeWitt equation

$$H\Psi = 0$$

where $H$ is the usual Hamiltonian constraint turned above into a quantum-mechanical operator applied to the wave function (of the universe) and (b) a law for a cosmological initial condition. Nb. There are many other features that can be invoked but I am taking the simple line to convey the set-up. But let us add a few comments about this statement. In a phase space description of cosmological dynamics, some features are strongly dependent on a specific choice of initial state. In addition, the subject of study is the whole universe, i.e., in cosmology the boundary conditions (necessary to solve the dynamical laws) for the evolution of the whole universe cannot be obtained from observations of a 'part' of the universe, 'outside' the ('sub') system under analysis, i.e., the universe itself.

Hence, the cosmological boundary must be formulated within a fundamental law. The establishment of such a law is the guiding focus of quantum cosmology. And is there a fundamental principle that assists in determining such a law for the initial condition of the universe? In other words, is there a selection rule for a boundary condition? This is supremely crucial: the necessary initial conditions for inflation must be implied by the boundary condition for the wave function of the universe; subsequently we must face the fact that a cosmological wave function must lead to a physical test, i.e., empirical distinctions. In summary, this means the necessary initial conditions for inflation must be implied by the boundary condition for the wave function of the universe.

In the last point, to retrieve predictions from quantum cosmology, a choice of boundary conditions has to be made (for the Wheeler–DeWitt equation or alternative), to select one of them from the class(es) of solutions. However, a few more (severe?) obstacles emerge. Yes, QC is the toughest beast in the research jungle and takes a few strong harnesses bones to investigate it… To begin with, there are quite a few boundary conditions to choose from and it is not yet possible (meaning novel and ground-breaking technological savviness) to distinguish between their observational differences. Another point is that, even in the case of the no-boundary proposal (that has been so much promoted by the late Stephen Hawking), it happens that not even a unique wave function is selected. Indeed, there are many such wave functions, which do not differ in their semiclassical predictions but some signatures could be desired to remain to be detected at some earlier layer in the fabric of space-time. To be more concrete, additional elements, surely of a more fundamental nature, are required to yield a unique solution. And this is where "observational" QC can provide hints on *what* to search for, *where, when, how*.

It is here that things can become more complicated. The Wheeler-DeWitt equation above and the wave function of the universe need to be 'expanded' in expressions that bear terms from where fluctuations on the fields (inc. geometry) can be associated. In a more easily to understand expression,

$$H = H_0 + H_2$$

the first term on the rhs conveys the background and the final one, the fluctuations on the fields. Accordingly, the wave function is then expanded as

$$\Psi = \exp(iS)\Pi_m \psi_m$$

where the exponential term is a traditional WKB wave function, from which the background can be recovered within a Hamilton-Jacobi equation, whereas from the product in '$m$' a Schrödinger equation plus quantum gravitational corrections can be extracted. It is from the two above combined equations that (within this very basic and simplified description) that falsifiable features concerning QC can be made.

At this point we may be wondering about the potential use of this semiclassical framework to provide QC with an "observational" viewpoint as indicated at the beginning of this note. The presence of $H_0, H_2$ in the above expressions allows us to estimate corrections and their importance with regard to potential observational effects. For a Friedmann universe (i.e., a universe simple in geometrical terms) with scale factor *a*, we can roughly estimate corrections, that for an energy *E = 700 GeV* and 70 km/(s Mpc) for the Hubble parameter, we obtain approximately $10^{-44}$ for the ratio between background and between corrections as well. Therefore, any quadratic (matter) Hamiltonian correction is not yet (!) an observable effect for our current standard of measurements. Hence, it still remains to initiate future investigations, which will deal with the context of observational consequences within QC, such as structure formation.

SKA implementation will provide an immensely beneficial plethora of new data, from new broadness of observations and 'visualizations' scope, much deeper and wider than current observatories and their detection range. Extreme events, triggered within extreme configurations would be, as we all are hoping for, the terrain to test novel ideas and proposals. Recent gravitational and visual observations have already allowed to set up new restrictions, selecting a set of remaining and more restricted models of gravity. If the sensitivity is improved, if the range and scope of detection is broaden, beyond the usual and so far taken limits, if the amount of data is much increased and then analysed with care, our probing capacity will be one (or more) level up. This is where SKA intervenes, as true XXIst century tool, which no other has yet been able to bring.

At CMA – UBI, either intrinsically regarding the theoretical physics team plus the data analysis for 'big data' from the Mathematics-Statistics group, we are fully able to meet the challenges in the above paragraphs. Moreover, there are in this region, data storage contributing conditions, technical experience as well as opportunities to grow that are not yet bounded by any regulatory constraint. In short, there is scientific leadership, team capacity and synergy, plus an intertwining framework that works, within faculties as well as other higher education institutions (viz. effective consortia).

There is absolutely no reason for discouragement for QC pondering at CMA – UBI within data analysis as well as setting up tests for 'observational' QC to be eventually considered, through the SKA assembled data associated to suitable extreme detected events. On the contrary, in contrast to inaccurate rumours, quantum cosmology still offers a few, quite tenacious. Technically speaking, on a mathematical sense, it may depend on the chosen complex contour where the integration is implemented. Technically speaking, it may also depend on SKA data delivery. This is why we need SKA and be a contributing team.

## ACKNOWLEDGMENTS

P. Moniz acknowledges the support from Fundação para a Ciência e a Tecnologia (FCT) under the grant UID/MAT/00212/2013.

## REFERENCES

[1] **Moniz, P.** "Quantum cosmology - the supersymmetric perspective: Vol. 1: Fundamentals" Lect. Notes Phys. 803 (2010) 1-351, DOI: 10.1007/978-3-642-11575-2

[2] **Moniz, P.** "Quantum cosmology - the supersymmetric perspective: Vol. 2: Advanced Topics" Lect. Notes Phys. 804 (2010) 1-283, DOI: 10.1007/978-3-642-11570-7

[3] Jalalzadeh, S., **Moniz, P.** "Challenging Routes in Quantum Cosmology", World-Scientific ISBN: 978-981-4415-06-4 (forthcoming)

[4] Jalalzadeh, S., Rostami, T., **Moniz, P.** "Quantum cosmology: From hidden symmetries towards a new supersymmetric perspective", Int.J.Mod.Phys. D25 (2016) no.03, 1630009 DOI: 10.1142/S0218271816300093

2020					Portuguese SKA White Book					Page 48 of 210

# Probing Particle Physics via Primordial Gravitational Waves at SKA

António P. Morais

CIDMA, Department of Physics, University of Aveiro, Campus Universitário de Santiago, 3810-193 Aveiro, Portugal

**ABSTRACT**

The Square Kilometer Array (SKA) will be leading in an era where a Pulsar Timing Array (PTA) based search for gravitational waves (GWs) will be able to use hundreds of well timed millisecond pulsars rather than a few dozens in existing PTAs. The ground-breaking nature of the SKA design will offer the possibility to search for both individual super-massive black hole binaries or rather seek for a stochastic background of GWs whose origin may be due to cosmological phase transitions in early Universe, particularly important for Baryogenesis. This neatly fits into our research where we are studying sources of primordial GWs such as the electroweak phase transition (EWPT) and where SKA will offer the possibility to access such details upon measurement of the peak-frequency and the peak-amplitude of the stochastic GW spectrum. Furthermore, the low frequency sensitivity of SKA makes it capable of probing inaccessible regions by, e.g. the space based LISA interferometers enhancing the possibility for a broader reach and a doorway to perform particle physics studies beyond collider experiments.

**Keywords:** Gravitational Waves, Electroweak phase transition, Higgs boson, Particle Physics, Cosmology

## 1    INTRODUCTION

Despite the great success of the Large Hadron Collider (LHC) in the discovery of the Higgs boson [1, 2], thus completing the Standard Model (SM), the current void of new phenomena either indicates that new physics can only be manifest at a larger energy scale than previously thought, or results from a lack of sensitivity of the current experiments measuring rare events. In fact, the weaker the interaction strength between the SM and new physics, the greater the challenge to probe it.

On the other hand, the recent discovery of a binary neutron star merger, firstly observed by the gravitational waves (GW) interferometers of the LIGO-Virgo collaboration [3, 4], a new era of multi-messenger astronomy has begun. Furthermore, the reach of GW observatories is by no means exhausted and broader sensitivities are designed for future space-based interferometers such as those of the LISA [5, 6], DECIGO [7], BBO [8] and SKA [9] collaborations. This opens up the door for a plethora of new studies including connections with both cosmology and particle physics (see e.g. Refs. [10–18]). In particular, the potential observation of a stochastic GW background produced by violent processes in the early universe, e.g. by expanding and colliding vacuum bubbles associated with strong cosmological phase transitions [19, 20], may well become a gravitational probe for beyond-the-Standard-Model (BSM) physics and a complement for recent and future collider experiments.

We propose a possibility to probe new particle physics models with non-trivial vacuum configurations by means of the GW data. We point out that the GW spectrum produced in multi-step strong EWPTs occurring at distinct temperatures in the early universe exhibits a non-trivial multi-peak shape which may be probed by future GW observatories such as SKA and LISA. Such an observation would be a signature of multiple symmetry breaking stages, a step forward in our understanding of the EWPT and offer gravitational probes to BSM physics.

## 2    DETECTABILITY OF RELIC GRAVITATIONAL WAVES

The PTA at SKA is an ongoing attempt to detect low frequency GWs with a sensitivity in the range of $10^{-9}$ Hz to $10^{-6}$ Hz. In particular, PTAs are typically most sensitive to the lowest frequencies in this range. The detectability of relic GW background produced by the electroweak phase transition can be determined in terms of (a) the released latent heat during the phase transition, $\alpha$ and (b) inverse time scale of each transition, $\tilde{\beta}$. Such quantities are the most important ones to characterize the shape of the gravitational wave spectrum and directly relate to a given particle physics model. Considering the case of bubble collisions, an example extracted from [21] and displayed in Fig. 1 shows different possibilities for GW signals. While the peak frequency is only controlled by $\tilde{\beta}$, the peak amplitude is determined by both $\alpha$ and $\tilde{\beta}$. On the other hand, the transition temperature has little effect in the signal as shown by each pair of blue and red lines for fixed $\alpha$ and $\tilde{\beta}$. In essence, the important message to retain is that there are parameter regions, which only SKA can observe at small frequencies, thus small $\tilde{\beta}$. This can be further seen in Fig. 2, also taken from [21], which highlights the importance of SKA for probing the regime of small frequencies.

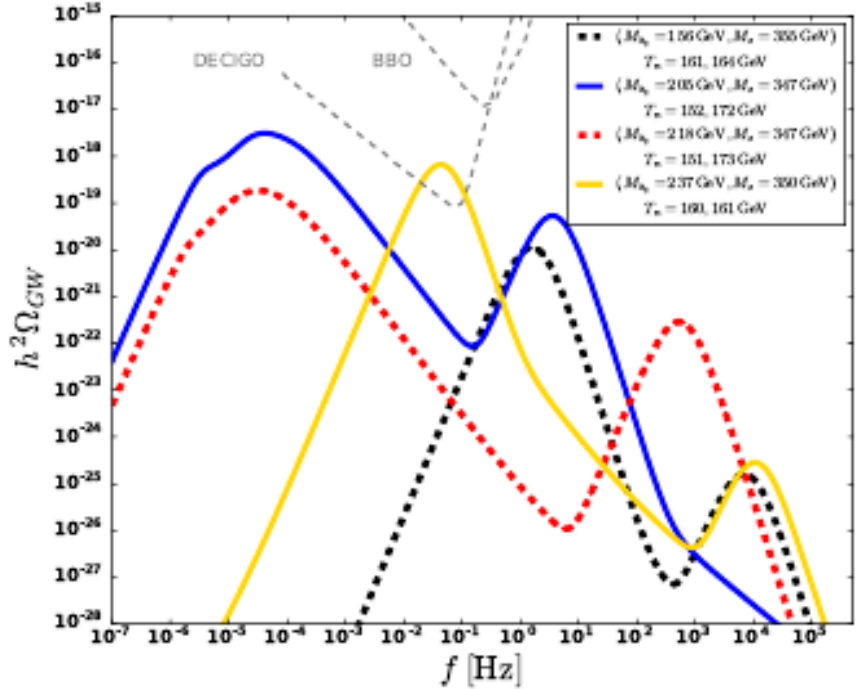

**Figure 1.** Signals of the relic gravitational wave background in case of the bubble collision. The red lines correspond to a transition temperature of 70 GeV while the blue lines correspond to 100 GeV.

## 3 MULTI-PEAKED SIGNATURES

Following our studies in [22], we may argue that the combined sensitivities of SKA and other GW observatories may become of crucial importance if the details of the EWPT are non-trivial. In particular, a generic BSM scenario typically contains a large number of scalar degrees of freedom which can be advantageous e.g. for EW baryogenesis. Even reducing the scalar sector to a few fields, new unexplored possibilities of transition patterns arise, in particular, transitions in several successive first-order steps. Therefore, a non-trivial EWPT is expected and multi-step transitions may well have occurred in the early universe. An important consequence of this yet unexplored ground is that we can have more than a single transition pattern for a particular point in the parameter space, which results in sequential nucleation of bubbles of different vacua. As a phenomenological probe, we suggest that multi-step transition pattern leave a characteristic signature in the spectrum of GWs exhibiting a multi-peaked shape. The main idea is that, provided that the properties of the bubble nucleation process are different for successive transitions between distinct phases, we can expect a superposition of GW signals with different frequency peaks, whose positions mostly depend on the inverse time scale of each transition. Typically, the larger the time scale, the smaller the frequency of the corresponding GW signal. Our studies in [22] show that, for a toy-model, the frequency spread of such multi-peaked GW spectra can be wide enough to demand a combined analysis of different experiments. An illustrative example can be seen in Fig. 3. This example indicates that the observation of such a pattern would provide us with a rather detailed information about dynamics of the EWPT, thus of the underlying particle physics. However, in the considered toy-model, such signals with well-separated peaks in the GW spectrum are too weak. One can expect to have distinct detectable peaks within reach of SKA and other GW observatories if more realistic models with a larger energy budget and an enhanced release of the latent heat are considered.

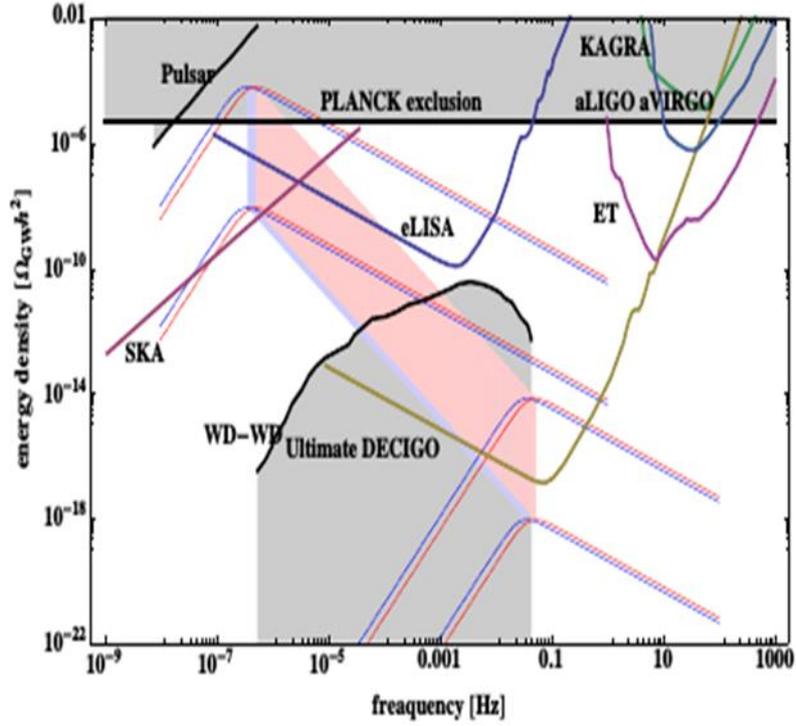

**Figure 2**. Detectabilities in the (α,β̃) plane for the signals sourced by the bubble collision

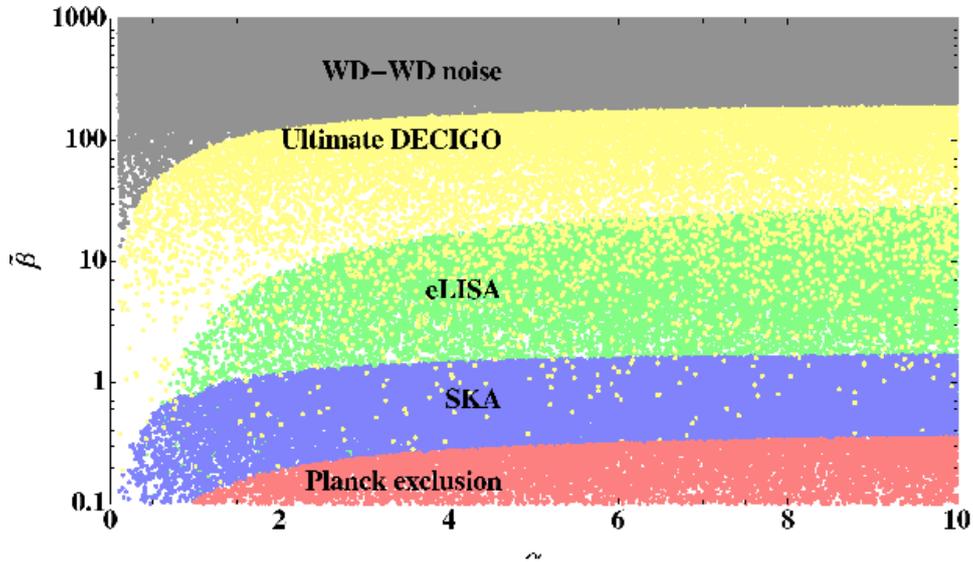

**Figure 3**. Examples of the net multi-peaked GW spectra emerging in two-step transitions. Detectabilities in the (α, β̃) plane for the signals sourced by the bubble collision. From the top to the bottom, the cases for Ultimate DECIGO, eLISA, SKA are plotted, respectively. The WD-WD noise means the region where signals are covered by the foreground noise by the WD-WD binaries. From [21].

## 4     SUMMARY

After the discovery of the Higgs boson, the particle physics community have been experiencing an increasing effort to tackle the details of the electroweak phase transition head-on and a number of opened questions related to the Higgs sector have been explored. It is then becoming attractive to study possible scenarios for the EWPT, and particularly non-trivial multi-step phase transitions, in order to investigate their role in the stochastic gravitational wave background. We point out that the role of SKA in this domain will be unique and crucial to search for particle physics models that predict low frequency peaks in the GW spectrum, strongly motivating further studies in this direction.

## ACKNOWLEDGMENTS

APM acknowledges support from CIDMA strategic project (UID/MAT/04106/2013), and is funded by the FCT grant SFRH/BPD/97126/2013.

## REFERENCES

[1] G. Aad et al. (ATLAS), Phys. Lett. B716, 1 (2012), 1207.7214.

[2] S. Chatrchyan et al. (CMS), Phys. Lett. B716, 30 (2012), 1207.7235.

[3] B. P. Abbott et al. (Virgo, LIGO Scientific), Phys. Rev. Lett. 116, 061102 (2016), 1602.03837.

[4] B. P. Abbott et al. (Virgo, LIGO Scientific), Phys. Rev. Lett. 116, 241103 (2016), 1606.04855.

[5] P. A. Seoane et al. (eLISA) (2013), 1305.5720.

[6] N. Bartolo et al., JCAP 1612, 026 (2016), 1610.06481.

[7] S. Kawamura et al., Class. Quant. Grav. 28, 094011 (2011).

[8] V. Corbin and N. J. Cornish, Class. Quant. Grav. 23, 2435 (2006), gr-qc/0512039.

[9] https://www.skatelescope.org/project/projecttimeline/.

[10] P. Huang, A. J. Long, and L.-T. Wang, Phys. Rev. D94, 075008 (2016), 1608.06619.

[11] P. Huang, A. J. Long, and L.-T. Wang, Phys. Rev. D94, 075008 (2016), 1608.06619.

[12] J. M. No, Phys. Rev. D84, 124025 (2011), 1103.2159.

[13] C. Grojean and G. Servant, Phys. Rev. D75, 043507 (2007), hep-ph/0607107.

[14] R. Apreda, M. Maggiore, A. Nicolis, and A. Riotto, Nucl. Phys. B631, 342 (2002), gr-qc/0107033.

[15] K. Hashino, M. Kakizaki, S. Kanemura, and T. Matsui, Phys. Rev. D94, 015005 (2016), 1604.02069.

[16] M. Kakizaki, S. Kanemura, and T. Matsui, Phys. Rev. D92, 115007 (2015), 1509.08394.

[17] P. S. B. Dev and A. Mazumdar, Phys. Rev. D93, 104001 (2016), 1602.04203.

[18] P. S. B. Dev, M. Lindner, and S. Ohmer, Phys. Lett. B773, 219 (2017), 1609.03939.

[19] M. Hindmarsh, S. J. Huber, K. Rummukainen, and D. J. Weir, Phys. Rev. Lett. 112, 041301 (2014), 1304.2433.

[20] M. Hindmarsh, S. J. Huber, K. Rummukainen, and D. J. Weir, Phys. Rev. D92, 123009 (2015), 1504.03291.

[21] Y. Kikuta, K. Kohri, and E. So (2014), 1405.4166.

[22] T. Vieu, A. P. Morais, and R. Pasechnik (2018), 1802.10109

# Witnessing the birth of galaxies with SKA and SKA-percursors

José Afonso[a,b], Stergios Amarantidis[a,b], Israel Matute[a,b], Ciro Pappalardo[a,b]

[a] Instituto de Astrofísica e Ciências do Espaço, Universidade de Lisboa, OAL, Tapada da Ajuda, PT1349-018 Lisboa, Portugal
[b] Departamento de Física, Faculdade de Ciências, Universidade de Lisboa, Edifício C8, Campo Grande, PT1749-016 Lisboa, Portugal

**ABSTRACT**

The epoch of first light in the Universe, or Epoch of Reionization (EoR), is one of the most exciting frontiers in current astrophysical knowledge. The SKA, with its revolutionary capabilities in terms of frequency range, resolution and sensitivity, will allow to explore the first Gyr of structure formation in the Universe, in particular, with the detection and study of the earliest manifestations of the AGN phenomenon. The recent detection of powerful quasars at those epochs implies an amazingly rapid growth, previously considered impossible, of supermassive black holes, and suggests that the observation of even earlier counterparts, in particular at radio wavelengths where detection has so far eluded us, is within our grasp. Not only would such detections be paramount to the understanding of the earliest stages of galaxy evolution, they are necessary for the direct study of neutral hydrogen in the Epoch of Reionization, through SKA observations of the HI 21cm forest against such background sources. In order to understand how SKA and SKA-precursors can be optimised to reveal these earliest AGN, we are exploring state-of-the-art models of galaxy formation and evolution. This work is showing us how to reach the highest redshifts, and, as a result, we have recently revealed what is likely the most distant radio-selected AGN ever found. This research also demonstrates the importance of exploring the most powerful telescope currently in operation, the Atacama Large Millimetre Array (ALMA), for the efficient determination of the redshift to very distant sources - in particular if they are not selected by their bright optical or near-infrared emission. The activity of PACE, the Portuguese ALMA Centre of Expertise, is presented, as an example of relevant scientific and technical capability the Portuguese astronomical community has already acquired.

**Keywords:** galaxies: active, galaxies: evolution, radio continuum: galaxies, radiotelescopes, surveys.

## 1 INTRODUCTION

The epoch of first light in the Universe, or Epoch of Reionization (EoR), is one of the most exciting frontiers in current astrophysical knowledge. When and how did the first galaxies, stars and supermassive black holes (SMBHs) form? How did the first light they originated rapidly (re)ionized the entire Universe? How did the neutral gas evolve throughout those initial few hundred thousand years? Understanding the earliest phases of galaxy evolution requires not only the deepest and widest observations of the Universe, but, at least as importantly, understanding how to fine-tune observations for the identification of these "holy-grail" objects - mastering not only the observational limitations and caveats, but also the theoretical framework that represents our best knowledge of the early Universe.

Surprisingly, we are currently starting to glance into the formation of the first stars and galaxies, a research topic that will witness a dramatic expansion over the next few years, with revolutionary observational capabilities that will feed (and be fed by) better theoretical simulations. Although the direct observation of the first sources of light hasn't been possible yet, we have targeted increasingly higher redshift, younger galaxies, by focusing on the processes that we know are fundamental to their evolution: star-formation and the infall of matter to a SMBH - an active galactic nucleus (AGN).

While the radiation from the first stars is often assumed to be the major culprit for the Reionization of the Universe, at $z$>6-7, recent work [1, 2] has raised some doubts over the contribution of accretion to early supermassive black holes (SMBHs). This is particularly relevant to radio observations with the upcoming Square Kilometre Array (SKA), as radio emission from the earliest AGN should be well within its reach. The detection of such very high redshift radio galaxies would even be more exciting, as it would then be feasible to consider the *direct* study of neutral hydrogen and its evolution *in the Epoch of Reionisation itself*, through observations of the HI 21cm forest against such background sources [3, 4].

However, the detection of such radio powerful AGN remains extremely challenging. Over the past few years, in the framework of international consortia, we have explored new radio selection techniques [e.g., 5, 6, 7, 8, 9, 10], and the problems associated with the emission of synchrotron radiation at an epoch where an extremely energetic Cosmic Microwave Background exists [11]. This has showed us how to reach the highest redshifts and find some of the earliest AGN in the Universe. In fact, our efforts have recently resulted in the selection of sources that likely include the most distant radio-selected AGN ever found. Such efforts have also been fundamental in increasing the Portuguese expertise in the use of one of the most powerful telescopes currently in operation, the Atacama Large Millimetre Array (ALMA), which is particularly relevant when considering the potential use of SKA by the Portuguese astronomical community.

## 2    WHERE ARE THE MOST POWERFUL RADIO GALAXIES?

Radio selection was, until the early 2000's, a powerful way to explore the highest redshift Universe. With huge areas of the sky surveyed by powerful radio telescopes (e.g., 3C, 4C, NVSS, FIRST), many radio galaxies were identified at increasingly higher distances, up to the (until recently) record holder TN J0924-2201, at a redshift of $z$ = 5.2 [12]. However, other wavelengths and selection techniques quickly picked up pace and overthrown radio as a high-redshift discovery machine. To understand the difficulties affecting radio-based high-redshift searches one should note: (a) powerful radio galaxies are rare objects, and their space density falls off substantially at $z>2$-3 [e.g., 13, 14], (b) their radio emission is decreased at very high redshifts [11] and (c) their redshift confirmation requires demanding optical and/or near-infrared and/or, more recently, sub-millimetre follow-up spectroscopic observations. It should thus be not surprising that it took ten years before another similarly high redshift radio-selected galaxy was found, at "only" $z$=4.9 (J163912 [15]), and almost 20 years until the redshift record was broken, with the discovery of TGSS1530, at a redshift of $z$=5.7 [16].

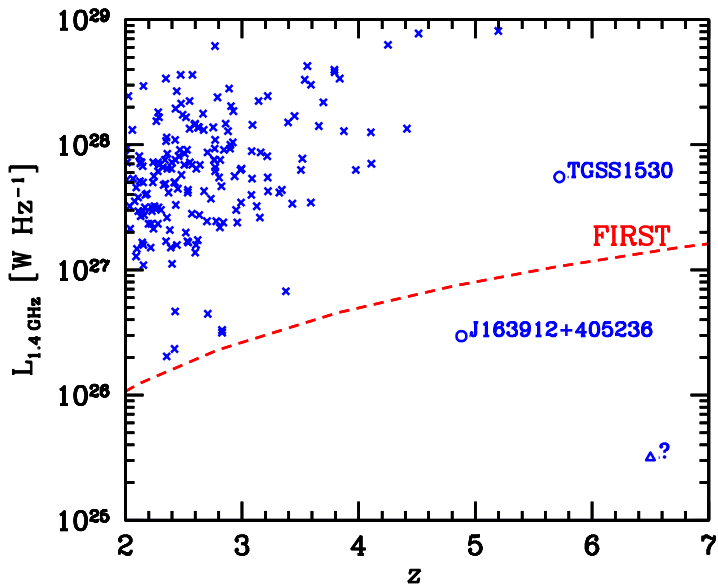

**Figure 1**. Compilation of high-z radio galaxies from [17], complemented by recent high-z sources (J163912, from [15] and TGSS1530, from [16]). The triangle at z~6.5 denotes a candidate high-z source from our own work, currently being observed with ALMA for a redshift confirmation. Also shown is the sensitivity limit from the FIRST survey.

In Figure 1 we show a compilation of high redshift (z>2) radio galaxies from [17], together with some recent discoveries of high(er) redshift radio sources, J163912 and TGSS1530, and one of our recent z>6 radio source candidates (marked as a "?"), which is currently being observed with the Atacama Large Millimetre Array (ALMA) for a final redshift confirmation (see below). The figure mirrors how difficult it has been to explore the highest-redshift regime, mostly due to the difficulty in successfully *confirm* the redshifts radio-selected very high redshift candidates. Unlike optical-to-near-infrared selection criteria, which have already been successful in revealing tens of optical/NIR selected QSOs at $z \sim 6-7$ [e.g., 18, 19, 20, 21, 22, 23], selection in the radio does not imply a bright optical or near-infrared magnitude, which would all but ensure successful spectroscopic follow-up and redshift confirmation. Only recently, with the advent of ALMA, it has become feasible to effectively follow-up radio-selected high-redshift candidates.

## 3    EXPLOITING STATE-OF-THE-ART GALAXY FORMATION MODELS

In order to understand how best to explore upcoming radio telescopes to reach the highest redshifts, it is important to explore what the most recent models of the early Universe can tell us about the expected abundance of radio-detectable AGN in the first $10^9$ years. In spite of the often surprisingly consistent picture that state-of-the-art galaxy formation and evolution semi-analytic and hydrodynamic simulations can already provide, they are frequently validated only through much lower redshift observations ($z<2$), since complete and detailed samples of galaxy populations in the Universe cannot be achieved at much higher distances. Far from perfect, the unavoidable uncertainties at the highest redshifts can nevertheless suggest how SKA and, even sooner, upcoming SKA-pathfinder telescopes can be successful (or be tuned to be successful) in directly exploring the galaxy populations at the Epoch of Reionisation.

Using some of the most advanced galaxy formation models, we have started exploring their predictions for the detectability of Active Galactic Nuclei (AGN) at radio wavelengths for $z = 6-10$ (Amarantidis et al, in prep.). Figure 2 shows the predictions to the mass of the most massive supermassive black holes (SMBH) at any given redshift, and its comparison to current (optical and NIR) observations. Strikingly, all models fail to reach the masses of the most massive SMBHs already observed. This effect can be attributed to the computational limitation in volume that all models face, which will prevent the existence, in the simulation, of the most extreme dark matter haloes and, consequently, of the most massive SMBHs. These simulations are, nevertheless, still able to place a lower limit to the number of $z>6$ AGNs that SKA will be able to detect – a number that is found to be at a few tens per square degree (Amarantidis et al., in prep.).

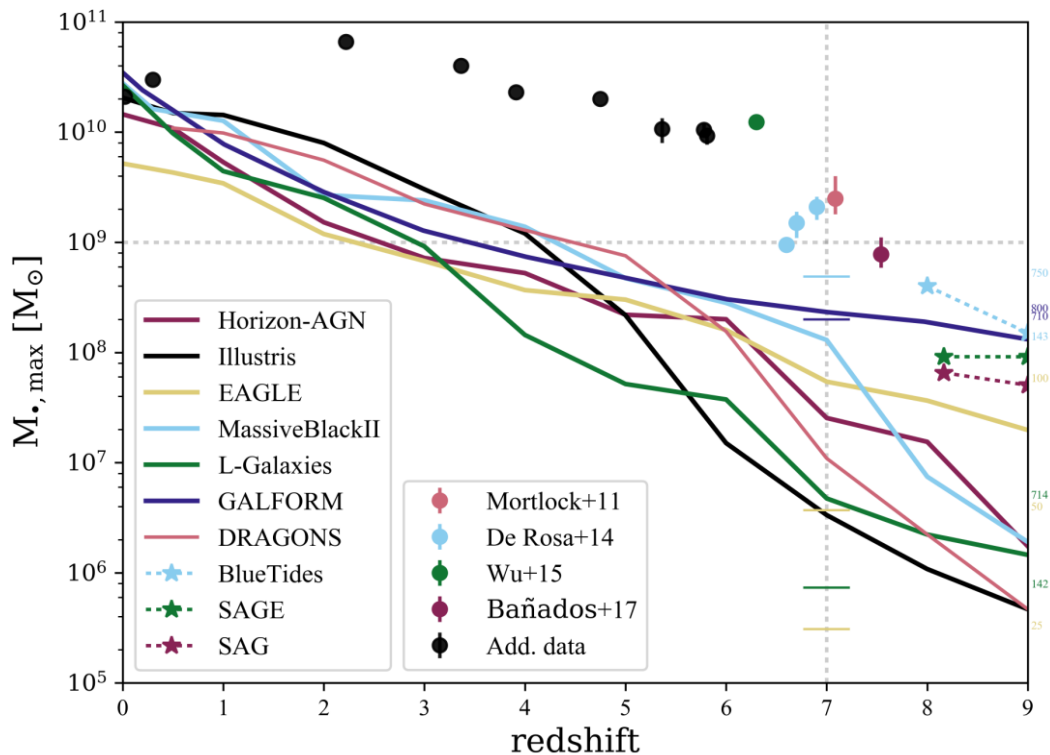

**Figure 2**. The most massive supermassive black holes (SMBH) predicted by recent state-of-the-art galaxy formation models (lines) at different redshifts, compared with observational data (points) for $z>6$ [22, 24, 25, 26] and for lower redshifts (additional data). For full details see Amarantidis et al., in preparation.

## 4     SCIENCE WITH AND FOR SKA-PRECURSORS

Besides being involved with the development of SKA through the (Extragalactic Continuum) Science Working Group, our team is also pushing the exploration of models of galaxy formation in the context of the development of several SKA-precursors [see also 9]: we are leading the activities in the Key Science Project "Radio AGN in the EoR" of the Evolutionary Map of the Universe survey (EMU: [27]), that will be performed with ASKAP, and also participating in WODAN [28], to be observed with WSRT-APERTIF, MIGHTEE [29] and MeerKLASS [30], both to be performed with MeerKAT. In all cases, different combinations of depth, area and resolution will require different approaches to maximize the efficiency in the search for the highest redshift radio sources, and potentially result in different degrees of success.

As part of our preparation for the scientific exploitation of these upcoming surveys, we have started a detailed analysis of current deep radio surveys, having already revealed some exciting candidates. Their high redshift

2020     Portuguese SKA White Book     Page 55 of 210

confirmation is, as expected, extraordinarily difficult, as these radio-selected sources have had no constraints in terms of optical and NIR brightness. In particular, one of the sources identified in the VLA survey of the COSMOS field was subsequently found to have red far-infrared Herschel fluxes, a tell-tail sign of a very high redshift nature. Figure 3 shows the multiwavelength view of the source, undetected in the optical and increasingly brighter at longer wavelengths, and the photometric redshift indication for a high-redshift nature. Being too faint for optical or near-infrared spectroscopic redshift confirmation, we have secured follow-up observations using what is arguably the most powerful telescope currently in operation, the Atacama Large Millimetre Array. If this source is confirmed to be at $z>6$, it will be the first radio-selected source in the EoR, providing exciting prospects for the science that SKA-precursors and, soon, also SKA, will be able to achieve.

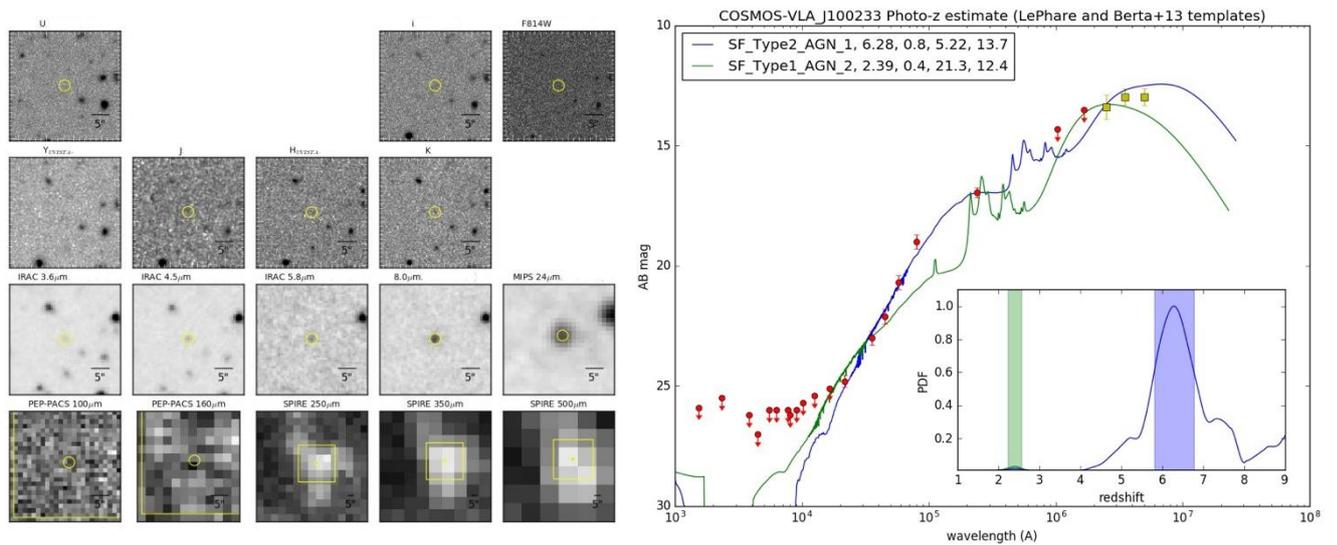

**Figure 3**. Left panel: Multiwavelength cutouts (90x90 arcsec across for SPIRE Herschel bands, 30x30 for the rest) for the high-redshift candidate CVLA-100233. From top to bottom and left to right, the bands go from u to the Herschel SPIRE bands. Right panel: spectral energy distribution of the best fit (blue line) for CVLA-100233, and the redshift probability

## 5 THE PORTUGUESE ALMA CENTRE OF EXPERTISE

As seen above, the confirmation of very high-redshift radio sources is extremely demanding. The most interesting candidates may remain undetected even with the deepest observations at optical and NIR wavelengths. The recent construction of ALMA, arguably the most powerful telescope in operations, opened what may well be the best possibility in a long time to finally discover high-redshift radio sources within the EoR.

Recognising the unparalleled capabilities of ALMA for the study of the highest redshift Universe, our team has pushed heavily for increasing the national expertise in its use. We were part of the core group behind the successful increase in Portuguese use of ALMA in recent years, quickly growing the number of successful ALMA proposals with Portuguese participation over the past 6 years. The effort lead to the creation of the Portuguese ALMA Centre of Expertise (PACE: see http://pace.oal.ul.pt), hosted by the Institute of Astrophysics and Space Sciences, and recognised by the European Southern Observatory as part of the ALMA Regional Centre Network in Europe.

PACE ALMA support activities include:

- Support to Portuguese ALMA users in proposal preparation, data reduction, and archive mining – With PACE, the improvement in terms of submitted/accepted proposals with Portuguese participation has been remarkable (from 4 in 2012 to more than 40 in 2017 for submitted proposals, and from 1 accepted proposal in 2012 to 8 in 2017), fully rewarding the huge effort that PACE staff placed in building a strong Portuguese ALMA community;

- ALMA data Quality Assurance – PACE contributes to the ALMA project, participating in data quality assessment (QA2), software development, and data archiving. Such tasks show clearly the level of expertise PACE has already achieved;

- Software Testing – The ALMA user community makes use of different software tools, which need to be tested following continuous upgrades. PACE has been involved in tests for the Common Astronomy Software Applications (CASA), pushing for improvements in terms of running speed for different software routines. Beyond this, PACE participated to the ALMA user Science Portal tests, in collaboration with ESO and NRAO, designed to facilitate the usefulness of the ALMA website to the scientific community;

- Events and Activities – To support Portuguese ALMA users, PACE organises ALMA events as a preparation for upcoming observing cycles. PACE organises periodic Portuguese ALMA Community days focused on the preparation of ALMA proposals for the upcoming Cycle. PACE also hosts, in collaboration with ESO and the remaining ALMA Regional Centre Network, more technical meetings focused on the operations and development of ALMA, like the "ALMA All-hands" meeting (2016) and the first "ALMA data Processing Workshop" (2017).

- Outreach – PACE contributes regularly to outreach events, promoting the ALMA project and educating the community in ALMA-related science and scientific results.

The growth in expertise towards ALMA, one of the most revolutionary telescopes in operation, shows the capability and strategic vision the Portuguese astronomical community has already achieved. This is particularly relevant when considering the support to SKA, which requires many of the competences already developed for ALMA. Looking both at the scientific capability and the technical expertise of the Portuguese astronomical community, one can only conclude that Portugal is ready and will be successful in continuing its contribution to the development of SKA and, soon, to its full scientific exploitation.

## ACKNOWLEDGMENTS

The authors gratefully acknowledge support from the Science and Technology Foundation (FCT, Portugal) through the research grants UID/FIS/04434/2013, SFRH/BPD/90559/2012 and SFRH/ BPD/95578/2013.

## REFERENCES

[1] Robertson, B. E., Furlanetto, S. R., Schneider, E., et al. 2013, ApJ, 768, 71

[2] Fontanot, F., Cristiani, S., Pfrommer, C., Cupani, G., & Vanzella, E. 2014, MNRAS, 438, 2097

[3] Carilli, C. L., Furlanetto, S., Briggs, F., et al. 2004, New Astron. Rev., 48, 1029

[4] Khatri, R. & Wandelt, B. D. 2010, Highlights of Astronomy, 15, 312

[5] Afonso, J., Mobasher, B., Koekemoer, A., Norris, R. P., & Cram, L. 2006, AJ, 131, 1216

[6] Norris, R. P., Afonso, J., Appleton, P. N., et al. 2006, AJ, 132, 2409

[7] Norris, R. P., Afonso, J., Cava, A., et al. 2011, ApJ, 736, 55

[8] Afonso, J., Bizzocchi, L., Ibar, E., et al. 2011, ApJ, 743, 122

[9] Norris, R. P., Afonso, J., Bacon, D., et al. 2013, PASA, 30, e020

[10] Smolcic, V., Padovani, P., Delhaize, J., et al. 2015, Advancing Astrophysics with the Square Kilometre Array (AASKA14), 69

[11] Afonso, J., Casanellas, J., Prandoni, I., et al. 2015, Advancing Astrophysics with the Square Kilometre Array (AASKA14), 71

[12] van Breugel, W., De Breuck, C., Stanford, S. A., et al. 1999, ApJ, 518, L61

[13] Rigby, E. E., Best, P. N., Brookes, M. H., et al. 2011, MNRAS, 416, 1900

[14] Rigby, E. E., Argyle, J., Best, P. N., Rosario, D., & Rottgering, H. J. A. 2015, A&A, 581, A96

[15] Jarvis, M. J., Teimourian, H., Simpson, C., et al. 2009, MNRAS, 398, L83

[16] Saxena, A., Marinello, M., Overzier, R. A., et al. 2018, submitted to MNRAS, arXiv:1806.01191

[17] Miley, G., & De Breuck, C. 2008, A&ARv, 15, 67

[18] Fan X., Carilli C. L., Keating B. 2006. ARA&A, 44:415

[19] Willott, C. J., Delorme, P., Omont, A., et al. 2007, AJ, 134, 2435

[20] Willott, C. J., Delorme, P., Reylé, C., et al. 2009, AJ, 137, 3541

[21] Jiang, L., Fan, X., Annis, J., et al. 2008, AJ, 135, 1057

[22] Mortlock, D. J., Warren, S. J., Venemans, B. P., et al. 2011, Nature, 474, 616

[23] Venemans, B. P., Findlay, J. R., Sutherland, W. J., et al. 2013, ApJ, 779, 24

[24] De Rosa, G., Venemans, B. P., Decarli, R., et al. 2014, ApJ, 790, 145

[25] Wu X. B., et al., 2015, Nature, 518, 512

[26] Bañados E., et al., 2018, Nature, 553, 473

[27] Norris, R. P., Hopkins, A. M., Afonso, J., et al. 2011, PASA, 28, 215

[28] Rottgering, H., Afonso, J., Barthel, P., et al. 2011, JApA, 32, 557

[29] Jarvis, M. J., Taylor, A. R., Agudo, I., et al. 2017, Proceedings of Science, "MeerKAT Science: On the Pathway to the SKA", arXiv:1709.01901

[30] Santos, M. G., Cluver, M., Hilton, M., et al. 2017, Proceedings of Science, "MeerKAT Science: On the Pathway to the SKA", arXiv:1709.06099

# Active Galactic Nuclei and High redshift Radio Galaxies in the SKA era

Sonia Antón[a,b], Andrew Humphrey[c], Tom Scott[c]

[a] CIDMA, Departamento de Física, Universidade de Aveiro, Campus Universitário de Santiago, 3810-193 Aveiro, Portugal
[b] Instituto de Telecomunicações, Campus Universitário, 3810-193 Aveiro, Portugal
[c] Instituto de Astrofísica e Ciências do Espaço, Universidade do Porto, CAUP, Rua das Estrelas, 4150-762 Porto, Portugal

## ABSTRACT

This is the era of the multi-messenger astronomy, where we receive news of phenomena across the Universe from electromagnetic emission (EM), neutrinos, cosmic-rays and gravitational waves, allowing the investigation of the events in a multi parameter space. Active galactic nuclei (AGN) galaxies are amongst the most extreme long-lived phenomena in the Universe, emitting over almost the entire EM spectrum and being among the best candidates for sources of neutrinos and high energy cosmic-rays. A large fraction of AGNs are radio emitters, and in this EM domain SKA will be the cornerstone of astronomy in the next decade, in terms of probed sky area, sensitivity, spatial and temporal resolution, which will have an enormous impact on our understanding of galaxy evolution. Extragalactic research is an area where the SKA in combination with other new instruments will boost knowledge by orders of magnitude. **Star formation and Supermassive Black Hole accretion history** are among the **priority scientific objective**s for the first SKA phase, and these are areas in which we are involved.

**Keywords:** galaxy evolution, AGN, star formation, multi-messenger astronomy

## 1     COSMIC BEASTS

Active Galactic Nuclei (AGN) galaxies are among the most extreme and long-lived phenomena in the Universe, emitting up to ~$10^{47}$-$10^{49}$ erg$s^{-r}$ from regions smaller than a tenth of the whole galaxy, covering almost the entire electromagnetic spectrum, from radio to X-ray (or even up to GeV) bands, and so far, sharing with the pulsars the detected TeV sky [1 for a review]. They are also strong candidates for neutrino [e.g. 2] and high energy cosmic rays [3] emitting sources and are foreseen as potential for gravitational waves emitters [e.g. 4]. It is broadly accepted that the main cause of such "activity" is related with the presence of a central supermassive black hole (SMBH; $M_{BH}$ ~ $10^7$-$10^{10}$ $M_{sun}$) surrounded by a viscous accretion disk. Interestingly, the scaling relations between the central SMBH and the host galaxy properties are well established [5], and those relations have triggered many questions about the role of AGNs in the framework of galaxy evolution [6], its origin becoming one of the most critical astrophysical problems. Observations so far argue for a Universe that first experienced high star formation rate and SMBH accretion rate, with both subsequently declining due AGN feedback or other mechanisms, resulting in quiescent supermassive black holes harboring star formation quenched galaxies. Yet, we still have no direct information of the process, due to the limited resolution and sensitivity of the current telescopes.

SKA sensitivity will permit to probe a new regime of radio emission at sub µJy level. This means that the number of known AGNs will grow by orders of magnitude, especially among the so-called radio-quiet population, but also new types of AGN may be discovered in this new parameter-space. Offering both resolution and significant increases in sensitivity, the SKA will allow a better understand of the interrelation between "normal" components of the galaxy (i.e. stellar and dust/gas content) and the active nucleus, which will be crucial to advance knowledge in areas such as AGN activity, star formation rate (SFR) histories, AGN feedback in galaxies, the origin of relativistic jets, among others. Star formation and Supermassive Black Hole accretion histories are among the priority scientific objectives for SKA1 (see Figure 1) and these are areas in which we are involved.

| | |
|---|---|
| CD/EoR | Physics of the early universe IGM – I. Imaging |
| CD/EoR | Physics of the early universe IGM – II. Power spectrum |
| Pulsars | Reveal pulsar population and MSPs for gravity tests and Gravitational Wave detection |
| Pulsars | High precision timing for testing gravity and GW detection |
| HI | Resolved HI kinematics and morphology of ~10^10 M_sol mass galaxies out to z~0.8 |
| HI | High spatial resolution studies of the ISM in the nearby Universe. |
| HI | Multi-resolution mapping studies of the ISM in our Galaxy |
| Transients | Solve missing baryon problem at z~2 and determine the Dark Energy Equation of State |
| Cradle of Life | Map dust grain growth in the terrestrial planet forming zones at a distance of 100 pc |
| Magnetism | The resolved all-Sky characterisation of the interstellar and intergalactic magnetic fields |
| Cosmology | Constraints on primordial non-Gaussianity and tests of gravity on super-horizon scales. |
| Cosmology | Angular correlation functions to probe non-Gaussianity and the matter dipole |
| Continuum | Star formation history of the Universe (SFHU) – I+II. Non-thermal & Thermal processes |

**Figure 1**. These are the highest priority scientific goals, in arbitrary order, for the first phase of SKA, i.e. SKA1, which represents 10% of the whole SKA in terms of collecting area. It is highlighted the relevant subject for the present contribution.

## 2 RADIO LOUD NLS1, A LOCAL LABORATORY OF THE HIGH-Z AGN EVOLUTION

One of the main problems when studying extragalactic objects is the lack of enough resolution permitting to investigate in detail the different processes occurring in the inner regions of the galaxies, a difficulty which only increases as one goes to higher redshifts. For example, at z~2-3, when both AGN and star formation rate show a peak in activity, in order to resolve ~ tens of pc one needs a resolution at milli-arcsec level, and that is not available from almost any telescope, see below. Interesting, there is a population of AGNs that may be regarded as the local Universe counterpart of the high-z AGNs, the so-called radio loud narrow line Seyfert 1 galaxies (RL-NLS1), see Figure 2. They are characterized by (1) having relatively low black hole mass (105-107 Msun) but are accreting central matter at an extremely high rate, close to the Eddington limit [ex. 7], and (2) being frequently hosted by rejuvenated, gas-rich galaxies, some of which show distorted morphology either due to past mergers or by belonging to interacting systems [8]. Figure 3 presents the energy output along the EM spectrum of one of best studied RL-NLS. The figure contains optical (NOT B image) and radio (VLA-A) of the same object, which harbors a complex system where star formation and AGN are comingled [8], as it is the case of many more RL-NLS1 systems. These systems, which are building up central mass, are experiencing the same AGN-galaxy evolution as the first high-z AGNS, so they constitute a superb laboratory to investigate some of the open questions on galaxy evolution as their relative low-z will permit a detailed investigation with the new generation of telescopes, prominent amongst them being the SKA.

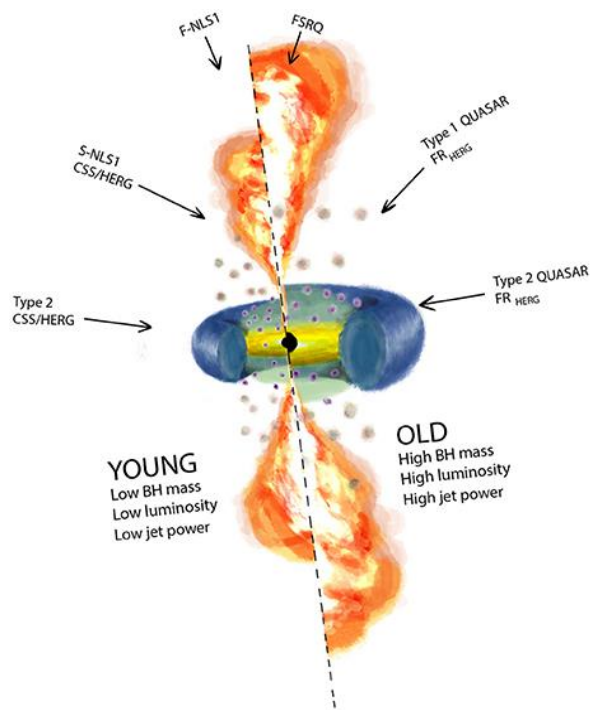

**Figure 2**. The cartoon presents a proposed unification scheme of two types of radio loud AGNs: (LHS) RL-NLS1 that are accreting matter at very high rate with respect to the Eddington limit, building up in central mass, and (RHS) high-z blazars, with a larger SMBH but which are accreting at a lower rate, from [9].

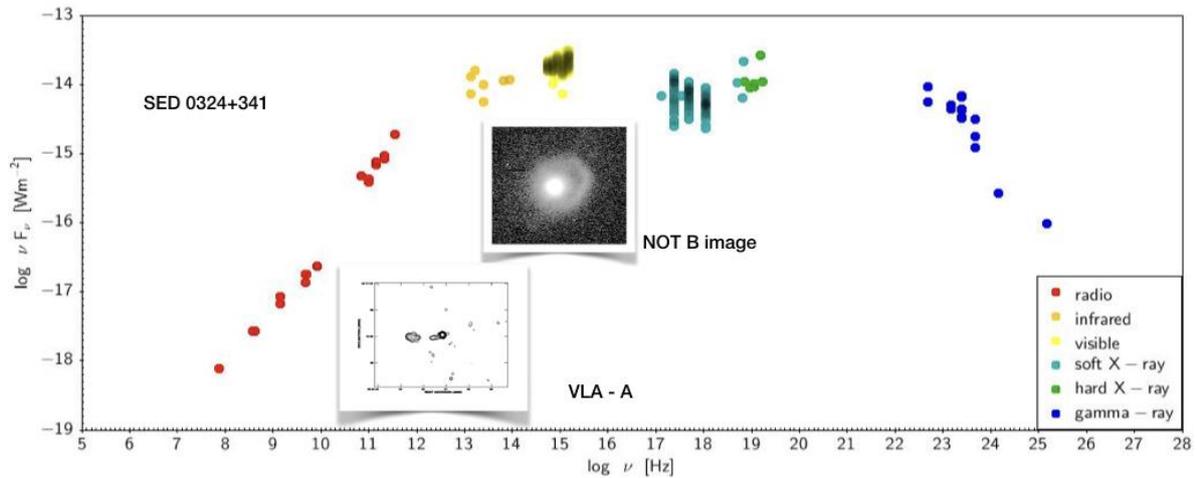

**Figure 3**. Spectral energy distribution of RL-NLS1 0324+341, from the radio band up to the gamma-ray regime, different colors correspond to the different bands. The figure includes an inset B-band image (Nordic Optical Telescope) and radio image (1.4 GHz, VLA, A configuration); from [8].

## Multiwavelength studies: the many views of phenomena

Multiwavelength studies are crucial to pinpoint the physical processes occurring in galaxies containing AGNs, including estimates of redshifts, i.e. distances. For example, our work on the mid-infrared properties of RL-NLS1s has shown that a large fraction of the objects has an "active" host galaxy with very high star formation rate, estimated between 10-500 $M_{sun}$/year [10], showing that also in this aspect RL-NLS1 objects are much more alike high-z systems than those found at their redshift range. Only by combining the information from the different energy domains, from the radio up to high energy regimes, is possible to pinpoint and study the physical processes in operation. The SKA will dedicate an important fraction of its time to all-sky surveys, which is a powerful way to detect transient phenomena, including relativistic jet flares, or transformational science such as FRBs or black hole mergers (putative place for gravitational wave emission). In the examples above, the identification of the phenomena needs characterization of the object at other wavelengths. The SKA-era will see the commissioning of new higher specification telescopes, covering other parts of electromagnetic spectrum, for example: EUCLID (ESA; infrared), JSWT (NASA, infrared-optical), ELT (ESO; optical), e-Rosita (X-rays), ATHENA (ESA, X-rays), CTA (gamma-ray), besides missions like Gaia that are already in place. Note that in some cases the resolution of these telescopes will match that of SKA, as for example that of SKA1-MID and EUCLID or JSWT.

## A question of resolution and sensitivity

In order to disentangle the different physical processes occurring in the RL-NLS1s (or more generally in any extragalactic object) it is necessary to spatially and (sometimes kinematically and temporally) resolve the emitting regions. Most of the energy output of these objects emanates from the central few kpcs, which for systems beyond z>0.1 translates to angular distances of tenth of an arc-second. Figure 4 is a composition of a radio image of the quasar 3C454.3 (that shows the core and extended regions of the jet), with a cartoon of the central AGN engine proposed by [11]. In order to investigate such regions, milli-arcsec resolution is needed. This resolution is currently offered only by the very large baseline radio interferometers (EVN/VLBA). SKA baselines will not be comparable with the later, for example SKA1-MID will have baselines up to only 150 km, similar to those of e-MERLIN. But will be possible to combine SKA with VLBI, in order to reach both ultraprecise astrometry and milliarcsecond resolution. Indeed, SKA-VLBI will offer the best of two worlds: the first offers collecting area, up to 1 km2, therefore sensitivity, the second offers the most exquisite resolution. SKA-VLBI, with baselines up to 10 000 km, will allow us for the first time to investigate sub-mJy sources at milli-arcsec resolution, and a panoply of new results may emerge from the exploration of this new parameter-space. In terms of RL-NLS1 research this means that on one hand there will be a major growth in the number of objects (currently there are few hundred of known sources), permitting a statistical study of the population, on the other hand the superb resolution will permit to disentangling of concurrent processes and enable significant progress in the knowledge of the relation between other galaxy components and AGN, also in the framework of galaxy evolution.

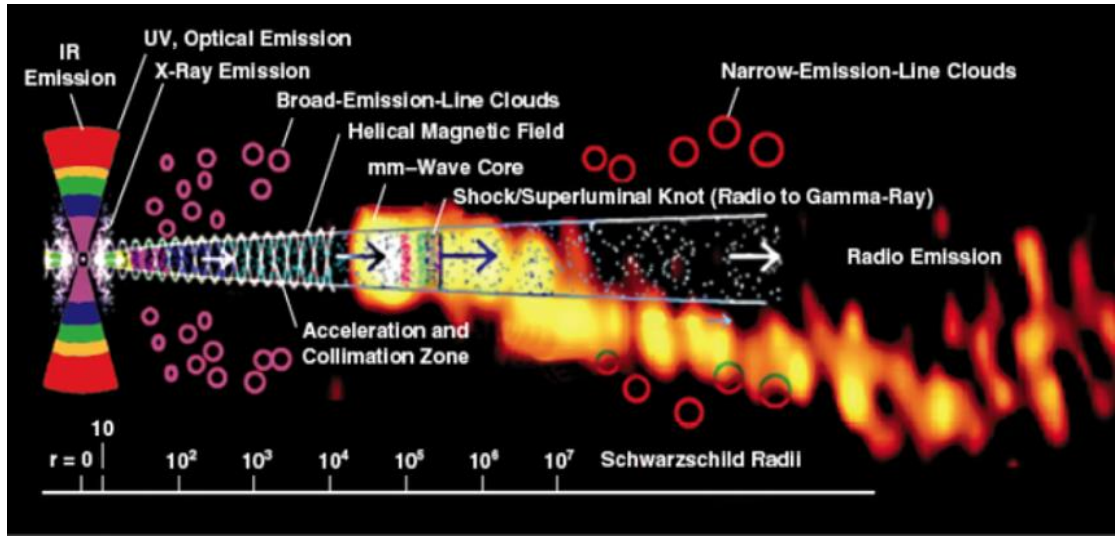

**Figure 4**. Overlay of the mm emission of AGN 3C454.3 and a model of a quasar, from [12] .The central engine is contained in a region smaller than 103 times Schwarzschild radii, which for a black hole of 1010 Msun translates to ~1 pc. In order to probe regions of ~100 pc, for a z> 0.1 object it is needed sub-arcsecond resolution and for z>1 milliarcsecond one.

### 3   High redshift radio galaxies: feedback physics and the first AGN

Radio galaxies in the high redshift Universe (HzRGs; z>2) offer an unrivalled view of a key phase in the evolution of objects that are destined to become the red, inactive massive galaxies we see in the nearby Universe. With the otherwise optically highly luminous AGN obscured from direct view by a dusty "torus", which acts as a "natural coronagraph", their host galaxies and immediate environments can be studied in great detail, using observations from right across the EM spectrum [see 13 for a review]. A notable feature of high redshift radio galaxies are powerful jets of radio-emitting plasma, formed in the active nucleus and propagating out through the galaxy and, in some cases, into intergalactic space. These jets provide not only a means for efficiently finding such galaxies, but also a window into the physics of the feedback activity that is thought to play a central role in the quenching of star formation activity in massive galaxies. The extraordinary sensitivity of SKA will afford a clearer view of the radio-jet activity in these galaxies and, crucially, will allow identification of samples of high redshift radio galaxies in much greater numbers, and out to higher redshifts -- potentially showing us the first radio-loud AGN in the Universe [14].

Around half of HzRGs show strong HI Lyman-alpha absorption lines in their ultraviolet spectra. The properties of the absorption features indicate that these particular HzRGs are surrounded by an expanding bubble of atomic gas, with huge sizes (>30 kpc) [15]. Although the precise origin of these giant bubbles remains unclear, it is thought that they are a remnant of powerful feedback processes which swept up a substantial fraction of the interstellar medium and carried it into the outskirts of the host galaxy. The SKA will allow us to detect these absorbing structures via the HI 21cm line, freeing us from the degeneracies involved in using solely the HI Lyman-alpha line in the ultraviolet [16]. More generally, the 21 cm absorption line will also allow us to probe the atomic gaseous environment of HzRGs, enabling us to assess the gas content and search for the presence of outflows and inflows.

### 4   SUMMARY

A large fraction of AGNs are radio emitters, and in this EM domain SKA will be the cornerstone of astronomy in the next decade, in terms of probed sky area, sensitivity, spatial and temporal resolution, which will have an enormous impact on different areas of galaxy evolution knowledge. Also, extragalactic related research is one of the areas where the exploration of the synergies with SKA will boost by orders of magnitude the current knowledge. The SKA-era will issue in other new telescopes, covering complementary parts of the electromagnetic spectrum with the best ever sensitivity and resolution.

## ACKNOWLEDGMENTS

SA acknowledges financial support from Centre for Research & Development in Mathematics and Applications (CIDMA) strategic project UID/MAT/04106/2013 and from Enabling Green E-science for the Square Kilometre Array Research Infrastructure (ENGAGESKA), POCI-01-0145-FEDER-022217, funded by Programa Operacional Competitividade e Internacionalização (COMPETE 2020) and FCT, Portugal. This work was supported by Fundação para a Ciência e a Tecnologia (FCT) through national funds (UID/FIS/04434/2013) and by FEDER through COMPETE2020 (POCI-01-0145- FEDER-007672). TS acknowledges the support by the fellowship SFRH/BPD/ 103385/2014 funded by FCT (Portugal) and POPH/FSE (EC).

## REFERENCES

[1] Padovani, P. et al, 2017, A&ARv, 25, 2

[2] Padovani, P. et al. 2018, MNRAS, 477, 3469

[3] Aab, A. et al., 2015, ApJ, 804, 15

[4] Enoki, M. et al., 2004, ApJ, 615, 19

[5] Ferrarese, L. & Merritt, D., 2000, ApJL, 539, L9

[6] Haehnelt & Kauffmann, 2000, MNRAS, 318, 23

[7] Boroson, T. A., 2002, ApJ, 565, 78

[8] **Antón, S.** et al. 2008, A&A, 490, 583

[9] Berton, M. et al., 2017, Frontiers in Astronomy and Space Sciences, 4, 8

[10] Caccianiga, A., **Antón, S.** et al., MNRAS, 2015, 451, 1795

[11] Marscher, A. et al., 2008, Nature 452, 966

[12] Wehrle, A. et al., 2009, White paper for Astro2010

[13] Miley, G. & De Breuck, C., 2008, A&ARv, 15, 67M

[14] Afonso, J. et al., Proceedings of Advancing Astrophysics with the Square Kilometre Array (AASKA14)

[15] van Ojik, R. et al., 1997, A&A, 317, 358V

[16] Silva, M., **Humphrey, A.** et al., 2018, MNRAS, 474, 3649S

# HI Gas in Star-Forming Dwarf Galaxies: The SKA Perspective

Mercedes E. Filho[a,b]

[a] Center for Astrophysics and Gravitation - CENTRA/SIM, Faculdade de Ciências da Universidade de Lisboa, Campo Grande, 1740-016 Lisboa
[b] Departamento de Engenharia Física, Faculdade de Engenharia, Universidade do Porto, Rua Dr. Roberto Frias s/n, P-400-465, Oporto, Portugal

**ABSTRACT**

Star-forming dwarf galaxies in the local Universe are unique in that they resemble, in many ways, the first galaxies ever formed in the early Universe; they are chemically and dynamically unevolved. As such, they are the ideal local laboratories to search for signals of cosmic web gas accretion that drives star formation and galaxy evolution. However, until now, signals of cosmic web gas accretion have been mostly indirect. With the advent of the SKA, the paradigm will shift; the high sensitivity of SKA will allow to detect and trace very faint gas at larger distances from the galaxies, and will allow to detect and trace gas in galaxies that are farther away. Thus, the knowledge of the gas cycle driving galaxy evolution will see a substantial increment with the advent of SKA.

**Keywords:** radio astronomy; HI gas; dwarf galaxies

E-mail: mfilho@fe.up.pt

## 1   THE GAS-CYCLE

Galaxy evolution is driven by the cycle of gas: star-forming regions in galaxies, appearing as bright blue regions in optical images (Fig. 1), are where the gas cools, collapses and fragments to make stars. At the end of their life the more massive stars explode as supernova, expelling a fraction of the gas back into the outer galaxy and, in some cases, expelling the gas outside of the galaxy. However, as star formation consumes the gas, gas needs to be replenished in order to support long-term star formation. There are several ways in which galaxies can acquire gas. One way is through mergers or interactions with other galaxies; the process disturbs the gas, the gas cools and then settles onto the galaxy disk where it can be used for star formation. In a similar process, gas can be acquired through the accretion of smaller satellite galaxies. Galaxies may also recycle the gas expelled from star formation processes; as hot gas is ejected from the galaxy disk by supernova explosions, a part of this gas cools and falls backs onto the galaxy disk where it can be used to form stars. The final gas accretion mechanism is cosmological gas accretion. When galaxies were first formed, not all of the gas was incorporated into galaxies; some of the gas ended up in the space between galaxies. Over time, that primordial gas slowly falls onto the galaxies through filaments. These filaments make up what is known as the cosmic web, a faint web of gas that permeates the entire Universe. Cosmological gas accretion is thought to be the main mechanism by which galaxies formed and evolved in the early Universe, and the main means by which galaxies in the local Universe acquire gas that drives star formation and galaxy evolution [1]. As the gas infalling from the cosmic web is expected to be very faint, small-scale and clumpy [2], cosmological gas accretion has been notoriously difficult to detect.

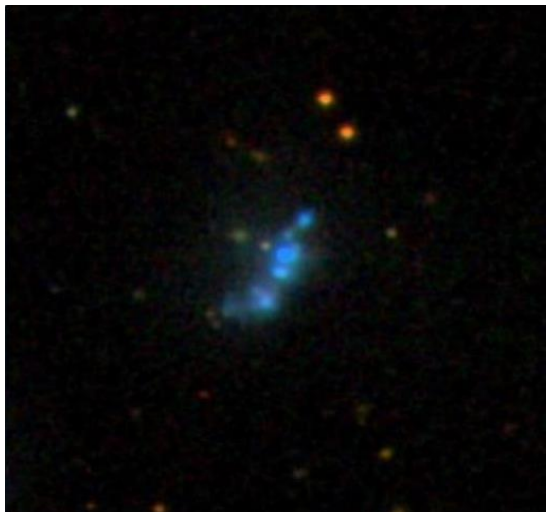

**Figure 1**. An optical image from the Sloan Digital Sky Survey of the star-forming dwarf galaxy CGCG007-025. The bright blue knots are the star-forming regions.

## 2 STAR-FORMING DWARF GALAXIES

Star-forming dwarf galaxies are compact, low-mass galaxies in the local Universe, known to be actively forming stars [3] (Fig. 1). They are of particular interest because they possess similar properties to the first galaxies ever formed, i.e., they are unevolved in terms of their dynamics and their chemistry. In the impossibility of directly studying the first galaxies and the processes responsible for galaxy formation and evolution such as star formation, local star-forming dwarf galaxies constitute the foremost proxies for the first galaxies; by studying star-forming dwarf galaxies it is possible to gain information as to how galaxies form and evolve across time. In particular, because star-forming dwarf galaxies are relatively isolated in space and actively forming stars [3], it is thought that their star formation and evolution is not significantly driven by mergers and interactions with other galaxies, but instead is driven by cosmological gas accretion, similarly to what occurs in the first galaxies. Nonetheless, the faintness and patchiness of the gas [2] poses challenges to its detection; evidence for cosmological gas accretion in star-forming dwarf galaxies has been mostly indirect[4].

## 3 EXAMPLE: SEXTANS A

Sextans A is a local star-forming dwarf galaxy. Figure 2 (far left) shows the gas [5] (black contours) as observed by the Parkes 64 metre radio dish (Fig. 2; middle left), superimposed on an image of the stars (greyscale) acquired with an optical telescope. The gas is shown centered around the stars, but extending further out. An image acquired with such a single radio dish allows to estimate the total amount of gas present in a galaxy. However, the detail with which the gas can be observed (i.e., resolution), and the ability to detect faint gas (i.e., sensitivity) depends on the diameter of the radio dish. In order to investigate how the gas is acquired and how it feeds the star-forming regions, higher sensitivity and more detail is required, i.e., a larger dish is needed, preferably a dish that is several kilometers in diameter. However, such a structure is physically impossible to construct. The alternative is to simulate a large dish using many smaller dishes separated in space, such as the VLA (Very Large Array), an array of twenty seven 25 metre dishes which can be moved on tracks (Fig. 2; middle right). In an array, the resolution is given by the largest distance between two antennas (i.e., maximum baseline), while the capacity to image faint structures depends on the individual dish diameters, the number of antennas and the maximum baseline. With Parkes, a large extended region of gas is observed (Fig. 2; far left), but the image fails to show detail and faint features [5]. With the VLA (Fig. 2; far right), the more extended gas regions are not observed (the size of the detected gas region is approximately the size of the stellar region; Fig. 2; far left; greyscale), but fainter gas and more detail are discernible [6].

Nevertheless, even with this sensitivity, signatures of infalling cosmological gas, such as tails, streams and gas clouds, have gone undetected. The direct detection of cosmological gas accretion signatures requires the detection capabilities only provided by an instrument such as the SKA.

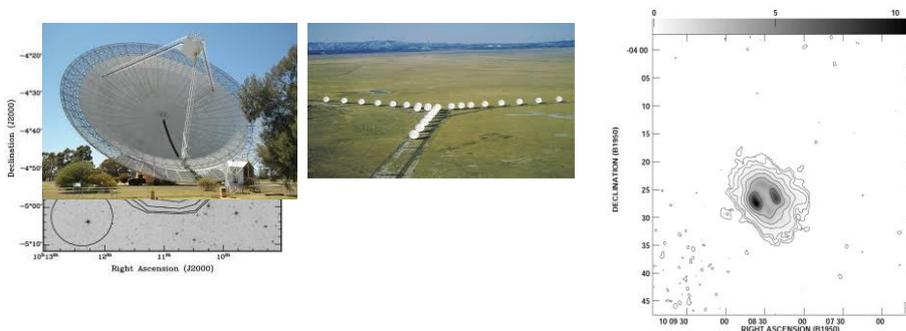

**Figure 2**. Far left: Parkes 64 metre radio dish image of the gas [5] (black contours) in the star-forming dwarg galaxy Sextans A, superimposed on an image of the stars (greyscale) taken with an optical telescope. Middle left: The Parkes 64 metre disk. Middle right: The VLA. Far right: VLA image of the gas (black contours), superimposed on the optical image of the stars (greyscale) [6]. The size of the detected gas region is approximately the size of the stellar region (far left; greyscale).

# 4  THE SQUARE KILOMETRE ARRAY

The SKA will provide an enormous gain relative to using a small single dish or arrays such as the VLA, as it will combine several advantageous properties of both: a billion observing elements, of at least two types, arranged in multiple spiral patterns with a large baseline range, the maximum of which will be over 3 000 km. The large number of antennas will provide a factor 50 increase in sensitivity. Because of such high sensitivity, the sky can be surveyed 10 000 faster. The resolution will be similar to some already existing arrays, and will be equivalent to being able to observe a one Euro coin atop the Eiffel tower all the way from New York City. In addition, parts of the array can be used independently to observe different parts of the sky simultaneously.

# 5  HI GAS IN THE SKA ERA

### Star-Forming Dwarf Galaxies

For only the nearest galaxies, and only for a handful of these, have signatures of cosmological gas accretion (clouds, tails and streams) been found [7] (Fig. 3; black contours). However, even in these cases, it is unclear whether the gas is truly infalling (cosmological gas accretion) or if it is outflowing (gas expelled from the star-forming regions by supernova explosions). Because of the SKA's sensitivity, the SKA will be able to detect faint gas with a high level of detail for a larger number of nearby galaxies, including star-forming dwarf galaxies, and galaxies that are farther away. In particular, the SKA will be able to distinguish between infalling (cosmological gas accretion) and outflowing (stellar processes) gas. In addition, gas can be detected at larger distances from the optical galaxy and at lower gas masses.

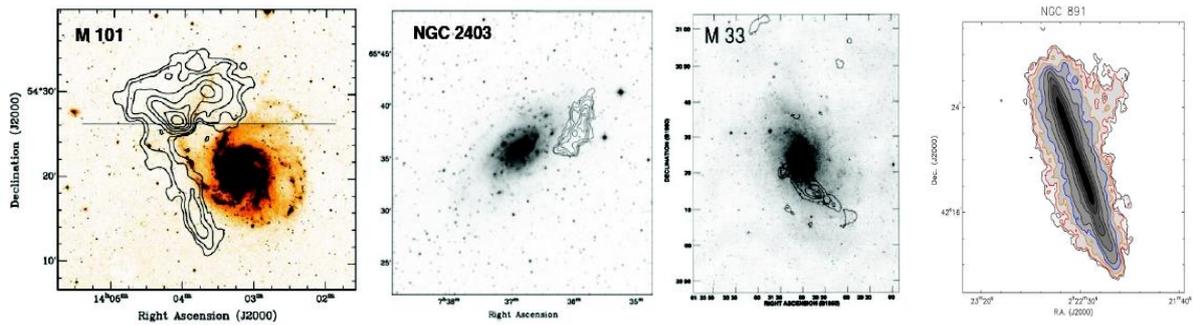

**Figure 3**. From left to right: M101, NGC2403, M33 and NGC891, nearby galaxies shown as examples for the future SKA detection of infalling/outflowing gas, and the detection of faint gas features (tails, streams and clouds) at large distances from the optical galaxy[7]. The gas is shown in black contours superimposed on the optical images of the galaxies in color or greyscale.

# 6  CONCLUSIONS

The advent of the SKA will bring significant advancements in gas detection and tracing, and its role in triggering star formation and galaxy evolution, particularly in star-forming dwarf galaxies. This is particularly important because, not only are star-forming dwarf galaxies surrogates for the first galaxies, the star formation and evolution in star-forming dwarf galaxies is thought to be driven by cosmological gas accretion. Hence, star-forming dwarf galaxies can provide evidence for how galaxies form and evolve over time. Due to its sensitivity, the SKA will allow to detect and trace fainter gas and at lower masses further out from the individual galaxies. In addition, it will be possible to study the gas in a larger number of galaxies, and galaxies at larger distances.

# ACKNOWLEDGEMENTS

M. E. F. gratefully acknowledges the financial support of the Fundação para a Ciência e Tecnologia (FCT – Portugal), through the research grant SFRH/BPD/107801/2015. M.E.F. was supported by ENGAGE SKA for the participation in the SKA days.

# REFERENCES

[1]   Dekel, A., Birnboim, Y., Engel, G., Freundlich, J., Goerdt, T., Mumcuoglu, M., Neistein, E., Pichon, C., Teyssier, R. & Zinger, E. 2009, Nature, 457, 451

[2]   Schaye, J., Dalla Vecchia, C., Booth, C. M., Wiersma R. P.C., Theuns, T. et al. 2010, MNRAS, 402, 1536

[3]   Morales-Luis, A. B., Sánchez Almeida, J., Aguerri, J. A. L. & Muñoz-Tuñón, C. 2011 ApJ, 743, 77

[4]   Sánchez Almeida, J., Caon, N., Muñoz-Tuñón, C., Filho, M. & Cerviño, M. 2018, MNRAS, 476, 4765

[5]   Barnes, D. G. & de Blok, W. J. G. 2004, MNRAS, 351, 333

[6]   Wilcots, E. M. & Hunter, D. A. 2002, AJ, 123, 1476

[7]   Blythe et al. 2015, Proceedings of Advancing Astrophysics with the Square Kilometre Array (AASKA14). 9-13 June, 2014, Giardini Naxos, Italy

# Probing the interior of Neutron Stars and the QCD phase diagram with SKA

Constança Providência*, Pedro Costa, Márcio Ferreira, Helena Pais, Renan C. Pereira

CFisUC, Department of Physics, University of Coimbra, 3004-516 Coimbra, Portugal

## ABSTRACT

We briefly review how the accurate measurement of neutron star masses may constrain the equation of state of dense matter and give information on the appearance of non-nucleonic degrees of freedom. It will also provide information on the structure of the QCD phase diagram, in particular, on the possible existence of a first order phase transition and a critical end point. The interpretation of observations requires the knowledge of the crust equation of state, and a brief discussion of its constitution is given. The SKA telescope will provide important information on the pulsar mass and frequency that will certainly allow the constraining of the high density equation of state of strongly interacting matter.

**Keywords:** neutron stars, equation of state, pasta phases, non-nucleonic degrees of freedom, QCD phase diagram

*Electronic address: cp@uc.pt

## 1 INTRODUCTION

Neutron stars are the most compact objects in the Universe, having in their interior nuclear matter under very extreme conditions of density and isospin asymmetry. These objects are a unique probe to test states of matter not attainable in the laboratory, including the low temperature and high density phase diagram of Quantum Chromodynamics (QCD), very asymmetric nuclear matter, nuclear superfluidity and superconductivity, colour superconducting phases of quark matter, hypernuclear matter or deconfined quark matter at very low temperatures, or even the possible formation of kaon or pion condensates [1].

In order to determine the structure of neutron stars, in particular, the mass-radius relation, it is necessary to know the equation of state (EoS) of the matter that constitutes the star, i.e. the relation between the pressure, energy density and temperature (see [2] for a recent review). A unified equation of state is obtained taking the same nuclear model from the outer crust to the centre of the star. This is usually a difficult task because the different types of matter in the crust, outer and inner, and the core require quite different approaches, and also because there is still no formalism that allows the description of hadronic matter and quark matter within the same model. There has, however, been put a large effort during the last decade to build a complete EoS constrained by laboratory measurements, *ab-initio* theoretical calculations of neutron matter and neutron star observations.

Among the observational constraints, the most stringent are set by the masses of the pulsars PSR J1614-2230 and PSR J0348+0432 of the order of two solar masses. These pulsars leave the question: which is the largest mass of a stable neutron star? Very large masses limit the possibility that non-nucleonic degrees of freedom exist in their interior. In fact, the presence of non-nucleonic degrees of freedom, also known as exotic degrees of freedom in the sense that the constituents of matter are not restricted to neutrons, protons, electrons and muons, will soften the EoS making more difficult for stellar matter to be able to counterbalance gravity and attain very large masses.

The fastest spinning neutron star detected has been the pulsar PSR J1748-2446ad which spins with a frequency of 716 times per second. Millisecond pulsars are recycled pulsars that have gone through a period of accretion of mass from their companion star in a binary system. This pulsar is not constraining much the EoS of dense matter, but if faster pulsars are detected, in particular a sub-millisecond pulsar, they may set real constraints by eliminating models that predict smaller Keppler frequencies, i.e. have smaller mass-shedding limits, see Fig. 2 left panel.

Other important constraints coming from neutron star observations are the ones referring to the so call glitches, i.e. irregular changes of the spinning rate, consisting most of the time of a sudden speed up. The explanation of these events is presently associated to the superfluidity state of neutrons in the inner crust of neutron stars. Measuring the glitch rise time and the post-glitch relaxation time would certainly bring important information on neutron superfluidity, in particular on the nuclear superfluidity parameters. Knowing these parameters will also have important implications on the understanding of the cooling evolution of neutron stars.

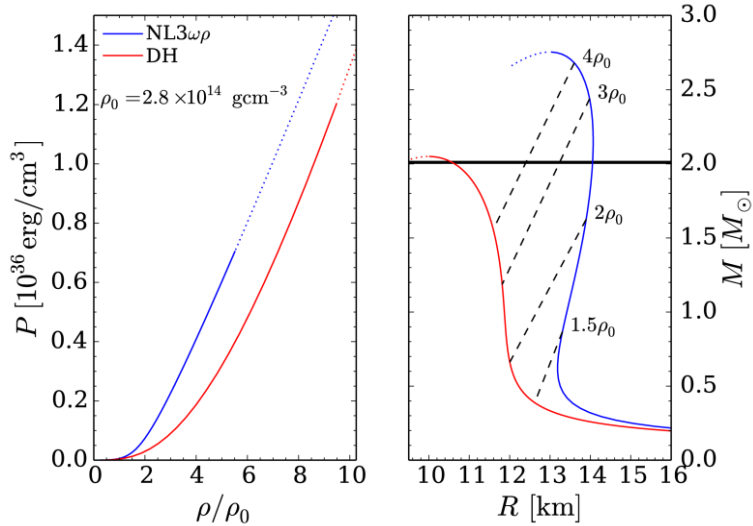

**Figure 1**. The EoS for two different models (left panel) and the corresponding mass-radius curves (right panel) obtained from the integration of the TOV equations. In the right panel we also show the central baryonic density of several stars.

The Square Kilometre Array (SKA) is a unique telescope that will allow the measurement of masses through its high-precision timing measurement. Through pulsar timing, it will be possible to account for every neutron star rotation over large periods of time, making it possible to determine, with large accuracy, the spin, orbital parameters, and, for some binary pulsars, some post-Keplerian (PK) parameters. The measurement of two PK parameters will allow the determination of the neutron star mass. If more than two PK parameters are obtained, then it will also be possible to test general relativity [5].

## 2 NEUTRON STARS: A LABORATORY OF DENSE MATTER PHYSICS

In the interior of neutron stars high densities are attained. Unfortunately, presently it is still not possible to describe matter at high density from first principles, using, for instance, the Lattice QCD formalism, due to the *sign problem*. It is, therefore, necessary to use phenomenological models that describe correctly the symmetries of the strong force, and that have their parameters constrained by experiment, first principle calculations and observations.

The simultaneous measurement of the mass and radius of a neutron star would provide a very important constraint on the high density equation of state. This is clearly seen from Fig. 1a). Here we show the correspondence between the pressure-energy density curve and the mass-radius curve obtained by integrating the Tolman-Oppenheimer-Volkov (TOV) equations, that describe a spherical symmetric object in hydrostatic equilibrium within general relativity. The mass-radius curve (M-R) represents a family of stars described within the same model and having different central baryonic densities: the larger the mass of the star, the larger the central density attained. The correspondence between both curves is a one-to-one relation and having simultaneous information on the mass and radius of a neutron star would impose very strict restrictions on the EoS of dense matter. The two different EoS shown in the left panel give rise to two quite different M-R curves. In the right panel, we identify a set of stars having the same central baryonic density and belonging to each one of the families. It is clear that if we would have access simultaneously to the mass and radius of a neutron star, we would be able to decide which model describes correctly the high density matter.

With SKA it will be difficult to measure the radius of a neutron star with precision, but the same is not true for the mass. Having the mass of massive neutron stars measured with high precision will impose strong constraints on the constitution of matter at high densities, and, therefore, may be used to constraint phenomenological models. The M-R curves plotted on the right panel of Fig. 2 have been calculated from models that include hyperons, see [6] for more details. The interaction between hyperons and nucleons and hyperons and hyperons is still weakly known. The curves shown have been obtained with different coupling constants. It is clear that the two solar mass constraint eliminates some models, but it is premature to say that massive stars do not contain hyperons in their interior.

One of the presently unsolved problems in particle physics is the determination of the QCD phase diagram, in particular, if there is a first order phase transition from confined hadronic matter to deconfined quark matter at high density and intermediate temperature. Lattice QCD predicts a crossover at zero chemical potential but cannot say anything on the high density behaviour. If the crossover goes to a first order phase transition at high densities, then the QCD phase diagram should include a critical end point (CEP). Unveiling the possible existence of a CEP in the QCD phase diagram is the objective of many future heavy ion collision programs. However, if a first order phase transition really exists then this will also have implication on the EoS of neutron stars. In Fig. 3 we show possible EoS that describe matter, including quark matter in the neutron star core.

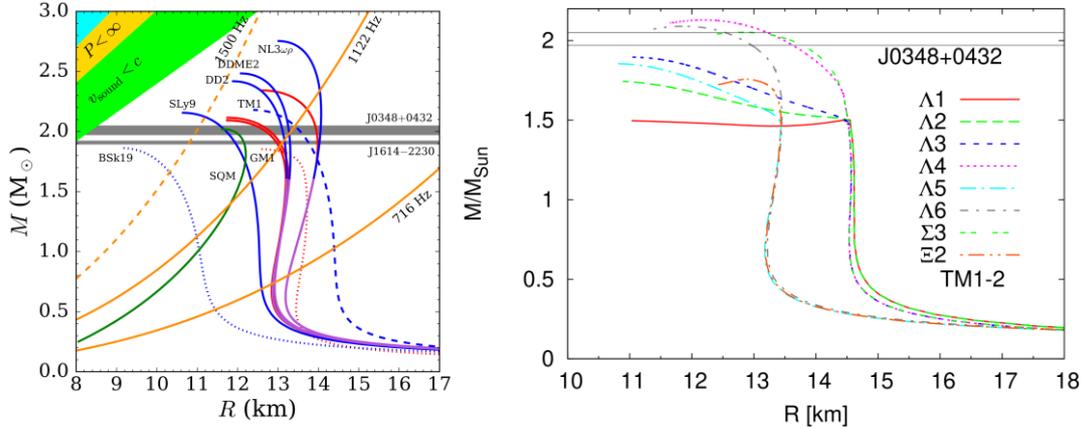

**Figure 2**. In the left panel the mass-radius curves for different nuclear models are plotted. The mass constraints set by the two solar mass pulsars PSR J1614-2230 and PSR J0348+0432 have been included. We also include the mass-shedding limit for pulsars with spinning frequencies 716 Hz, 1122 Hz and 1500 Hz and the causality limit $v_{sound} < c$. M-R curves for EoS including hyperons at high density are plotted in the right panel, from [6]. The two solar mass pulsars constraint is also shown.

The EoS includes a hadronic outer core and a quark core. Quark matter is described within a model with chiral symmetry [7] and a vector interaction is included in order to allow the description of massive stars. This exemplifies the role of neutron star observations in constraining phenomenological models.

We have recently shown that in order to have a reliable estimation of the neutron star radius, it is important to have a unified crust-core EoS. At least the inner crust constituted by a lattice of clusters in a neutron and electron background should be calculated within the same model, or uncertainties as large as 1km may arise, see [8, 9]. In the inner crust, exotic clusters, known as pasta phases, that result from the competition of the Coulomb interaction and the surface tension, giving rise to a frustration phenomenon, are formed. Different approaches have been used to study these phases, and, in particular, in our group, we are applying both the compressible liquid drop model and a Thomas Fermi approach to minimize the free energy of the system [10–12]. At finite temperature it is very important to take into account the presence of light clusters [14] that have an important role in the evolution of a supernova explosion [13], and perhaps on neutron star mergers.

SKA will not be able to provide direct information on the radius, and even indirect information through the moment of inertia will be extremely difficult to get [5]. However, calculating a realistic EoS for the inner crust and determining the crust thickness is important for the explanation of glitches. It is presently not clear if the contribution for these astrophysical events comes only from the crust or whether the core also contributes. With the pulsar timing process operating in SKA, we will be able to collect information on the slow post-glitch relaxation and possibly the glitch rise, and it will be important to have reliable information on the crust size in order to get information on the nuclear superfluidity parameters.

## 3  CONTRIBUTION OF CFISUC TO THE NEUTRON STAR PROBLEM

One of the problems addressed by the Group Hadronic Physics and Fundamental Interactions of the Centre of Physics of the University of Coimbra has been the constraining of the neutron star EoS. In parallel, members of the group have also put a lot of effort on calculating the QCD phase diagram. Our research is strongly integrated in a European effort to join the communities of particle and nuclear physics, astrophysicists and gravitational theory with the main objective of understanding neutron stars through the COST actions NewCompStar (a project that finished on November 2017) and PHAROS (a project started on November 2017). We highly support the creation of a Portuguese network interested on the SKA telescope, from the multifaceted interests that this project arises.

For CFisUC, the SKA observations are very important to constrain the EoS of dense strongly interacting matter and learn about the QCD phase diagram. Our main areas of action are:

- Constraining the EoS of strongly interacting matter.

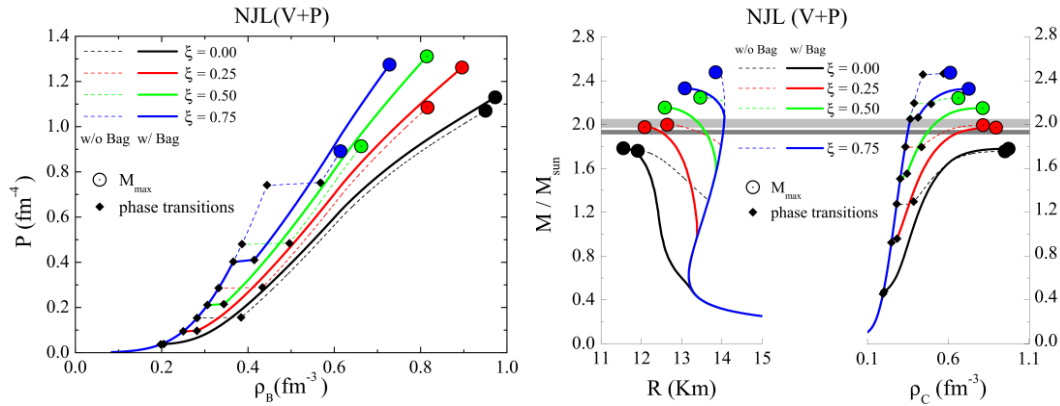

**Figure 3**. EoS for hybrid stars with a quark core in their interior (left panel) and the corresponding M-R curves. In the right panel it also shown the central baryonic density as a function of the star mass. The constraints set by the two solar mass pulsars PSR J1614-2230 and PSR J0348+0432 are also shown.

- Build constrained models that include hyperon degrees of freedom and quark degrees of freedom.
- Calculate the EoS of the inner crust, taking into account the pasta phases, and the effect of strong magnetic fields.
- Calculate the non-homogeneous equation of state of warm matter including the effect of light clusters and pasta phases.
- Contribute to the CompOSE database that provides open access data tables for different state-of-the-art equations of state ready for further usage in astrophysical applications and nuclear physics.
- Study the effect of strong magnetic fields on the EoS of hadronic matter, which is important for the study of magnetars, neutron stars with extremely large magnetic fields.
- Calculate the QCD phase diagram and identify the properties that signal the presence of a CEP.

## ACKNOWLEDGMENTS

This work was supported by the "Fundação para a Ciência e Tecnologia", Portugal, under the projects No. UID/FIS/04564/2016, POCI-01-0145-FEDER-029912 (with financial support from POCI, in its FEDER component, and by the FCT/MCTES budget through national funds (OE)), and under the Grants Nos. SFRH/BPD/102273/2014 (P.C.), SFRH/BPD/95566/2013 (H. P.), PD/BD/128234/2016 (R. P.) and CENTRO-01-0145-FEDER-000014 (M.F.) through CENTRO2020 program.

## REFERENCES

[1] N. K. Glendenning, Compact Stars: Nuclear Physics, Particle Physics, and General Relativity (Springer, New York, 2000).

[2] F. Özel and P. Freire, Annual Review of Astronomy and Astrophysics, 54, 401 (2016).

[3] P. B. Demorest, T. Pennucci, S. M. Ransom, M. S. E. Roberts, and J. W. T. Hessels, Nature (London) 467, 1081 (2010); E. Fonseca, T. T. Pennucci, J. A. Ellis *et al*., The Astrophysical Journal, 832, 167 (2016).

[4] J. Antoniadis *et al*., Science 340, 1233232 (2013).

[5] A. Watts *et al.*, PoS AASKA14 (2015) 043.

[6] M. Oertel, **C. Providência**, F. Gulminelli, A. R. Raduta, J. Phys. G42 (7), 075202 (2015).

[7] **R. C. Pereira, P. Costa, C. Providência**, Phys. Rev. D 94, 094001 (2016).

[8] M. Fortin, **C. Providência**, A. R. Raduta, F. Gulminelli, J. L Zdunik, P. Haensel, M. Bejger, Phys. Rev. C 94, 052801 (2016).

[9] **H. Pais** and **C. Providência**, Phys. Rev. C 94, 015808 (2016).

[10] D. G. Ravenhall, C. J. Pethick, J. R. Wilson, Phys. Rev. Lett. 50, 2066 (1983).

[11] S. S. Avancini, S. Chiacchiera, D. P. Menezes, **C. Providência**, Phys. Rev. C 82, 055807 (2010).

[12] H. Pais and J. R. Stone, Phys. Rev. Lett. 109 (2012).

[13] A. Arcones *et al.*, Phys. Rev. C 78, 015806 (2008).

[14] S. S. Avancini, **M. Ferreira, H. Pais, C. Providência**, G. Röpke, Phys. Rev. C 95, 045804 (2017).

# Astrophysical Transients

Valério A. R. M. Ribeiro[a,b], João G. Rosa[c], Sonia Antón[a]

[a] CIDMA, Departamento de Física, Universidade de Aveiro, Campus Universitário de Santiago, 3810-193 Aveiro, Portugal
[b] Instituto de Telecomunicações, Campus Universitário de Santiago, 3810-193 Aveiro, Portugal
[c] CFisUC, Department of Physics, University of Coimbra, 3004-516 Coimbra, Portugal

**ABSTRACT**

The SKA will be a game changer for astrophysical transients with its unprecedented sensitivity, resolution and ground-breaking nature. Broadly, the SKA will unveil a population of astrophysical transients which will allow the exploration of the unknown and test a number of theories. In this contribution, we identify two very different types of astrophysical transients, novae and fast radio bursts (FRBs), which will undoubtedly take advantage of all the capabilities of the SKA in very different ways. The sensitivity of the SKA will allow for the discovery of all novae in the Galaxy allowing for a complete study of the population. While for FRBs we will be able to probe them at much further distances, in both cases testing a number of different theories. The resolution of the SKA will allow us to localise FRBs as well as probe the nature of their dispersion measure (for example, whether FRBs arise within a supernova remnant or at cosmological distances). For novae, the resolution will allow us to trace the ejecta in detail and pin point where gamma-ray emission arises – if from internal shocks on the ejecta or somewhere else. Tracing the ejecta will allow us to model the geometry and hence retrieve an accurate measure of the ejected mass, which can be compared to the accretion rate to determine if the mass of the white dwarf is growing to the Chandrasekar limit. Synergies between the SKA and other up-and-coming ground- and space-based telescopes will play a crucial role for our fundamental understanding of the nature of astrophysical transients.

**Keywords:** Astrophysical radio transients, novae, fast radio burst, kilonovae, supernovae

E-mail: valerio.ribeiro@ua.pt; joao.rosa@ua.pt; santon@ua.pt

## 1       INTRODUCTION

The South African and Australian Square Kilometre Array (SKA) projects will light up the sky with astrophysical transient events. These astrophysical transients may be broadly divided into two groups: incoherent synchrotron sources and coherent bursts (Figure 1). The former is associated with sources of low brightness temperature ($T_B \leq 10^{12}$ K) and most luminous events where their emission vary on timescales of days to years. These events include merger of compact objects, collapsing stars and explosive events. Furthermore, these incoherent synchrotron sources are associated with outflow material and are the main contributors of heavy elements in the universe.

Coherent bursts on the other hand, are characterised by very high brightness temperatures (in excess of $10^{30}$ K) and very short duration (order of milliseconds to seconds). In particular, coherent bursts probe objects with extreme densities and allow us to test fundamental physics. In this paper, we describe a biased account of some of the research areas which will take advantage of the SKA, both at low- and mid-frequencies. For a broader account the reader is invited to numerous contributions on the SKA science book [1].

The SKA will provide two orders of magnitude jump, over the current best radio telescopes, in sensitivity during its lifetime. This means that in the next decade, or two, our rate of discovery of astrophysical sources will vastly increase. Notwithstanding, output from the SKA will only be improved with a full study of the electromagnetic spectrum of these sources, shifting the paradigm on how we do astronomy. Given the large data rates, this paradigm shift means that as scientists we no longer hold the data on our laptops or university computer clusters. We will only be provided with information to which we will then require to use the knowledge obtained at other electromagnetic wavelengths in order to characterise in detail the observations. As eloquently described in [2], *"The SKA as currently designed will be a fantastic and ground-breaking facility for radio transient studies, but the scientific yield will be dramatically increased by the addition of (i) near-real-time commensal searches of data streams for events, and (ii) on occasion, rapid robotic response to Target-of-Opportunity style triggers."*

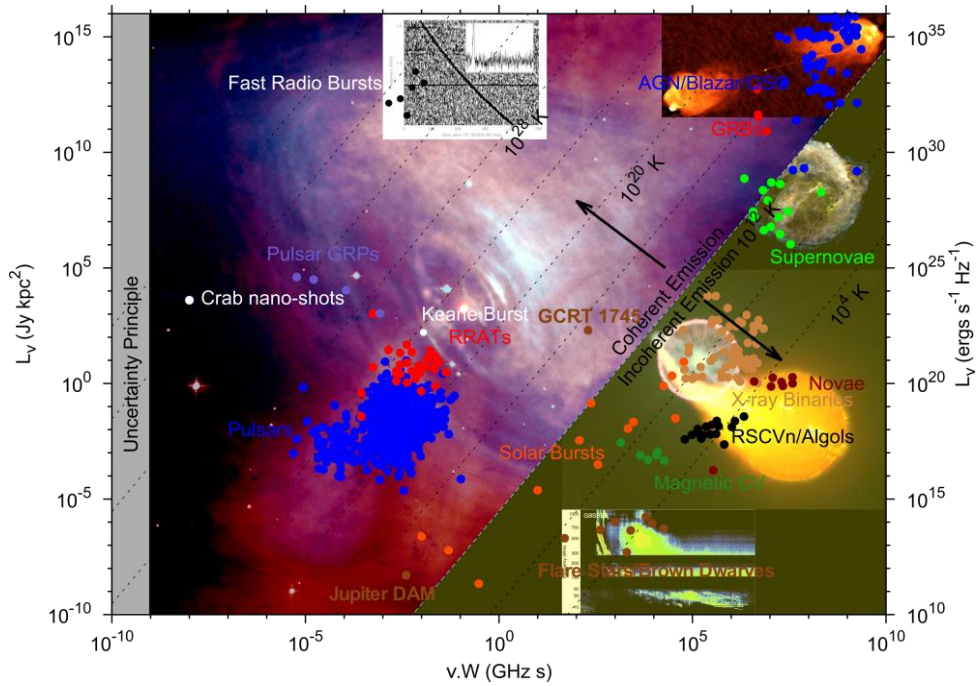

**Figure 1**. Astrophysical radio transient parameter space, demonstrating the wide variety of known sources ([3]; see also [4]).

In the next sections we describe some astrophysics transient sources where some research has been performed and where the SKA will provide a significant contribution towards. In Table 1, we also provide a non-exhaustive list of some of these sources and how the SKA will contribute.

## 2 NOVAE

Classical and recurrent novae are the most common thermonuclear eruption (essentially a nuclear bomb in space) event, in the local universe, that occurs on the surface of a white dwarf star following extensive accretion of hydrogen rich material from a less evolved donor star (see [5,6] for recent contributions). The white dwarf star is the end point of the evolution of a star similar to our Sun once the nuclear burning has stopped at the core. Depending on the evolution of these white dwarf stars, the end product is either a carbon-oxygen or oxygen-neon core. The donor star can either be a main sequence, sub-giant or red-giant star [7]. Due to the nature of nova, where matter is accreted and subsequently ejected, various scenarios have been put forward as to the ratio of accreted versus ejected matter. If less matter is ejected then the mass of the white dwarf star grows to the Chandrasekar Limit, and depending on the underlying white dwarf, either explodes as a Type Ia Supernova or collapses into a neutron star.

The thermonuclear eruption expels a few to tens Earth masses worth of hydrogen, helium and heavy elements[1] all the way to iron. In particular, it is well known that the Universe only produced hydrogen and helium and trace amounts of heavy elements. However, the universe we observe today has a larger amount of these heavy elements. How did they come to be larger? It turns out from exploding stars.

Novae are also excellent test beds for a number of astrophysical processes such as shock physics, the formation of dust, the evolution of material into the interstellar medium, the contribution of heavy elements to life, among others. Furthermore, since novae occur regularly in our Galaxy we are able to observe these in much greater detail than any other erupting star. For these reasons, novae can play a crucial role as they behave essentially as a supernova but they happen much closer (a few 100s to 1,000s light-years vs 10s to 1,000s million light-years in supernovae) and evolve at a slower rate (1,000s of kilometres per second vs 10,000s of kilometres per second in supernovae).

---

[1] heavy elements here mean chemical elements above the atomic number of Helium.

Novae emit radiation at all wavelengths from gamma-rays to radio. In particular, radio observations of novae are useful as the Galaxy is transparent at these wavelengths and the thermal radio emission that arises from the ejecta is a fundamental tracer of outflows from stellar systems. This allows us to derive fundamental quantities such as the ejected mass, kinetic energy and density profile of the ejecta. However, with the upgrade of the Karl Jansky Very Large Array (VLA), New Mexico, USA, and e-MERLIN, Manchester, United Kingdom, our understanding of the nova phenomena has required us to return to the drawing board. These new instruments revealed a complex process of mass ejection in novae.

However, radio solely does not provide the full picture of a nova event. For example, optical observations are able to provide crucial parameters of the nova ejecta [8] which can then be applied to radio observations to derive further parameters [9] and understand the origin of gamma-ray emission [10]. However, at the moment we are dealing with small size statistics as we are only able to follow 1-2 novae per year, at radio wavelengths, which are specifically selected given a number of criteria. In comparison, at optical wavelengths we discover 10 or so nova a year in our Galaxy. The SKA will be a game changer in nova science.

**Novae and SKA phase 1**

The sensitivity afforded by SKA1 will allow us to leap from the 1-2 novae that are currently followed to discovering all novae that explode in the Galaxy, accessible to the SKA – we expect ~35 novae per year in the Galaxy [11]. Here we make the distinction between followed and discovering from the stand point that the former requires an external trigger (this has largely been the amateur astronomy community) to actually having the SKA discovering new novae inaccessible at optical wavelengths due to dust obscuration. The discovery of a larger sample of novae will permit us to study novae not as individual objects but as statistically complete populations. Novae are brightest at the highest frequencies which allows us to work commensally with other projects that will be scanning the Galaxy, for example, for signs of extra-terrestrial life [12]. We can take advantage of these discoveries to then utilise the full suite of SKA1 LOW and MID instruments to get a fuller picture of the event. Particularly, at low frequencies we expect the shocks to be more prevalent and may provide information on the underlying binary system (see, e.g., [13]).

The ~100 km baseline of the SKA1 will provide a better resolution than currently available on the VLA and similar to the e-MERLIN. In tandem with SKA1 the African Very Long Baseline Interferometry Network (AVN) is being prepared across the partner countries. The AVN will have a similar resolution to the current Very Long Baseline Array. It is noteworthy that this baseline configuration will be available for 12 months of the year compared with only a quarter of the year for the VLA longest baseline.

**Novae and SKA Phase 2**

The improved sensitivity of SKA2 will permit us to detect, *for the first time*, an extragalactic nova, at radio frequencies, in the Magellanic Clouds as well as efficiently survey novae throughout the Galaxy. The very long-baselines, from the African Partner countries, will permit us to have very high resolution allowing us to resolve the expanding ejecta much earlier in eruption and at even greater distances.

# 3  FAST RADIO BURSTS

Short duration radio bursts produce some of the most exotic events in the Universe. Fast radio bursts (FRBs), first discovered serendipitously in archival data from 2001 [14], are amongst the most mysterious astrophysical transients observed today. These extremely bright events are very short in duration, ranging from less than a millisecond to a few milliseconds, with large fluencies (time-integral of the received flux) in the range 1-10 Jy-ms (see, e.g., [15]). Over thirty such events have been detected so far at GHz frequencies, mainly with the Parkes 64 metre and Arecibo 305 metre radio telescopes, in New South Wales, Australia, and Arecibo, Puerto Rico, United States of America, respectively. Observed FRBs have extremely high brightness temperatures, up to $10^{37}$ K. This is, of course, a sign of a non-thermal emission mechanism rather than thermal objects emitting at such high physical temperatures. Such high brightness temperatures are associated with coherent emission by "bunches" or coherent radiation involving a large number of particles. The duration of the events also places an upper bound on the size of the progenitor around the size of a neutron star or stellar mass black hole, i.e. up to a few kilometres. However, little is known about the sources of such bright events, despite the plethora of theoretical models that have been proposed in the literature over the last decade.

FRBs are, furthermore, characterised by high dispersion measures (DM), which indicate that the signals have been dispersed by intervening ionized plasma before reaching us, possibly due to dense plasma in the vicinity of the progenitor and/or the lower density intergalactic medium itself. The latter case is the favoured possibility, particularly following the recent discovery of linear polarization and Faraday rotation in FRBs [16], suggesting an extra-galactic origin for the bursts.

Most of the observed FRBs have been identified only once, however, FRB 121102 detected at Arecibo [17,18], has been observed to repeat at irregular intervals. These repeats included, for example, four in a space of twenty minutes, with varying fluence. The repeat nature of FRB 121102 coupled with radio interferometric observations allowed for sub-arcseconds localisation, and the association of the FRB to a faint dwarf galaxy with a low-luminosity active nucleus, emitting a continuous radio signal and a faint optical counterpart close to the location of the repeating bursts [19,20].

Models to explain the FRBs can be roughly divided into either cataclysmic events or sources capable of generating multiple bursts. The above-mentioned repeating FRB 121102 rules out cataclysmic scenarios as sources of all FRBs, taking also into account that repetitions of other known FRBs cannot conclusively be ruled out. Such cataclysmic one-time events are typically associated with mergers of compact objects, including binary neutron stars [21,22,23], binary black holes [24,25], black hole-neutron star binaries [26] and binary white dwarfs [27]. In the latter case the FRB may be the result of coherent emission in the rapidly rotating and magnetized white dwarf resulting from the merger. Non-repeating FRBs could also be the result of gravitational collapse, for example of a supra-massive neutron star into a black hole, with an intermediate stage where the neutron star's magnetosphere remains outside the black hole for a short period, generating a *blitzar* with a sufficiently large luminosity to be observed as an FRB [28].

Several non-cataclysmic FRB progenitor models have also been proposed. For example, 1) giant pulses from extra-galactic pulsars [29] are known to occur in nearby pulsars such as the Crab pulsar, although these might be too atypical. 2) Young energetic pulsars embedded in supernovae remnants [30], which may explain the significant contribution to the observed DM, potentially ruling out cosmological distances as the origin of at least some of the FRBs. 3) Galactic centre magnetars (highly magnetized neutron stars), as well as the associated soft gamma repeaters [31], have also the right properties to account for FRBs however, their population should be substantially smaller than that of young pulsars.

The mechanism driving FRBs may also be the result of novel, more exotic, physics, and a plethora of such scenarios can be found in the recent literature. Amongst these, we highlight the "BLAST" scenario co-proposed by an *Engage SKA* team member [32], where FRBs are the result of coherent laser emission in the vicinity of a spinning primordial black hole. A black hole suffers from the so-called super-radiant instability when it spins too fast and sheds its surplus angular momentum by producing a dense cloud of axions in its vicinity. Such axions are amongst the leading dark matter candidates in extensions of the Standard Model of particle physics, and their stimulated decay into photon pairs can lead to extremely powerful radio laser bursts, powered by the black hole's rotation and repeating every few hours or less.

Distinguishing between these very different scenarios requires, naturally, an increase in statistics, and SKA has the necessary features (frequency, multi-beam capabilities, wide field-of-view) to increase the number of FRB detections up thousands or even tens of thousands [33]. The expected detection rate for SKA1 is about 200 times larger than that of the Parkes radio telescope, the latter detecting on average one FRB every 10 days of telescope time. SKA1-MID will observe at frequencies close or in the range of known FRBs, but SKA1-LOW may also be very efficient in an initial low-frequency survey. The addition of a spectral index, which will only be possible with the sensitivity of the SKA instruments, will rule out a number of models.

The resolution afforded by the SKA long-baselines will be instrumental in providing accurate FRB localisation. For events with redshift z>1, localisation within 0.1-0.5 arcseconds is crucial for an association with a host galaxy, allowing for a comparison with measured DM and polarization of the signal. Furthermore, the increase in both population statistics and localisation accuracy will also allow one to use FRBs as cosmological rulers, for example, to constrain the dark energy equation of state at redshifts beyond those of Type Ia supernovae or to probe primordial magnetic fields and associated turbulent phenomena (see, e.g., [33]).

Finally, the SKA may also be used to indirectly probe the nature of FRBs, since in many scenarios the coherent emission mechanism behind the bursts is accompanied by a continuous signal, not necessarily localised in the FRB vicinity. For example, the BLAST mechanism mentioned above is associated with exotic axion particles that may be the dominant component of dark matter in all galaxies. The very same process responsible for stimulated axion decay around spinning black holes also leads to axion-photon conversion in galactic magnetic fields. Detecting such a signal in the Galactic centre has been shown to be within the reach of SKA [34], providing a combined probe of dark matter and of its possible connection with FRBs.

# 4   SYNERGIES

The characterisation of individual astrophysical transients provides us with useful information however, we require a comprehensive and multi-wavelength study of a wide sample. These astrophysical transients in the SKA era will quickly become a big data challenge at radio frequencies. A number of optical telescopes are either online or about to come online where large areas of the sky can be observed in one snapshot, these include, for example:

- All-Sky Automated Survey for Supernovae (ASAS-SN); [35].
- The Intermediate Palomar Transient Facility (iPTF); [36].
- Zwicky Transient Facility (ZTF); [37].
- Public ESO Spectroscopic Survey of Transient Objects (PESSTO); [38].
- SkyMapper Southern Sky Survey; [39].
- MeerLICHT and BlackGEM; [40].

The various surveys above have a combination of surveying large parts of the sky and being spread throughout the world – some have multiple telescopes – in order to achieve optimal coverage. Noteworthy are the MeerLICHT and BlackGEM project, which use identical optical telescope design, however for different science goals. BlackGEM will be an array of optical telescopes in La Silla Observatory, Chile, with the aim of detecting and characterizing optical counterparts of gravitational wave events detected by Advance LIGO and Virgo. While MeerLICHT, hosted at the South African Astronomical Observatory, in Sutherland, South Africa, will co-point with the MeerKAT, the mid-frequency radio telescope precursor to the SKA. Essentially, wherever MeerKAT points MeerLICHT will point in the same direction[1]. Given that a number of the key science projects on MeerKAT will stare at the same patch of sky for extended periods of time (a few thousands of hours), this will allow for a unique characterisation of deep fields for transient phenomena. In the planned SKA phase 2 expansion to other African partner countries, the MeerLICHT consortia is looking into expanding to the same countries. This will allow for a unique and unprecedented view of astrophysical sources. Furthermore, the SKA sensitivity and resolution will allow us to compare radio images with, for example, the James Webb Space Telescope, probing the nova structure revealing unprecedented detail.

A key driver for understanding the nature of FRBs is their localisation. Associating an FRB with a host galaxy and a counterpart will necessarily involve synergies with optical telescopes and/or higher frequency telescopes – for example, the above mentioned MeerLICHT optical telescope which will be twinned to the MeerKAT radio telescope and the concept of simultaneous radio-optical observations is undoubtedly scalable to the SKA. The associating will prove essential in order to understand the nature of FRBs since many FRB models do not involve counterparts at higher frequencies. Furthermore, the multi-frequencies capabilities of the SKA will provide an unprecedented means to study the spectrum of FRBs at lower frequencies than those currently observed, which will be essential in determining the matter content of the source and may be used to rule out several of the proposed emission mechanisms.

# 5   SUMMARY

Undoubtedly, the SKA will be a game changer for both coherent and incoherent astrophysical transients. Both Novae and FRBs explored here will take full advantage of the SKA that will provide us with unprecedented complete and well understood samples of these sources (see Table 1 for an overview of some further sources). In novae science we will be able to tackle questions such as: can we determine the dominant mechanism of mass loss in novae? Do novae adhere to theoretical expectations? Does the mass of the white dwarfs really grow by accreting non-degenerate matter? While for FRBs we will answer questions such as: what is the underlying progenitor model? What type of population of astrophysical sources create FRBs? Do all FRBs repeat?

---

[1] with the caveat that this will only work at night time for the optical telescope.

Table 1. A few identified coherent and incoherent synchrotron sources where SKA will vastly improve on our current knowledge. Where we provide a question mark is due to the fact that we do not fully understand the underlying population therefore, these numbers are to be taken with caution specially towards SKA2. However, undoubtedly, following SKA1 we will be able to better constrain the rates at which these objects occur.

|  | **Currently** | **SKA1** | **SKA2** |
|---|---|---|---|
| *Core-collapse Supernovae* | ~50 (over 30 yrs) <br><br> Improved understanding of supernova properties | 100s per yr <br><br> Uncover larger supernova population | 1,000s per yr <br><br> Star-formation rate of massive stars and locate missing baryons |
| *Type Ia Supernovae* | 0 <br><br> Only upper limits on the circumstellar gas | 1 per 13 yrs? <br><br> Deeper radio limits to circumstellar gas | 2 per yr? <br><br> Radio and X-ray can discriminate models |
| *Kilonovae* | 1 GW counterpart <br><br> Start testing our theories | 10s – 100s per yr? <br><br> Deeper searches and population studies | 100s – 1,000s per yr? <br><br> Completer population studies |
| *Novae* | 1-2 per yr <br><br> Can determine individual fundamental quantities | ~35 per yr <br><br> This will unveil statistically complete populations | Similar studies will be possible to the Magellanic Clouds |
| *FRB* | ~30 | 1000s – 10000s per year <br><br> A more complete statistical study is possible | Better localisation and larger volume to explore cosmological origins |

## ACKNOWLEDGMENTS

VARMR acknowledges financial support from the Fundação para a Ciência e a Tecnologia (FCT) in the form of an exploratory project of reference IF/00498/2015. JGR is supported by the FCT Investigator Grant No. IF/01597/2015 and partially by the H2020-MSCA-RISE-2015 Grant No. StronGrHEP-690904. VARMR, JGR and SA acknowledges financial support from the Center for Research & Development in Mathematics and Applications (CIDMA) strategic project UID/MAT/04106/2013 and support by Enabling Green E-science for the Square Kilometre Array Research Infrastructure (ENGAGESKA), POCI-01-0145-FEDER-022217, funded by Programa Operacional Competitividade e Internacionalização (COMPETE 2020) and FCT, Portugal.

## REFERENCES

[1] Bourke, T. L. et al. "Advancing Astrophysics with the Square Kilometre Array," in Proceedings of Advancing Astrophysics with the Square Kilometre Array (AASKA14), 14 (2015).

[2] Fender, R. et al. "Transient Astrophysics with the Square Kilometre Array," in Proceedings of Advancing Astrophysics with the Square Kilometre Array (AASKA14), 14, 051 (2015).

[3] Pietka, M., Fender, R. P., Keane, E. F. "The variability time-scales and brightness temperatures of radio flares from stars to supermassive black holes," Monthly Notices of the Royal Astronomical Society, 446, 3687 (2015).

[4] Cordes, J. M., Lazio, T. J. W., McLaughlin, M. A. "The dynamic radio sky," New Astronomy Reviews, 48, 1459 (2004).

[5] Woudt, P. A. and *Ribeiro, V. A. R. M.* ed. "Stella Novae: Past and Future Decades," in Astronomical Society of the Pacific Conference Series, 490 (2014).

[6] O'Brien, T., Rupen, M., Chomiuk, L., *Ribeiro, V.* et al. "Thermal radio emission from novae & symbiotics with the Square Kilometre Array," Advancing Astrophysics with the Square Kilometre Array (AASKA14), 14, 062 (2015).

[7] Darnley, M. J., *Ribeiro, V. A. R. M.* et al. "On the Progenitors of Galactic Novae," The Astrophysical Journal, 746, 61 (2012).

[8] *Ribeiro, V. A. R. M.*, Munari, U., Valisa, P. "Optical Morphology, Inclination, and Expansion Velocity of the Ejected Shell of Nova Monocerotis 2012," The Astrophysical Journal, 768, 49 (2013).

[9] Linford, J. D., *Ribeiro, V. A. R. M.* et al. "The Distance to Nova V959 Mon from VLA Imaging," The Astrophysical Journal, 805, 136 (2015).

[10] Chomiuk, L., Linford, J. D., Yang, J., O'Brien, T. J., Paragi, Z., Mioduszewski, A. J., Beswick, R. J., Cheung, C. C., Mukai, K., Nelson, T., *Ribeiro, V. A. R. M.* et al. "Binary orbits as the driver of γ-ray emission and mass ejection in classical novae," Nature, 514, 339 (2014).

[11] Darnley, M. J. et al "Classical novae from the POINT-AGAPE microlensing survey of M31 - II. Rate and statistical characteristics of the nova population," Monthly Notices of the Royal Astronomical Society, 369, 257 (2006).

[12] Hoare, M. G. et al. "The Cradle of Life with the SKA," in Proceedings of Advancing Astrophysics with the Square Kilometre Array (AASKA14), 14, 115 (2015).

[13] Kantharia, N. G. et al. "Insights into the evolution of symbiotic recurrent novae from radio synchrotron emission: V745 Scorpii and RS Ophiuchi," Monthly Notices of the Royal Astronomical Society, 456, L49 (2016).

[14] Lorimer, D. R. et al. "A bright millisecond radio burst of extragalactic origin," Science 318, 777 (2007).

[15] Katz, J. I. "Fast radio bursts - A brief review: Some questions, fewer answers," Modern Physics Letters A, 31, no. 14, 1630013 (2016).

[16] Masui, K., et al. "Dense magnetized plasma associated with a fast radio burst," Nature 528, 523 (2015).

[17] Spitler, L. G. et al. "A Repeating Fast Radio Burst," Nature 531, 202 (2016).

[18] Scholz, P. et al. "The repeating Fast Radio Burst FRB 121102: Multi-wavelength observations and additional bursts," The Astrophysical Journal, 833, 117 (2016).

[19] Chatterjee, S., Law, C. J., Wharton, R. S. et al. "The direct localization of a fast radio burst and its host," Nature, 541, 58 (2017).

[20] Tendulkar, S. P., et al. "The Host Galaxy and Redshift of the Repeating Fast Radio Burst FRB 121102," Astrophysical Journal, 834, L7 (2017).

[21] Piro, A. L. "Magnetic Interactions in Coalescing Neutron Star Binaries," The Astrophysical Journal, 755, 80 (2012).

[22] Totani, T. "Cosmological Fast Radio Bursts from Binary Neutron Star Mergers," Publications of the Astronomical Society of Japan, 65, L12 (2013).

[23] Wang, J.-S., Yang, Y.-P., Wu, X.-F., Dai, Z.-G., Wang, F.-Y. "Fast Radio Bursts from the inspiral of double neutron stars," The Astrophysical Journal Letters, 822, L7 (2016).

[24] Zhang, B. "On the afterglow and progenitor of FRB 150418," The Astrophysical Journal Letters, 822, L14 (2016).

[25] Liebling, S. L., Palenzuela, C. "Electromagnetic Luminosity of the Coalescence of Charged Black Hole Binaries," Physical Review D, 94, 064046 (2016).

[26] Mingarelli, C. M. F., Levin, J., Lazio, T. J. W. "Fast Radio Bursts and Radio Transients from Black Hole Batteries," The Astrophysical Journal Letters, 814, L20 (2015).

[27] Kashiyama, K., Ioka, K., Mészáros, P. "Cosmological fast radio bursts from binary white dwarf mergers", The Astrophysical Journal Letters, 776, L39 (2013).

[28] Falcke, H., Rezzolla, L. "Fast radio bursts: the last sign of supramassive neutron stars," Astronomy and Astrophysics, 562, A137 (2014).

[29] Cordes, J. M., Wasserman, I. "Supergiant pulses from extragalactic neutron stars," Monthly Notices of the Royal Astronomical Society, 457, 232 (2016).

[30] Pen, U.-L., Connor, L. "Local Circumnuclear Magnetar Solution to Extragalactic Fast Radio Bursts," The Astrophysical Journal, 807, 179 (2015).

[31] Popov, S. B., Postnov, K. A. "Millisecond extragalactic radio bursts as magnetar flares," arXiv:1307.4924 [astro-ph.HE] (2013).

[32] ***Rosa, J. G.***, Kephart, T. W. "Stimulated axion decay in superradiant clouds around primordial black holes," arXiv:1709.06581 [gr-qc], to appear in Physical Review Letters (2018).

[33] Macquart, J.-P. et al. "Fast Transients at Cosmological Distances with the SKA," Advancing Astrophysics with the Square Kilometre Array (AASKA14), 14, 055 (2015).

[34] Kelley, K. and Quinn, P. J., "A Radio Astronomy Search for Cold Dark Matter Axions," The Astrophysical Journal, 845, L4 (2017).

[35] Shappee, B. J. et al. "The Man behind the Curtain: X-Rays Drive the UV through NIR Variability in the 2013 Active Galactic Nucleus Outburst in NGC 2617," The Astrophysical Journal, 788, 48 (2014).

[36] Law, N. M. et al. "The Palomar Transient Factory: System Overview, Performance, and First Results," Publications of the Astronomical Society of the Pacific, 121, 886, 1396 (2009).

[37] Bellm, E. "The Zwicky Transient Facility," in The Third Hot-wiring the Transient Universe Workshop, 27 (2014).

[38] Smartt, S. J. et al. "PESSTO: survey description and products from the first data release by the Public ESO Spectroscopic Survey of Transient Objects," Astronomy and Astrophysics, 579, 40 (2015).

[39] Wolf, C. et al. "SkyMapper Southern Survey: First Data Release (DR1)," Publications of the Astronomical Society of Australia, 35, e010 (2018).

[40] Bloemen, S. et al. "MeerLICHT and BlackGEM: custom-built telescopes to detect faint optical transients," Proceedings of the SPIE, 9906, 990664 (2016).

# Exoplanets and the Cradle of Life with the SKA Telescope

Alexandre C. M. Correia[a], Tjarda C. N. Boekholt[a], Alan J. Alves[a], Ema Valente[a]

[a] CFisUC, Department of Physics, University of Coimbra, 3004-516 Coimbra, Portugal

## ABSTRACT

Life as we know it requires a proper environment to develop. In particular, we expect to find it when similar conditions to those on Earth can be found elsewhere in the universe. The Square Kilometre Array Telescope (SKA) is a perfect instrument to search for life outside the Solar System, since it is able to detect the radio signatures of a large number of phenomena that are directly or indirectly related to the building blocks of life. In this paper, we focus on the formation and detection of exoplanets. SKA observations will provide unique constraints on proto-planetary disks, composition of exoplanets through their magnetic fields emissions, and first detection of exomoons interacting with these fields

**Keywords:** Radioastronomy, SKA, Protoplanetary disks, Magnestopheric Radio Emisions, Exoplanets, Exomoons

## 1    INTRODUCTION

The Square Kilometre Array (SKA) will yield new and unique insights into the different stages of the origin and existence of life elsewhere in the Universe, which were assembled in a SKA working group entitled "Cradle of Life". There is a large number of different topics that are included in this SKA field of research such as

- Searching for Extraterrestrial Intelligence;
- Galactic Radioastronomy: continuum observations;
- Maser Astrometry;
- OH masers in the Milky Way and Local Group galaxies;
- Radio Jets in Young Stellar Objects;
- Tomography of Galactic star-forming regions and spiral arms;
- Complex organic molecules in protostellar environments;
- The ionised, radical and molecular Milky Way: spectroscopic surveys;
- Studies of Anomalous Microwave Emission (AME);
- Protoplanetary disks and the dawn of planets;
- Magnetospheric Radio Emissions from Exoplanets;
- Detection of Exomoons.

For a comprehensive description of all these items we recommend the review work by Hoare et al (2015). Planets appear to be the most favourable cradle of life in the context of current astrobiology. For this reason, in this paper, we focus on the last items listed above. SKA's high frequency receivers will be able to study the early phases of the formation of terrestrial planets, from small grains to rocks. They can directly monitor the growth of dust grains through the important cm-sized regime. For the nearest proto-planetary disk systems, the SKA can study the mechanism of grain growth within the snow line, which is at the basis of the formation of rocky planets. The SKA can also detect the presence of organic molecules, such as amino acids, which are the building blocks of life. These prebiotic molecules can be found in the outer regions of proto-planetary discs, which are free from strong dust emission in millimetre-wave bands. It is in these cold outer disk regions that organic molecules can be incorporated onto comets, which in turn may deliver them to the rocky planets in the inner Solar System. SKA's low-frequency receivers will be able to study the magnetic fields of exoplanets through aurorae emissions, providing a direct detection of these planets and important constraints on their habitability. Magnetic fields provide clues about the composition of planetary interiors, but they also protect planetary surfaces against high energy stellar wind particles. In addition, the modulations of these radio emissions will allow us to extend the field of comparative magnetic physics to a wider range of planet-star interactions. Finally, these modulations can also detect the presence of potentially habitable exo-moons, as well as their compositions.

## 2      PROTO-PLANETARY DISKS

This section is based on the review paper by Testi et al (2015). Proto-planetary (or accretion) discs are a natural consequence of the conservation of angular momentum during the star formation process (eg, Shu et al., 1987). The presence of proto-planetary discs around young stars was initially inferred from the emission in excess of the stellar photosphere at infrared and sub-millimeter wavelengths (Beckwith et al., 1990). These conclusions were later confirmed by the HST silhouette optical absorption images (O'Dell et al., 1993). After the initial observations, proto-planetary discs around young solar analogues were studied extensively in regions of star formation. During the early stages, the disk serves as an intermediary between the hydrogen molecular cloud and the proto-star in formation, while in the later stages, the remaining material is used in the formation of the planets. Therefore, direct observation of the properties and evolution of the proto-planetary disks is a way of studying the initial conditions of planet formation. The physical and chemical evolution of proto-planetary disks is fundamental to understanding the formation of planetary systems in general, and of our own Solar System in particular. The SKA will enable unique observations for constraining the physics of planet formation and disk dissipation mechanisms.

In the continuum, the main emission mechanisms are related to three distinct phenomena that can be understood by observations at different wavelengths in the range of 1-15 GHz: thermal emission from dust, thermal emission from ionized winds, and strongly variable thermal and non-thermal emission arising close to the stellar surface from accretion and magnetic activity. The main observations of SKA on proto-planetary discs will then be centered on i) the growth of dust grains towards the planetesimals, which can be constrained only by gaining access to cm-wave emission from large dust grains and pebbles; ii) the disk cold gas component and its possible pre-biotic molecular species; iii) the disk-star interaction, traced by the emission of photoevaporative and disk winds, which regulate angular momentum transport in disks and the origin of outflows, and of energetic stellar flares that may drive some of the chemical processing of both solid and gaseous material in the disk (Testi et al., 2015).

For spectral lines, the emission of complex and pre-biotic molecules is potentially detectable in proto-planetary disks at the high end of the SKA frequency range. Given the cold conditions of disk midplanes, the organic molecules may be hidden in ice sheets of dust and remain undetectable. However, cosmic ray induced secondary uv-photons can desorb in heavily shielded regions, as suggested by recent Herschel measurements of water in proto-planetary disks (eg Hogerheijde et al., 2011). This process allows the release of organic molecules into the gas phase. Jiménez-Serra et al. (2014) have shown that complex prebiotic molecules (such as glycine) may be detectable if desorbed together with the water molecules. We expect that this result can be extended to the cold midplanes of proto-planetary disks. Direct gas-phase detection and water-abundance measurements of complex organic molecules would be a significant milestone in the study of our cosmic heritage and the ability of the interstellar medium and disk chemistry to produce the raw material necessary for development of life in exoplanets

## 3      MAGNETOSPHERIC RADIO EMISSIONS

This section is based on the review paper by Zarka et al (2015). After more than twenty years of exoplanet detections, we now know almost four thousand of these bodies, covering a wide range of sizes, masses and orbital parameters. It is increasingly desirable to determine the physical properties of exoplanetary systems in order to perform deeper comparative studies. Some topics of interest include magnetospheric dynamics, star-planet interactions, planetary rotation, and planetary dynamos. Large-scale planetary magnetic fields play an important role in protecting the planet's surface and atmosphere. Accelerated charged particles in a magnetic field generate radio emissions. The most intense ones are produced at high magnetic latitudes (called auroral, circumpolar regions) by a well-known non-thermal coherent process, involving keV electrons: the Cyclotron Maser Instability. This process widely operates on all magnetized planets, which are therefore strong radio emitters, and thus observable targets for the SKA (Zarka, 1998).

Planetary magnetic fields are a window into a planet's interior and magnetospheric radio emissions are a probe of planetary magnetic fields. Most of the exoplanet detection techniques are indirect (based on the effect of the planet on the star), so the simple detection of their magnetic fields already provides a direct confirmation of the planet itself. Moreover, magnetospheric radio emissions are a unique tool for probing exoplanet's inner structure (composition, thermal state) and dynamics (including the effect of spin-orbit locking for hot Jupiters), leading to a better understanding of the planetary dynamo process. Ground-based detection of Jupiter's decametric radio emission (Burke & Franklin, 1955) provided the first proof of existence and the first measurement of the Jovian magnetic field. Its monitoring allowed to define precisely the rotation of Jupiter's interior, and to discover the existence of a strong interaction between the moon Io and the Jovian magnetosphere (Bigg 1964). Synchrotron emission maps also revealed the tilt of the magnetic field relative to the rotation axis (Berge & Gulkis 1976).

SKA low-frequency capabilities will have a sensitivity around 10 µJy in the 50-350 MHz range. The SKA is unlikely to be able to detect magnetic fields of terrestrial planets if they have a similar strength to the Earth's, since they will emit at too low a frequency. Much more likely are detections of Jovian planets. Jupiter bursts in the 30-40 MHz range may exceed 107 Jy on Earth, which corresponds to about 40 µJy in the 10-pc range (Zarka et al., 2015). There is, therefore, interest in performing observations within 10 pc on already known exoplanets that orbit heavily magnetized stars, or for which there is a hint of planet detection. For the Jovian planets there is also an additional interest related to the possibility of detecting exo-moons, which are difficult to detect in any other way. The Jovian moons leave an imprint on the auroral emission pattern, which is then revealed as periodic modulation of the radio emission. This may be the case whether the exo-moon itself has a magnetic field (eg, Ganymede) or not (eg, Io). If the exo-moons are icy and magnetic, then they also hold the perspective of hosting biological activity in a deep ocean beneath the cold surface.

## 4  SYNERGIES

The characterisation of proto-planetary discs and magnetic fields of exoplanets provides us with invaluable information. However, we require a multi-wavelength study of these astrophysical objects in order to get a more comprehensive and global understanding.

Concerning the proto-planetary discs, for smaller dust particles at the early stages of their formation, it is important to observe proto-planetary discs also in smaller radio waves (mm), such as those reachable by the ALMA telescope (ESO). In addition, the hot parts of the discs that are close the parent stars also emit thermal radiation in the infrared and visible wavelengths, which can be observed by the James Webb Space Telescope (NASA).

Concerning the magnetic fields, only the SKA wavelengths can study them, but the targets must first be identified using other observational techniques. The GAIA space telescope (ESA) will provide a large list of the giant exoplanets in the neighbourhood of the Solar System that may have significant magnetic fields. Moreover, in order to better characterize these planets, spectroscopic measurements like those taken by the ESPRESSO spectrograph (ESO) or photometric measurements from CHEOPS and PLATO (ESA) will also be very useful. Finally, planetary missions like the JUNO spacecraft (NASA) and the Cassini–Huygens spacecraft (NASA/ESA) to the outer Solar system are also very important to update information on typical magnetic fields properties of giant planets like Jupiter and Saturn, respectively. The JUICE spacecraft (ESA) will visit the icy satellites of Jupiter and hence explore the emergence of habitable exo-moons around gas giants that can be also detected by SKA.

## 5  CONCLUSIONS

SKA will provide unique observations to constrain the physics of planet formation and composition in the frequency range below 15 GHz. SKA observations of proto-planetary disks will focus on the growth of dust grains towards planetesimals, which may be constrained by gaining access to cm-wave emission from large dust grains and pebbles. Grain settling and growth in proto-planetary discs is considered to be the initial step of the formation of the rocky cores of the planets. Planetary-scale magnetic fields are a window into the interior of a planet and provide shielding of the planet's atmosphere and surface for life. These emissions are produced by all the magnetized planets of the solar system in the MHz band. The detection of similar emissions from exoplanets will provide constraints on the thermal state, composition and dynamics of their interior, understanding of the planetary dynamo process, and the physics of star-planet plasma interactions. The Jovian moons also leave an imprint on the auroral emission pattern, which is then revealed as periodic modulation of the radio emission.

## ACKNOWLEDGMENTS

We acknowledge support from CFisUC strategic project (UID/FIS/04564/2019), ENGAGE SKA (POCI-01-0145-FEDER-022217), and PHOBOS (POCI-01-0145-FEDER-029932), funded by COMPETE 2020 and FCT, Portugal.

## REFERENCES

[1] Beckwith, S. V. W., Sargent, A. I., Chini, R. S., & Guesten, R. 1990, Astron. J., 99, 924

[2] Berge, G. L. & Gulkis, S. 1976, in Jupiter, T. Gehrels ed., Univ. Arizona Press, 621

[3] Bigg, E. K. 1964, Nature, 203, 1008

[4] Burke, B. F., & Franklin, K. L. 1955, J. Geophys. Res., 60, 213

[5] Hoare, M. G., et al. 2015, The Cradle of Life and the SKA. In Advancing Astrophysics with the Square Kilometre Array - AASKA14 (pp. 1-11)

[6] Hogerheijde, M. R., Bergin, E. A., Brinch, C., et al. 2011, Science, 334, 338

[7] Jiménez-Serra, I., Testi, L., Caselli, P., & Viti, S. 2014, Astrophys. J. Lett., 787, L33

[8] O'dell, C. R., Wen, Z., & Hu, X. 1993, Astrophys. J., 410, 696

[9] Shu, F. H., Adams, F. C., & Lizano, S. 1987, ARA&A, 25, 23

[10] Testi, L., et al. 2015, Protoplanetary disks and the dawn of planets with SKA. In Advancing Astrophysics with the Square Kilometre Array - AASKA14 (pp. 1-11)

[11] Zarka, P. 1998, J. Geophys. Res., 103, 20159

[12] Zarka, P., et al. 2015, Magnetospheric Radio Emissions from Exoplanets with the SKA. In Advancing Astrophysics with the Square Kilometre Array - AASKA14 (pp. 1-18)

# Solar and space weather science with the SKA

Dalmiro Jorge Filipe Maia

Faculdade de Ciências da Universidade do Porto, Rua do Campo Alegre, s/n 4169-007 Porto, Portugal

**ABSTRACT**

Due to its close proximity the Sun is the brightest radio object visible from the Earth. The high levels of flux density from the Sun and the wide variety of mechanisms for solar and heliospheric radio emission provide us with unique information required for understanding fundamental problems of plasma astrophysics. The high spectral, time and spatial resolution, and the unprecedented sensitivity of the SKA, will provide new insights and results in topics of fundamental importance, such as the physics of impulsive energy releases like flares, the early development of coronal mass ejections, the dynamics of post-eruptive processes, energetic particle acceleration, the structure of the solar wind and the development and evolution of solar wind transients at distances up to and beyond the orbit of the Earth. Radio observations with the SKA, incorporated within current and planned activities in space weather monitoring will provide a solid foundation towards predictive capacity, allowing for better forecasting of extreme space weather events that can deleterious effects in made infrastructures and equipment both on Earth and in outer space.

## 1 INTRODUCTION

In this paper we present several research topics that will specifically benefit from the high resolution and sensitivity to be provided by the SKA, and which are expected to bring us new results of transformative nature. The SKA will enable us to address many of the open questions in solar and heliospheric physics, using its extremely high sensitivity, spectral coverage, and high time cadence. Answering those questions will also have important implications for other astrophysical transients including stellar flares, which have been shown to be common across many types of stars [1].

A strong additional motivation for the intensive development of solar and heliospheric radio physics is of a more practical nature: high-energy radiation, non-thermal energetic particles and bulk plasma mass motions generated as a result of transient episodes of energy release in the solar atmosphere are directly relevant to geophysical challenges such as space weather. Space weather events can have significant effects on technological systems on Earth and on satellites in the terrestrial environment. They are increasingly identified as a major societal risk, with high economic impact.

## 2 EMISSION MECHANISMS AND CLASSIFICATION OF SOLAR RADIO BURSTS

Discovered many decades ago [2,3] solar radio bursts remain obscure to most radioastronomers outside the field of solar radio science. For the uninitiated the classification scheme depending on their spectral shape, with designations given by order of historical discovery from type I to type V, may look somewhat opaque. Nonetheless, these categories are rather informative and identify the presence in the corona or the interplanetary medium of phenomena like electron beams (type III), coronal shocks (type II), and particles accelerated inside loops (type IV).

The emission mechanisms behind those bursts, or at least some of the specifics of those mechanisms, are also not very familiar to many radioastronomers. Types II and III, as well as types I and V (and perhaps some classes of type IV radio bursts), are produced by the so-called "plasma emission" mechanism, one of only four known collective emission processes. This mechanism involves accelerated electrons generating high levels of Langmuir waves with frequencies near the plasma frequency, fp, that are then converted into electromagnetic emission near the fp and its first harmonic [2,3]. Decimetric events within the type IV burst category involve incoherent processes, gyrosynchrotron and synchrotron emission from energetic electrons [3], but even those well-known mechanisms will present spectra that will look somewhat atypical for those outside solar radio science, since Razin-Tsytovich suppression will be quite evident in most spectra.

### 3  SOLAR PHENOMENA PRODUCING RADIO EMISSIONS

The radio bursts, in the categories II to IV that we have alluded to, are not just interesting in themselves, they are also important tracers of the development of large scale coronal disturbances since they provide strong non-thermal signatures at heights were other wavelengths like X-ray are not useful due to the low density in the corona. The most important disturbances associated to radio emissions in the Sun are flares and coronal mass ejections (CMEs).

Flares almost certainly represent the release of stored magnetic energy through the process of magnetic reconnection [4] and are known to efficiently accelerate electrons in large numbers [5], but the detailed physics of the process is not known. Although we have a basic physical picture of electron transport from the Sun to the Earth, there are many unsolved questions concerning energetic electron acceleration, storage, and release in the corona, and transport in interplanetary space [6]. Radio observations are often the only means by which to observe the escaping particles [7] and new observations [8] suggest that radio emission can even be used as a unique tool to diagnose the region where energetic electrons are accelerated when traditional X-ray techniques are insensitive.

CMEs are the white light signature associated with the release of large volumes of plasma and magnetic fields from the Sun into the heliosphere, which can travel at speeds of up to 3,000 km s$^{-1}$ and have average mass about $2 \times 10^{15}$ g. CMEs are the most important driver of space weather and during solar maximum they can occur several times per day [9]. Radio observations provide important diagnostics about CME initiation, early development, propagation and the restructuring low in the corona in the aftermath of the CME. A full review is outside of the scope of this document, see for example [10].

### 4  SOLAR OBSERVATIONS WITH THE SKA

The impact and usefulness of SKA for coronal observations will depend on the imaging capabilities available and their adaptation for solar observations. Analyses are complicated among other factors by the extreme variations in brightness of the different radio sources, with some of them barely seen above the quiet Sun while other very compact sources may contribute with a flux exceeding that of the quiet Sun by several order of magnitude. An extremely good calibration, fast modes of image acquisition, and adequate coverage at short and long baselines, providing both a large field and a large resolution, are important to insure an adequate dynamic range. SKA should outperform current instruments in that regard. We will discuss those aspects briefly in this section.

The first criterion to consider is the frequency coverage and the type of instrument. A major improvement with SKA is that it will be able to simultaneously image at near continuous frequencies, instead of the reasonably low number of discrete frequencies achieved with most instruments operating in the same frequency range currently. Solar emissions cover a wide range of frequencies but many of the features are relatively narrow in frequency, often escaping detection in more than one of the discrete frequencies available. With SKA we will be able to track these spectral features along a substantial fraction of their frequency range. This will provide a much needed clarification on the event progression and a much better association with features seen in other wavelengths, such as γ-rays, X-rays, white light and EUV. In particular, we will be able to study the frequency range from 450MHz to 1GHz where there has been a noticeable lack of imaging on the Sun. That alone would be a major contribution for solar physics from SKA. One exciting development during solar cycle 23 was the observation of very large extent radio loops due to synchrotron emissions from MeV electrons trapped inside the expanding CME loop [11]. Since these features are subject to Razin suppression these observations can be used to constrain values for magnetic fields inside CMEs low in the corona [12]. Being able to reconstruct the synchrotron spectra at every point inside the loop will be a major advancement possible with the nearly continuous frequency coverage from SKA (alone or in conjunction with LOFAR).

The spatial scales of solar emissions vary widely, from the thermal emission from the corona exceeding the size of the solar disk, and loop gyro-synchrotron and synchrotron emissions reaching even larger sizes, while the emission from type I noise storms is close to the minimum size defined by scattering in the solar atmosphere, e.g.

the effect of plasma turbulence in the meter wave range [7]. One of the features distinguishing SKA from previous instruments observing the Sun in the decimeter to meter wavelength range is that it combines a large field of view with high resolution. The interest that an instrument combining both high resolution and large field of view would have been shown by combining Nançay Radioheliograph (NRH) and Giant Metrewave (GMRT) observations at 327 MHz [13]; GMRT provided the long baselines (up to 26 km) while NRH the short baselines (dense coverage below 1 km). Whole sun imaging combining GMRT and NRH was thus possible with a resolution of 49"; the full resolution possible with GMRT was not achieved due to several calibration issues and mainly due to the fact that the coverage of GMRT at the longest baselines is very sparse for instantaneous imaging. A resolution on the order of 10 to 20" can probably be achieved if there is a sufficiently dense coverage for baselines at least up to 20 km. Even considering only the 6 km core, SKA will most likely outperform in terms of resolution all the existing and even the upcoming new generation of state-of-the-art specialized solar radio interferometers.

Regarding the amount of flux, handling the Sun is a major departure from the sub-Jansky modes typical of non-solar observations. Solar radioastronomers like to define their flux in terms of solar flux units (SFU) such that 1SFU=10,000Jy. Solar fluxes in the decimeter wavelength range are often close to 100 SFU. During the really big events, like the Halloween 2003 events, the solar emission in the decimeter to meter wavelength can exceed $10^5$ SFU [14]. One might think that, given these numbers improved sensitivity would not be of paramount importance, but given the immense dynamic range of solar phenomena it is in fact one the most highly anticipated SKA features. During major outbursts at the Sun the flux can be dominated by very spatially localised sources and simultaneously there can co-exist elongated features whose brightness temperature over the same spatial extent as the narrow source could be nine orders of magnitude lower. With SKA increased sensitivity we will be able to simultaneously track the progression of both the weaker and stronger sources. Even during quiet Sun periods increased sensitivity will be of relevance. During solar maximum the flux at decimetre/metre wavelengths is dominated by noise storms and associated type I bursts [15]. The noise storm is a reasonably stable broadband continuum (on a scale from hours to several days) that can have associated with it narrow frequency short bursts (< 1 sec). The level of fluctuations in the noise storm flux, though small, often exceeds the contrast of quiet sun features and as such preclude the use of aperture synthesis even during quiet times. Hence, during solar maximum instantaneous imaging would be required to study even quiet Sun features.

The temporal scales of phenomena at the Sun also vary widely. The thermal emissions can be stable on the order of hours or days, but many outbursts or quasi-periodic pulsations require better than 0.1-second time resolution, and the typical eruptive event will develop through a series of outbursts during a time interval of less than ten minutes [16]. This means that although aperture synthesis could be used for the quiet Sun during solar minimum, it is not an option for solar radio burst science, for which instantaneous imaging is required. The major issue with coronal mass ejection studies based on solar imaging is the short time scale of the associated emissions. In the most extreme events very strong nonthermal emissions (on the order of $10^8$ K at each point over the source region) can rapidly cover nearly the whole Sun at meter wavelengths [17,14]. This extremely rapid response to CME initiation, huge variations in brightness, and short duration means that aperture synthesis techniques cannot be used, and also that an observing program cannot be based on infrequently assigned observing times. During the adequate period in the solar cycle (from the rise in activity following the onset of a new cycle to the decay phase of the cycle) a substantial fraction of the daytime period should thus be assigned to solar observations in order to be able to observe these events.

# 5    CLOSING REMARKS

By providing simultaneously high spectral and spatial resolution unavailable with current instruments, the SKA will radically (by two orders of magnitude) improve on their sensitivity, allowing for the detection of a number of physical phenomena predicted theoretically. The SKA offers the ability to address some of the many fundamental and important issues in solar physics and new and unanticipated discoveries in solar radio physics are confidently expected from SKA that will advance solar and heliospheric physics, fundamental plasma astrophysics, and space weather. The breakthrough potential of SKA in solar and heliospheric studies in the low frequency band has already been demonstrated by MWA [17] and LOFAR [18], both of which are SKA pathfinder projects. These instruments include solar and heliospheric physics, and space weather among their key science objectives.

Finally, we note that solar radio astronomy has a long tradition in Portugal: the first Portuguese radio-telescope was a solar radiospectrograph built in the late 1970s [19—22].

# ACKNOWLEDGMENTS

DJFM acknowledges support from the Enabling Green E-science for the Square Kilometre Array Research Infrastructure (ENGAGESKA), POCI-01-0145-FEDER-022217, funded by Programa Operacional Competitividade e Internacionalização (COMPETE 2020) and Fundação para a Ciência e a Tecnologia (FCT), Portugal.

# REFERENCES

[1] Maehara, H., Shibayama, T., Notsu, S., et al. 2012, Nature, 485, 478

[2] Cairns, I. H. 2011, Coherent Radio Emissions Associated with Star System Shocks, 267

[3] White, S. M., Benz, A. O., Christe, S., et al. 2011, Space Science Reviews, 159, 225

[4] Shibata, K. & Magara, T. 2011, Living Reviews in Solar Physics, 8, 6

[5] Holman, G. D., Aschwanden, M. J., Aurass, H., et al. 2011, Space Science Reviews, 159, 107

[6] Kontar, E. P. & Reid, H. A. S. 2009, ApJ Let., 695, L140

[7] Bastian, T. S., Benz, A. O., & Gary, D. E. 1998, An. Rev. A&A, 36, 131

[8] Fleishman, G. D., Kontar, E. P., Nita, G. M., & Gary, D. E. 2011, ApJ Let., 731, L19

[9] Schwenn, R., Raymond, J. C., Alexander, D., et al. 2006, Space Science Reviews, 123, 127

[10] Pick, M., Forbes, T. G., Mann, G., et al. 2006, Space Science Reviews, 123, 341

[11] Bastian, T. S., Pick, M., Kerdraon, A., Maia, D., & Vourlidas, A. 2001, ApJ Let., 558, L65

[12] **Maia, D. J. F.**, Gama, R., Mercier, C., et al. 2007, ApJ, 660, 874

[13] Mercier, C., Subramanian, P., Kerdraon, A., et al. 2006, A&A, 447, 1189

[14] Pick, M., Malherbe, J.-M., Kerdraon, A., & **Maia, D. J. F.** 2005, ApJ Let., 631, L97

[15] Kai, K., Melrose, D. B., & Suzuki, S. 1985, Storms, 415–441

[16] **Maia, D.**, Vourlidas, A., Pick, M., et al. 1999, Journal of Geophys. Research, 104, 12507

[17] Oberoi, D., Matthews, L. D., Cairns, I. H., et al. 2011, ApJ Let., 728, L27

[18] Carley, E. P., Long, D. M., Byrne, J. P., et al. 2013, Nature Physics, 9, 811

[19] Magalhães, A. & Carneiro, J. 1998, Astrophysics and Space Science, 261, 211

[20] Raoult, A., Pick, M., Magalhaes, A., & Carneiro, J. 1996, Solar Physics, 165, 201

[21] **Maia, D.**, Pick, M., Hawkins, III, S. E., Fomichev, V. V., & Jiřička, K. 2001, Solar Physics, 204, 197

[22] Wang, S. J., Maia, D., Pick, M., et al. 2005, Advances in Space Research, 36, 2273

# Radioastronomy and Space Science in Azores: Enhancing the Atlantic VLBI Infrastructure cluster ; I - Scientific opportunities

Domingos Barbosa[1], Sonia Anton[2], Miguel Bergano[1], Tjarda Boekholt[1], Bruno Coelho[1], Alexandre Correia[5], Dalmiro Maia[3], Valério Ribeiro[1,2], Jason Adams[3,6]

[1]Instituto de Telecomunicações, 3810-193 Aveiro, Portugal;
[2]CIDMA, Department of Physics, University of Aveiro, 3810-193 Aveiro, Portugal;
[3] University of Aveiro, 3810-193 Aveiro, Portugal;
[4] CICGE, Faculdade de Ciências da Universidade do Porto, Porto, Portugal;
[5]CFisUC, Department of Physics, University of Coimbra, 3004-516 Coimbra, Portugal
[6]AMSL Semiconductors, Bladel, the Netherlands

**ABSTRACT**

Radioastronomy and Space Infrastructure in the Azores have a great scientific and industrial interest because they benefit from a UNIQUE geographical location in the middle of the North Atlantic, allowing a vast improvement in the sky coverage. This fact obviously has a very high added value to connect the scientific and space infrastructure networks in Africa, Europe and USA using Very Large Baseline Interferometry (VLBI) techniques. Recently, Azores has started VLBI observations mainly for space geodesy, motion of Earth tectonics, global change research studies, complementing the activities of the future AIR Centre. The construction of the Square Kilometre Array project (SKA) and in particular its SKA1-MID development in South Africa and the development of the Africa VLBI Network (AVN) has setup a new large-scale South Atlantic dynamics that may be greatly enhanced by the inclusion of Azores as a region capable of world-class infrastructure for advanced VLBI services connecting the African and Eurasia infrastructures, as a component to future space infrastructures. There are infrastructures like the large 32-m SATCOM antenna in S. Miguel in the region that could integrate such advanced VLBI networks thus complementing the RAEGE radiotelescopes or we may even foresee the deployment of SKA1 dishes providing very long VLBI baselines to SKA1, EVN and AVN. Such infrastructure is an opportunity for a world-class infrastructure for radioastronomy and space exploration and would constitute a key technological facility for data production, promoting local digital infrastructure investments and the testing of cutting-edge Information technologies**; This paper I explores the scientific synergies of VLBI offered by an upgrade of S. Miguel 32-metre SATCOM antenna.**

**Keywords:** Radioastronomy, VLBI, geodesy, Space Science, Space debris, Space tracking, Atlantic connections, global change

## 1  INTRODUCTION:

Azores has a unique location across the Atlantic: spread over 600Km, the archipelago has islands dispersed across three tectonic plates: American, European and African plates. In fact, the plate junctions pass through its central group (Islands of Faial, Pico, S. Jorge, Terceira and Graciosa). In the 1990's, NASA used the Azores as an important point for the International Reference System to improve the spatial navigation with GPS. Furthermore, since the early 2000's there has been considerable interest for a station in the Azores for science and innovation to provide a unique radioastronomical and space navigation facility.

Azores is now developing a new infrastructure: Atlantic International Research Center (AIR), which is an international platform integrating research and innovation on multi-related areas such as climate change, earth observation, energy, space and oceans. The Florianopolis Declaration [26] led to the confirmation of this new Atlantic interaction level [25, 26, 27] which will aggregate complementary aspects of a new critical infrastructure towards high impact space sciences.

Radioastronomy and space navigation are also areas that can take advantage of the unique location of Azores, and projects on either the installation of new stations in St. Maria or Flores islands or the update and retrofit of a space communications (SATCOM) dish in S. Miguel, have been proposed. One of the projects that has already began is the 13.2 metre radiotelescope with VLBI capabilities of the Rede Atlantica de Estações Geo-Espaciais (RAEGE), deployed at St. Maria island (Azores Western group), a project led by the IGN-Spain and the Region of Azores [16].

On the other hand, this is the era of great developments in the South Atlantic: (a) the Square Kilometre Array (SKA) will become the largest scientific facility of the XXIst century, will yield new and unique insights into the different stages of Cosmic history up to the quest for the Origins (b) and the African VLBI Network (AVN), a network of radio telescopes throughout Africa, mostly from upgrades of SATCOM Intelsat Standard-A 32-m antennas, that will have the capability of extending the VLBI worldwide network. SKA and AVN are also expected to provide during their lifetime important transformational telemetry support and data downlink to the new generation of Deep space probes [28]. With this contribution we aim to address the potential of Azores location for multi-disciplinary research, in the framework of existing infrastructure (RAEGE dishes, putative updated SATCOM dish, SATCOM for Cubesats, AIR) and new high sensitivity facilities like SKA, AVN and EVN.

The impact and visibility are leveraged for instance by the European Commission (EC) and European Parliament (EP). This area is defined as a priority area for funding in EU-Africa relations, in particular, in collaborative radio astronomy studies, as evidenced by the global Square Kilometer Array (SKA) project, the European Parliament Written Declaration 45/11 and the decision of the Heads of State of the African Union "Assembly/AU/Dec.407 CXVIII", for radio astronomy to be a priority focus area for Africa—EU cooperation.

## 2 EXISTING INFRASTRUCTURES IN AZORES: THE EMERGENCE OF LOCAL VLBI CAPABILITIES

The RAEGE consists of a network of four Geodetic Fundamental Stations in Spain (Yebes and Canary Islands), and Portugal (Azores Islands of Santa Maria and Flores), as part of the developments required to set up a VLBI Geodetic Observing System, VGOS [17,18]. VGOS is part of the Global Geodetic Observing System (GGOS) of the International Association of Geodesy (IAG), which integrates different geodetic techniques to provide the geodetic infrastructure necessary for monitoring the Earth system and for global change research. Besides the existent station in St. Maria (African tectonic plate; see previous section), it is expected that a second RAEGE radiotelescope will be installed in Flores Island (American tectonic plate) by 2021.

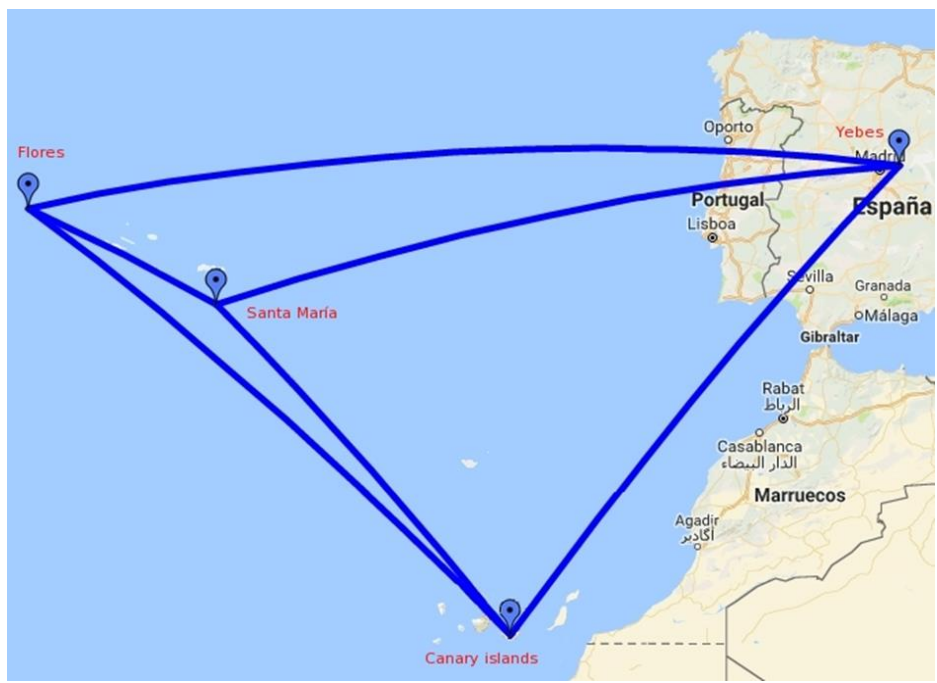

**Figure 1.** The RAEGE network, covering most of Macaronesia, or the central and north-west Atlantic Ocean. When fully deployed, RAEGE will have VLBI stations spread over the American, European and African plates.

# 3 FROM SATCOM TO VLBI NETWORKS

These RAEGE telescopes have a very stringent optical design, optimized for geodesic VLBI observations enabling the measurement of the Earth reference frame with 1 mm accuracy and are designed to enable observations up to or above Ka band (40 GHz). Due to their size optimized for geodesy applications the RAEGE telescopes lack the capability for very high resolution astronomy and Deep space experiments with planetary missions to the outer Solar System that require telescopes with bigger areas and higher sensitivity.

The conversion of SATCOM antennas for scientific purposes and the establishment of Space infrastructure to support space navigation, space tracking and science operations is gaining momentum and offers an opportunity for higher sensitivities as seen with the development of the Africa VLBI Network (AVN). Satellite communications (SATCOM) were the main carrier of telephone (audio), data and TV signals to peripheral regions, thus supplementing undersea cables service. The radio bands allocated for this service is mainly the frequency range known as C-band, ie 5.925 to 6.425 GHz for uplink and 3.700 to 4.200 GHz for downlink.

To provide support to these services, Intelsat Standard-A antenna Earth stations were designed and deployed with large dishes 32 m in diameter around the world, in particular in the 70's and 80's years. From mid-80s the new satellite communication technologies and the ever increasing coverage and bandwidth of modern digital ground optical fiber networks paved the way for much smaller ground station apparatus and the largest stations became redundant with time.

Remarkably, there are a number of examples across the world of transforming redundant Intelsat Standard A dishes into scientific world class facilities. Those Intelsat dishes share a similar design, have focal distances similar to radiotelescopes and therefore are much easier and cheaper to upgrade and to be transformed into scientific or space service stations, as illustrated by some examples: the 30-m dish at Ceduna in Australia [3] was transformed into a radiotelescope in 1985 by the University of Tasmania, in Atlanta (USA), a 30m SATCOM antenna, was acquired and fully renovated as radiotelescope by the Georgia Institute of Technology [6]; in Japan, near Yamaguchi, a 32-m antenna [7] was upgraded to science operations by the National Astronomical Observatory of

| Station | Dish Size | Freq | Lat | Long | Capabilities |
|---|---|---|---|---|---|
| RAEGE St. Maria (in operations) | 13.6 m | 2-40 GHz | 36.985175 | -25.125878 | Geodetic VLBI, Geodesy, global change studies, GNSS |
| RAEGE Flores (in preparation; for ~ +2022) | 13.6 m | 2-40 GHz | ~39.45 | ~-32.2 | Geodetic VLBI; Geodesy, global change studies |
| SMA St. Maria ESA Tracking (in operations) | 5.5 m | S, C band, X band | 36° 59' 50.10 | -25° 08' 08.60" | Kourou launch tracking; CleanSeaNet, satellite detection of oil slicks |
| VLBI SATCOM Station S.Miguel (planned) | 32 m | 5-10 GHz | 37º 47.36 'N | 25º 39.85' W | VLBI; Very High resolution Astronomy + Space VLBI; Deep Space Network Doppler Tracking; space debris |
| FCT Space station (for installation; 2019) | 15 m | C, X band | | | Proba-3 DSN; LEO polar + helio synchronous orbits $ |
| LEO station (planned) | 2.4 / 4.5 m | S, X band | SMA | SMA | LEO tracking; Rx-Tx GGSN for cubesat / formation flying * |

**Table 1.** Summary of Existing and planned Space and Radioastronomy Infrastructure in Azores.; $ - based on FCT/Thales-Edisoft;

Japan (NAOJ); a SATCOM transformation in Elfordstown near Cork in the Republic of Ireland [8] was considered to enhance the resolution and uv-coverage of both the Multi-Element Radio-Link Interferometer Network (e-MERLIN) and the European Very Long Baseline Interferometer (VLBI) Network (EVN). Additionally, in Latvia, the Ventspils International Radio Astronomy Center (VIRAC) has recovered a 32-metre old Soviet radar station to an operating radiotelescope, now a pivotal facility of the Baltic research in space science with a particular focus in planetary sciences (comets, NEOs, etc)..

In Africa, the AVN is leveraging the constitution of radioastronomical and space science communities across Africa through the conversion of former SATCOM dishes. In fact, there are about 29 documented 30-m class telecommunications antennas in 19 African countries though some no longer exist. Ghana just saw the first successful African example of an Earth station in Kutunse transformed into a world class radioastronomical and space facility [9,10,11], with Zambia and Kenya to follow.

The sensitivity contribution of a VLBI radiotelescope to existing networks can be perceived by what is called the UV- plane (sparing details, the FFT of image sky brightness). The better we fill-in the UV-plane, the higher the quality and fidelity of the images of the sources (astronomical objects, spacecrafts). Figure 2, 3 and 4 demonstrate the UV coverage of European VLBI network and the upcoming Africa VLBI Network. Adding an Azores dish alone contributes to a much needed East-West baseline relative to the main stations in Eurasia and Africa (which provide a broad North-South baseline). Long baselines detect the most compact, otherwise unresolved structures. This remarkable information highlights the great scientific potential of such a contribution. **The Azores is the only place in the North Atlantic that can provide the much necessary East-West baselines.**

**Networking, timing and e-services for e-VLBI**

As it is expected in VLBI networks like the EVN and AVN, large radio telescopes regularly participate in VLBI and electronic VLBI (e-VLBI) observations. This requires good quality of service delivery for data transfer across carrier digital networks. The Azores Intelsat Standard-A antenna station compound in S. Miguel possesses good digital connectivity and has besides the 32-metre station, a 12-metre station for ca SATCOM communication link with the Azores Western group (Flores and Corvo Islands). The compound is served by a dedicated 1-Gbps point of presence (Giga-PoP) link connecting to the Azores fiber loop at 10Gpbs. The Azores itself have very good connectivity having access to the new submarine cables nearby and the archipelagos is connected to mainland Portugal. This should enable e-VLBI operations with 1-Gbps speed (i.e. 16 channels each with 16-MHz bandwidth and 2-bit resolution). In fact, the RAEGE station in St. Maria already participates in e-VLBI experiments with Yebes radiotelescope (OAN-Spain) with interferometric fringe analysis by the JIVE ERIC[1] SXFC correlator in the Netherlands. We note that operators and NRENS are already implementing 100Gbit/sec enabling large data streams between Africa or Australia and Europe thus opening great prospects for added capacity and real time correlations.

Precise Timing is necessary to enable phase coherency when correlating signals. A maximum permitted coherence loss of 2% equates to 0.2 radians of phase error, which, at a maximum observing frequency of 6 GHz, corresponds to accuracies of ~10 ps. Also, high precision long-term timing is necessary for astrophysical phenomena such as pulsars and transients, nowadays major scientific cases. In particular, pulsar monitoring experiments require timing accuracies of 10 ns over time periods of 10 years. Overall, synchronization to an absolute time provider is required for system management, antenna pointing, beam steering, time stamping of data and producing regular timing ticks. From the experimental point of view, the clocks should enable a frequency standard for the Local Oscillators (LOs), digitizer clocks and other devices.

Timing and synchronization to the required stability should be provided by a local hydrogen maser (in the VLBI case) or from a distributed frequency reference signal, locked to a central 'master' hydrogen maser frequency standard. Since Azores islands are connected by their Azorean fibre ring loop, the SATCOM antenna could also be synchronized after time transfer over fibre from the operating "clock master" at St. Maria RAEGE station with White Rabbit technologies, a framework for Ethernet-based network for general purpose data transfer and sub-nanosecond accuracy time transfer being used in radioastronomy and space, for instance by EVN [22,23]. This is also the case for the SKA1, where White Rabbit is being considered as a prime option for accurate time transfer across the SKA1 network [24]. As a spin-off, it has been shown recent optical fibre submarine cable scan be explored as supplementary seismograph infrastructure, therefore also enhancing the scope of the scientific instruments infrastructure spread over the Azores. **Indeed, the reinforcement of the Atlantic radioastronomical and space cluster favours the emergence of new testbeds on time distribution technologies across digital networks and new science applications.**

---

[1] https://www.jive.nl/

# 4   LARGE RADIOTELESCOPES AND VLBI SCIENCE GOALS

Scientifically, the use of VLBI with large radiotelescopes enables unsurpassed high resolution of compact sources. In particular, Imaging with high angular resolution of compact, bright radio sources, with very high science impact, currently also as part of multi-messenger astronomy. We can cite as most important science cases:

- Quasars – physics of relativistic jets, , calibrators, astrometry (also in the framework of Gaia mission)
- Microquasars – behavior and parallaxes – distances
- Masers – star-forming (e.g. methanol) – understand variability and measure the distances to star-forming regions in the Milky by parallax, to map spiral arms
- Pulsars – proper motions and parallax, interstellar scattering, emission region size
- Transient sources, including Novae, FRBs and GW triggered events
- Supernovae – behaviour of exploding star remnants
- Interacting binary star behaviour
- E-VLBI and TOO VLBI through internet – rapid response to new events (triggered by Gravitational waves or cosmic rays' events).
- Near-Earth object tracking capability gained by the combined optical and VLBI methods

Additionally, VLBI capable telescopes can also operate in "single dish" mode, since e-VLBI experiments run only for some weeks in a row. An Azores SATCOM antenna has the advantage of being closer to the Equator than other EVN stations. It therefore has access to a larger area of the sky and can observe a much bigger fraction of the Milky Way compared with EVN radiotelescopes based in the Northern Hemisphere. The high sensitivity of such a large dish can be used for these "single dish" opportunities:

- Spectroscopy with narrowband multi-channel receiver
- Monitor masers in star-forming regions eg for periodic variations (methanol at 6668MHz)
- Survey formaldehyde absorption in Milky Way dark clouds (4829 MHz)
- Pulsar observing with wideband multi-channel receiver
- Monitor pulsars for glitches, long term behaviour, proper motion
- Search / monitor for intermittent pulsars and transients (RRATs)
- Radio continuum flux measurement with wideband multi-channel receiver
- Monitoring of Gamma-ray flare sources
- Planetary sciences: Single dish observations of NEOs (Space Situational awareness) or comets at OH maser radiation line;

# 5   SKA AND VLBI

The Atlantic scientific dynamics will greatly accelerate with the deployments of the SKA and the Africa VLBI Network (AVN). Since the SKA 1 (MID) will behave as a big very sensitive station, the expansion of its VLBI capabilities (SKA-VLBI) with addition of stations across Africa and eventually the EVN will greatly broaden the science of the SKA, see [65] for an extensive analysis. In fact, the VLBI will be much necessary to increase substantially spatial resolution. For the future Phase 2, it is expected that the outer SKA stations will greatly enhance the VLBI capabilities through a distributed baseline configuration of up to a few thousands of Km, eventually merging it with the existing single sited VLBI station networks.

This high spatial resolution capability has long been considered an essential part of the SKA [66-70]. The science goals are best achieved with SKA1 by forming phased-array elements from SKA1-MID observing together with existing VLBI arrays in the 1-15 GHz frequency (L to C- Band).

| SWG | Science Objectives | SWG Rank | VLBI with: |
|---|---|---|---|
| CD/EoR | *Physics of the Early Universe IGM I. Imaging* | 1/3 | |
| CD/EoR | *Physics of the Early Universe IGM II. Power Spectrum* | 2/3 | |
| Pulsars | *Reveal pulsar population and MSPs for Gravity Tests and GW detection* | 1/3 | |
| Pulsars | *High Precision timing for testing Gravity and GW detection* | 1/3 | **LOW/MID** |
| HI | *Resolved HI Kinematics and morphology of ~$10^{10}$ Msol galaxies out to z~0.8* | 1/5 | **LOW/MID** |
| HI | *High spatial resolution I the nearby Universe* | 2/5 | |
| HI | *Multi- resolution studies of the ISM in our galaxy* | 3/5 | |
| Transients | *Solve missing baryon problem at z~2 and determine Dark Energy Equation of State* | =1/4 | **MID** |
| Craddle of Life | *Map dust grain growth in the terrestrial planet forming zones at a distance of 100pc* | 1/5 | **MID** |
| Magnetism | *The resolved all-Sky characterization of interstellar and intergalactic magnetic fields* | 1/5 | |
| Cosmology | *Constraints on primordial non-Gaussianity and tests of Gravity on super-horizon scales* | 1/5 | |
| Cosmology | *Angular correlation functions to probe non-Gaussianity and the matter dipole* | 2/5 | |
| Continuum | *Star Formation history of the Universe (SFHU) I+II. Non-Thermal and Thermal processes* | 1+2/8 | **MID** |

**Table 2. –** SKA High Priority Science Objectives and VLBI service planning. Legends: SWG – Science Working Group; CD/EoR – Cosmic Dawn/Epoch of Reionization.

Furthermore, VLBI astrometry will remain a very important tool for astrophysics enhancing and surpassing Gaia astrometry products, with several applications to navigation and Reference Systems. For example, pulsar parallax measurements using SKA-VLBI will play an essential role in several high impact areas, including strong field tests of gravity in relativistic binary systems, tomographic mapping of the Galactic magnetic field and mapping the ionized interstellar plasma in the Galaxy, and the physics of neutron stars, as well as detecting the gravitational wave background.

SKA-VLBI will provide very sensitive, milliarcsecond (mas) resolution imaging that is important for the study of the physics of jets in Active Galactic Nuclei (AGNs), the formation of the very first generation of MBs in the Universe and in connection with the Cherenkov Telescope Array (CTA) or the Pierre Auger it will help reveal the nature of the large population unidentified high-energy sources [71].

Since the VLBI configuration with SKA will be rather conventional and similar to other tied beam arrays, an addition of a station to cover the Atlantic gap may greatly enhance the scientific retour of SKA-VLBI.

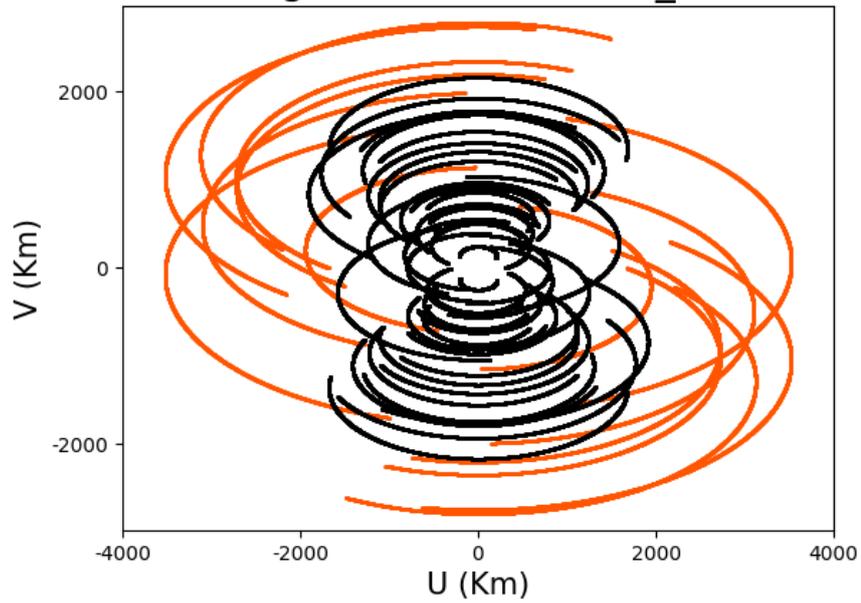

**Figure 2.** U-V plane Coverage of European VLBI Network EU seven stations + Azores (red). Azores provides the Atlantic unique and fantastic coverage and enhances the dynamic range of observations. Source used: QSO 0234+285.

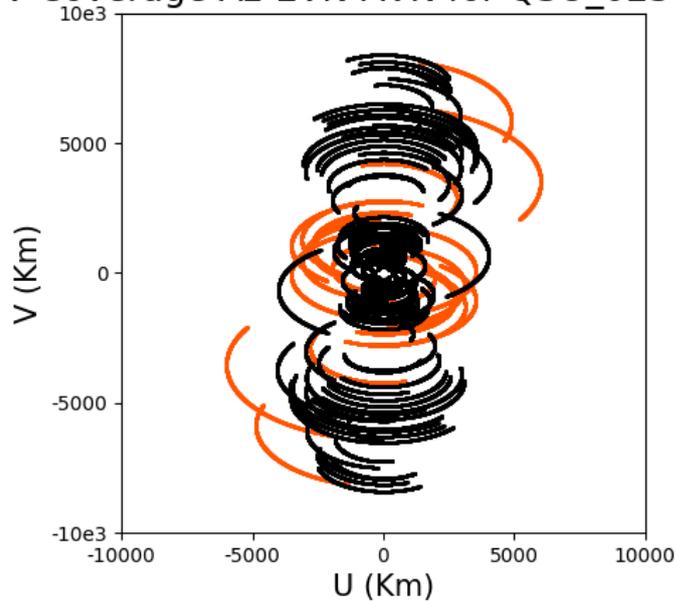

**Figure 3 .** U-V Coverage contributions of European VLBI Network + Africa VLBI Network (Ghana – Kuntunse; Longonot – Kenya, HART – South Africa) + Azores (red). Azores Atlantic coverage is unique. Source used: QSO 0234+285.

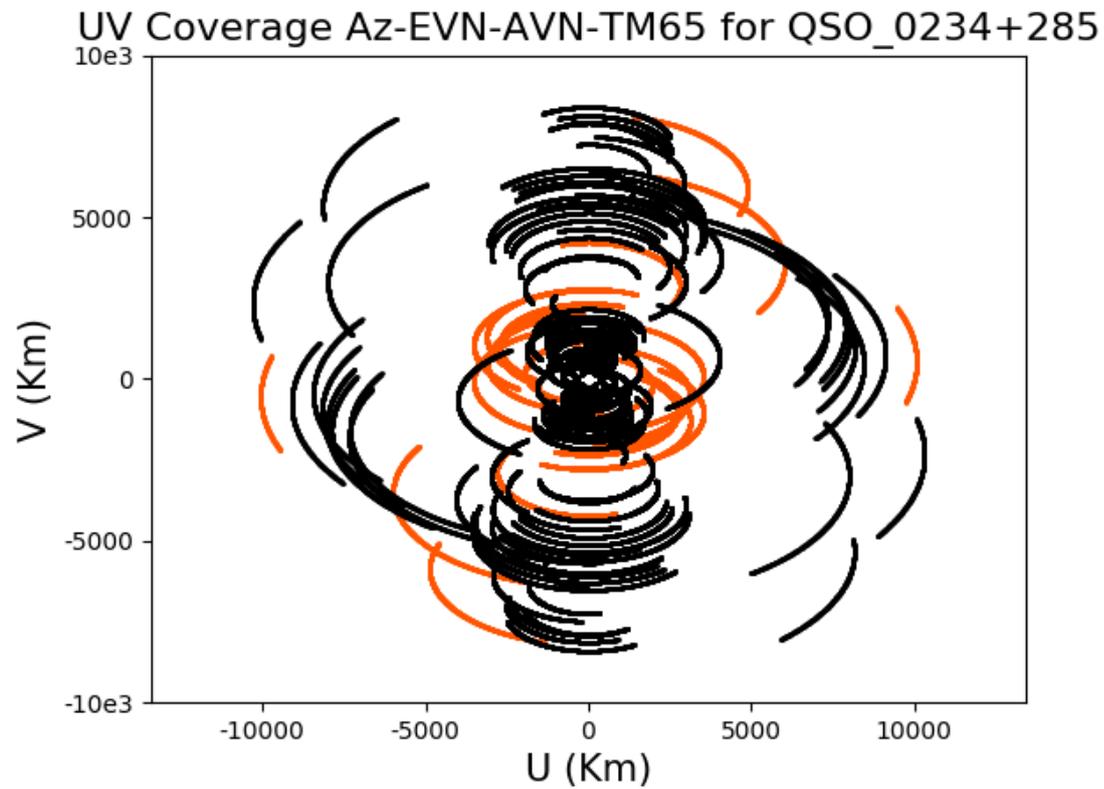

**Figure 4.** U-V Coverage contributions of existing Eurasia EVN + AVN + China VLBI network (CVN) Shangai TIANMA65 (China) + Azores (red). Azores provides unique coverage, and enhances dynamic range of observations of EVN + AVN + Shangai (China) stations added. Source used: QSO 0234+285.

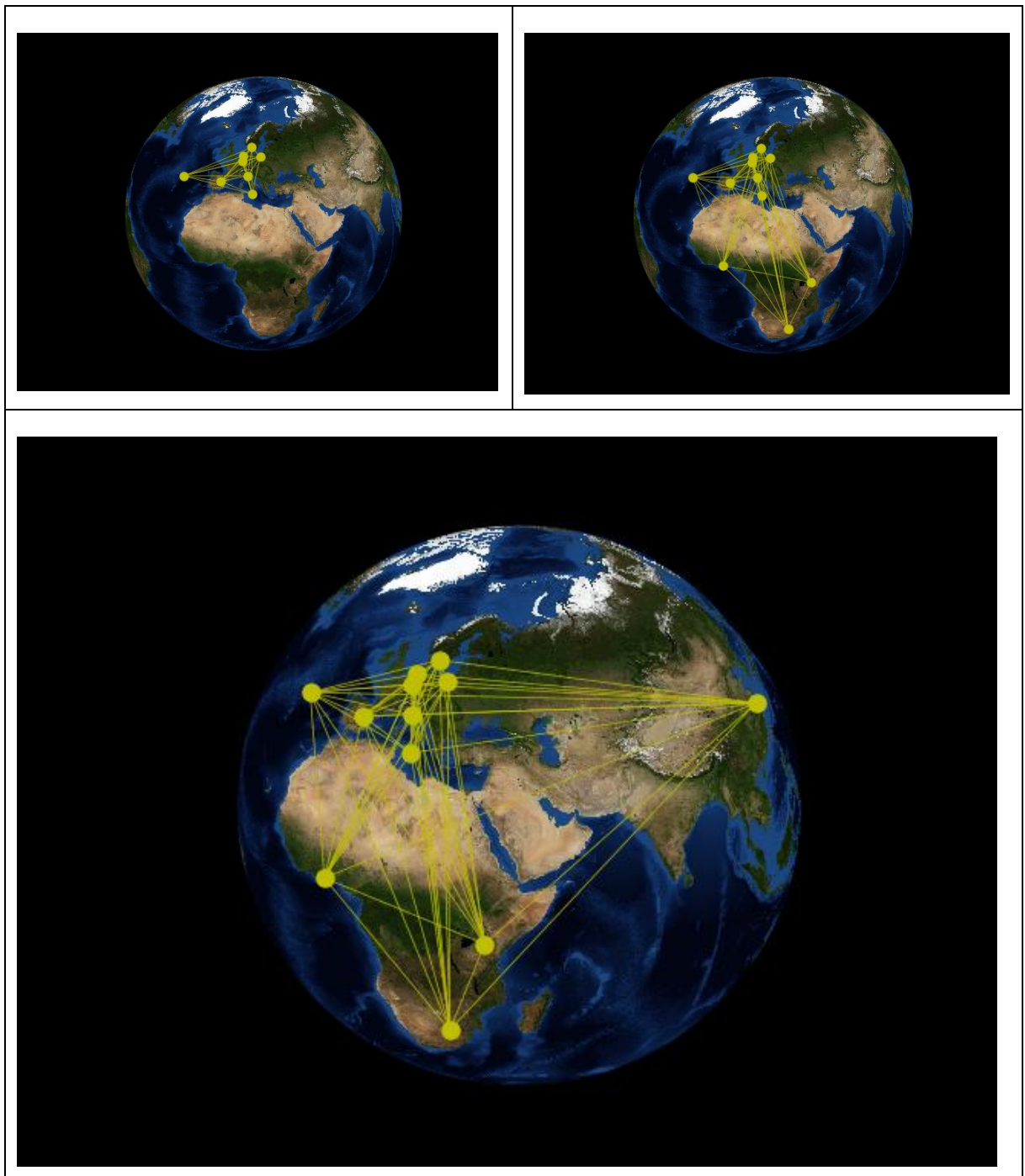

**Figure 5.** Baseline corresponding to the simulated UV plane considered in Figures 2-4.

## 6 CONCLUSIONS

Azores has a unique location across the Atlantic for space and radioastronomy infrastructure of world-class capabilities: its location covers the so called "Atlantic gap". An Astronomical VLBI capability in Azores added through the conversion of the former 32-metre SATCOM antenna in S. Miguel (or even the installation of a couple of MeerKAT/SKA1-MID dishes) would offer long and unique East-West visibilities to SKA and existing VLBI networks (AVN and EVN). This is an opportunity for cutting-edge research and for the development of a world-class infrastructure for radioastronomy and space explorations at a relatively modest investment.

Radioastronomy techniques enable ground-breaking studies of the widest variety of cosmic phenomena: VLBI is the only technique enabling the imaging of the horizon of Black Holes amplifying the impact of larger projects like SKA; it enables unsurpassed precision in the tracking and reception of planetary probes in the deep Solar system beyond Mars. On itself, the SKA-VLBI mode will significantly improve the connection between the celestial reference frames defined in the optical and radio bands and will have a profound effect on a large number of fields within astronomy.

The benefits from astronomical VLBI through the conversion of the STACCOM dish in S. Miguel are impactful:

- Continued use of very expensive installation that is now or is becoming redundant, at relatively lower cost, with insertion in international networks with access to funding programs
- Reinforcement of links to global radio astronomy networks through VLBI and international usage of a national facility
- In country training in practical radio astronomy with synergies/aligned with RAEGE/AIR goals:
- Single-dish research
- Very Long Baseline Interferometry (VLBI) for high angular resolution imaging and space
- Create a pool of astronomers and experimentalists able to use current and future large scale radio telescope arrays:
    - Spinoffs: reinforcement of Azores and Portuguese digital infrastructures
    - Opportunities for training, research and development in engineering and technology
    - low-noise microwave feeds and receivers
    - analogue and digital electronics
    - digital signal processing, Big Data software engineering
- Stimulate interest in science, engineering and technology through outreach program connected to very high impact science.
- Promote/reinforce the establishment / development of a national Space Agency and highly synergistic interactions with/within the Atlantic International Research Centre (AIR)

## ACKNOWLEDGMENTS

We warmly thank Francisco Colomer (JIVE - Joint Institute for VLBI in Europe ERIC) and Leonid Gurvitz (JIVE and Astrodynamics and Space Missions, TU Delft) for the long time and comprehensive support to the Azores astronomical VLBI cluster. We acknowledge support from CIDMA strategic project (UID/MAT/04106/2013), ENGAGE SKA (POCI-01-0145-FEDER-022217), and PHOBOS (POCI-01-0145-FEDER-029932), funded by COMPETE 2020 and FCT, Portugal.

## REFERENCES

[1] Gaylard, M.J., Bietenholz. M.F., Combrinck, L., Booth, R.S., S.J., Buchner, Fanaroff, B., MacLeod, G., Nicolson, G., Quick, J., Stronkhorst, P., Venkatasubramani, T.L., "An African VLBI network of radio telescopes", South African Institute of Physics 56th Annual Conference, 2014, arXiv:1405.7214

[2] Barbosa, D., Paulo, C., Ribeiro, V., Loots, A., Thondikulam, V.L., Gaylard, M., van Ardenne, A., Colafrancesco,S., Bergano, J., Amador, J., Maia, R., Melo, R., "Design, Environmental and Sustainability Constraints of new African Observatories: The example of the Mozambique Radio Astronomy Observatory", Proceedings of the URSI BEJ Session 'Large Scale Science Projects: Europa-Africa Connects', IEEE Africon 2013 Conference Mauritius (9-12 Sep) 2013, IEEE Xplorer, Nov 2013, arXiv:1311.4464

[3] McCulloch, P. M., et al. 2005, ApJ, 129, 2034

[4] Woodburn, L., Natusch, T., Weston, S., Thomasson,P., Godwin, M., Granet, C., and Gulyaev, S., "Conversion of a New Zealand 30-Metre Telecommunications Antenna into a Radio Telescope", Publications of the Astronomical Society of Australia (PASA), Vol. 32, e017, 14 pages (2015), doi:10.1017/pasa.2015.13

[5] Heywood, I. 2011, Expanding e-MERLIN with the Goonhilly Earth Station

[6] DeBoer, D. R., & Steffes, P. G. 1999, RaSc, 34, 991

[7] Fujisawa, K., Mashiyama, H., Shimoikura, T., & Kawaguchi, N., 2002, in Proc. of the IAU, 8th Asian-Pacific Regional Meeting, Vol. II, ed. S. Ikeuchi, J. Hearnshaw, & T. Hanawa (Tokyo, Japan), 3

[8] Gabuzda, D., Golden, A., & ARTI Consortium 2005, in ASP Conf. Proc., Vol. 340, Future Directions in High Resolution Astronomy, ed. J. Romney & M. Reid (San Francisco: ASP), 566

[9] Gaylard, M. J., et al. 2012, in Proc. of SAIP2011, ed. I. Basson & B. A. E. (Univ. of South Africa, Pretoria), 473

[10] Nordling, L. 2012, Nature, 488, 571

[11] Perks, S. 2012, PhyW, 25, 9

[12] Paragi, Z., et al., "Very Long Baseline Interferometry with the SKA", Proceedings of Science, "Advancing Astrophysics with the Square Kilometre Array", June 8-13, 2014, Giardini Naxos, Italy, PoS(AASKA14)143

[13] Duev, D. et al.., "Planetary Radio Interferometry and Doppler Experiment (PRIDE) technique: A test case of the Mars Express Phobos fly-by", A&A 593, A34 (2016)

[14] Bocanegra-Bahamón, T.M., et al., "Planetary Radio Interferometry and Doppler Experiment (PRIDE) Technique: a Test Case of the Mars Express Phobos Fly-by. 2. Doppler tracking: Formulation of observed and computed values, and noise budget", A&A 609, A59 (2018)

[15] Dirkx, D, Gurvits, L., Laineyc, V., Larid, G., Milani, A., Cimòa,G., Bocanegra-Bahamon, T.M., "On the contribution of PRIDE-JUICE to Jovian system ephemerides", Planetary and Space Science, Volume 147, 1 November 2017, Pages 14-27 ISBN: 978-989-20-6191-7

[16] Proceedings of the 22nd European VLBI Group for Geodesy and Astrometry Working Meeting, edited by R. Hass and F. Colomer, 17-21 May, 2015, ISBN: 978-989-20-6191-7

[17] J. Gomez–González, L. Santos, J. A. López, Fernandez, F. Colomer, "Status of the Spanish-Portuguese RAEGE Project", Proceedings of the 22nd European VLBI Group for Geodesy and Astrometry Working Meeting, edited by R. Hass and F. Colomer, 17-21 May, 2015, ISBN: 978-989-20-6191-7

[18] "RAEGE: An Atlantic Network of Geodynamical Fundamental Stations", IVS 2010, General Meeting Proceedings, p.101–105, 2010

[19] Yajima, M., Tsuchikawa, K., Murakami, T., Katsumoto, K., Takano, T., "Study of a Bistatic Radar System Using VLBI Technologies for Detecting Space Debris and the Experimental Verification of its Validity", Earth, Moon, and Planets, April 2007, Volume 100, Issue 1–2, pp 57–76

[20] Nechaeva, M., Antipenko, A., Bezrukovs, V., Bezrukov, D., Dementjev, A., Dugin, N, A experiment on radio location of objects in the near-Earth space with VLBI in 2012, Baltic Astronomy, Vol. 22, p. 35-41

[21] Mahdi, M.C., Study the Space Debris Impact in the Early Stages of the Nano-Satellite Design, ARTIFICIAL SATELLITES, Vol. 5, No. 4 – 2016, DOI: 10.1515/arsa-2016-0014

[22] J. Serrano, P. Alvarez, M. Cattin, E. G. Cota, P. M. J. H. Lewis, T. Włostowski et al., "The White Rabbit Project", in Proceedings of ICALEPCS TUC004, Kobe, Japan, 2009.

[23] Namneet Kaur, Florian Frank, Paul-Eric Pottie, Philip Tuckey, "Time and frequency transfer over a 500 km cascaded White Rabbit network", 2017 Joint Conference of the European Frequency and Time Forum and IEEE International Frequency Control Symposium (IFTF/IFCS), Besançon, France, 2017

[24] Grange , K., Alachkar, B., Amy, S., Barbosa, D., et al., "Square Kilometre Array: the radio telescope of the XXI century", Astronomy reports, Volume 61, Issue 4, 1 April 2017, Pages 288-296

[25] Florianopolis Declaration, Towards AIR Centre, http://www.atlanticinteractions.org/documents/

[26] "Atlantic Interactions White paper - A Science and Technology Agenda, for an integrative approach to the Atlantic: Integrating Space, Climate, Oceans and Data Sciences through North-South / South-North

Cooperation", Towards the Atlantic International Research Center (AIR Center), FCT, July 2017, available at: http://www.atlanticinteractions.org/wp-content/uploads/2017/07/AIR-white_paper-July-2017_VF_BOM.pdf

[27] Atlantic_Interactions_Book, A PROCESS OF SCIENTIFIC DIPLOMACY, Integrating Space, Climate, Oceans and Data Sciences through North-South / South-North Cooperation, Towards the Atlantic International Research Centre (AIR Centre), 3rd High-level Industry-Science-Government Dialogue Interactions, Praia, Cape Verde, May 2018, available at http://www.atlanticinteractions.org/wp-content/uploads/2018/05/Atlantic_Interactions_Book_2_May_2018_Web.pdf

# Radioastronomy and Space Science in Azores: Enhancing the Atlantic VLBI Infrastructure cluster; II- Space synergies

Domingos Barbosa[1], Sonia Anton[1,2], Miguel Bergano[1], Tjarda Boekholt[1], Bruno Coelho[1], Alexandre Correia[5], Dalmiro Maia[3], Valério Ribeiro[1,2], Jason Adams[3,6]

[1] Instituto de Telecomunicações, 3810-193 Aveiro, Portugal;
[2] CIDMA, Department of Physics, University of Aveiro, 3810-193 Aveiro, Portugal;
[3] University of Aveiro, 3810-193 Aveiro, Portugal;
[4] CICGE, Faculdade de Ciências da Universidade do Porto, Porto, Portugal;
[5] CFisUC, Department of Physics, University of Coimbra, 3004-516 Coimbra, Portugal
[6] AMSL Semiconductors, Bladel, the Netherlands

## ABSTRACT

Radioastronomy and Space Infrastructure in the Azores have a great scientific and industrial interest because they benefit from a UNIQUE geographical location (longitude and latitude) in the middle of the North Atlantic allowing a vast improvement in the sky coverage. This fact obviously has a very high added value for: i) the establishment of space tracking and communications networks for the emergent global small satellite fleets ii) it is invaluable to connect the scientific infrastructure networks in Africa, Europe and USA using Very Large Baseline Interferometry (VLBI) techniques, iii) it allows excellent potential for monitoring Space debris and Near Earth Objects (NEOs). There are putative infrastructures in the region like the large 32-m SATCOM antenna in S. Miguel Island that could integrate advanced VLBI networks and be capable of additional Deep Space Network ground support. Such infrastructure is an opportunity for a world-class infrastructure for radioastronomy and space exploration and would constitute (a) a key technological facility for data production, promoting local digital infrastructure investments and the testing of cutting-edge Information technologies; (b) would enable participation in space debris monitoring experiments up to Geostationary Objects (GEO), NEOs monitoring experiments and (c) would greatly enhance recent VLBI infrastructure developments like the RAEGE project d) would complement the near future 15.5-metre FCT antenna for Low Orbit Objects (LEO) tracking by enabling space tracking for Deep Space missions; **This paper II explores the space science synergies offered by the upgrade of S. Miguel 32-metre SATCOM antenna.**

**Keywords:** Radioastronomy, VLBI, geodesy, Space Science, Space debris, Space tracking, Atlantic connections, global change

## 1 INTRODUCTION:

Azores has a unique location across the Atlantic: spread over 600Km, the archipelagos has islands dispersed across three tectonic plates: American, European and African plates. In fact, the plates junctions pass through its central group (Islands of Faial, Pico, S. Jorge, Terceira and Graciosa). In the 1990s years of the XX century, NASA used the Azores as an important point for the International Reference System to improve the spatial navigation with GPS. Furthermore, since the early 2000's there has been considerable interest for a station in the Azores for science and innovation to provide a unique radioastronomical and space navigation facility.

Azores is now developing a new infrastructure: Atlantic International Research Center (AIR), which is an international platform integrating research and innovation on multi-related areas such as climate change, earth observation, energy, space and oceans. The Florianopolis Declaration [26] led to the confirmation of this new Atlantic interaction level [25, 26, 27] which will aggregate complementary aspects of a new critical infrastructure towards high impact space science.

Radioastronomy and space navigation are also areas that can take advantage of the unique location of Azores, and projects on either the installation of new stations in St. Maria or Flores islands or the update and retrofit of a space communications (SATCOM) dish in S. Miguel, have been proposed. One of the projects that has already began is the 13.2 metre radiotelescope with VLBI capabilities of the Rede Atlântica de Estações Geo-Espaciais

(RAEGE), deployed at St. Maria island (Azores Western group), a project led by the IGN-Spain and the Region of Azores [16] and discussed in paper I on the Azores radioastronomical potential.

## 2   EXISTING INFRASTRUCTURES IN AZORES

In paper I, we discussed current plans for Azores infrastructures, including the RAEGE developments. In this paper (paper II) we will focus on the space capabilities and support to Deep Space network or space tracking activities. See Table 1, from previous Paper I.

Furthermore, the Portuguese Science and Technology Foundation (FCT) has commissioned the deployment of a 15-metre station for space communication in Azores St. Maria Island. This will potentially enhance downlink coverage in the Atlantic for Leo satellites, in particular for ESA support. However, support to Deep Space missions to very distant targets in the Solar system may require larger antennas, of the class 25-32 metres. Such large dish infrastructure already exists in S. Miguel[1] for availability, that were dedicated previously to SATCOM activities, and are currently redundant.

With part of the Square Kilometre Array (SKA) in Africa, there is now a considerable interest in radio astronomy across the African continent with the launch of the Africa VLNI network (AVN). AVN is accelerating the constitution of radioastronomical and space science communities across Africa by leveraging the conversion of former SATCOM dishes in Africa. In fact, there are about 29 documented 30-m class telecommunications antennas in 19 African countries that could benefit AVN. Ghana just saw the first successful African example of an Earth station in Kutunse transformed into a world class radioastronomical and space facility [9,10,11], with Zambia and Kenya to follow.

A very interesting example is the transformation of the famous and former British Telecom SATCOM facilities at the Goonhilly Earth Station (GES) in Cornwall, UK [5]. GES[2], currently a private company, provides Deep Space tracking and ground support servicing to the UK Space Port Cornwall activities in Newquay , Southwest England and its main activities  include space navigation and telemetry support to near future cubesat constellations in LEO orbits and support to private space operators like Virgin Orbit. GES is now ready to perform radioastronomy observations with a conversion of a 25-metre dish to radiotelescope in collaboration with a British University Consortium.

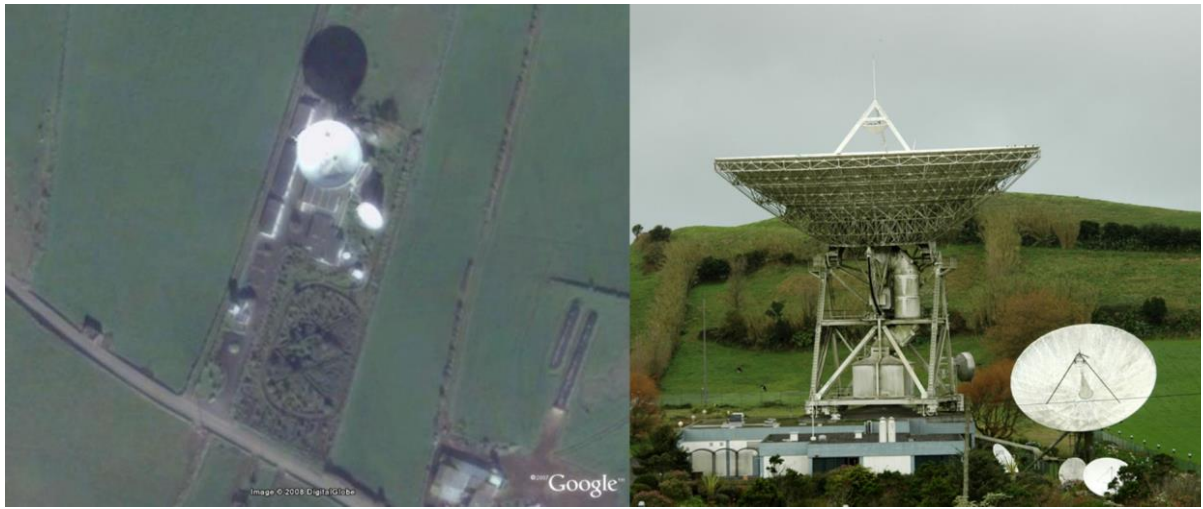

**Figure 1**.  Earth Stations in São Miguel (Azores) – Intelsat Standard A 32-metre parabolic antenna. We can see the waveguide system below the dish.

---

[1] 32-metre Earth station from former Marconi/PT Comunicações, currently at Altice Portugal.

[2] Goonhilly Earth Station - Gateway to SPACE - http://www.goonhilly.org/

# 3 FROM SPACE TRACKING TO SPACE PORTS

Recent developments in Azores target the emergent global small satellite markets that are rapidly growing, especially in the nano and micro-satellites mass ranges. The rapidly decreasing costs of such small satellites coupled to the ever increasing capabilities enabled by many COTS developments and availability of state-of-art technologies make the new generation of satellites more performant than previous more massive generations at a fraction of their cost. As such, small satellites constellations or swarms become an interesting concept enabling innovative missions that were not previously possible, and requiring as well innovative infrastructures to provide ground support. Those range from LEO cubesats for Earth observations (climate change, ocean monitoring, etc) to science missions like the Chinese Chang'e-4 mission to the far side of the Moon that will include a pair of microsatellites to be placed in lunar orbit to test low frequency radio astronomy and space-based interferometry [30].

Notably, "small sats" (cubesats) swarms in LEO orbits are commercially and "environmentally" sustainable: high atmosphere drag self-cleans constellations with time and they do not contribute for the long term space debris field affecting higher orbits (they simply fall down and disintegrate with time). Their number is also growing very fast and the number of expanding large constellations does indeed represent a data challenge in terms of the aggregated data transmission and in orbit occupancy.

Typically, smaller and fast tracking stations are the prime choice for low frequency-band communications. However, these smaller stations do not possess the sensitivity for ground support to missions requiring Deep Space Network capabilities because these missions are very faraway or they rely on communications using higher frequency bands (from C, X-band or even Ka bands). Fortunately, Azores also possess a decommissioned 32-metre Intelsat Standard-A SATCOM in S. Miguel island that can cover the Deep Space segment and open Azores to the high impact contributions of Deep Space missions, enhancing the AIR footprint to potential collaborations with major space agencies (NASA, ESA).

Deep-space missions largely rely on the use of radio tracking for their orbit determination and the associated parameter estimation, using Doppler data (closed loop) obtained by, for example, NASA's Deep Space Network (DSN) and ESA's TRACKing station network (ESTRACK). VLBI measurements using the Planetary Radio Interferometry and Doppler Experiment (PRIDE) technique (open loop) have been used on a large number of past and current planetary missions to provide unsurpassed precision about spacecraft accurate angular positions in the sky, ie spacecraft lateral position (right ascension α and declination δ) measurements with an uncertainty of approximately 1.0 nrad (~200 μarcsec) or about 50 metre precision at a distance of 1.4 AU. PRIDE techniques are used by the large radio telescopes from the European VLBI Network (EVN, EurAsia) and the Very Large Baseline Array (VLBA, USA) in collaboration with major space agencies.

PRIDE does not require special capabilities from the mission's on-board instrumentation and it can be applied to almost any radio signal from a spacecraft, provided the spacecraft signal is minimally powerful and phase-stable. In short, PRIDE techniques exploits the radio (re-)transmitting capabilities of spacecraft from the most modern space science missions and can become the core of mission external experiments. Additionally, DSN/ESTRACK and VLBI networks are well known for their long standing close technological collaboration sharing much of the design concepts for equipment and data acquisition software. Although, the characteristics and capabilities of the receivers may be similar, the post-processing techniques have different software processing pipeline customizations.

PRIDE/VLBI have been used as a multi-purpose, multi-disciplinary enhancement of planetary missions science return on a large number of past and current planetary missions : VEGA Venus atmosphere balloons, Ulysses solar orbiter [35], the Huygens Probe during its descent to the surface of Saturn's moon Titan, from the Cassini-Huygens mission and the VLBI tracking of the Cassini spacecraft at Saturn [36,37,38], Chang'E−1 flight to the Moon [39], VLBI tracking with the European VLBI Network (EVN) antennas of the controlled impact of ESA's Smart-1 probe on the surface of the Moon [40], VLBI tracking of NASA's Mars Exploration Rover B spacecraft during its final cruise phase [41], VLBI tracking of the solar sail mission IKAROS [42], ESA's Venus EXpress (VEX) VLBI spacecraft observations [43], the ESA Mars Express (MEX) Phobos-flyby [32,44, 45] monitored by the EVN radio telescopes and the SKA precursor Murchinson Radio Observatory (MRO), and NASA's Juno flyby of Jupiter [38].

The information provided with PRIDE on the ultra-precise spacecraft state vectors can be used for a variety of scientific applications including planetary science (measurements of tidal deformations of planetary moons, atmosphere dynamics and climatology as well as seismology, tectonics, internal structure and composition of planetary bodies), ultra-precise celestial mechanics of planetary systems, gravimetry and fundamental physics (e.g. tests of general relativity or study of the anomalous accelerations of deep-space probes like the study on the Pioneer Anomaly [47]), interplanetary scintillations [32], Spacecraft Doppler tracking as the only possible way of detecting low-frequency (10−5−1 Hz) gravitational waves [48]. PRIDE can also provide diagnostics of deep-space missions ("health check") and direct-to-Earth delivery of a limited amount of critical data (e.g. Friedman et al. 2008) and PRIDE enabled radio science experiment on the interaction of the spacecraft signal with planetary bodies and interplanetary media as it propagates from the spacecraft to Earth [49-56].

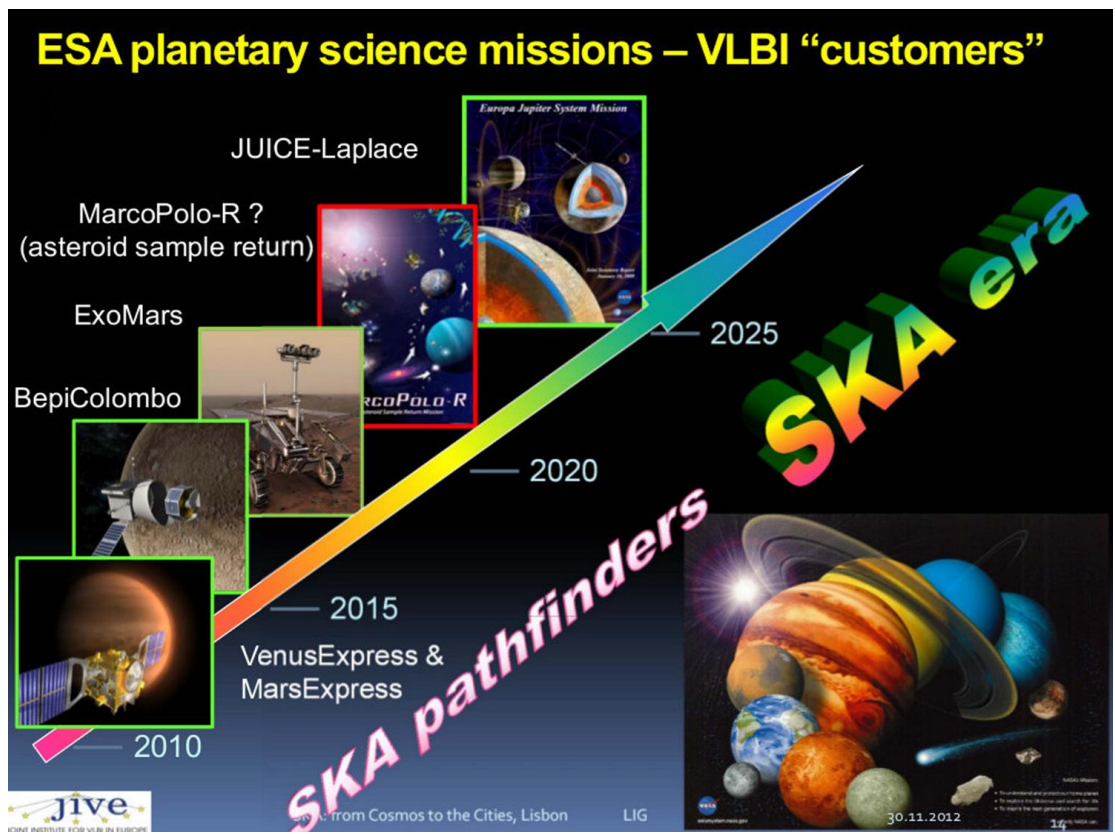

**Figure 2.** ESA Planetary Science missions and VLBI Customers; From L. Gurvits, EU-AFR ICT 8Th Partnership side event, Lisbon 2012.

Furthermore, PRIDE was also included in a number of other space mission Design studies like European Venus Explorer (EVE) [56], Titan and Enceladus Mission (TandEM) [57], Kronos, to explore the depths of Saturn with probes and remote sensing [58]. Therefore, it is not surprising that Deep Space exploration with VLBI support is actively recognized by ESA**: it is worth noting PRIDE was selected by ESA as one of the experiments of its L-class JUpiter ICy moons Explorer mission (JUICE) [59], enhancing radio science synergies and addressing the quest of Icy Moons interior and Jovian system ephemerides.**

Other notable example is Chang'e program – the Chinese Lunar Exploration Program [60, 61]: this long term program develops a range of lunar orbiters, landers, rovers and sample return spacecraft, and the dedicated ground segment infrastructure that includes the Chinese VLBI network; Although not using VLBI, the descent of EXOMARS mission Landing module Schiaparelli telemetry was followed by the Giant Metrewave Radio Telescope (GMRT) [62], also an SKA Pathfinder located near Pune, India, and operated by the National Centre for Radio Astrophysics, part of the Tata Institute of Fundamental Research. GMRT comprises an array of 30 radio telescopes, each with a dish diameter of 25 m, and it is one of the world's largest interferometric arrays.

These examples just show the multipurpose character, low cost and capabilities of modern radiotelescopes and radiointerferometres to provide advanced ground segment support to Deep Space Planetary exploration missions and enhance the scientific retour trough important radio science experiment with very high science impact. Additionally, programs like SKA, with their very wide band frequency coverage, do provide a technology and science development path. **These examples just emphasize the importance and the scientific enhancements provided by VLBI based techniques to planetary missions.**

# 4  FUTURE TELEMETRY CONSIDERATIONS WITH SKA1 AND RADIOASTRONOMY NETWORKS

Of course, besides spatial resolution, telemetry considerations will much benefit from the enhanced arrays. The near future SKA availability show SKA 1 sensitivity for telemetry may dramatically change the landscape of deep Solar System exploration telemetry support for specific missions.

The SKA1 baseline design covers several bands widely in space exploration. In particular for SKA-MID :

- Band 1 covers VHF/UHF frequencies, used by planetary atmospheric probes;
- Band 3 covers standard S-band (2.3 GHz), used for deep space downlink frequency allocation.
- Band 5 covers standard X-band (8.4GHz), used for deep space downlink allocation.

As in Jones and Lazio [73] *"the assumptions of a 10X increase in sensitivity and 20X increase in angular resolution for SKA result in a truly unique and spectacular future spacecraft tracking capability"*. This gain in sensitivity may allow for more complex space probing experiments enhancing the scientific exploration and planetary resource mapping (see Figure 3). This sensitivity leads to two transformational capabilities for future deep space missions Schutte, [28].:

- SKA1-Mid will be able to increase the data volume returned to earth by more than an order of magnitude for traditional deep space missions enabling live HD video from beyond the orbit of Saturn.
- SKA1-Mid sensitivity is such that small interplanetary descent and landing probes could be tracked without requiring relay orbiter. This would enable an entirely new class of low-cost in-situ missions (lower launch mass).
- Coupling SKA-MID telemetry capabilities with VLBI super-resolution could become a game change in spacecraft tracking, by providing "GPS" accuracy in deep space navigation.

Of course, SKA will most likely be a very highly demanded telescope, and therefore pending discussions over time availability will require future high level agreements between the SKA Observatory and Space Agencies. As a token, the range of radio science and data transmission technologies are just the pathfinders of data and tracking technologies to be used routinely in the new space economy targeting space resource exploration (asteroid mining for instance or Earth resources) whose technological and methodological synergies with ocean exploration are great.

# 5  SPACE SECURITY AND ASSET PROTECTION: SPACE DEBRIS SURVEYING

Securing space assets, in particular the emergent satellite constellations is of major economic and strategic importance and considered along three main vectors of what is best known as Space Situational Awareness (SSA) programs: Space Weather phenomena, Near-Earth Objects and Space debris. Accurate measurements of the position and trajectory of the space debris fragments is indeed of primary importance for the characterization of the orbital debris environment The main techniques for monitoring space debris require a combination of optical (telescopes) and radar detection, requiring in the case of radar large-size instrument infrastructures. Thus, there are plenty of opportunities for new large radiotelescopes with radar capabilities, either as monostatic radars equipped with emitters in high frequency (X or Ka band) or as part of bi-static radar network using large radiotelescopes to receive the echo or backscattering of radar pulses emitted by one or two large antennas emitters.

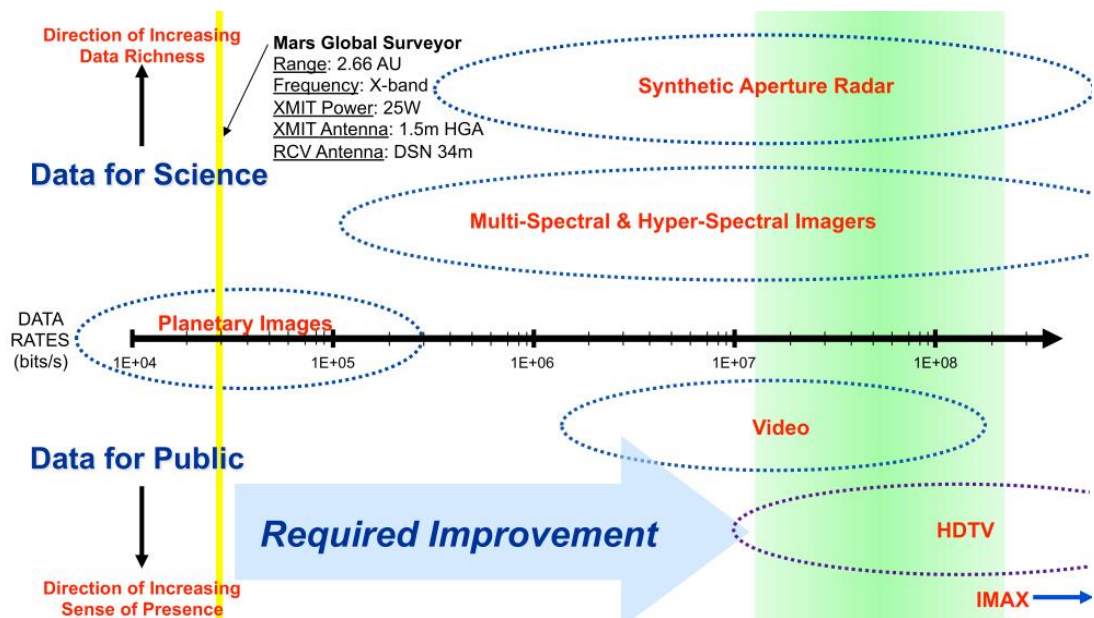

**Figure 3.** Different types of observation possible with different data rates. The Voyager data rates were were ~10 kb/s, while observing other planets as we do Earth requires data rates > 10 Mb/s. From Jones & Lazio, arXiv:1412.6006v1; PoScience, 2014.

Of particular interests for the SSA is the European Space Survey and Tracking (SST) segment program fostering the creation of a network infrastructure of new sensors (optical, radio) capable of monitoring accurately the space debris field affecting commercially interesting orbits. Radiotelescopes and radio interferometers are examples of ground based passive radio sensors capable of providing important ancillary debris information if they are set in the appropriate configuration required for this.

There have been considerable interests in promoting radar sensors in Portugal mainland and the Azores (in particular at its Eastern group) since both Portugal mainland or the Azores could operate providing an interesting connecting between sensors like the Haystack Auxiliary Radar, operated by the Massachusetts Institute of Technology Lincoln Laboratory (MIT LL) for the NASA Orbital Debris Program Office (ODPO) and the European sensors.

Larger antennas could also be used for piggy-back space debris in GEO orbits through in bistatic or multistatic surveying configurations. In 2001 the Medicina parabolic antenna was also used together with the Goldstone (USA) and the Evpatoria (Ukraine) transmitters to perform observations of Near-Earth Asteroid 1998 WT24 [63]. In February 2008 the malfunction of the American military satellite USA-193 gave the Northern Cross the opportunity to be tested for the first time as a receiving element of a space surveillance bistatic radar system [64].

Therefore, the large S. Miguel SATCOM dish could also contribute if appropriately configured with piggy back observations or monitoring of space debris in geostationary orbits, which due to distance are more difficult to observe after receiving echos of a faraway transmitter in a bistatic configuration. The bistatic radar capabilities of the VLBI radiotelescopes in space debris detection and tracking is unsurpassed in combination with facilities in other radio bands.

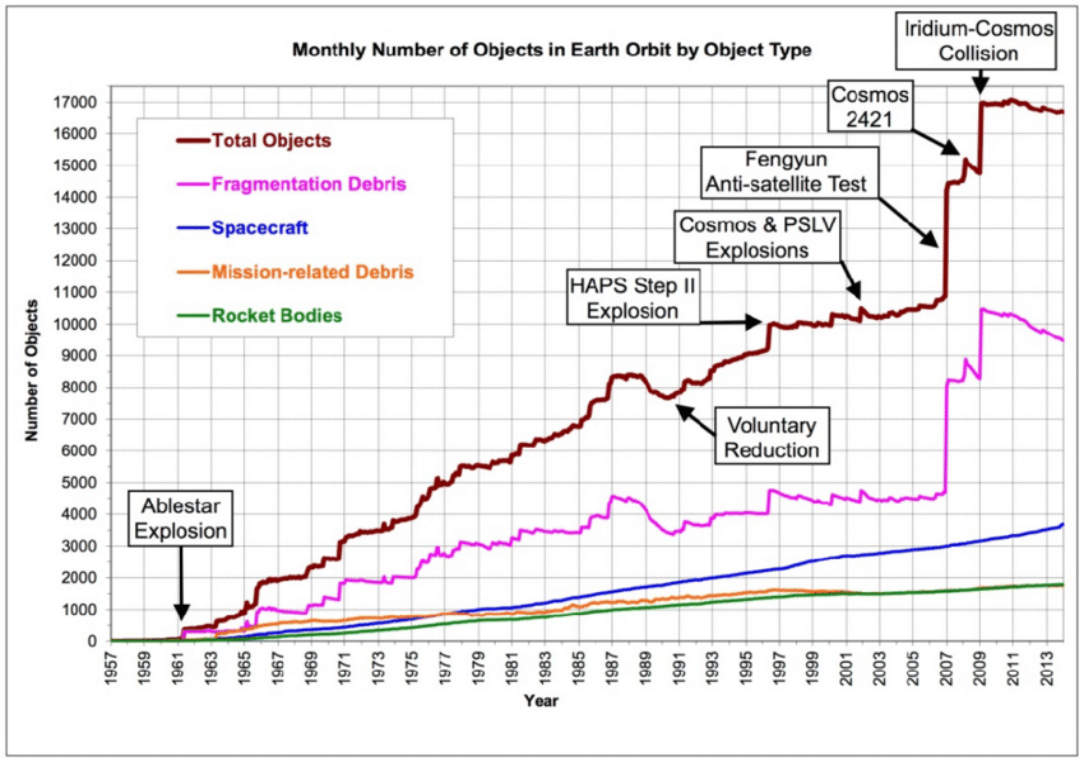

**Figure 4 .** Summary of all objects in Earth orbit officially catalogued by the U.S. Space Surveillance Network as a function of time.

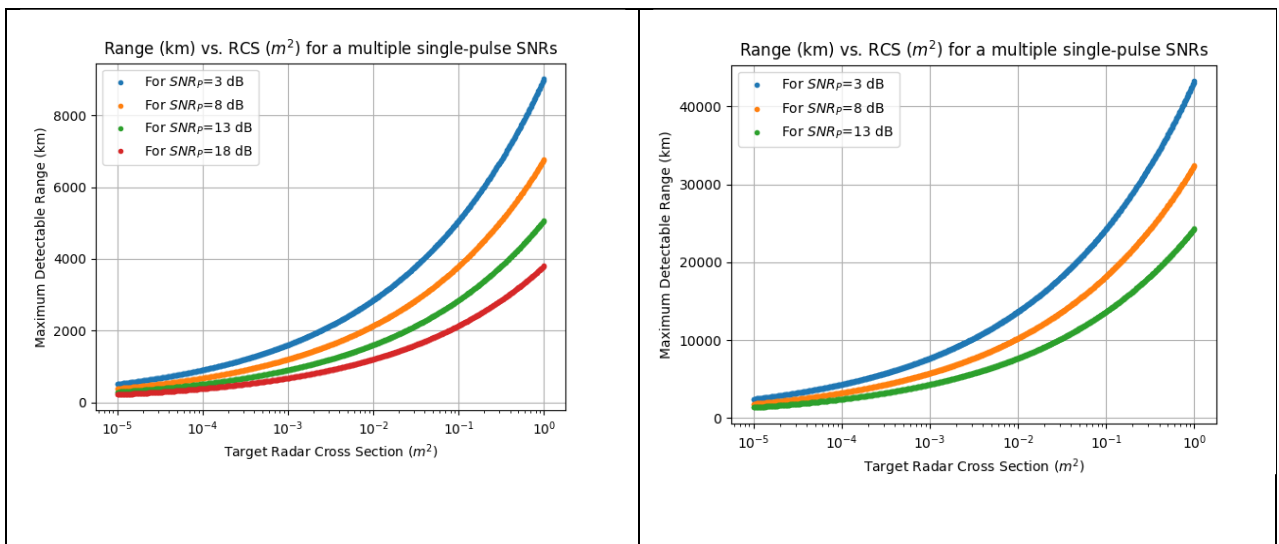

**Figure 5 .** Signal-to-Noise ratio in dB versus target range in kilometers for multiple target radar cross sections. Left: monostatic RAEGE configuration; Right: piggy back bistatic using S. Miguel SATCOM – it can detect GEO debris. From [72].

**CONCLUSIONS**

As already mentioned in paper I, Azores has a unique location across the Atlantic for space and radioastronomy infrastructure of world-class capabilities: its location covers the so called "Atlantic gap". The conversion of former SATCOM antennas into operational radio telescopes with capabilities for space sciences applications would also offer unique opportunities for cutting-edge research and a world-class infrastructure for radioastronomy and space exploration and open capabilities for Deep Space support at a relatively modest investment.

Portugal can benefit the on-going space science investments either through conversion of S. Miguel SATCOM or even through the installation of MeerKAT/SKA-MID dishes into a networked radio telescope connected to large distributed infrastructures. Azores can greatly enhance its VLBI cluster with a large dish that can benefit the international scientific community and could be strategic to: i) provide added value services to Space navigation; ii) become a sensor contributing to the Portuguese Space Survey and Tracking (SST) program for space debris monitoring up to GEO orbits; iii) contribute to the provision of "stellar GPS" accuracy to space tracking and telemetry of probes sent to the outer Solar system, from landing modules in Mars or missions to Jupiter like the JUICE spacecraft.

Such an asset would complement and enhance the recent infrastructure developments like the RAEGE project and the near future 15-metre FCT antenna deployed for Proba-3 and future Low Orbit Objects (LEO) tracking. The large STACOM availability is a great asset to enhance the investments currently being considered for the establishment of the Atlantic International Research Centre (AIR Centre) and the newly created Portugal Space Agency.

## ACKNOWLEDGMENTS

We warmly thank Francisco Colomer (JIVE - Joint Institute for VLBI in Europe ERIC) and Leonid Gurvitz (JIVE and Astrodynamics and Space Missions, TU Delft) for the full and comprehensive support about the international interest on the Azores VLBI cluster. We acknowledge support from CIDMA strategic project (UID/MAT/04106/2013), ENGAGE SKA (POCI-01-0145-FEDER-022217), and PHOBOS (POCI-01-0145-FEDER-029932), funded by COMPETE 2020 and FCT, Portugal.

## REFERENCES

[1] Gaylard, M.J., Bietenholz. M.F., Combrinck, L., Booth, R.S., S.J., Buchner, Fanaroff, B., MacLeod, G., Nicolson, G., Quick, J., Stronkhorst, P., Venkatasubramani, T.L., "An African VLBI network of radio telescopes", South African Institute of Physics 56th Annual Conference, 2014, arXiv:1405.7214

[2] Barbosa, D., Paulo, C., Ribeiro, V., Loots, A., Thondikulam, V.L., Gaylard, M., van Ardenne, A., Colafrancesco,S., Bergano, J., Amador, J., Maia, R., Melo, R., "Design, Environmental and Sustainability Constraints of new African Observatories: The example of the Mozambique Radio Astronomy Observatory", Proceedings of the URSI BEJ Session 'Large Scale Science Projects: Europa-Africa Connects', IEEE Africon 2013 Conference Mauritius (9-12 Sep) 2013, IEEE Xplorer, Nov 2013, arXiv:1311.4464

[3] McCulloch, P. M., et al. 2005, ApJ, 129, 2034

[4] Woodburn, L., Natusch, T., Weston, S., Thomasson,P., Godwin, M., Granet, C., and Gulyaev, S., "Conversion of a New Zealand 30-Metre Telecommunications Antenna into a Radio Telescope", Publications of the Astronomical Society of Australia (PASA), Vol. 32, e017, 14 pages (2015), doi:10.1017/pasa.2015.13

[5] Heywood, I. 2011, Expanding e-MERLIN with the Goonhilly Earth Station

[6] DeBoer, D. R., & Steffes, P. G. 1999, RaSc, 34, 991

[7] Fujisawa, K., Mashiyama, H., Shimoikura, T., & Kawaguchi, N., 2002, in Proc. of the IAU, 8th Asian-Pacific Regional Meeting, Vol. II, ed. S. Ikeuchi, J. Hearnshaw, & T. Hanawa (Tokyo, Japan), 3

[8] Gabuzda, D., Golden, A., & ARTI Consortium 2005, in ASP Conf. Proc., Vol. 340, Future Directions in High Resolution Astronomy, ed. J. Romney & M. Reid (San Francisco: ASP), 566

[9] Gaylard, M. J., et al. 2012, in Proc. of SAIP2011, ed. I. Basson & B. A. E. (Univ. of South Africa, Pretoria), 473

[10] Nordling, L. 2012, Nature, 488, 571

[11] Perks, S. 2012, PhyW, 25, 9

[12] Paragi, Z., et al., "Very Long Baseline Interferometry with the SKA", Proceedings of Science, "Advancing Astrophysics with the Square Kilometre Array", June 8-13, 2014, Giardini Naxos, Italy, PoS(AASKA14)143

[13] Duev, D. et al.., "Planetary Radio Interferometry and Doppler Experiment (PRIDE) technique: A test case of the Mars Express Phobos fly-by", A&A 593, A34 (2016)

[14] Bocanegra-Bahamón, T.M., et al., "Planetary Radio Interferometry and Doppler Experiment (PRIDE) Technique: a Test Case of the Mars Express Phobos Fly-by. 2. Doppler tracking: Formulation of observed and computed values, and noise budget", A&A 609, A59 (2018)

[15] Dirkx, D, Gurvits, L., Laineyc, V., Larid, G., Milani, A., Cimòa,G., Bocanegra-Bahamon, T.M., "On the contribution of PRIDE-JUICE to Jovian system ephemerides", Planetary and Space Science, Volume 147, 1 November 2017, Pages 14-27 ISBN: 978-989-20-6191-7

[16] Proceedings of the 22nd European VLBI Group for Geodesy and Astrometry Working Meeting, edited by R. Hass and F. Colomer, 17-21 May, 2015, ISBN: 978-989-20-6191-7

[17] J. Gomez–González, L. Santos, J. A. López, Fernandez, F. Colomer, "Status of the Spanish-Portuguese RAEGE Project", Proceedings of the 22nd European VLBI Group for Geodesy and Astrometry Working Meeting, edited by R. Hass and F. Colomer, 17-21 May, 2015, ISBN: 978-989-20-6191-7

[18] "RAEGE: An Atlantic Network of Geodynamical Fundamental Stations", IVS 2010, General Meeting Proceedings, p.101–105, 2010

[19] Yajima, M., Tsuchikawa, K., Murakami, T., Katsumoto, K., Takano, T., "Study of a Bistatic Radar System Using VLBI Technologies for Detecting Space Debris and the Experimental Verification of its Validity", Earth, Moon, and Planets, April 2007, Volume 100, Issue 1–2, pp 57–76

[20] Nechaeva, M., Antipenko, A., Bezrukovs, V., Bezrukov, D., Dementjev, A., Dugin, N, A experiment on radio location of objects in the near-Earth space with VLBI in 2012, Baltic Astronomy, Vol. 22, p. 35-41

[21] Mahdi, M.C., Study the Space Debris Impact in the Early Stages of the Nano-Satellite Design, ARTIFICIAL SATELLITES, Vol. 5, No. 4 – 2016, DOI: 10.1515/arsa-2016-0014

[22] J. Serrano, P. Alvarez, M. Cattin, E. G. Cota, P. M. J. H. Lewis, T. Włostowski et al., "The White Rabbit Project", in Proceedings of ICALEPCS TUC004, Kobe, Japan, 2009.

[23] Namneet Kaur, Florian Frank, Paul-Eric Pottie, Philip Tuckey, "Time and frequency transfer over a 500 km cascaded White Rabbit network", 2017 Joint Conference of the European Frequency and Time Forum and IEEE International Frequency Control Symposium (IFTF/IFCS), Besançon, France, 2017

[24] Grange , K., Alachkar, B., Amy, S., Barbosa, D., et al., "Square Kilometre Array: the radio telescope of the XXI century", Astronomy reports, Volume 61, Issue 4, 1 April 2017, Pages 288-296

[25] Florianopolis Declaration, Towards AIR Centre, http://www.atlanticinteractions.org/documents/

[26] "Atlantic Interactions White paper - A Science and Technology Agenda, for an integrative approach to the Atlantic: Integrating Space, Climate, Oceans and Data Sciences through North-South / South-North Cooperation", Towards the Atlantic International Research Center (AIR Center), FCT, July 2017, available at: http://www.atlanticinteractions.org/wp-content/uploads/2017/07/AIR-white_paper-July-2017_VF_BOM.pdf

[27] Atlantic_Interactions_Book, A PROCESS OF SCIENTIFIC DIPLOMACY, Integrating Space, Climate, Oceans and Data Sciences through North-South / South-North Cooperation, Towards the Atlantic International Research Centre (AIR Centre), 3rd High-level Industry-Science-Government Dialogue Interactions, Praia, Cape Verde, May 2018, available at http://www.atlanticinteractions.org/wp-content/uploads/2018/05/Atlantic_Interactions_Book_2_May_2018_Web.pdf

[28] Schutte, A., "The Square Kilometre Array Radio Telescope: Transformational Capabilities for Deep Space Missions", Reinventing Space Conference 2016, BIS-RS-2016-15, 2016

[29] P L.I. Gurvits, T.M. Bocanegra Bahamon, G. Cimò, D.A. Duev, G. Molera Calvés, S.V. Pogrebenko, I. de Pater, L.L.A. Vermeersen, P. Rosenblatt, J. Oberst, P. Charlot, S. Frey, V. Tudose, "Planetary Radio Interferometry and Doppler Experiment (PRIDE) for the JUICE mission", EPSC Abstracts Vol. 8, EPSC2013-357, European Planetary Science Congress, 2013

[30] Ye, Peijian; Sun, Zezhou; Zhang, He; Li, Fei (2017). "An overview of the mission and technical characteristics of Change'-4 Lunar Probe", Science China Technological Sciences. 60 (5): 658, 2017

[31] Yingzhuo Jia, Yongliao Zou, Jinsong Ping, Changbin Xue, Jun Yan, Yuanming Ning, The scientific objectives and payloads of Chang'E−4 mission. Planetary and Space Science. 21 February 2018. doi:10.1016/j.pss.2018.02.011

[32] Molera Calvés et al., "Observations and analysis of phase scintillation of spacecraft signal on the interplanetary plasma", A&A, 564 (2014) A4, 7

[33] Crisham et al., " A decade of the Super Dual Radar Network (SuperDARN): scientific achievements, new techniques and future directions", Surveys in Geophysics, 28 (1): 33-109, 2007

[34] J. Utzmann, et. al., "ARCHITECTURAL DESIGN FOR A EUROPEAN SST SYSTEM", Proc. 6th, European Conference on Space Debris, Darmstadt, Germany, 22-25 April, 2013.

[35] Folkner et al., "Determination of Position of Jupiter From Very-Long Baseline Interferometry Observations of ULYSSES", Astronomy Journal, 112, 1294F, 1996

[36] Pogrebenko, S. V., Gurvits, L. I., Campbell, R. M., Avruch, I. M., Lebreton, J.-P., and van't Klooster, C. G. M. (2004). "VLBI tracking of the Huygens probe in the atmosphere of Titan", In A. Wilson, editor, Planetary Probe Atmospheric Entry and Descent Trajectory Analysis and Science, volume 544 of ESA Special Publication, pages 197–204

[37] Lebreton, J.-P., Witasse, O., Sollazzo, C., Blancquaert, T., Couzin, P., Schipper, A.-M., Jones, J. B., Matson, D. L., Gurvits, L. I., Atkinson, D. H., Kazeminejad, B., and Pérez-Ayúcar, M. (2005), "An overview of the descent and landing of the Huygens probe on Titan", Nature, 438:758–764.

[38] Jones, D., Romney, J., Dhawan, V., Folkner, W., Jacobson, R., Jacobs, C., and Fomalont, E. (2017). A decade of astrometric observations of Cassini: Past results and future prospects, In 2017 IEEE Aerospace Conference, pages 1–8

[39] Jianguo, Y., Jinsong, P., Fei, L., Jianfeng, C., Qian, H., and Lihe, F. (2010), "Chang'E-1 precision orbit determination and lunar gravity field solution", Advances in Space Research, 46:50–57.

[40] Pogrebenko, S.V. , Gurvits, L.I., Wagner, J. et al. 2006, First results of the First EVN VLBI Practice Run on the Smart-1, in Cassini PSG meeting, 21-23 June 2006, Nantes, France

[41] Lanyi, G., Bagri, D. S., and Border, J. (2007), "Angular Position and Determination of Spacecraft by Radio Interferometry", IEEE Proceedings, 95(11):2193–2201.

[42] Takeuchi, H., Horiuchi, S., Phillips, C., Edwards, P., McCallum, J., Ellingsen, S., Dickey, J., Ichikawa, R., Takefuji, K., Yamaguchi, T., et al. (2011). Vlbi tracking of the solar sail mission Ikaros. In General Assembly and Scientific Symposium, 2011 XXXth URSI, pages 1–4. IEEE.

[43] Duev, D. A., Molera Calv´es, G., Pogrebenko, S. V., Gurvits, L. I., Cimó, G., and Bocanegra Bahamon, T. (2012), "Spacecraft VLBI and Doppler tracking: algorithms and implementation", Astronomy and Astrophysics, 541:A43.

[44] Duev, D., Pogrebenko, S., Cim, G., Calvs, G. M., Bahamn, T. B., Kettenis, L. G. M., Kania, J., Tudose, V., Rosenblatt, P., Marty, J.-C., Lainey, V., de Vicente, P., Quick, J., Nicola, M., Neidhard, A., Kronschnabl, G., Ploetz, G., Haas, R., Lindquist, M., Orlatti, A., Ipatov, A., Kharinov, M., Mikhailov, A., Gulyaev, S., Weston, S., Natush, T., Zhang, W., Wang, W., Bo, X., Yang, W., Hao, L., and Kallunki, J. (2016)., "Planetary Radio Interferometry and Doppler Experiment (PRIDE) technique: a test case of the Mars Express Phobos fly-by", Astronomy and Astrophysics, 593:A34, 2016

[45] Park, R. S., Folkner, W. M., Jones, D. L., Border, J. S., Konopliv, A. S., Martin-Mur, T. J., Dhawan, V., Fomalont, E., and Romney, J. D. (2015)., "Very Long Baseline Array Astrometric Observations of Mars Orbiters",  The Astronomical Journal, 150:121., 2015

[46] Grasset, O., Dougherty, M. K., Coustenis, A., Bunce, E. J., Erd, C., Titov, D., Blanc, M., Coates, A., Drossart, P., Fletcher, L. N., Hussmann, H., Jaumann, R., Krupp, N., Lebreton, J.-P., Prieto, Ballesteros, O., Tortora, P., Tosi, F., and Van Hoolst, T. (2013), "JUpiter ICy moons Explorer (JUICE): An ESA mission to orbit Ganymede and to characterise the Jupiter system",  Planetary and Space Science, 78:1–21

[47] Turishev, S., Toth, V., "The Pioneer Anomaly", Living Reviews in Relativity,  13, 4 (2010)

[48] Kopeikin, S., Shafer, G., "Lorentz Covariant Theory of Light Propagation in Gravitational Fields of Arbitrary-Moving Bodies", Phys.Rev.D60:124002,1999

[49] Tyler, G., Sweetnam, D., Anderson, J., et al. 1989, Science, 246, 1466

[50] Howard, H., Eshleman, V., Hinson, D., et al. 1992, Space Sci. Rev., 60, 565

[51] Kliore, A., Anderson, J., Armstrong, J., et al. 2004, Space Sci. Rev., 115, 1

[52] Pätzold, M., Neubauer, F., Carone, L., et al. 2004, in Mars Express: The Scientific Payload, 1240, 141

[53] Häusler, B., Pätzold, M., Tyler, G., et al. 2006, Planet. Space Sci., 54, 1315

[54] Iess, L., Di Benedetto, M., James, N., et al. 2014, Acta Astron., 94, 699

[55] Bocanegra-Bahamón, T.M., Molera, G., Gurvits, L., Duev, D., Pogrebenko, S.V., Cimò, C., Dirkx, D. and Rosenblatt, P., "Planetary Radio Interferometry and Doppler Experiment (PRIDE) technique: A test case of the Mars Express Phobos Flyby, II. Doppler tracking: Formulation of observed and computed values, and noise budge", A&A, Vol. 609, A59, 2018,

[56] Wilson, C.F. et al., "The 2010 European Venus Explorer (EVE) mission proposal", Experimental Astronomy, Springer Link, 2012, 33 (2-3), pp.305-335.

[57] Coustenis, A., Atreya, S.K., Balint, T. et al., "TandEM: Titan and Enceladus mission", Exp Astron (2009) 23: 893

[58] Marty, B., Guillot, T., Coustenis, A. 2009, Exp. Astron. 23, p. 947

[59] Grasset, O. et al., "JUpiter ICy moons Explorer (JUICE): An ESA mission to orbit Ganymede and to characterise the Jupiter system", Planetary and Space Science, Volume 78, April 2013, Pages 1-21

[60] Huixian, Sun; Shuwu, Dai; Jianfeng, Yang; Ji, Wu; Jingshan, Jiang, "Scientific objectives and payloads of Chang'E-1 lunar satellite", Journal of Earth System Science, vol. 114, issue 6, pp. 789-794

[61] Yong Huang, Shengqi Chang, Peijia Li, Xiaogong Hu, Guangli Wang, Qinghui Liu, Weimin Zheng, Min Fan,"Orbit determination of Chang'E-3 and positioning of the lander and the rover", Chinese Science Bulletin 59(29-30):3858-3867, August 2014

[62] Ormston, T.(18 October 2016). "Listening to an alien landing". European Space Agency, http://blogs.esa.int/rocketscience/2016/10/18/listening-to-an-alien-landing/

[63] Di Martino, M., Montebugnoli, S., Cevolani,G., et al. 2004, Planet. Space Sci., 52, 325

[64] Pupillo, G., Salerno, E., Pluchino, S., Bartolini, M., Montebugnoli, S., Di Martino, M , "A potential Italian radar network for NEO and space debris observations" Pupillo, G., Salerno, E., Pluchino, S., Bartolini, M., Montebugnoli, S., Di Martino, M., Journal: Memorie della Societa Astronomica Italiana Supplement, v.16, p.59 (2011)

[65] Paragi, Z. et al., "Very Long Baseline Interferometry with the SKA", Proceedings of "Advancing Astrophysics with the Square Kilometre Array", PoS(AASKA14)14

[66] Garrett, M. A. 2000, in M. P. van Haarlem (ed.) Perspectives on Radio Astronomy: Science with Large Antenna Arrays, 139 (Astron:Dwingeloo)

[67] Gurvits, L. I. 2004, New Astronomy Reviews, 48, 1211

[68] Godfrey, L. E. H., Bignall, H., Tingay, S. et al. 2012, PASA, 29, 42

[69] Fomalont, E., & Reid, M. 2004, New Astronomy Reviews, 48, 1473

[70] Schilizzi, R. T., Alexander, P., Cordes, J. M. et al. 2007, SKA Memo No. 100

[71] Giroletti, M., Orienti, M., D'Amando, F. et al. 2014, "The connection between radio and high energy emission in black hole powered systems", in proceedings of "Advancing Astrophysics with the Square Kilometre Array", PoS(AASKA14)153

[72] Adams, J., "Analysis and Simulation of a Ground-based Radar for Space Debris Detection", MsC Thseis, University of Aveiro, 8th October 2018

[73] Jones & Lazio, arXiv:1412.6006v1; PoScience, 2014.

# Big Data and Computing

# Computational Astrophysics and SKA

Tjarda C. N. Boekholt[b,c], Valério A. R. M. Ribeiro[a,b], Alan J. Alves do Carmo[c] and

Alexandre C. M. Correia[c]

[a] CIDMA, Departamento de Física, Universidade de Aveiro, Campus Universitário de Santiago, 3810-193 Aveiro, Portugal
[b] Instituto de Telecomunicações, Campus Universitário de Santiago, 3810-193 Aveiro, Portugal
[c] CFisUC, Department of Physics, University of Coimbra, 3004-516 Coimbra, Portugal

**ABSTRACT**

The computer has become an indispensable tool for astronomers. Not only for storing, data mining and visualizing large data sets obtained from large astronomical missions, but also for theoretical modelling of astrophysical systems in the universe. Both hardware and software with increasing levels of performance and modularity have to be developed in order to be able to construct sufficiently realistic models that can be compared directly to upcoming observations from missions such as the SKA.

**Keywords:** computational astrophysics, hardware and software, numerical simulations, SKA, gravitational dynamics, hydrodynamics, astrophysical transients

## 1 COMPUTER SIMULATIONS IN ASTRONOMY

Up to about the 1960s the theoretical study of most astrophysical phenomena relied on mathematical analysis using pencil and paper. This still remains the preferred method of research due to its rigor and ability to derive mathematical relations. However, with the rapid development of chaos theory it was recognized that most astrophysical phenomena exhibit some degree of chaos, which is usually related to the presence of non-linear equations and some form of feedback. Also, the multi-physics aspect adds to the complexity of the analysis. For example, one can imagine a dense stellar system in which many stars interact gravitationally, where some of them collide, where the most massive stars explode to become supernovae, and where the ejecta and strong radiation interact with the primordial gas out of which the stellar system was formed. Another illustrative example, but closer to home, is the origin and longevity of Jupiter's great red spot. The combination of non-linearity and multi-physics in most astrophysical phenomena has led astronomers to resort to discretized, numerical models that can be solved on a computer using a variety of integration techniques.

Two common types of systems in the universe are dynamical systems and gaseous systems, which are often represented by an ensemble of particles interacting either through Newton's equations of gravity and motion (N-body simulations) or through the equations of smoothed particle hydrodynamics (SPH codes). In both cases a very large particle number is preferred in order to reach the desired resolution that allows a direct comparison to observations. In order to achieve this, it is required to have two key elements at one's disposal: 1) state of the art computer resources, i.e. clusters of central processing units (CPUs) and/or graphical processing units (GPUs), and 2) modular simulation software that allows for experimenting with theories and implementations. There are great synergies between computational science and astrophysics, and the SKA will contribute to these synergies through new radio observations of a wide variety of astronomical systems and the development of new computing infrastructures.

## 2 MODELLING OBSERVATIONS FROM THE SKA

The time scale on which astrophysical systems change is often comparable or longer than a human lifetime. Those systems therefore provide us with a mere snapshot in time. It is the aim of the computational modeller to extend our knowledge of the system by calculating possible dynamical histories and futures that are consistent with the observational snapshot. It is invaluable here to have at one's disposal several independent observations of the system that the model should reproduce simultaneously. For example, multi-wavelength studies will prove crucial for understanding the nature of stellar novae and fast radio bursts (see contribution by Ribeiro et al.). The SKA will provide such complementary data through high resolution radio observations.

To give a specific example, SKA will reveal the radio signatures from the accretion of gas onto young stellar objects [1]. These observational constraints will help to improve our models, which include gravitational dynamics, hydrodynamics and stellar feedback. In this way the SKA observations will further our understanding of star formation and the origin of the stellar initial mass function. Besides providing complementary constraints, the SKA will also present modellers with new challenges. In order to reproduce the new observations from the SKA, it might be necessary to implement new physical processes into the model and to increase the level of precision. An illustrative example and part of the Cradle of Life program is the modelling of the magnetic fields of exoplanets (see contribution by Correia et al.). The interaction between energetic electrons from the stellar wind and the planet's magnetic field can give rise to circularly polarized cyclotron radio emissions. Another example is the modelling of classical and recurrent novae [2] (see also contribution by Ribeiro et al.). The complex interplay between the primary and the secondary star of a binary system has to be resolved in both space and time, including multiple phases of mass eruptions, mass transfer, interaction of the ejecta with the secondary and changing orbital elements. These complexities are well demonstrated in radio observations of a nova that also showed spots of synchrotron emission on the surface of the ejecta photosphere. The emission was suggested to be due to interaction between slow and fast winds, which would have also given rise to gamma-ray emission [3]. This can only be truly tested hydrodynamically and, in particular, such new physical ingredients are best implemented in a modular fashion so that the new module can be coupled to existing codes, and where the modeller can experiment with different numerical implementations of the module.

Other computational applications directly related to the SKA key science program are simulations of the epoch of reionization [4] and formation of the first galaxies [5]. In both cases the size of the problem and the enormous range in scales prevent us from performing realistic direct simulations. It is an art to construct semi-analytical recipes that speed up the calculations and at the same time preserve the precision of the solution. Meanwhile, Moore's law for computing power [6] predicts a steady increase in the forthcoming years. The usage of clusters of GPUs to solve massively parallel problems in astrophysics will become increasingly important and should also be pushed forward by both modellers and computer scientists within the ENGAGE-SKA community.

## 3 STATE OF THE ART HARDWARE AND SOFTWARE

Current state of the art hardware for solving scientific problems include clusters of GPUs in which the same set of operations is solved in parallel over many cores, and which are subsequently combined in an efficient manner to evolve the full system in time. This technique has recently made it possible to model the evolution of globular clusters consisting of millions of stars on a star by star basis [7]. With the current fastest treecode called Bonsai it is possible to model our Galaxy with billions of stars with a performance of 24.77 PFLOPS (PFLOPS = $10^{15}$ floating-point operations per second)[1] on the ORNL Titan supercomputer in the USA, and 33.49 PFLOPS on the Piz Daint supercomputer in Switzerland [8]. There is an active development of new supercomputers by a variety of countries, which is mirrored in the annual competitive ranking of most powerful supercomputers around the globe[2]. Whereas Titan was at the head of this list in 2015, it has now been surpassed by the Chinese machines Sunway TaihuLight and Tianhe-2.

On a somewhat smaller scale, but nevertheless very useful scientifically, is the development of the Little Green Machine II in the Netherlands. This €200,000 machine consists of four worker nodes each with four NVIDIA Tesla P100 GPUs with NVLink, reaching a peak performance just shy of 90 TFLOPS (double-precision). Such medium-sized machines are ideal for scientists as they are local and provide easy access. As the following back of the envelope calculation shows, performances of the order 1-10 TFLOPS are ideal for medium-sized scientific simulations.

We consider a typical simulation of a modest globular cluster. To save the full phase space information we would need to store 7 numbers: mass, three position coordinates and three velocity components. A modest globular cluster consists of $10^5$ stars. Storing a number in double-precision usually takes 64 bits or 8 bytes. One snapshot of our globular cluster is thus about $10^5$ x 7 x 8 = 5.6 MB. The total simulation time is of the order the lifetime of the universe (13.7 Gyr) with snapshot intervals every dynamical crossing time (of order 1 Myr). This would result in about $10^4$ snapshots with a total storage space of about 56 GB. In N-body simulations, the computing time is dominated by the calculation of the accelerations, which scales with the number of stars squared. On average, there are about $10^2$ x $N^2$ FLOP per integration step where N is the total number of stars. In order to resolve the dynamics accurately, we require time steps smaller than the dynamical time, say one hundredth of a dynamical crossing time. The total number of time steps is then around $10^6$. We estimate that a full simulation performs of the order $10^6$ x $10^2$ x $(10^5)^2 = 10^{18}$ FLOP. In order to complete such a simulation within a reasonable time, assuming one week, this would require a performance of about 1.7 TFLOPS.

---

[1] In comparison, a household computer will have a few tens of GFLOPS ($10^9$ floating-point operations per second)
[2] https://www.top500.org/lists/top500/

The current and upcoming computing facilities in Portugal are well matched to these constraints. For an overview of computer resources in Portugal we refer the reader to the contribution by M. Avillez. In particular, these include computer clusters at the University of Aveiro, University of Évora, the national HPC Infrastructure LCA-PRACE cluster at University of Coimbra, and the planned Minho Advanced Computing Centre (MACC), in the North of Portugal[1].

The development in hardware should go hand in hand with the development of modular and open-source software. A giant leap forward in this respect has been made by the Astrophysical MUlti-physics Software Environment[2] consortium [9]. This software framework collects community codes for gravity, hydrodynamics, stellar and binary star evolution, stellar mergers and radiative transfer. The main novelty is the Python interface given to each code which allows the different codes to communicate and exchange data. Rather than building one kitchen-sink code, one can construct multi-physics scripts by gluing together the required physical ingredients to model the system of interest. Two illustrative applications of modular scripting are the dynamics of stars in star forming regions [10] and the collisional growth of massive black holes [11].

## 4 SUMMARY

The SKA provides the computational astrophysicist with exciting challenges and opportunities. First of all, the new observations in the radio of a wide variety of astronomical systems, ranging from planets to stellar explosions to large scale structure, will provide new targets for modellers to reproduce. We will test our current models and numerical implementations of physical ingredients such as stellar evolution, stellar dynamics, (magneto-)hydrodynamics and radiative transfer. Multi-wavelength observations will require multi-physics simulations in order to explain the fine details in the observations. It is through the unprecedented high-quality observations of SKA that the field of computational astrophysics will receive a great boost.

## ACKNOWLEDGMENTS

We acknowledge support from the FCT (SFRH/BPD/122325/2016), CIDMA strategic project (UID/MAT/04106/2013), ENGAGE SKA (POCI-01-0145-FEDER-022217), and PHOBOS (POCI-01-0145-FEDER-029932), funded by COMPETE 2020 and FCT, Portugal.

## REFERENCES

[1] Fuller, G. et al., Star and Stellar Cluster Formation: ALMA-SKA Synergies, Proceedings of Advancing Astrophysics with the Square Kilometre Array (AASKA14). 9 -13 June, 2014. Giardini Naxos, Italy, 04/2015.

[2] O'Brien, T. Rupen, M., Chomiuk, L., **Ribeiro, V. A. R. M.** et al., Thermal radio emission from novae & symbiotics with the Square Kilometre Array, Proceedings of Advancing Astrophysics with the Square Kilometre Array (AASKA14). 9 -13 June, 2014. Giardini Naxos, Italy, 04/2015.

[3] Chomiuk, L., Linford, J. D., Yang, J., O'Brien, T. J., Paragi, Z., Mioduszewski, A. J., Beswick, R. J., Cheung, C. C., Mukai, K., Nelson, T., **Ribeiro, V. A. R. M.** et al. "Binary orbits as the driver of γ-ray emission and mass ejection in classical novae," Nature, 514, 339 2014.

[4] Iliev, I. et al., Epoch of Reionization modelling and simulations for SKA, Proceedings of Advancing Astrophysics with the Square Kilometre Array (AASKA14). 9 -13 June, 2014. Giardini Naxos, Italy, 04/2015.

---

[1] INCoDe.2030 NO CIÊNCIA'17, Fundação para a Ciência e a Tecnologia, ISBN 978-972-667-348-4, Aug 2017, Portugal, http://www.incode2030.gov.pt/en/home/

[2] AMUSE, http://www.amusecode.org/

[5] Santos, M. et al., HI galaxy simulations for the SKA: number counts and bias, Proceedings of Advancing Astrophysics with the Square Kilometre Array (AASKA14). 9 -13 June, 2014. Giardini Naxos, Italy, 04/2015.

[6] Bedorf, J. and Portegies Zwart, S. F., A pilgrimage to gravity on GPUs, The European Physical Journal Special Topics, Volume 210, 2012, pp.201-216, 08/2012.

[7] Wang, L. et al., NBODY6++GPU: ready for the gravitational million-body problem, Monthly Notices of the Royal Astronomical Society, Volume 450, Issue 4, p.4070-4080, 07/2015.

[8] Bedorf, J. et al., 24.77 Pflops on a Gravitational Tree-Code to Simulate the Milky Way Galaxy with 18600 GPUs, Proceedings of the International Conference for High Performance Computing, Networking, Storage and Analysis, p. 54-65, 11/2014.

[9] Portegies Zwart et al., A multiphysics and multiscale software environment for modeling astrophysical systems, New Astronomy, Volume 14, Issue 4, p. 369-378, 05/2009.

[10] **Boekholt, T. C. N.** et al., Dynamical ejections of stars due to an accelerating gas filament, Monthly Notices of the Royal Astronomical Society, Volume 471, Issue 3, p.3590-3598, 11/2017.

[11] **Boekholt, T. C. N.** et al., Formation of massive seed black holes via collisions and accretion, Monthly Notices of the Royal Astronomical Society, Volume 476, Issue 1, p.366-380, 05/2018.

# High Performance Computing in Astrophysics

Miguel A. de Avillez

Department of Mathematics, University of Evora, Portugal
Zentrum für Astronomie und Astrophysik, Technische Universität Berlin, Germany

**ABSTRACT**

Owing to the availability of supercomputer resources and the increase in computing power the astrophysical community has benefited over the last two decades of unprecedented means to model with great detail the Universe and its components, and at the same time include more physical processes into the existing models. With the Square Kilometer Array (SKA) demands on high performance computing a further development is expected towards exascale computing.

**Keywords:** plasmas – hydrodynamics – magnetohydrodynamics – atomic processes – radiation mechanisms – galaxies – Milky Way – interstellar medium – high performance computing – numerical simulations

## 1   INTRODUCTION

Over the last decade multi-wavelength space and ground-based observations have provided the Astrophysical/Astronomical community with an unprecedented insight of the Universe, cluster of galaxies, galaxies, the Milky Way, the interstellar medium, the Local Bubble, stars and planets. However, observations are meaningless if there is no sound theoretical basis to explain the observed features. There has been a large effort by the theoreticians to develop models and tools to explain the observed radiation from the individual objects. Based on the physical processes theoreticians often develop complex models that in turn rely on the solutions of non-linear equations coupling together macro and microscopic processes (e.g., gas dynamics, atomic and molecular physics, magnetic fields, acceleration and propagation of high energy particles, such as cosmic-rays, etc).

Owing to the increase in computing power and access to supercomputer facilities, for instance the Tier 0 and 1 facilities in Europe available under the PRACE[1] research infrastructure service, has made possible the development of codes that can tackle the evolution of an astrophysical system and at the same time have the adequate dimensionality, increasing resolution and inclusion of more physical processes. Accessing such facilities requires the community develops software that must take advantage of capabilities provided by the new generation CPUs, e.g., vectorization, and the availability of multi-cores multi-node, accelerators (e.g., GPGPUs) and or coprocessors (MICs). This requires the development, testing and refactoring of software and the adoption of programming models involving homogeneous or heterogeneous architectures.

The increasing need of computing power per core to handle the massive data sets resulting from the observations generated with the SKA will drive further development of low-energy consumption multi-cores CPUs with larger L3 caches, faster memory access and larger floating point operations per second per core. Consequently, the astrophysical community will take advantage of the new resources to improve models and accelerate calculations striving for a better description of the reality.

## 2   MODELING THE EVOLUTION OF ASTROPHYSICAL SYSTEMS

The dynamical evolution of astrophysical systems is in general described by means of a single or multi-fluid calculations (the latter is used in cases where the joint evolution of ions and plasma is required; [1]) using either a Lagrangean or an Eulerian approach. While in the former the fluid is divided into an ensemble of fluid elements whose history is traced independently, in the Eulerian approach the fluid's history is traced as a whole from a reference frame.

---

[1] http://www.prace-ri.eu

Lagrangean calculations are suitable to trace the evolution of, e.g., (i) ions in a plasma taking into account the recombination, ionization and emission processes – often this calculation is followed with a remap to the Eulerian grid, (ii) evolution of the shock structure is compressible gases, or (iii) in gravitationally dominated systems where the fluid is represented by an N-body system, with each particle enclosing a large mass in the case of cosmological simulations each particle encloses millions of solar masses. The Eulerian approach is in general used in gas dynamics simulations where shocks are dominant and great detail is required during the dynamical evolution. This approach is often used in, e.g., galaxy-scale simulations, evolution of disk and halo of galaxies, interstellar medium, star formation, or even in turbulence simulations.

Hybrid schemes coupling the two approaches are also used, e.g., in the study of the time-dependent joint thermal (ionic) and dynamical evolution of plasmas in the interstellar medium of galaxies, or track the dynamics of individual systems (by particles) embedded in a fluid, e.g., evolution of newly formed stars (represented by particles) within a molecular cloud and later their drifting from the parental cloud.

### Dimensionality and Resolution

The success of the calculations require a balance between dimensionality, size of the computational domain, resolution and adequacy of the physical processes for the adopted resolution (see discussion in [2]). Dimensionality is paramount in the solution of any astrophysical problem. Many of the astrophysical systems, and in particular the interstellar medium in galaxies, are turbulent [3] one can not rely on their two-dimensional representation. In fact two and three-dimensional turbulences are quite different. Three-dimensional turbulence is characterized by the dissipation of energy through vortex stretching, a property that is not present in two-dimensional turbulence (see, e.g., [4,5]). Thus, any two-dimensional solution of a system, even it it reproduces the system's observations, is by definition wrong.

Resolution is fundamental for the convergence of the solutions for the physical processes at hand. If the adopted resolution is lower than the length scales for the relevant physical processes (e.g., cooling, atomic and molecular processes) the simulation does not catch these processes. Thus, their feedback into the system evolution is unnoticed and therefore affects the solution and thus its convergence. The smallest physical structures can resolved using the adaptive mesh refinement [6,7] which consists in the creation on-the-spot and on-the-fly of higher resolution grids. Due to the complexity of its implementation only a few Astrophysical codes have this capability.

### Thermal history of the plasma

Often is assumed that the all the gas parcels in the plasma have the same ionic and radiative histories. This means that the plasma does not have a recollection of its history and thus the ionic evolution of the gas parcels is exactly the same, but this is not quite true. In fact the different ions of an element have different recombination time scales which in turn are longer than the cooling time of the plasma. Thus, shock heated gas cools faster than it recombines and it becomes over ionized. Similarly, if a gas is suddenly heated, say, by a shock wave, it heats up but its ionization lags behind - the plasma is under ionized. Consequently the ionization structure, emission spectra (line and continuum) and cooling of the plasma are quite different from place to place even at the same temperature (see discussions in, for example, [8—10]).

A good example is the finding that x-ray emission can occur at temperatures of $10^4$ K contrary to what is expected for a plasma in collisional ionization equilibrium, in which the soft x-ray emission occurs around $10^6$ K. The emission at low temperatures is in fact a result of the delay recombination of the plasma [11]. A further effect was shown by [12] who using for the first time to date a time-dependent multi-fluid hydrodynamical and atomic physics evolution of the interstellar gas in the Milky Way found that the X-ray emission at low temperatures reflects the the feedback between heating, cooling and the dynamics of the system. Even with the same initial conditions the gas parcels will have a different ionic/radiative history with the corresponding emissivities having an order of magnitude differences.

Contrary to what has been the dominant thinking the plasma may not be in thermal equilibrium and thus has not reached the Maxwell-Boltzmann distribution. In fact there is mounting evidence that non-thermal distributions (see, e.g., [13—16]) distributions that deviate from the Maxwell-Boltzmann - are present in many astrophysical systems [17—19] characterized by low densities and having high temperature or density gradients exist leading to the deposition of energy into the tail of the distribution at a rate high enough to prevent relaxation towards thermal equilibrium.

A further complexity that comes into these simulations is the fact that the adiabatic parameter (the ratio of the specific heats at constant pressure and volume) depends on the plasma internal energy, which includes the contributions from the thermal, ionization and excitation energies (see discussions in, e.g., [20—22] and evolves according to the ionic state of the plasma. Hence, its value in the case of a monotatomic plasma varies between 1.05 and 5/3 over the full range of temperatures (Figure 1).

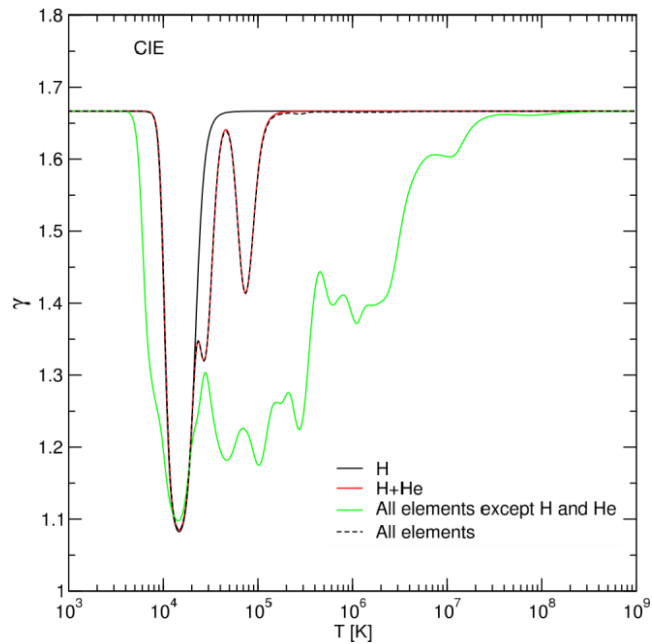

**Figure 1**. Evolution of the $\gamma$ parameter in different plasmas composed only of H (black line), H and He (red line), metals (C, N, O, Ne, Mg, Si, S, and Fe; green line), and all elements (H, He and metals; black dashed line). Note that H and He determine the evolution of $\gamma$ for the calculations involving all the ten elements because the abundances of the metals are much smaller than those of H and He. From de [22].

The inclusion of the atomic and molecular evolutions of the plasmas into the dynamical simulations of the different astrophysical systems implies an increase in complexity and an increasing demand in computing resources. If the electron distribution function is traced over time and atomic calculations have to be carried out on-the-spot and on-the-fly. This implies a large database for the cross sections associated to the different ionization, recombination and excitation/deexcitation processes.

## 3     HIGH PERFORMANCE COMPUTING AND THE COMMUNITY

Much of the ongoing and future HPC efforts are related to the implementation of hybrid numerical schemes exploiting multi-cores heterogeneous architectures (comprising a combination of CPUs, GPUS or MICs) using well established programming models that rely on MPI, OpenMP, OpenACC, CUDA or OpenCL and their combined use providing it is efficient to accelerate the calculations. This approach to HPC implies the joint work of scientists, software developers, and HPC experts. While the former two understand the problems at hand and have insight on their physics and on the numerical algorithms to be more efficient and adequate, the HPC experts have the knowledge and skills on the novel hardware architectures, programming models and software acceleration and refactoring.

This joint effort, which has been the core business of work packages under the PRACE service, will also be at the core of the SKA HPC developments that are expected to occur. This will drive the astrophysical community to better define what are the schemes and approaches that are more adequate to their calculations using grid or particle-based codes.

## 4     FINAL REMARKS - HPC AT THE UNIVERSITY OF ÉVORA

The University of Évora has a long tradition in HPC applied to research in theoretical and computational Astrophysics and in the training of human resources. For instance the University established the first international research and training networks in Computational Astrophysics funded by the ERASMUS program (e.g., Supercomputing and Numerical Methods in Astrophysical Fluid Flow Modelling; 2008 and 2009) and by the European Science Foundation, e.g., the Supercomputing 2008 and 2009 Schools, and the *a la carte* program "European Network of Computational Astrophysics - ASTROSIM" (2006-2011). The university lead the astrophysical efforts in the development and deployment of parallelisation models for heterogenous systems composed of accelerators and coprocessors under the PRACE Second Implementation Phase (PRACE-2IP) Work Package 8.

Besides the current HPC resources, the University is acquiring, under the ENGAGE SKA research infrastructure, a new supercomputer with up to 11200 cores, 500 TB in storage and 260 TFLOPS. This new facility combined with the existing expertise (prototyping, development of paralellization models and software refactoring) will be an adding advantage for the HPC efforts associated to the SKA.

## ACKNOWLEDGEMENTS

Projects PRACE fifth implementation phase (PRACE-5IP) RI-730913, *DeutscheForschungsgemeinschaft*, DFG project ISM-SPP 1573, "Hybrid Computing Using Accelerators & Coprocessors", In Alentejo program, CCDRA, Portugal, and Enabling Green E-science for the SKA Research Infrastructure (ENGAGE SKA), reference POCI01-0145-FEDER-022217, funded by COMPETE 2020 and FCT, Portugal.

## REFERENCES

[1] de Avillez, M. A., Breitschwerdt, D., Asgekar, A., and Spitoni, E. 2015, Highlights of Astronomy 16, 606

[2] de Avillez, M. A. and Breitschwerdt, D. 2004, A&A 425, 899

[3] von Weizsäcker, C. F. 1951, ApJ 114, 16

[4] Davidson, P. A. 2004, Turbulence: an introduction for scientists and engineers, Oxford University Press

[5] Tennekes, H. and Lumley, J. L. 1972, First Course in Turbulence, MIT Press

[6] Berger, M. J. and Colella, P. 1989, Journal of Computational Physics 82, 64

[7] Berger, M. J. and Oliger, J. 1984, Journal of Computational Physics 53, 484

[8] Gnat, O. and Sternberg, A. 2007, ApJS 168, 213

[9] Shapiro, P. R. and Moore, R. T. 1976, ApJ 207, 460

[10] Sutherland, R. S. and Dopita, M. A. 1993, ApJS 88, 253

[11] Breitschwerdt, D. and Schmutzler, T. 1994, Nature 371, 774

[12] de Avillez, M. A. and Breitschwerdt, D. 2012, ApJ 756, L3

[13] Berezhko, E. G. and Ellison, D. C. 1999, ApJ 526, 385

[14] Druyvesteyn, M. J. 1930, Zeitschrift fur Physik 64, 781

[15] Hares, J. D., Kilkenny, J. D., Key, M. H., and Lunney, J. G. 1979, Physical Review Letters 42, 1216

[16] Vasyliunas, V. M. 1968, in R. D. L. Carovillano and J. F. McClay (eds.), Physics of the Magnetosphere, Vol. 10 of Astrophysics and Space Science Library, p. 622

[17] Dzifčzková, E. and Dudík, J. 2013, ApJS 206, 6

[18] Humphrey, A. and Binette, L. 2014, MNRAS 442, 753

[19] Karlický, M., Dzifčáková, E., and Dudík, J. 2012, A&A 537, A36

[20] Cox, J. P. and Giuli, R. T.: 1968, Principles of stellar structure, Vol. 1, Gordon and Breach

[21] D'Angelo, G. and Bodenheimer, P.: 2013, ApJ 778, 77

[22] de Avillez, M. A., Anela, G. J., and Breitschwerdt, D.: 2018, A&A (in Press)

# Virtualization technologies for the SKA data challenge: high availability and reproducibility requirements[1]

J. B. Morgado[*a], D. Maia[a], D. Barbosa[b], João Paulo Barraca[b],
J. Bergano[b], D. Bartashevich[b], Nuno Silva[c]

[a] CICGE, Faculdade de Ciências da Universidade do Porto, 4430-146 V. N. Gaia, Portugal
[b] Instituto de Telecomunicações, Campus Universitário de Santiago, 3810-193 Aveiro, Portugal
[c] Critical Software, Parque Industrial do Taveiro, lote 49, 3045-504 Coimbra, Portugal

## ABSTRACT

The Square Kilometre Array (SKA) Telescope, is an ongoing project set to start its building phase in 2019 and achieve first light in 2022. The first part of the project, SKA1, will produce a raw data rate of ~10 Tb/s, requiring a computing power of 100 Pflop/s and an archiving capacity of ~300 PB/year. The next phase of the project, SKA2, will increase these demands by a factor of 10. The SKA has a very high availability key requirement of 99.9% for its operations. Other requirements are computing scalability and scientific outcome reproducibility. Focusing on the SKA Telescope Manager requirements and architecture, we propose an approach to enforce these requirements – with an optimal use of resources – by using highly distributed computing and virtualization technologies.

**Keywords:** SKA, Telescope Manager, HPC, Virtualization, Radio Astronomy, Reproducibility

[*] E-mail: jorge.morgado@fc.up.pt; phone +351 227 861 290

## 1    INTRODUCTION

The Square Kilometre Array (SKA), is set to start its building phase in 2019, achieve first light by 2022, be fully operational by 2025 and collect data for at least 20 years thereafter. The SKA will gather radio signals from the whole sky in the frequency range from 50 MHz to 15.5 GHz. The 1st phase of the SKA, SKA1, will consist of 130000 low frequency radio antennas – operating in the 50 MHz to 350 MHz band, and ~200 mid frequency radio antennas - operating in the 350 MHz to 15.5 GHz band, incorporating the South African MeerKAT 64 antenna precursor. Phase 2 of the SKA, SKA2, will expand in frequency and spread its stations throughout the African continent and Australia and New Zealand. In Africa, the stations will cover a range over ~6000 kilometers – from Ghana to Madagascar/Mauritius, increasing the number of antennas by a factor of 10. This will increase all the computing, data transfer and storage requirements accordingly.

SKA1 will produce a raw data rate of ~10 Tb/s, require a computing power of 100 Pflop/s and archive ~300 PB/year of science products, solely for acquiring, processing the signal, transferring the data products to data centers and store them. The computing requirements for doing science with these data products is expected to be several orders of magnitude higher. For comparison, in 2016 worldwide total internet traffic was ~212.8 Tb/s, present worlds' fastest supercomputer has a capacity of 93 Pflop/s and in 2012 Facebook stored ~100 PB of photos and videos. Due to the nature of the data gathered by the SKA project and due to its extended operating lifetime, it is also of major importance to take into consideration provision for future uses of the data which currently show no immediate scientific value. These data could, on first approach, be discarded in order to keep the data storage, processing and transfer costs down. For a practical example on the analysis on the use of data previously discarded as noise, that ended up having scientific interest going beyond the original purpose of the instrument, see [1].

The SKA observing operations, data transfer, computing and storage follow a very high availability key requirement of 99.9%. Another requirement, not only for the SKA, but one that we expect to be transversal to scientific research in the future, is that of *reproducibility*. Presently, when more researchers start to have access to the data used in refereed publications, scientists expect to be able to reproduce results from those publications independently, but are failing at it in what is being aptly named a "*Reproducibility Crisis*" [2]. In the SKA era, the issue goes even beyond that. The SKA scientific outcomes will rely on such a massive amount of data involving a demanding amount of computational resources, that we cannot afford to re-run specific computations or algorithms, in order to check if a result was actually achieved using any specific parameter configuration, algorithm or even a specific workflow that connects information from well-known astronomical databases [3, 4]. As such, the SKA, reproducibility must be an enforced feature throughout the all systems.

---

[1] Based on article published on SPIE Telescopes+Instrumentation 2018, [7].

The SKA Telescope Manager (TM) is the element responsible for the architecture design of the Observations Planning and Operations and the Monitoring and Control system of the SKA. This architecture includes the TM's interaction through Internal Interfaces with compute intensive elements like the Science Data Processor (SDP) and Central Signal Processing (CSP). Within the SKA TM, its Local Infrastructure Architecture (TM LINFRA) and Services (TM SER) are solving part of these technological hurdles with the help of virtualization technologies.

Virtualization technologies have been under heavy development during the last decade accompanying the change in computing paradigm from increasingly fast single processing units to an increasingly bigger number of processing units. This meant a convergence between High Performance Computing (HPC) and Cloud based environments paving the way to a renewed effort in parallel computing algorithms and an interest in virtualization and its ability to run several isolated computational threads across various computing units, including geographically disparate sites.

Looking at the current state of the art, we chose to base the design of our system in the OpenStack[1] platform taking into account its modularity and the open source development that is being accomplished. Still, all the rationale presented in our work is platform agnostic and can be adapted and used with other virtualization technologies. We propose a system philosophy that can take into account the requirements for scalability, availability and reproducibility needed in the SKA.

## 2   REQUIREMENTS

Considering that the SKA will have several levels of interaction, we are building a platform that will provide different tools for different key-holders while minimizing duplication of efforts. Our main work so far has been done in the TM, which purpose and scope are presented at great length in [5]. Of special importance in the TM, is the nature of its deliverables. Identified as TM Apps, TM's deliverables may be stand alone or exist together with other TM Apps. TM components constitute one of the main critical system elements in the telescope. The SKA requirements align the TM with a Tier 3 system as defined by the TIA-942 standard and the SKA TM LINFRA is designed according to the ruling principles of such systems.

Similar principles apply to the SKA Science Data Processor (SDP) element [6] – with the role of processing science data from the correlator or non-imaging processor into science data products – and the future storage and provision of data through regional digital infrastructures that are currently in the planning phases. For the latter, and as an example, the H2020 project Advanced European Network of E-infrastructures for Astronomy with the SKA (AENEAS) – is developing the design of the SKA's data mining infrastructure federating the underlying large-scale e-infrastructures (compute, connectivity), to enable the scientific community at large future access and exploitation of collected data products.

The key requirements for any TM Application (Apps), from the point of view of the SKA1 TM infrastructure and the TM service layer, or any Apps running in a virtualization layer in SDP or any future AENEAS infrastructure federations is then three fold:

- *Modularization* – An App should be defined by smaller, independent parts: Apps should not be monolithic, but rely on smaller modules. This will allow TM LINFRA to migrate modules in real time between hardware resources, allowing for failsafe implementations and optimal use of computing resources.

- *State based* – An App should depend only on the state of the system in order to function: Relying on a set of configuration files, databases and temporary state description variables, an App can be easily migrated during computation and be able to pick up exactly where it left. This operating principle allied to a redundancy of the underlying computation layer, will ensure the high availability of the system and its scalability.

---

[1] https://www.openstack.org/

- *Parallelization* - An App should, preferably, be able to carry out its function while running in parallel with other instances of the same App: If an App is capable of performing in parallel - and this requirement is especially important for systems like the SKA SDP element and future federation of e-infrastructures efforts like AENEAS – the underlying orchestration system is capable of launching several threads of the same App if the available computing resources at a given time permit it, greatly speeding up the computation in some cases.

## 3 METHODS

To allow for the previously exposed requirements to integrate with different systems – namely at the TM level as showcase – we propose that the access and configuration of the various parts of the system, built upon a Virtualization layer, rely on a three layered approach. At base level, the systems configuration would be defined by SKA1 TM infrastructure. A middle layer with access and computing resources agreed upon by the various key players. Finally a top layer where the Apps would be able to perform computation and require access between them and beyond them.

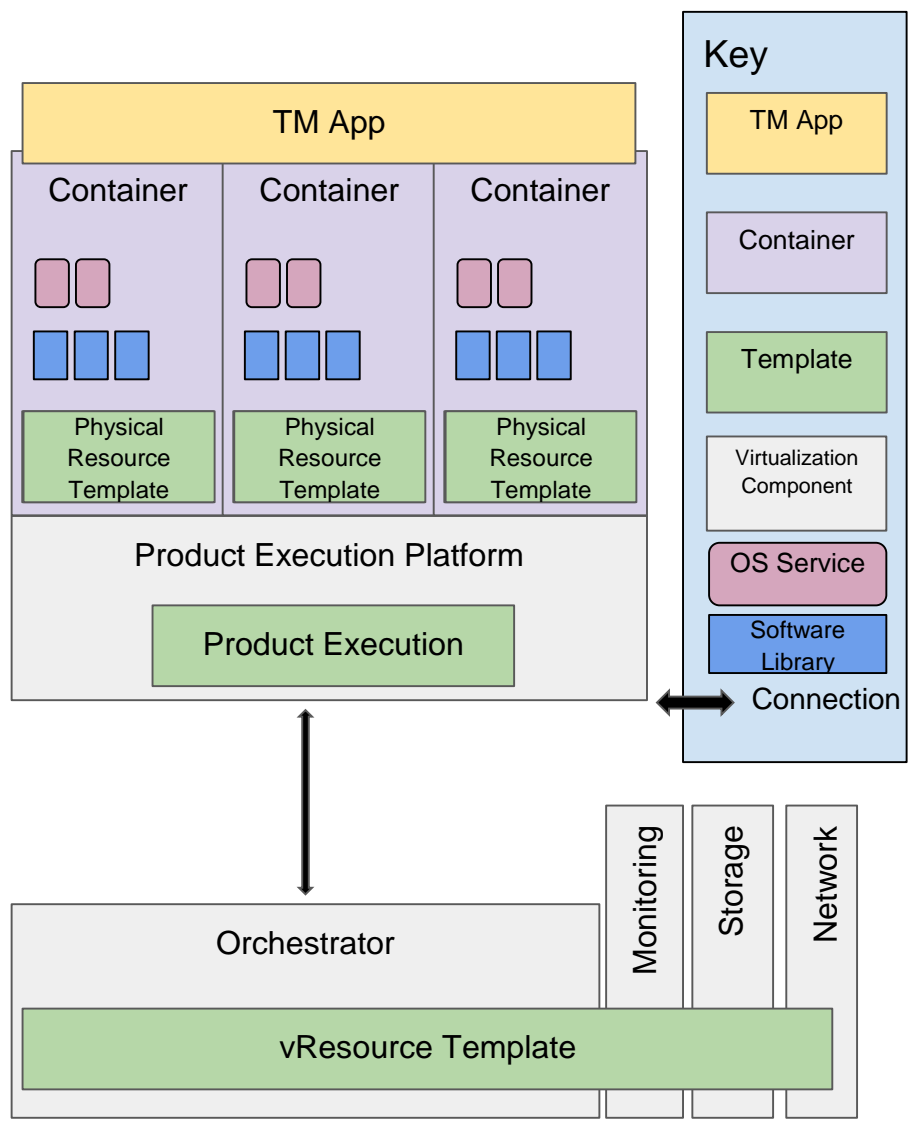

**Figure 1.** Template and Instance presentation.

We predefined these three levels of complexity within the system and divided the needed configurations for each one. Figure 1 presents a view of this approach and it is specified as follows:

- *Physical Resource Layer* – providing a uniform hardware view for all software: SKA1 will include tens of thousands of devices for controlling the telescopes and acquire data. We aim to provide all the key-holders of the project with the means to build tools that will work independently of the hardware. We also want to ensure that advances in computing hardware or changes in the computing resources providers can be met easily by the project. This avoids any adaptation process outside of the work done by the SKA TM LINFRA and the SKA TM SER.

- *Product Execution Layer* – consisting of a distributed environment operating towards the provisioning of highly available products: SKA1 services will be composed of a multitude of software code components that need to observe certain key points. One of those key points is to ensure that the adequate computing resources get allocated to each of these services and the SKA TM LINFRA will be able to automatically balance the hardware resources available to each service according to their immediate specific needs and the hardware resources available at the moment. Another key point is to ensure a high availability of services. In order to do so, the SKA TM LINFRA will detach the running hardware from a specific machine meaning that in case of hardware failure the service will continue to run on other available machines.

- *Virtualized Resource Layer* – consisting of virtual machines, containers and other hardware that has a logical representation to the Product Execution Layer and is part of a template or is available to be used by future templates: SKA services need a high degree of interoperability and communication. This part of the software abstraction will provide a deliverable product, modular in its nature, that will already have defined its hardware access, resource usage and communication capabilities with other parts of the SKA. By ensuring that it resides in the top layer, it means that the specifics of the SKA product can be changed at a lower level and later deployed and put into production without disruptions to the service being provided.

What this architecture achieves, is that the interface for the virtualization system – that will be the basis upon which all the SKA Telescope Manager software will run – will be constrained by exposing template actions externally according to the aforementioned three layers. The end objective is then to provide high availability, abstracting the underlying hardware infrastructure, and allowing software defined failover and horizontal scalability.

# 4  CONCLUSION

The SKA, due to its massive scale, has extremely demanding requirements set in place at various levels in its infrastructure. These requirements mainly focus on three key points: availability, scalability and reproducibility.

Implementing from the start of the project a paradigm built upon virtualization and expressing a set of requirements enforcing: modularization, state based logic and parallelization; allows for an architecture relying upon a three level access and control system, capable of optimally managing available computing resources in order to deliver TM's products. It also sets similar scenarios for other SKA elements such as the SDP and future post-SDP e-infrastructures federations such as AENEAS. Moreover, this same paradigm when implemented in a system with redundant hardware resources, ensures a high availability of services while simultaneously fully utilizing all available computing resources.

Finally, since all the various components of the system will necessarily have to be described through configuration files that define their building blocks, resource access and computing steps needed to obtain a product, the system becomes, by design, reproducible at any given time in the future. Thereby, any data or scientific deliverable created, could have associated the full history – from source data gathering to the final result – on how that deliverable came to exist.

### ACKNOWLEDGEMENTS

This research was supported by the project Enabling Green E-science for the SKA Research Infrastructure (ENGAGE SKA), reference POCI-01-0145-FEDER-022217, funded by COMPETE 2020 and FCT, Portugal. This work has been made possible thanks to the financial support by the Italian Government (MEF - Ministero dell'Economia e delle Finanze, MIUR - Ministero dell'Istruzione, dell'Università e della Ricerca). BM, DB and JPB acknowledge support from the European Commision through H2020 AENEAS project, Grant 731016. We thank discussions on reproducibility with Lourdes Verdes-Montenegro and Suzana Sanchez from IAA-CSIC, Spain.

# REFERENCES

[1] **J. B. Morgado** et al., "The low energy magnetic spectrometer on Ulysses and ACE response to near relativistic protons," Astronomy & Astrophysics 577, 61 (2015).

[2] B. Monya, "1,500 scientists lift the lid on reproducibility" Nature 533, 452–454 (2016).

[3] J. E. Ruiz et al., "AstroTaverna - Building workflows with Virtual Observatory services", Astronomy and Computing 7-8, 3-11 (2014).

[4] K. Hettne et al., "Structuring research methods and data with the Research Object model: genomics workflows as a case study," Journal of Biomedical Semantics 5, 41 (2014)

[5] S. Natarajan et al., "SKA Telescope Manager (TM): Status and Architecture Overview," Proc. of SPIE 9913, 991302 (2016).

[6] P. Alexander et al., "SKA Data Processor Architecture," SKA--TEL--SDP--0000013, (2016).

[7] **J. B. Morgado, D. Maia, D. Barbosa, J. P. Barraca, J. Bergano, D. Bartashevich, Nuno Silva**, M. Di Carlo, M. Canzari, M. Dolci, R. Smareglia , "Large scale high performance computing, instrument management for high availability systems and scientific reproducibility through the use of virtualization at the Square Kilometre Array (SKA) telescope," Proceedings of SPIE, AS18-AS110-103, Austin-Texas, accepted for publication (2018).

# SKA Slow Transients Pipeline Prototype and compute platforms: towards multi-messaging Big data pipelines

Luis Lucas[a], Nuno Silva[a], João Esteves[a], Domingos Barbosa[b], Dalmiro Maia[c],
João Paulo Barraca[b,c], Miguel Bergano[b], Diogo Gomes[b,c]

[a] Critical Software SA, Taveiro, Coimbra, Portugal
[b] Instituto de Telecomunicações, Campus Universitário de Santiago, 3810-193 Aveiro, Portugal
[c] Universidade de Aveiro, Campus Universitário de Santiago, 3810-193 Aveiro, Portugal
[d] Faculdade de Ciências da Universidade do Porto, Rua do Campo Alegre, s/n 4169-007 Porto, Portugal

**ABSTRACT**

Object detection in astronomical images, generically referred to as source finding, is often the final stage of image analysis in astrophysical processing work flows. In radio astronomy, source finding has historically been performed by bespoke offline systems; however, modern data acquisition systems as well as those proposed for upcoming observatories such as the Square Kilometre Array (SKA), will make this approach unfeasible. One area where a change of approach is particularly necessary is in the design of fast imaging systems for transient studies. This requires a number of advances in accelerating and automating the source finding in such systems to cope with the huge data rates involved and trigger when necessary a complementary Virtual Observatory (VO) follow-up. CRITICAL's experience with the Software Data Processor (SDP) and the telescope manager (TM) elements of SKA allowed us to identify the biggest bottlenecks and limitations of STP algorithms; to investigate more efficient solutions for processing that may significantly automate and boost the source extraction pipelines; optimize science exploration of high impact transient events and elaborated a TM Reliability, Availability, Maintainability and Safety (RAMS) analysis. The technologies used have a wide spectrum of application from radioastronomy to space and medical imaging.

**Keywords:** radioastronomy, transients, CLEAN, multi-messaging, Big Data, RAMS

E-mail: ricardo.armas@criticalsoftware.com

## 1   INTRODUCTION

The Square Kilometre Array (SKA) is one of the most ambitious scientific projects in the first quarter of this century. The data to be produced by the SKA will have unprecedented sensitivity, spatial and spectral resolution. From the scientific point of view, the instrument will open the door to research fields in topics that include fundamental aspects of cosmology, such as the study of the distribution of matter and dark energy, the study of gravitational waves and aspects not yet demonstrated of general relativity, mapping dynamics of cosmic magnetism, etc. From a technological-industrial point of view, the design is no less impressive. Once in operation, the SKA will produce a raw data stream close to the Exabyte/month[1], with pre-processing and local compression (see [9]), and sending more than 300 Petabytes of data per year – actually, it may be almost twice this value for redundancy reasons - for a federation of Data Centres and High Performance Computing in Europe and Africa orchestrated with the major European Digital Infrastructures (GEANT, EOSC, PRACE, among others).

The basis of operation of an instrument as the SKA (an interferometer) is the van Cittert-Zernike theorem, which relates the distribution of brightness in the sky to the amount obtained after correlation of the intensity measured by two interferometer elements (the so-called complex visibility). This relation is a Fourier transform. From a practical point of view, the important aspect to be retained is that the complex visibilities measured by an interferometer provide the information necessary to determine the position and brightness levels of the various cosmic sources in the field of view of the instrument, i.e. from visibilities, it is possible in principle to obtain an "image" of objects in the sky. However, obtaining such an image is not as simple as reversing a Fourier transform. The algorithms traditionally used in radio astronomy to move from a set of visibilities to images are iterative processes based on variants of the so-called CLEAN, which is essentially a numerical deconvolution method, must be optimized depending on its intended usage. The astronomical phenomena to be observed may have respective durations or flux variations in the order of the sub-seconds (like FRBs or magnetars) or days and months, presenting also great dynamic ranges of signal. In order to meet the scientific objectives, the corresponding chains have to be run at intervals one or several orders of magnitude shorter than the phenomena to be observed - a requirement that determines the need for high performance computing beyond the current state of the art.

---

[1] 1 Exabyte = $10^3$ Petabytes = $10^6$ Terabytes = $10^{12}$ Gigabytes

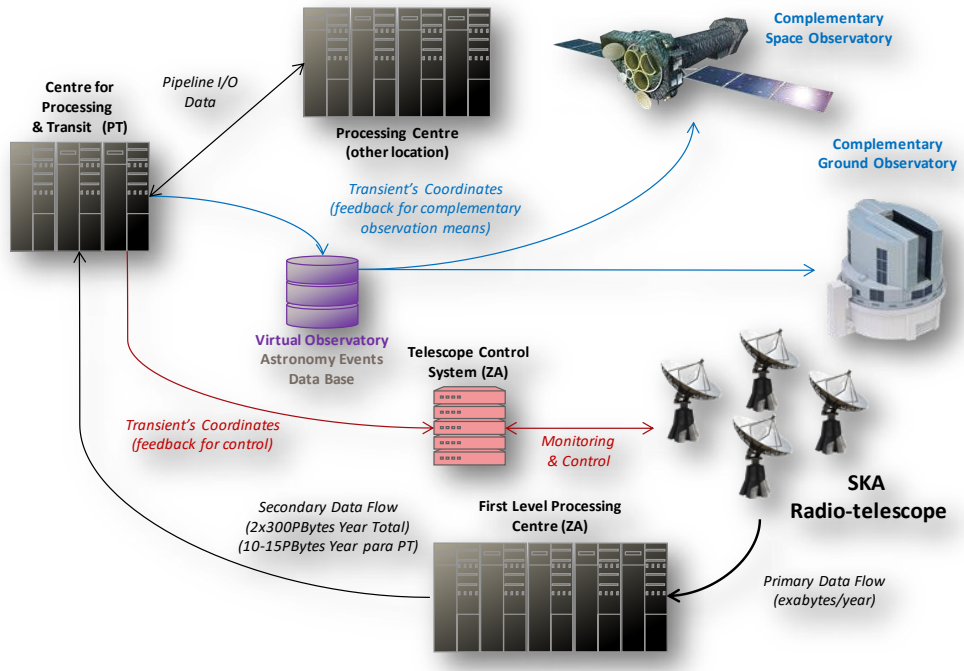

**Figure 1**. SKA Data Flows and Control Overview

Within the Science Data Processor (SDP), represented by the consortium of SDP, one of the eleven elements of the SKA project, CRITICAL has been directly involved in the development of the necessary software for the pipelines related to obtaining images from the visibilities, and its conversion into useful products for scientists.

CRITICAL is considering the specific case in which it is sought to obtain images at rapid rates (in the order of one image per second) in near real time. The need to obtain images in real time is motivated by the possibility of detecting phenomena such as novae, explosions of supernovae, jet flares; AGN accretion disk instabilities; novae, supernovae, kilonovae TDE (tidal disruption events, ie mass accretion into black holes), cataclysmic variable events, that make it necessary to change the modes of operation of the telescope to follow this particular object and study its development, as well as to provide information to complementary means such as space-based X-ray telescopes to point to regions of the sky where SKA has discovered a new phenomenon. This fast processing chain has the misleading designation of "Slow Transient Pipeline" (STP). The algorithms developed for STP thus transform temporal series of data into transient (transient) source catalogues, that is, sources whose brightness varies with respect to a global sky model. STP is therefore intended to identify and catalogue transients. CRITICAL's experience in SDP / SKA allowed us to identify the biggest bottlenecks and limitations of STP algorithms and to investigate more efficient solutions for processing.

Transients will be only a small part of the sources observed by the STP; most sources are expected to be relatively stable. Stable does not mean, however, that they do not suffer from glare disturbances due to factors extrinsic to the cosmic source, due to processes that are called scintillation. These scintillations are apparent variations of the flux density which are observed in very compact radio sources due to propagation of the signal through an ionized medium with a refractive index having irregularities. These scintillations may have their origin in the interstellar medium, the heliosphere (interplanetary medium), or the ionosphere. For the normal operation mode of the STP these scintillations are considered noise, and for the STP it is important only to define a threshold from which something is considered to be a true transient, or is simply affected by scintillation and not included in the list. But what for some is a source of noise for others this can be a valuable source of data. The various types of scintillation have distinct temporal characteristics. The most common, especially for night-time observations, will be ionospheric perturbations that are observed at the time of the order of seconds or tens of seconds. Interplanetary scintillations (IPS) due to phenomena in the heliosphere are blurred when considering integration times greater than one second, however, recent work with the LOFAR telescope, a precursor of SKA, has shown that it is possible to identify IPS even in data at two seconds, during the night and due to perturbations beyond the Earth's orbit ([1]). This means that these secondary data can be a potential source of interest for applications related to heliophysics and in particular space weather.

## 2    SLOW TRANSIENTS PIPELINE

The component for the implementation, optimization and integration of processing chains begins with the reuse/refactoring of a pipeline developed for the SKA SDP - a work re-created by CRITICAL for SDP in partnership with the University of Manchester [2]. This pipeline, shown in Figure 2, implements a slow transient processing chain and is prepared to single-node parallelisation - i.e., on a single multi-core CPU machine.

As a sum-up, the pipeline has as input the observation data from other upstream pipelines and information about the Global Sky Model (GSM). In the first step, the information contained in GSM is subtracted from the visibilities to begin isolating the transient sources that are the end result of the chain. Processing is done over several stages as shown in Figure 2, hence the name pipeline or processing chain. CRITICAL contributes to the implementation and optimization of the pipeline, resulting in source code and analysis reports, which include the architecture report and performance analysis.

The pipeline is implemented in C++11, uses the Thread Building Blocks (TBB) to parallelize it and the main mathematical libraries used are: Armadillo, OpenBLAS, LAPACK and FFTw. The pipeline includes a complete battery of tests including performance tests, which addresses the central points of the proposed solution.

The processing capability goal is to support input visibility data at a rate of about one second for $2^{16}$ x $2^{16}$ images. In tests made on a standard machine (4 cores with 32GB of RAM) this limit is respected for images of size $2^{13}$ x $2^{13}$ – i.e. images of this dimension are processed, in the current version of the prototype, in 0.75s for simple precision numerical values. A $2^{15}$ side image runs at 10.6s in simple precision. Tests have not yet been run on a 4x4 core machine - a substantially linear performance increase on a logarithmic scale is expected. For images with $2^{16}$ side, the 32GB RAM test machine is not sufficient. The effect of switching from single-precision to double-precision is roughly double the amount of memory consumption and runtime. The runtime measured from the first version of the prototype to the current version, for images of $2^{14}$, reduced from 44s to 2.89s. The optimization achieved is very significant, but it is to be hoped that further increases in performance will progressively become more difficult from now on.

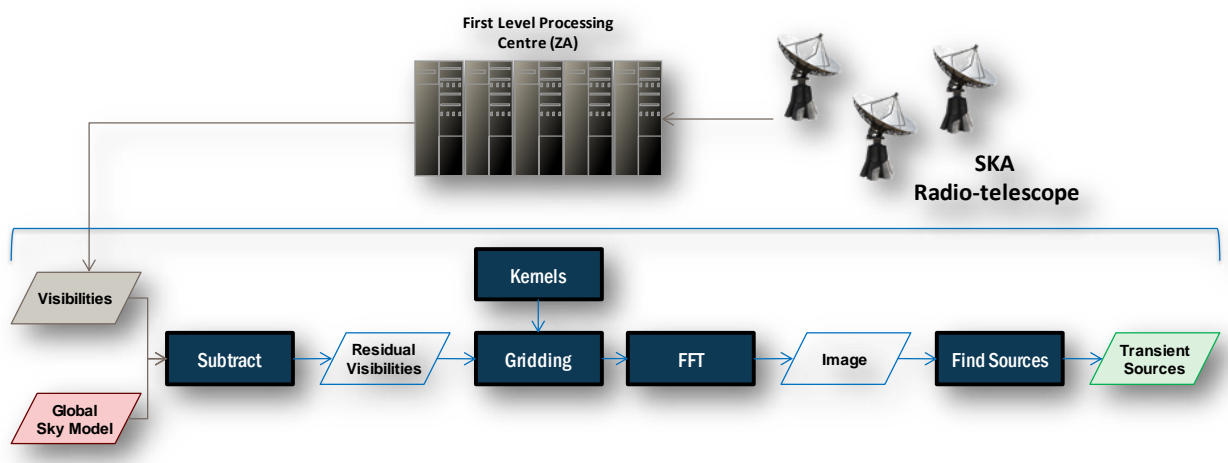

**Figure 2.**  Slow Transients Pipeline

# 3 PARALLEL COMPUTATION AND SOFTWARE OPTIMIZATION

One of the main points of the work developed in the implementation of processing chains is performance optimization. This point incorporates the use of technology for parallel computing, high-performance mathematical libraries, and a software architecture consistent with these principles. The technologies to analyse and validate in real situation include mathematical libraries such as: Armadillo, OpenBLAS, LAPACK and FFTW, and technologies of parallel and distributed computing such as TBB, OpenMP, MPI, Hadoop, Spark, etc. These technologies, even the most recent ones like TBB and Spark, are well established. Part of these technologies were used in the prototype of the Slow Transient Pipeline developed by CRITICAL SOFTWARE.

The fundamental challenge that persists is the design of software architectures for parallel computing - i.e. architectures that take full advantage of almost universally available multiprocessing capability. Concepts that previously could only be exploited having access to a supercomputer or large server are now available to virtually everyone, as illustrated below.

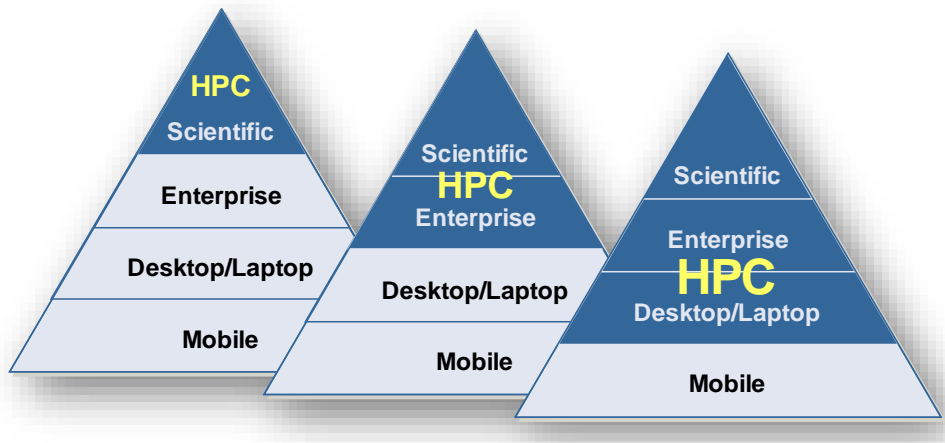

**Figure 3**. Evolution of the Ability to Explore Parallel Computing

After the emergence of multi-core processors Intel placed great emphasis on the dissemination of technology and parallel programming techniques. The TBB library, now available under the Apache 2.0 license, results from this effort. Still, the adoption of parallel / HPC techniques found few enthusiasts.

# 4 POTENTIAL MOBILIZER EFFECT FOR THE REPETITIVITY OF THE APPLICATION OF THE TECHNOLOGY TO VALIDATE/DEMONSTRATE OTHER ORGANISATIONS AND/OR ACTIVITY SECTORS

Viewed from a very restricted perspective, the first level of repeatability of the application of this technology is precisely within the SKA radio telescope. However, other system may actually benefit from such developments and be a market for such products. For instance, some of such systems are also presented in below.

**Table 1**. SKA and Recent Systems with Equivalent Needs

| Entity | System(s) and/or Target Program(s) | Application | Main Countries |
|---|---|---|---|
| SKA | SKA radio telescope | Astronomy | UK, SA, AUS, IT, NL, CHI, IN,... |
| ESA | *Copernicus, Living Planet* | Earth Observation | UK, FR, DE, SP, CAN, PT… |
| ESO | E-ELT, ALMA | Astronomy | FR, UK, DE, USA, JP,… |
| EUMETSAT | MTG, MetOp-SG, Jason-CS, etc. | Meteorology | EU… |

Despite this extremely restricted view, processing platforms and chains have high repeatability: radio astronomy and space science are currently developing strongly, with the need for high-performance processing chains in a context of great international dynamics. The catalogues of information on these transient cosmic sources produced by these processing chains are an essential product in the ingestion of information from the Virtual Observatory (VO) systems. This pan-European digital infrastructure correlates different spatial databases (managed by ESA and NASA), astronomical scans of large astronomical observatories (ESO, VLBI, Keck, SLOAN, SUBARU, ALMA, etc.) and databases of scientific literature, centred on the Strasbourg Astronomical Data Centre (CSD). The VO can trigger discovery alerts after ingestion of transient source catalogues and request spatial missions to quickly and automatically follow other wavelengths, reinforcing the impact of the discovery. The processing of information in this process is similar to that of the processing chains developed. It should be noted that the range of application of this technology is not limited to scientific fields, but also is seen in service oriented programs. For instance, we find the application of these technologies in meteorological data processing chains from an increasingly broader set of satellites and ground monitoring stations. In conclusion, several of the techniques used in STP demonstrator chains, such as gridding, are also used in fields as diverse as medical imaging [15].

Broadly, the activates performed by CRITICAL in collaboration with several partners from some SKA elements (namely SDP, TM, etc) can be summarized as:

### Slow Transients Pipeline Prototype

The development of the Slow Transients Pipeline (STP) prototype consists of two main areas:

1. Efficient implementation of the pipeline algorithms in modern C++ on Linux using appropriate coding techniques for performance optimisation leveraged by well-established and mature libraries for core functions, such as linear algebra and parallelisation; and

2. Algorithm selection and improvement by comparing different approaches to solving any given problem. Competing implementations are benchmarked for CPU and memory usage after which possible changes to the algorithms are equated to improve the hardware utilisation (max CPU, min running time, min memory). Examples of such changes include task parallelisation, memory reuse, and changing the in-memory data layouts to avoid or minimise CPU cache misses.

The original goals of the project were largely surpassed, granting an extension to the project with additional challenges.

### Systems Safety Engineering Services

The main objectives of the Systems Safety Engineering Services provided by CRITICAL Software were the following:

- Systematic audit of system requirements based on both interviews with the various international entities participating in the SKA Telescope Manager (TM) consortium and on documentation analysis;

- Perform an independent analysis of the SKA Telescope Manager (TM) element system requirements, as well as to elaborate a TM Reliability, Availability, Maintainability and Safety (RAMS) analysis.

- Systematic audit of system requirements based on both interviews with the various international entities participating in the TM consortia and on documentation analysis;

- Provide recommendations to address the issues encountered during the audit;

- Elaboration of a RAMS plan for SADT and TM elements to address the availability requirements of the telescope.

## 5 CONCLUSIONS

The convergence of Big data tools and HPC techniques paves ground for high gains of efficiency in parallelizing algorithms developed for astronomical source extraction by SKA SDP pipelines. CRITICAL's experience with the Software Data Processor (SDP) element of SKA allowed us to identify the biggest bottlenecks and limitations of STP algorithms and we have investigated more efficient solutions for processing in order to implement reliable, automated and efficient pipelines coping with vast amounts of data. Better design improves also science exploration, enhancing the discovery window of phenomena such as novae, supernovae, jet flares; AGN accretion disk; novae, supernovae, kilonovae, etc. In complement, systems engineering analysis was developed for the TM, the element controlling the operation of the SKA telescopes in order to identify failure modes and provide independent RAMS recommendations. Furthermore, some of these technologies and solutions have high repeatability: they can be used in several domains from space, meteorology and medical imaging.

## ACKNOWLEDGMENTS

DB, DM, MB, JPB, DG acknowledge support by the project Enabling Green E-science for the SKA Research Infrastructure (ENGAGE SKA), reference POCI-01-0145-FEDER-022217, funded by COMPETE 2020 and FCT, Portugal.

## REFERENCES

[1] R. A. Fallows, M. M. Bisi, B. Forte, T. Ulich, A. A. Konovalenko, G. Mann e C. Vocks, "Separating Nightside Interplanetary and Ionospheric Scintillation with LOFAR," The Astrophysical Journal Letters, vol. 828, nº 1, 2016.

[2] A. Scaife, "PDR.02.05.03 Imaging Pipeline," SKA Science Data Processor Consortium, 2015.

[3] CRITICAL Software, "Slow Transients Pipeline Prototype - C++ source," CRITICAL Software, 2017. [Online]. Available: https://github.com/SKA-ScienceDataProcessor/FastImaging.

[4] T. Staley, "Slow Transients Pipeline Prototype - Python interface implementation," 2017. [Online]. Available: https://github.com/SKA-ScienceDataProcessor/FastImaging-Python.

[5] L. Lucas, C. Lourenço, G. Almeida and A. Scaife, "Slow Transients Pipeline Prototype," SKA Science Data Processor Consortium, 2017.

[6] Intel, "Threading Building Blocks," Intel, [Online]. Available: https://www.threadingbuildingblocks.org/.

[7] C. Sanderson e R. Curtin, "Armadillo: a template-based C++ library for linear algebra," Journal of Open Source Software, vol. Vol 1, p. 26, 2016.

[8] OpenBLAS Contributors, "OpenBLAS - An Optimized BLAS Library," [Online]. Available: http://www.netlib.org/blas/.

[9] LAPACK Contibutors, "LAPACK - Linear Algebra Package," [Online]. Available: http://www.netlib.org/lapack/.

[10] FFTW Contributors, "FFTW - Fastest Fourier Transform in the West," [Online]. Available: http://www.fftw.org/.

[11] CRITICAL Software, "RAPPTOR - Radio Astronomy Pipeline Processor Demonstrator", CRITICAL Software, 2017.

[12] J. Dursi, "HPC is dying, and MPI is killing it," [Online]. Available: https://www.dursi.ca/post/hpc-is-dying-and-mpi-is-killing-it.html.

[13] R. Nijboer, "PDR.02.05 Pipelines Element Subsystem Design," SKA Science Data Processor Consortium, 2015.

[14] M. Johnston-Hollitt, C. Hollitt, S. Dehghan e M. Frean, "PDR.02.05.04 Science Data Analysis Pipeline Reference Document," SKA Science Data Processor Consortium, 2015.

[15] D. Rosenfeld, "An optimal and efficient new gridding algorithm using singular value decomposition," Magnetic Ressonance in Medicine, 1998.

# Turning radioastronomy and future SKA infrastructure into an Internet of Things (IoT) System: enabling Smart Farming and Water Quality Monitoring

José Jasnau Caeiro[a], João C. Martins[a], Domingos Barbosa[b,c], João Paulo Barraca[b,c], Diogo Gomes[b,c], Dalmiro Maia[d]

[a] Instituto Politécnico de Beja, Portugal
[b] Instituto de Telecomunicações
[c] DETI, Universidade de Aveiro, Portugal
[d] Faculdade de Ciências da Universidade do Porto, Porto, Portugal

## ABSTRACT

Radioastronomy has moved towards massive deployment of sensor arrays across wide geographical areas. The cyberinfrastructure deployed to transport and process the vast amounts of data into dedicated processing centres, presents an opportunity for piggy-back sensor applications enabling for instance monitoring of environmental parameters, important for other domains like the agribusiness or even water quality control. Furthermore, new technologies based on Internet of Things (IoT) systems are starting to be developed for applications in a number of fields that may benefit from radiotelescopes digital infrastructures. Here we present some of the work done in this area and also describe some systems being developed by the authors for the Alentejo region under the Engage SKA Research Infrastructure. A general architecture for water quality monitoring system is discussed. The important issue of computer security is considered and related to blockchain technology. Data transmission technology and protocols for IoT, micro web frameworks and cloud IoT services are also discussed as potential spin-off piggy-back applications.

**Keywords:** radiotelescopes, Sensors, Networks, Blockchain, Data Protocols, Cloud, Low Power, Smart farming, Water management, IoT, machine learning

## 1 INTRODUCTION

Radioastronomy has entered an era of increasing impact of Information, Computing and Telecommunications technologies (ICT). Motivated by the need of exquisite spatial resolution and higher sensitivity, interferometers in the radio domain spread thousands of antennas over vast areas. A particular example is the Low Frequency ARay (LOFAR), a radio interferometer and Square Kilometre Array (SKA) Pathfinder that spread initially over a large area of the Netherlands [16]. LOFAR evolved into a large international collaboration called the International LOFAR Telescope (ILT) [17] with many stations spread across Central and Northern Europe over +2000Km. The LOFAR digital infrastructure has also enabled a number of new smart sensor applications sharing the radiotelescope infrastructure. LOFAR_Argo [18, 19] was the first large scale experiment in precision agriculture in the Netherlands. They were able to predict the presence of phytophthora (a fungus affecting potato, fruits and other vegetables) plague in the crops after merging information from 150 sensor nodes, each node measuring the soil temperature and humidity. This pioneering project just showed that once sensors are deployed across an appropriate collecting digital infrastructure, the data produced can be used in many scenarios. Similarly, the rolling towards construction of the SKA Phase 1 in South Africa and Australia and its Phase 2 future expansion to SKA African Partner countries offers a window for potential piggy-back or serendipitous agro or environmental applications that may well contribute to the local socio-economic development.

Water is the main component of Earth's oceans, rivers and lakes. A large amount of Human activities is dependent of fresh water reserves. Not surprisingly, the sustainable management of these reserves is nowadays a political priority in many parts of the world. Water is a critical resource and will become even more important because of the pressure exerted by climate change that will affect Mediterranean regions. Therefore, new sustainable water management systems may benefit from Internet of Things (IoT) systems in several ways. The Alentejo region in Southern Portugal, is home to one of the largest dams and artificial lakes in Western Europe—the Alqueva dam—, that constitutes a strategic water reserve. This lake is a very important component of the irrigation system of Alentejo: it guarantees the water supply to the population and industry in the region and represents a major boost to the agribusiness, enabling a vast improvement in crop productivity. The Engage SKA (Enabling Green E-Sciences for the Square Kilometre Array[1]) scientific infrastructure in Beja is home to some of the IoT based water monitoring system pilots hereby briefly presented.

Interestingly, the Alentejo region is one of the few regions of Europe with similar characteristics to the places where the SKA telescope will be installed thus providing a good testing site for the deployment of systems connected to this project. For instance, over the last years some pilot testing of Aperture Array technologies were performed for the purpose of testing radioastronomy sensors with solar power provision to study aperture arrays power sustainability [20].

This contribution will describe how IoT can be used for several aspects of water management, namely starting form sensor networks dedicated to the acquisition of water quality and quantity related data, to aggregator microcomputer systems, security issues and centralisation of the information with further high-level processing. The importance of novel technologies such as blockchain and machine intelligence for the IoT area is stressed in this chapter. The Institute of Electrical and Electronics Engineers created a dedicated site to theme[2] and in 2015 published a report trying to define what IoT is ([1], IEEE Internet Initiative, 2015). A short definition attributed to IEEE in March 2014, and mentioned in the report, is that IoT is "*A network of items, each embedded with sensors, which are connected to the Internet*". It may be included under the broader definition of ubiquitous computing and sometimes we may consider the definitions indistinguishable. The water quality and resources monitoring systems described are limited to those that mention themselves as IoT proposals.

## 2    BACKGROUND

IoT applications for water management in rural environments tackle two main areas: irrigation water quality management and water resources management. In Portugal, information about the water quality and water resources is presented online[3]. This data is collected using traditional chemical and physical analysis from samples collected in the field. Unfortunately, due to cost issues, the data at many sites is not collected anymore. IoT systems are typically low cost and could present an alternative for the problem of updating old networks of water quality monitoring systems. During the last few years some proposals for IoT based water quality monitoring systems have appeared in the scientific literature. A short review follows.

- An architecture using web services for real time water quality data acquisition is presented by [2]. The proposal is centred around the web services concept, namely tackling the transmission of data collected by the hardware developed by the authors— a water quality sensor node using the NeoMote wireless sensing platform. A set of three different web services were implemented on three separate devices and programmed using three different programming languages. The web services are developed for the Xively IoT platform[4].

- A real-time water quality monitoring system using IoT is described in [3]. The system is designed with data collected from three sensors: pH; electrical water conductivity and temperature. The sensors collect analogue signals and use ADCs to convert to digital format. Zigbee communication modules send the data to a processing module microcontroller (LPC2148). After some processing the data is sent to a central server managed by The MathWorks, Inc., (the same company that owns MATLAB®), dedicated to IoT applications: ThingsSpeak[1].

---

1    http://engageska-portugal.pt/
2    https://iot.ieee.org/
3    http://snirh.pt/
4    https://www.xively.com/

- An online measurement and reporting system for the quality of water based on a wireless sensor network [4]. The objectives are the implementation of a low cost, IoT based system, measuring: temperature; electrical conductivity; acidity (pH) and turbidity. An Arduino Uno microcontroller is used for data acquisition and sending the processed data to a server with a Wi-Fi connection. Afterwards, the data is sent to a cloud-based server using a Google Drive spreadsheet.

- A proposal of an IoT platform with sensing, data processing and wireless communications, for remote aquatic environmental monitoring is presented by [5]. The quality of water is evaluated on its physical, chemical and biological parameters, namely: temperature; pH, dissolved oxygen, electrical water conductivity and oxidation-reduction potential. The conversion of sensor signals to sensor readings is carried out by a mote (Waspmote with the ATmega1281, manufactured by Libelium Communicaciones Distribuidas, S.L.). Data is processed by a Raspberry Pi 3 and then transmitted by Wi-Fi or Zigbee radio transmitter. The Base Station (BS) is built around a Linux based router. The authors present algorithms for characterization of a study area that they call a Hexagonal Grid-Based Survey Planner. The sampling locations are generated by adoption of a hexagonal cell decomposition and a spanning-tree based path planning algorithm is used to travel between the sampling location of interest. A normalized Online Water Quality Index is produced by the system.

- A surface water IoT system for measurement and monitoring of sensor data is presented in [6]. It uses an ARM microcontroller (LPC1768) with several sensors that measure: temperature and humidity; carbon monoxide; pH and water level. The data is saved into a data logger and sent to an Internet cloud system (IBM IoT Watson platform) for processing and analysis.

- An Intel Galileo Gen2 is used as the core controller for a water quality monitoring system, in the proposal of [7]. An Arduino microcontroller is connected to sensors for measurement of: temperature; pH and turbidity. The data collected is observed by connecting to the core controller.

- An implementation of a real-time water quality monitoring system is described by [8]. It measures temperature, acidity and dissolved oxygen. Sensors are placed underwater and data is sent using a wireless Zigbee communication to a central system. The system adopts MQ Telemetry Transport (MQTT) which is a lightweight, ISO-standard publish-subscribe protocol, intentionally developed for open and simple device communication at a premium network bandwidth and/or small code footprint. The communication nodes are built with Arduino boards. The public MQTT broker adopted for this implementation is HiveMQ[2].

- Another real time water quality monitoring system is proposed by [9]. The system measures the most typical parameters: temperature; turbidity; acidity (pH); dissolved oxygen and electrical conductivity, and also some chemical parameters, such as: sulphate; ammonia and nitrate. The sensors are supported by a WaspMote board. The system is powered by a solar panel and supports the communication of data with: LoRaWan; 4G; Wi-Fi and Zigbee protocols. The data from the WaspMote boards is transmitted to the base stations and afterwards sent from these base stations to Water Quality Monitoring stations where the concerned water authorities analyses the data and issue alerts. The IoT Azure platform[3], was chosen for the storage of data and ASP.NET for the web application.

- An efficient water quality monitoring system built using Field Programmable Gate Arrays (FPGA) is presented in [10]. An ADC is connected between the FPGA and a set of sensors that measure: acidity (pH); temperature; carbon dioxide; turbidity and water level. A Zigbee communications module sends the data from the FPGA system to a base station where the user can consult the acquired information.

- Work developed by some member of the national consortia, in collaboration with specific companies, also addressed the issue of water monitoring, by considering the development of a standards aligned (OneM2M) IoT platform for management, acquisition and processing of data from sensors. The Next Generation Smart Water Grids project [21] developed systems capable of accessing the quality of water resources, in the context of public water distribution grids. The online systems, communicating through LoRa, were deployed in pilots where they acted for early detection of degraded water quality. The sensors integrated included temperature and turbidity, acquired by novel fiber optical sensors.

---

1      https://thingspeak.com/

2      https://www.hivemq.com/

3      https://azure.microsoft.com/en-us/services/iot-hub/

# 3 WATER QUALITY MONITORING SYSTEM ARCHITECTURE

The majority of the systems described in the water quality monitoring literature follow the architecture in the diagram shown in Fig. 1. The system is divided into four major subsystems: data acquisition; data aggregator; data server and cloud server. The data acquisition subsystem incorporates the sensors, which communicate to the microcontroller the physical and chemical parameters of water under the form of electrical signals. These may be analogue or digital. In the case of analogue electrical signals, the microcontroller must have some data acquisition and conditioning electronics. The signals may be filtered and scaled and afterwards converted to a digital format by sampling with an analogue to digital converter (ADC). Many sensors have an associated circuitry that converts the analogue signal into a digital signal and permit an easy communication with the microcontroller via I2C (Inter Integrated Circuit) which is a synchronous, single ended computer bus, via SPI (Serial Peripheral Interface bus) or UART (Universal Asynchronous Receiver-Transmitter) protocol.

The microcontroller unit is a low power processing device that collects the sensor data, provides the first level of signal processing and sends the information to a data aggregator subsystem. The communication module is commonly wireless since the distances between the data acquisition subsystem and the data aggregator subsystem are usually high. Low power consumption is a concern for the data acquisition subsystem and it is common to use low power microcontrollers and low power long range communication protocols such as LoRaWan or SigFox or low power small range communication protocols such as Bluetooth Low Energy (BLE) or ZigBee.

The data aggregator provides another layer of computer processing power. It is composed of a low power microcomputer tailored for IoT applications. It usually runs a Linux based operating system with low RAM requirements and some sort of flash memory storage. The most typical example is the Raspberry Pi family of microcomputer boards. The data aggregator may perform some sort of signal processing, data encryption and temporary local storage. This subsystem will have an increasingly important role in what concerns computer security since it has enough processing power and software resources for this purpose. These microcomputers usually receive data using serial communication protocols from the microcontroller based data acquisition subsystem and send data to central data servers using the Internet. Sometimes the data aggregator subsystem sends directly the information to cloud based IoT services. It is common to use web services to send the data. Central data servers may use traditional relational databases to keep the information or NoSQL databases, such as MongoDB. In a foreseeable future, new types of databases, using some sort of blockchain representation, are to be expected.

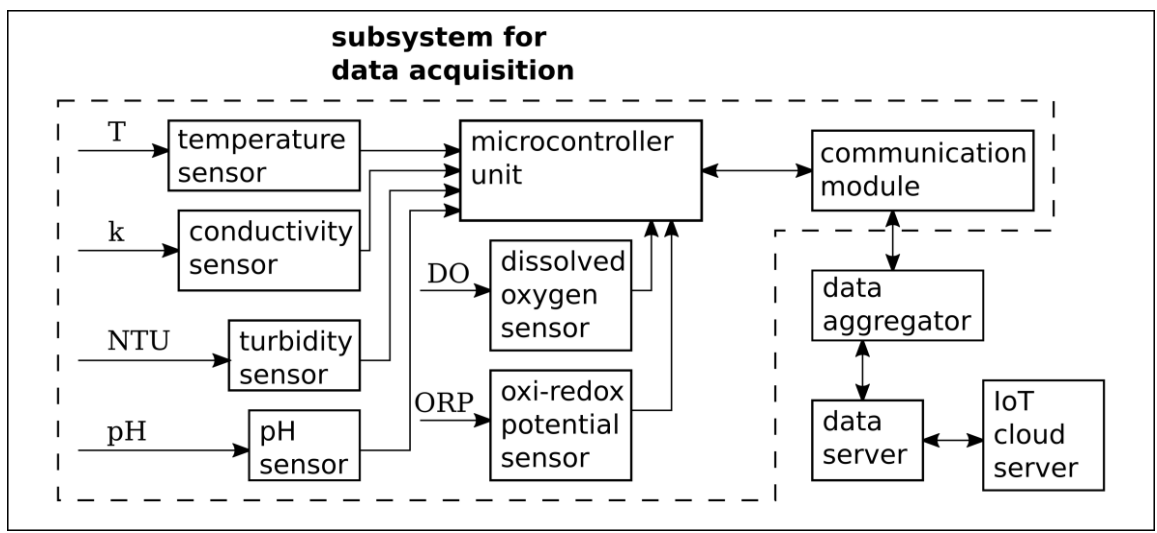

**Figure 1**. Generic architecture for water quality monitoring systems using an IoT approach. The sensors set is chosen according to the specific monitoring need of each system.

# 4 COMPUTER SECURITY IN WATER MONITORING SYSTEMS

One major concern on the IoT area is related to the computer security issues for IoT systems. Due to the connections to the Internet or to the wireless sensor networks, these systems are subject to attacks that usually plague other network connected computer-based systems. The large number of IoT devices and the low energy and low-cost requirements pose a challenge in terms of hardware protection and computer security policies.

Water is one of the most valuable and critical resources available to mankind and is usually included in the definition of strategic resources. Security is one major concern in terms of water quality monitoring systems. For a secure IoT system deployment some aspects must be considered: information privacy, confidentiality and integrity; authentication, authorization and accounting; services availability; etc.

Blockchain technology is emerging as one of the main constituents of computer security and the IoT area will certainly benefit from it [11]. A traditional approach to data security is focused on the implementation of computer security management policies for the database servers and the data transmission channels. Blockchain technology is focused on the security of the data itself. Each block of data is guaranteed not to be subject of attack therefore assuring that a large set of computer security requirements are fulfilled, namely: privacy; confidentiality; integrity, to mention a few. Since IoT devices have low computational power and storage, blockchain validation is a difficult problem. A model of the impact on block chain synchronization and wireless connectivity is studied by [12]. Another proposal for the application of blockchain technology to IoT is presented by [13]. Again, the low computing power available to IoT devices is seen as a challenge to the adoption of blockchain technology. Central to the security of blockchain is the proof of work (POW) since it is this mechanism that typically prevents the hacking of the data in the blockchain. The authors replace the POW with a distributed trust mechanism thus eliminating the high computational demand of the blockchain. The problem of control and configuration of IoT devices is studied in [14] and they propose using the Ethereum[1] blockchain platform for the device management. Smart contracts are used to write code that defines the behavior of IoT devices. The proof of concept is presented with Raspberry Pi microcomputers.

## 5     WEB SERVICES AND DATA TRANSMISSION

IoT systems may support one or more data transmission protocols. High latency, low energy resources and long transmission distances are typical for water quality management systems. These specialised data protocols are designed for, or adequately support, machine-to-machine (M2M) applications.

The first example of such a data transmission protocol, and perhaps the most widely adopted, is Message Queuing Telemetry Transport (MQTT) that enables a publish/subscribe messaging model[2]. The Constrained Application Protocol (CoAP) is an application layer protocol that is intended for use in resource constrained internet devices[3], and is being used in one of the most important joint efforts towards standardization of IoT devices, the OneM2M initiative[4]. The Simple (or Streaming) Text Orientated Messaging Protocol (STOMP) provides an interoperable wire format enabling the communication from STOMP clients with any STOMP message broker, providing easy and widespread messaging interoperability among programming languages, computational platforms and brokers[5]. A free and open standard is the Extensible Messaging and Presence Protocol (XMPP) available since 1999 and claims security features interesting for IoT[6]. Other protocols worth mentioning are: Mihini/M3DA; the Advanced Message Queuing Protocol (AMQP); the Data-Distribution Service for Real-Time Systems (DDS); the Lightweight Local Automation Protocol (LLAP); the Lightweight M2M (LWM2M); the Simple Sensor Interface (SSI); Representational State Transfer (REST); Hypertext Transfer Protocol Version 2 (HTTP/2); Simple Object Access Protocol (SOAP); Websocket, developed as part of the HTML5 initiative; etc.

Micro web frameworks such as the Python programming language based Flask web framework, provide an example of a software development platform for the creation of a web services server that runs in low power machines such as the Raspberry Pi[7]. Web sockets are implemented with the flaskSocketIO package[8]. A MQTT package for Flask is available at (https://github.com/neubatengog/FlaskMqtt) supporting the Eclipse PAHO implementation for IoT[1]. The Raspberry Pi microcomputer and the Flask micro web framework integrate easily. The Raspberry Pi gives strong support to the Python programming language development tools and packages. Flask is very light on resources and thus widely adopted in IoT systems.

---

1        https://www.ethereum.org/
2        https://mqtt.org/
3        http://coap.technology/
4        http://www.onem2m.org/
5        http://stomp.github.io/
6        https://xmpp.org/
7        http://flask.pocoo.org/
8        https://flask-socketio.readthedocs.io/en/latest/

# 6     CLOUD IOT PLATFORMS

Cloud IoT platforms provide a conceptually centralized storage for the information acquired from the water quality sensors. Several options are available with a large range of different services provided: cloud hosting; computing resources, message brokers, among others[2]. Most of companies with mature solutions are however outside Portugal, with the exception of some national telecom providers that are addressing that area, but not always following standardized approaches. Some cloud IoT providers have complete systems with dashboards, storage, and the means for batch or stream processing (e.g., Amazon), some also have System Development Kits available for several different programming languages whilst others only provide means to store and retrieve the data. Most support HTTP and MQTT protocols. IBM Watson and Google Cloud provide machine learning services that in a foreseeable future will become very important for water quality monitoring systems. Relevant for the choice of IoT cloud services is whether one has an open source license or not. Cloud IoT providers will soon face the challenge of blockchain technology and the InterPlanetary File System (IPFS) is an example for a distributed peer to peer method of storing data[3].

Still, there is a lack of computational power available in national clouds that can enable the operation of reliable IoT systems. A critical aspect of using IoT Cloud platforms is the communication latency between endpoints. Scenarios requiring real time alarms, or remote actuation of devices (e.g., water sprinklers and valves) greatly benefits from reduced latency. In some cases, there are hard boundaries for latency, limiting the use of public Cloud IoT platforms. The 5G initiative from 3GPPP and the 5G PPP, where the members of the Engage SKA have a seat in the board, already identified these challenges, and the benefit of low latency for scenarios of IoT, in particular tactile IoT, critical systems, future factories and precision production systems. In their view, which is being backed by actual solutions, trials and deployments, the cloud will burst to the network edge, so that cloud based services can be used in a wider range of applications. SKA has latency challenges to be solved, in order to provide consistent operation over thousands of kilometres, and we envision that some solutions can spill over to IoT systems.

# 7     IRRIGATION AND DRAINAGE NETWORK MONITORING SYSTEM

In the context of IoT applications for the Alentejo region, within the framework of the Engage SKA scientific infrastructure, a system is presented that monitors water drainage networks for monitoring irrigation channels in agriculture, or for monitoring an urban drainage network. The system detects the rise of water levels within the channels due to obstruction or due to heavy rainfall. Irrigation and drainage channels may become clogged which can result in serious damages and incur large amount of monetary loss. In the case of wastewater, when the regular drainage channel becomes obstructed, contaminated water is normally directed to alternative bypass channels or, in more extreme situations, it can expose the drainage system, invading open-air areas, with the resultant problems with the inhabitants, in addition to the public health problems that this situation may raise. Particularly, with the increasingly frequent situations of extreme climatic phenomena, which include high precipitation over a short period of time, it is important to have a way to get real-time information on the state and levels of water in the drainage channels. This data can also be useful in the monitoring of irrigation channels and water streams since it allows to take appropriate decisions for civil protection and to follow prevention measures to minimize the impact of these phenomena on populations or in cultures. It is known that the sooner a flood alert is issued, the less damage it will cause [15].

The usual solution for monitoring the channels is to inspect them, which in many situations implies the allocation of teams to inspect the channels on a regular basis, and in the case of sudden rainfall it is frequent to have no information at all. Therefore, a system that collects and sends information on the status of the channels, allows the taking of actions in a timely and informed way, which can avoid serious material damages, safeguards the people and cultures and, at the same time, allows the reduction of operational maintenance costs of these channels.

---

1     https://www.eclipse.org/paho/
2     https://www.postscapes.com/internet-of-things-platforms/
3     https://github.com/ipfs/ipfs

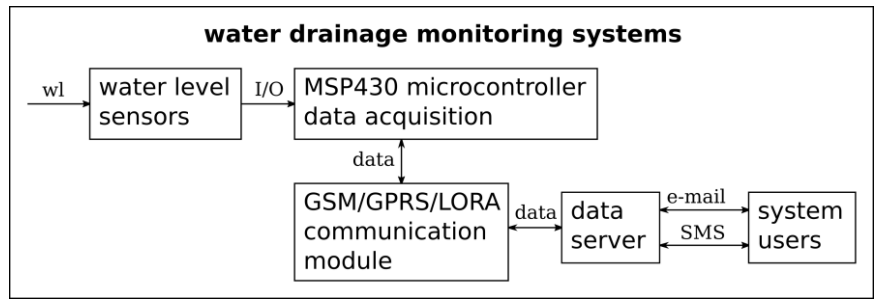

**Figure 2**. Architecture for water drainage monitoring systems using an IoT approach.

The system presented in Fig. 2 is a subset of the general architecture presented in Fig. 1 and is composed by two distinct modules. A remote module, based on a microcontroller, to which are associated level sensors and a communications module, which sends the information to a server. The second component of the system is the server application, which receives the information sent by the remote modules installed in the various critical points of the drainage channels, compiles and treats the information and makes its presentation, sending alerts to the stakeholder. The remote module is intended to monitor the drainage channels and report changes in the water level. The system presents the following constraints: since the modules can be installed in underground channels, the system needs to be exclusively powered by a battery and it is not possible to charge it by means of a solar panel, for example. Therefore, the energy consumption of the system must be low to preserve the battery charge and reduce maintenance costs.

Due to the previous restrictions the remote module is based on a low-power Texas Instruments microcontroller of the MSP430 family. The MSP430 is a 16-bit microcontroller family with limited processing capacity but with extremely low power consumption and versatile at the same time. The microcontroller is responsible for analysing changes in the sensor states and communicating with the server via SMS messages using a GSM/GPRS module. The set of sensors used are switch level sensors. For the communications module, an ITEAD's GPRS/GSM platform is used, based on the SIMCOM's SIM900 integrated circuit. Communication with this module is done using AT commands. The drawback of this module is that in the process of sending SMS messages, the current may reach a peak of 2 A. There are more suitable solutions, such as those offered by SIGFOX and LoraWAN, that allow communications over long distances with a reduced energy consumption, but which have the problem of not covering much of the territory and need access to local hubs. Therefore, GSM / GPRS communications are used in this system since it is deployment is more widespread.

The remote module software is programmed in the C language and implements several independent state machines, one for detecting changes in water levels that are triggered by interruptions sent by level switches, a state machine for management and reporting information, and another state machine that deals with communications. The various state machines place the system in a low power mode, which typically consumes a current of 0.9 $\mu A$ at a frequency of 1 MHz, which corresponds to one of the lowest energy consumption chips among microcontrollers on the market. The main goal of the server module is to receive and process the received data from the remote modules, and to notify specific user groups about change in the water levels. A full stack web framework built with the Python programming language: Django[1] is used for the data presentation. Django follows the traditional Model-View-Controller software paradigm and allows a flexible approach in terms of choice of the underlying database because it adopts an Object Relational Model (ORM) for dealing with the persistent data. The robustness, scalability, type of open source license, and a high level of reliability and availability, led to the selection of the PostgreSQL database[2].

The server module can use a GSM modem with a SIM card, or in the case of a great volume of messages the stakeholder can subscribe a SMS service from a mobile communications operator. In the case of direct use of a modem to receive SMS, a software package, like SMSTools[3], can be used to read the incoming SMS and send notifications by SMS to users.

---

1 https://www.djangoproject.com/
2 https://www.postgresql.org/
3 http://smstools3.kekekasvi.com/

The main functions of the server application are: reception and processing of data about the channels' water level received by SMS using the GSM communication technology; register and track the channels being monitored. A mapping of the positioning of the remote monitoring modules is done and the number of alerts that occurred on each channel is displayed in a map using the Google Maps API[1]. It registers and manages the remote modules: battery state, module version, communication modules, SIM cards, etc., as well as the recording and visualization of installations and maintenance performed on them. It registers and manages the maintenance interventions made to the modules and to the channels to eliminate the flood warnings. It generates reports with the statistical information about floods, water levels, for example, and channels visited and directly intervened. These reports can be personalized to display only the selected information, where the user selects the data that intended to export and select the time intervals. The server application also permits users to send messages between each other and schedule interventions, via email or by SMS. The server can also be connected to a weather forecast server so that it can predict the rainfall for the next five days and, based on history, it can also predict the channels that will flood with a given likelihood, that are shown with a red pin on the map. This forecast uses statistical data about the correlation between precipitation volume and previous flood occurrences. Due to its open source adoption, other modules can easily be added to the server application, namely the mapping of critical geographical areas in terms of floods, planning maintenance interventions and channels cleaning, a module to provide information to citizens in a city, including the sending of alerts, and the configuration and sending of automatically generated reports of several kinds.

# 8    WATER QUALITY MONITORING SYSTEM

An IoT system for monitoring the water quality of dams, lakes and ponds is presented. The monitoring of water quality is very important, especially if it is used for human or animal consumption, but also when it is used for agriculture or aquaculture. To measure the water quality parameters, it is common to send a human operator to the location and measure directly, possibly in several different places and with specific instrumentation, the values relevant to the water quality, and then return to the laboratory, download all the data to a computer and process the gathered information. Since this is a laborious procedure, the water quality is not measured with the most adequate frequency, and it can happen that certain physical and chemical parameters— for example, optimal for the emergence of given harmful algæ that demand to take a certain set of measures— are not detected on time. Again, IoT has opened the possibility to have systems that are real-time and uninterruptedly monitor the water quality, sending the acquired data to a server that manages and processes information, and extract patterns so that supervisors can take the necessary actions. This system also follows the general architecture shown in Fig. 1 but presents another subset of distinct modules as can be perceived by looking at Fig. 3.

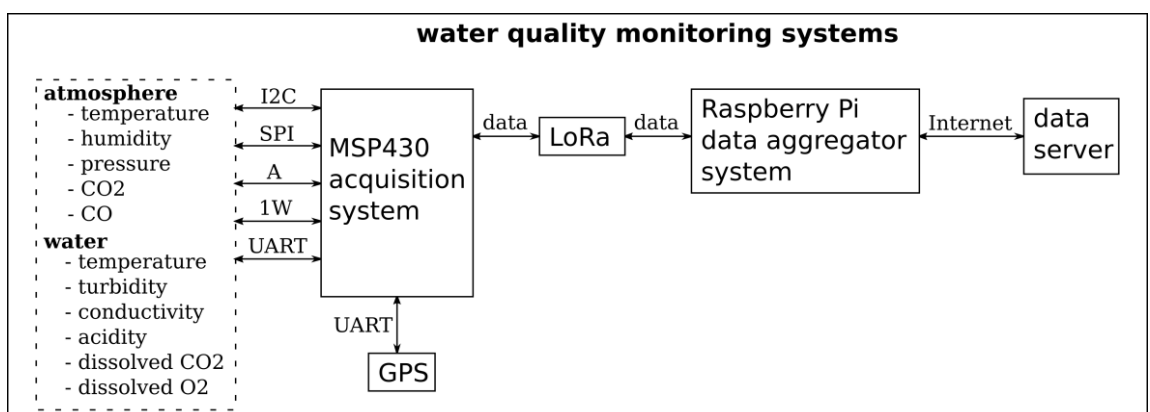

**Figure 3**. Architecture for water quality monitoring systems using an IoT approach.

The remote microcontroller based module acquires data from different types of sensors for each relevant physical and chemical parameters, that are sent to the server by the communications system. The server application retrieves the information sent by the remote modules located along the water mass and processes the information that is made available for analysis and visualization. The main functions of the remote system are the collection and transmission of the sensor data to a server where it is analysed. The sampling periodicity is flexible and adaptable to application needs. A higher sampling rate may increase the power consumption significantly. The system collects the following signals relevant to water quality: temperature of water (DS1820 waterproof sensor); pH (Atlas Scientific pH sensor); dissolved oxygen (O2) (Atlas Scientific Dissolved Oxygen sensor); dissolved carbon dioxide (CO2) (SKU: SEN0219); electrical conductivity (Atlas Scientific Conductivity Probe K 1.0 sensor); turbidity (Atlas Scientific Embedded NDIR CO2 Meter) and pressure (to measure depth) (SKU: SEN0257). Although less costly sensors are available, these offer a good trade-off between cost, performance, and accuracy.

---

1    https://developers.google.com/maps/

The remote module is complemented with sensors to measure air quality, like temperature (DHT22), humidity (DHT22), carbon monoxide (MQ-7 sensor). In many situations it is useful to have a GPS module thus providing the module with geographical positioning data. In dams and big water reservoirs, the water parameters are frequently measured at different depths, therefore the system can be equipped with a motor to lower and raise the sensors platform. Alternatively, one could opt to have several sensor groups positioned at different depths, however, since the sensors can be quite expensive, the first configuration would be preferred. The sampling interval can go from several minutes, to twice or three times a day, according to literature. This remote system is more demanding in terms of the number of sensors and interfaces than the previous one described. The microcontroller chosen to be used for this platform is again a MSP430 family microcontroller. These microcontrollers come in different varieties, and several types and number of peripherals are available, it is possible to choose the most adequate for the application with the necessary number of I/O pins, and the change between microcontrollers of the MSP430 family can be done with a minimum effort. The communications module should be chosen depending on the amount of data to be transmitted and on the terrain. Because of the amount of data to be transmitted, a low bandwidth technology, like SIGFOX, is not appropriate. Therefore, LoRa or a mobile communications operator that provides data communications is chosen. The LoRa system is particularly suited for a place where several locations are being simultaneously monitored and the installation cost of a LoRa gateway base station is justified. This communication base station is associated with the aggregation system, based on a more powerful computational platform, like the Raspberry Pi 3, that forms part of the blockchain processing system, and sends the processed data to a cloud server using a wired Internet connection or through a large bandwidth wireless mobile communications technology, like the 3G/4G mobile communications network.

Another important issue to consider for this system is the power supply. Normally, these kind of systems are installed in anchored floating platforms and they need considerable power, especially if a motor is used to measure water quality at different depths. The most appropriate power source is the use of a battery complemented by a solar panel, that powers the system and charges the battery at the same time.

The server gathers all the data from the remote modules, and saves it on a database for further processing and analysis. Besides information processing and analysis, the server can also provide a framework to manage the installation and maintenance of the remote stations. Again, a web framework to take into consideration is Django, based on the Python programming language, and for the database management system the scalability and availability of the open source database PostgreSQL is a convenient solution. The association of the blockchain with a conventional database improves the processing times. The web server application provides the following capabilities: the immediate geographically referenced visualization of the collected data in different formats like graphs but also as lists and tables for further study by external experts, with an easy tool to export those lists to different file formats like CSV, XML, JSON and PDF, for example. The server application is also able to configure alerts based on different threshold values for the relevant parameters and send it automatically to the stakeholders.

# 9 FROM THE SKA TECHNOLOGIES TO IRRIGATION AND LARGE SCALE IOT DEPLOYMENTS IN THE ALENTEJO REGION

The solution that is described mostly focus on the operational aspects found in the farming context, and the data acquisition methods, these are to be seen as the first step towards a unified monitoring solution for the Alentejo region. Considering its wide area, and the challenges presented by a dense mesh of sensors and monitoring stations, the scalability required for such system, greatly benefits from the expertise acquired in SKA. The number of sensors that are required for an effective deployment can easily surpass a thousand. When the practices for enhanced irrigation control through the use of ICT prove their value, other areas will also consider these approaches. In really, pilot trials are already being developed with focus on monitoring of individual animals. In this case, the number of devices will reach many thousands. The datarates will also be considerable, requiring the existence of a public cloud with the resources for managing all devices. While frequently works focus on the data rate as a result of sensed data and actuations, there is an actual burden to be considered due to dynamic software updates and basic telemetry. Other studies, and the expertise of our teams, have shown that simply handling metadata (connection, authentication, uptime notifications, and so on) requires dedicates resources. This is of particular relevance when considering the use of standardized approaches, such as the ones based on Lightweight M2M, or OneM2M, that provide enhanced management interfaces, complete device lifecycle management, and an object-oriented representation of devices, at the cost of complexity and overhead.

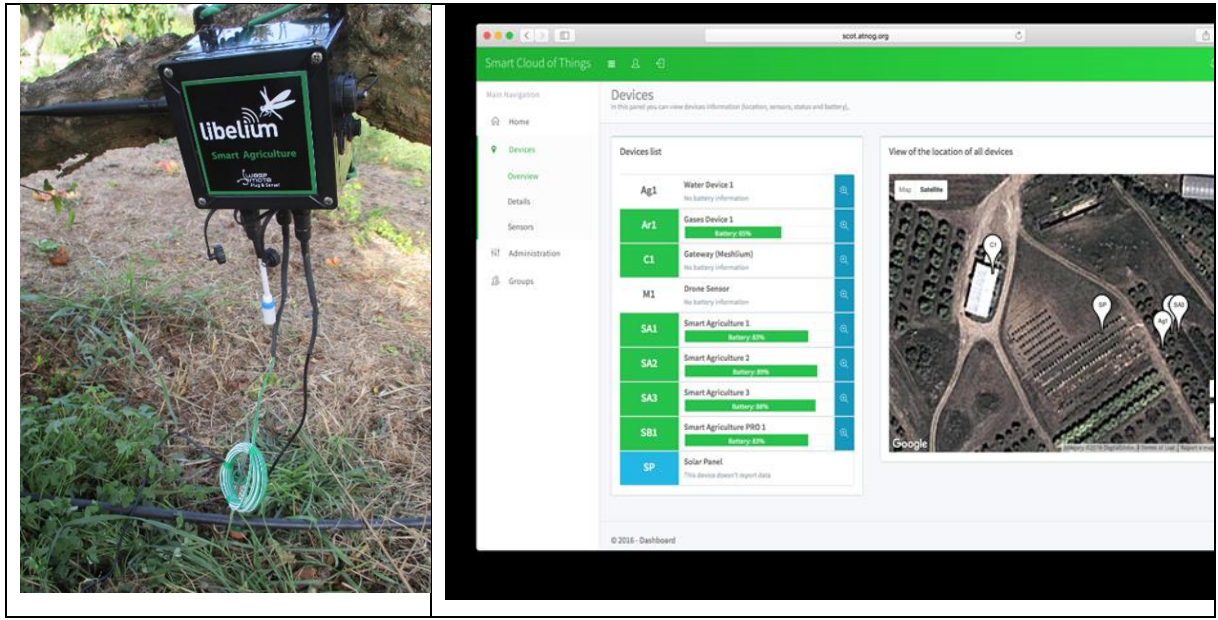

**Figure 4.** left) Smart sensor deployed at Herdade do Outeiro, IP Beja; right) Web interface for sensor control.

Data mining mechanisms, which are the current state of the art and that we envision to pursue as they have proven to have a strong value in the identification of patterns in sensed data, as well as in forecasting crop productivity and crop plagues bursts, require massive amounts of computational power. Solutions are frequently based on Deep Neural Networks, composed by hundreds to thousands of layers, and a flow that can have recursive characteristics may require specialized hardware in order to effectively provide results in near real time. These same category of dedicated devices (Intel Xeon Phi, Nvidia Tesla/Quadro, FPGAs) are broadly used in radioastronomy and SKA related projects and pathfinders. Hence, the acquired know how in integrating a large number of devices in SKA, may have a strong impact and spill over to these areas. It should not be forgotten that the Portuguese Consortia has demonstrated to have very strong skills in aspects related to critical operations and orchestration for virtualized environments, both vital for the effective implementation of large scale IoT solutions.

## 10  CONCLUSIONS

The knowledge obtained in SKA and the methods for management of extended processing capacity offer, with support for a large number of IoT devices in virtualized environments - especially if this capacity is available in public clouds - great prospects for an increasingly effective management of water, as well as crops and other related activities.

In fact, water quality and water resources monitoring and management IoT based systems present a viable low cost alternative to traditional monitoring and management methods. The introduction of blockchain technology into IoT solves a large number of computer security issues, namely those related to data tampering, i.e. critical to water quality and resources management since these are critical infrastructures. The pilot studies for Alentejo region including water resources management, namely flood monitoring are here described along a more detailed analysis of its general IoT background.

# ACKNOWLEDGMENTS

This research was supported by the ENGAGE SKA Project, [POCI-01-0145-FEDER-022217], financed by the Programa Operacional Competitividade e Internacionalização (COMPETE 2020), under the FEDER component, and by the Fundação para a Ciência e Tecnologia, I.P., under its national component.

# REFERENCES

[1] Minerva, R., Biru, A., and Rotondi, D., "Towards a definition of the Internet of Things (IoT)," tech. rep. (May 2015).

[2] Wong, B. P. and Kerkez, B., "Real-time environmental sensor data: An application to water quality using web services," *Environmental Modelling & Software* 84, 505–517 (2016).

[3] Das, B. and Jain, P. C., "Real-time water quality monitoring system using internet of things," in [*2017 International Conference on Computer, Communications and Electronics (Comptelix)*], 78–82 (July 2017).

[4] Parameswari, M. and Moses, M. B., "Online measurement of water quality and reporting system using prominent rule controller based on aqua care-IOT," *Design Automation for Embedded Systems* (Oct. 2017).

[5] Li, T., Xia, M., Chen, J., Zhao, Y., and de Silva, C., "Automated water quality survey and evaluation using an IoT platform with mobile sensor nodes," *Sensors* 17, 1735 (July 2017).

[6] Kafli, N. and Isa, K., "Internet of things (IoT) for measuring and monitoring sensors data of water surface platform," in [*2017 IEEE 7th International Conference on Underwater System Technology: Theory and Applications (USYS)*], 1–6 (Dec. 2017).

[7] Salunke, P. and Kate, J., "Advanced smart sensor interface in internet of things for water quality monitoring," in [*2017 International Conference on Data Management, Analytics and Innovation (ICDMAI)*], 298–302 (Feb. 2017).

[8] Pranata, A. A., Lee, J. M., and Kim, D. S., "Towards an IoT-based water quality monitoring system with brokerless pub/sub architecture," in [*2017 IEEE International Symposium on Local and Metropolitan Area Networks (LANMAN)*], 1–6 (June 2017).

[9] Menon, G. S., Ramesh, M. V., and Divya, P., "A low cost wireless sensor network for water quality monitoring in natural water bodies," in [*2017 IEEE Global Humanitarian Technology Conference (GHTC)*], 1–8 (Oct. 2017).

[10] Myint, C. Z., Gopal, L., and Aung, Y. L., "Reconfigurable smart water quality monitoring system in IoT environment," in [*2017 IEEE/ACIS 16th International Conference on Computer and Information Science (ICIS)*], 435–440 (May 2017).

[11] Khan, M. A. and Salah, K., "IoT security: Review, blockchain solutions, and open challenges," *Future Generation Computer Systems* 82, 395–411 (may 2018).

[12] Danzi, P., Kalør, A. E., Stefanovic, C., and Popovski, P., "Analysis of the communication traffic for blockchain synchronization of IoT devices," *CoRR* abs/1711.00540 (2017).

[13] Dorri, A., Kanhere, S. S., and Jurdak, R., "Towards an optimized blockchain for IoT," in [*2017 IEEE/ACM Second International Conference on Internet-of-Things Design and Implementation (IoTDI)*], 173–178 (April 2017).

[14] Huh, S., Cho, S., and Kim, S., "Managing IoT devices using blockchain platform," in [*2017 19th International Conference on Advanced Communication Technology (ICACT)*], IEEE (2017).

[15] Pappenberger, F., Cloke, H. L., Parker, D. J., Wetterhall, F., Richardson, D. S., and Thielen, J., "The monetary benefit of early flood warnings in Europe," *Environmental Science & Policy* 51, 278–291 (aug 2015).

[16] van Haarlem, M., et al., "LOFAR: The LOw-Frequency Array", A&A 556, A2 (2013)

[17] Vermeulen, R.C., van Haarlem, M. "The international LOFAR telescope (ILT)", 2011 XXXth URSI General Assembly and Scientific Symposium, 13-20 Aug. 2011, DOI: 10.1109/URSIGASS.2011.6051244

[18] K. Langendoen, A. Baggio, and O. Visser, "Murphy loves potatoes: experiences from a pilot sensor network deployment in precision agriculture", In 20th International Parallel and Distributed Processing Symposium, 2006, IPDPS 2006, p. 8, April (2006).

[19] J. Thelen, D. Goense, and K. Langendoen, "Radio wave propagation in potato fields", 1st Workshop on Wireless Network Measurements, Vol. 2, pp. 331–338, (2005).

[20] Barbosa, D., et al., "BIOSTIRLING-4SKA: A cost effective and efficient approach for a new generation of solar dish-Stirling plants based on storage and hybridization; An Energy demo project for Large Scale Infrastructures", Proceedings of the 7th International Workshop on Integration of Solar Power into Power Systems, 2C_5_SIW17_299, Ed. Uta Betancourt, Thomas Ackermann, Berlin, Germany, 24-25th October 2017

[21] Lúcia Bilro et al, "Next generation Smart Water Grids", Research Project, Online: https://www.it.pt/Projects/Index/1998

# Excellence and Society, Education and Dissemination

# ENAbling Green E-science for SKA - Capacitation and Sustainability of Portuguese participation in the SKA with radioastronomy as an Innovation Open Living Lab[1]

Domingos Barbosa[a], Dalmiro Maia[d], Bruno Morgado[d], Helder Ribeiro[d], Sonia Anton[b], Valério Ribeiro[a,b], Alexandre Correia[e], Miguel Avillez[f,g], João Paulo Martins[h], Joao Paulo Barros[h], José Jasnau Caeiro[h], Miguel Bergano[a], Tjarda Boekholt[a,e], Bruno Coelho[a], João Paulo Barraca[a,c], Diogo Gomes[a,c], Rui Aguiar[a,c]

[a] Instituto de Telecomunicações, Campus Universitário de Santiago, Aveiro, Portugal,
[b] CIDMA, University of Aveiro, Campus Universitário de Santiago, Aveiro, Portugal
[c] DETI, University of Aveiro, Campus Universitário de Santiago, Aveiro, Portugal
[d] FCUP, Rua do Campo Alegre, Porto
[e] CFisUC, University of Coimbra, Coimbra, Portugal
[f] University of Évora, Évora, Portugal,
[g] Technical University of Berlin, Berlin, Germany
[h] Instituto Politécnico de Beja, Beja, Portugal

## ABSTRACT

ENGAGE SKA is a Research Infrastructure of National Roadmap of Strategic Relevance fostering the development of radioastronomy and the Portuguese participation in the Square Kilometre Array (SKA) Observatory. ENGAGE SKA sets up a capacitation and sustainability plan for a Green e-Science Infrastructure fostering Portugal participation in the SKA project along the Big Data and Green Power axis: it will act as a driver for a smart and sustainable growth taking radioastronomy as an Innovation Open Living Lab. SKA is a global, unparalleled project not only because of its ambitious scientific goals, its overwhelming infrastructure (spreading through 3 continents, capable of outputting more data than the entire World Internet traffic) but also because it will be built through phases, each giving the opportunity for transformational science with cutting-edge technology. In 2016, the European Commission classified SKA as an ESFRI[2] Landmark project and Portugal has meanwhile secured key contributions to SKA consortia, that translate in a unique opportunity to effectively participate in the construction and the scientific exploration phases. ENGAGE SKA is involved in three main SKA related domains: (1) scientific research with SKA pathfinders and precursors, pipelines and simulations for multi-messenger astrophysics, (2) SKA design by participating in SKA Design and Pre-Construction Consortia (3) Innovation activities by exploring science and technology test-beds of moderate to high technology readiness level (TRL), including remote sensing, Internet of Things (IoT) applications and renewable power sources for SKA. With ENGAGE SKA we promote national cohesion by (a) connecting in a network three NUTII regions, namely Centro, Norte and Alentejo, foster collaboration with the Autonomous Region of the Azores, the national ICT TICEP.PT cluster (b) congregate know-how from different communities: theoreticians and observational astronomers with different backgrounds from solar physics and planetary sciences to cosmology, scientists, engineers and industry from Information and Computing field, from Power and Renewable Resources areas; c) aligning its activities to major national programs like INCoDE 2030 and SPACE.

**Keywords**: radio-astronomy, information and communication technologies, digital infrastructures, sustainable energy, smart farming

---

[1] https://engageska-portugal.pt/
[2] European Science Forum on Research Infrastructures

# 1 INTRODUCTION

SKA is an unparalleled project not only because of its ambitious scientific goals, its overwhelming infrastructure but because it will be built through phases, each giving the opportunity for transformational science with cutting-edge technology -- and that is quite unique! Since 2012, the EU Parliament has recommended radio-astronomy as a focus area for H2020 and in 2016, the European Commission (EC) granted SKA with the ESFRI Landmark status. ENGAGE SKA just fits this evolving global picture of radioastronomy by focusing national efforts towards critical mass aggregation and leveraging on new e-science skills in a close partnership with the dynamic ICT industry landscape. At each stage, the requirements of the SKA will push toward new research areas in e-Science, data handling, digital electronics, sustainable energy systems. These areas are central to several societal challenges, as defined in international framework programs like Horizon 2020, the near future Horizon Europe, its national counterpart framework program focusing on Innovation, Portugal 2020, and more recently on national flagship programs like the INCoDe2030 focusing on Information and Communication Technologies and the exploration of synergies for Atlantic interactions.

ENGAGE SKA aims to accomplish excellence research through the creation of a national Infrastructure addressing some of the most important quests of Fundamental Science (The Universe's history and Dark Content, the quest for Life Emergence), Skill Training and promotion of qualified Employment through an Industry-Academia partnership (big companies and SMEs), Technological Leadership aggregating Innovation in ICT, Energy, Advanced Manufacturing and contributing to respond to P2020 and H2020 Societal Challenges, promoting the Sustainability and Energy Efficiency of Mega-Science Infrastructures and placing Europe in a changing world through connections to a wider global context, including Asia, North America, Oceania and emergent Africa through Innovation and welfare creation.

ENGAGE SKA represents an opportunity to foster radioastronomy to higher levels, to pursue the design, deployment, and operation of the necessary research infrastructure for future SKA science at a National and European level and in close coordination with the SKA project, the SKA host and partner countries, and other international partners. The plan is to develop hubs for real-scale S&T testing in Portugal, that involve (1) Development of local skills, including prototype demonstrators and prepare for the news paradigms of multi-messaging science with a background on Information Technologies 2) contribute and upgrade of PT radio-astronomical facilities, compute and demonstration platforms to world-class levels(3) Scientific and engineering participation in SKA Design and Construction Phase endeavors with industrial leadership coupling ICT and Green Power, with smart agro-environment sensing spin-offs. Further, Engage SKA also collaborates with the Azores for the enhancement of the VLBI[1] RAEGE[2] cluster node.

# 2 THE PLAN

EngageSKA main scientific objectives are closely coupled with the SKA project. The scientific issues tackled throughout this organization are of two orders: engineering-oriented research, and radio-astronomy & astrophysics oriented research. Specially the former will be closely connected to the design of the SKA itself via the interfaces set from early participation of the team in FP7-funded proposals like PrepSKA and BIOSTIRLING 4 SKA, the H2020 AENEAS proposal for the SKA e-Infrastructure, as an Integrated Team in several SKA Pre-Construction Consortia and now contributing to the SKA Bridging period prior to Construction.

EngageSKA congregates know-how from different communities: theoreticians and observational astronomers with different backgrounds (e.g., solar physics, planetary sciences, transients, interstellar medium, galactic structure, galaxy evolution, cosmology and computational astrophysics), scientists, engineers linked to the industry of Information and Computing fields, and Power and Renewable Resources areas. Ultimately, our ambition is to ensure the scientific community has the resources it will need to achieve the truly transformational science potential of the SKA, implement demo pilots relevant for SKA in national soil, and anchor data curation in facilities in Portugal, in partnership with other Research Infrastructures from the national Roadmap like the RCTS (led by the national NREN FCCN/FCT) and the UC-LCA High performance Computing PRACE node.

Some of the objectives outlined in the Plan of Action are the following:

Radio-astronomy & Astrophysics: (a) time resolution astronomy & pulsars and all-sky continuum radio surveys (b) space weather observations (c) computational astrophysics and high performance computing (d) installation and operation of SKA prototypes (e) space situational awareness applications and facilities, using radio-astronomy tools for the SST and VLBI programs in Azores.

---

[1] Very Large Baseline Interferometry
[2] Rede Atlântica de Estações Geo-Espaciais

Engineering: (a) Pioneer self-sustainable green energy systems for SKA Aperture Array units (b) lead scientific imaging and associated digital hardware requirements for the SKA design (c) develop efficient software and firmware techniques to exploit SKA pathfinders observations (d) design and trial efficient cloud processing and storage techniques for swift and energy-efficient processing of the SKA pathfinders data, and management of Telescope facility operations.

Consolidation of the Cooperation agenda: members of our RI were founding members of the African European Radioastronomy Platform (AERAP); ENGAGE SKA is core to the Doppler[1] project supported by the FCT and the Aga Khan Development Network; participate in H2020 AENEAS project (Advanced European Network of E-infrastructures for Astronomy with the SKA) that will design and specify the distributed, European Science Data Centre (ESDC) to support the pan-European astronomical community and more broadly the science, technology and industries to achieve the scientific goals of the SKA.

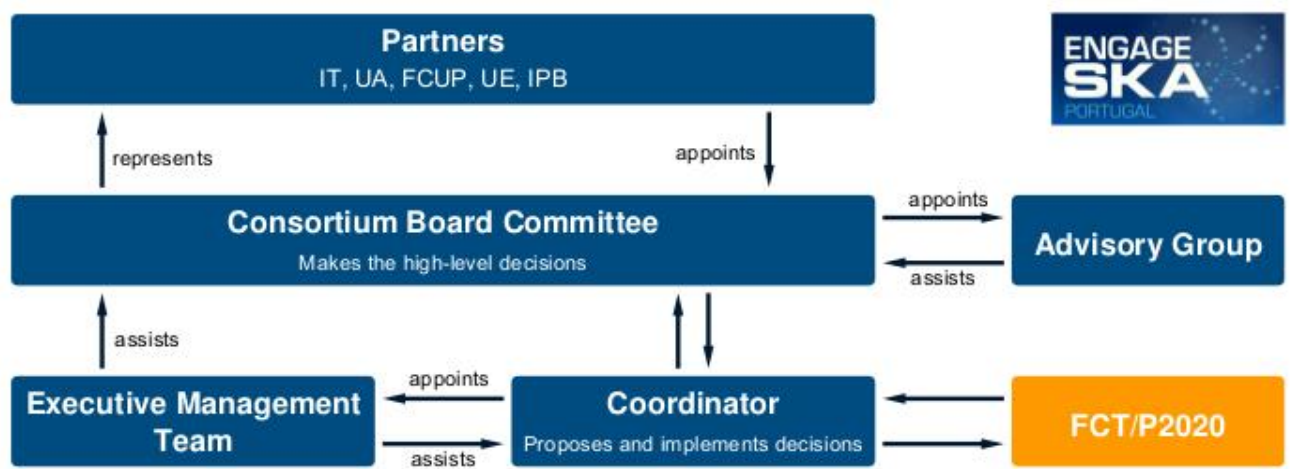

**Figure 1.** ENGAGE SKA Organization structure

ENGAGE SKA goals, if successful, will have an impact in different areas:

**Academia:** first contributing with their solid backgrounds in this field, will benefit from the infrastructure to improve their research profiles and visibility (SKA is an iconic project in the world), through consolidation of their presence in SKA Science Groups, creating critical mass and enabling future access to SKA data, thus maximizing scientific production. Furthermore, emphasis on computing and data curation of SKA data and ancillary projects in Portugal fills a gap currently in the astro science community, enhancing the required formation of skilled scientists with a strong know-how on Information Technologies and the industrial trends, specifically on Big Data processing technologies for large scale infrastructures. This infrastructure, with its Global impact, can attract researchers from other countries, in order to develop top level research in Portuguese universities and institutes.

**Industry:** (1) Telecommunications: the evaluation of novel architectures to meet the major goals of Cloud Computing, convergence scalability and upgradeability without the need for the deployment of large field trials. The use of the infrastructure to get fast results may have impact on the deployment strategies and spur new business models. (2) Equipment manufacturers: additional and reliable benchmark for the development of solutions targeting including Solar Energy systems and Smartgrid that are building blocks of the major sustainable systems, including Science infrastructures. The technologies required for SKA include major improvements to the state of the art and present clear application to many other scenarios (3) Service providers, who will have the opportunity to tune the main characteristics of their offered services according (ie, Smart Grid software and critical/risk analysis for fail-proof systems; PV testing; Cloud & Network software providers; high-performance circuits design; Internet of Things (IoT) sensor services for farming; equipment for new service programs like the Space Survey and Tracking (SST) using radioastronomical techniques).

---

[1] DevelOpping PaloP KnowLEdge on Radioastronomy

**Economy:** sourcing on the Innovation Agenda in partnership with the cluster TICE.PT for regional and national development: (1) sensor applications and radio-astronomical techniques for Earth monitoring enabling Forest management agribusiness. (2) SKA working package promote and support national ICT companies to develop and supply the building blocks for the SKA cyberinfrastructure. (3) regional data centers, at the cornerstone activities of data-mining & testing for mega-science, create revenue and qualified training (4) high performance heterogenous computing using CPUs and Coprocessors for SKA prototyping and Big Data handling (5) spinoffs and deployment of parallel computing solutions to small and medium service companies. (6) Astro tourism spin-off using ENGAGE SKA facilities as demo regional programs.

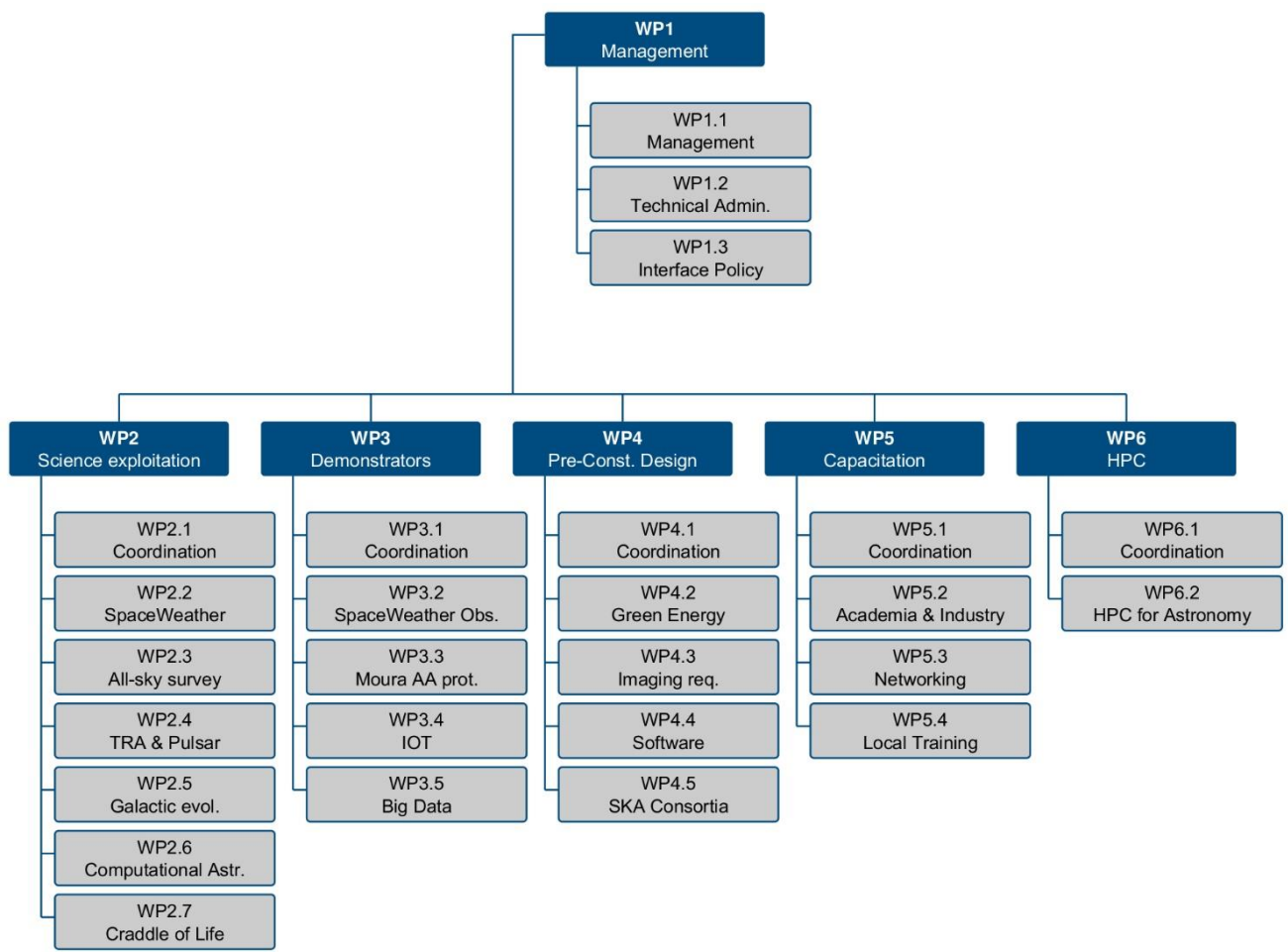

**Figure 2** . Operational WBS - Capacitation and Activities.

Assets:

Engage SKA aggregates a number of important assets that will be further enriched by the Infrastructure and will impact SKA national participation, besides contribution to the formation of a new breed of scientists and engineers. EngageSKA has some of the Portuguese radio-telescopes for advanced training in radio-astronomy and radio-frequency engineering, computing centers, data curation, IoT pilots, below are some of the resources:

- @ 9-meter radioetelescope (IT), 2.5-meter dish(IT) , 5-meter radio-spectroheliographic (FCUP)

- @ ICT facilities, currently involved in Pre-Construction/Bridging activities: "ALASKA"-IT based PaaS (300 cores); GPU based PaaS (FCUP,5000 cores NVidea).

- @ HPC node with dense Intel Xeon technologies with more than 10000 cores, to be included when commissioned in the national ICT strategy, contributing to the HPC AIR Data Centre network, as expected under INCoDE 2030.
- @ IoT pilots for remote sensing and agro-environmental monitoring (IT, FCUP, IPB)
- @ Equipped laboratories (all partners) enabling system integration and testing
- @ Aperture Array sensor technology prototypes - Field testing platforms for SKA (Herdade do Outeiro, Contenda).

# 3 CONCLUSIONS

ENGAGE SKA is part of the Research Infrastructures National Roadmap. As the only astronomy RI in the national Roadmap, ENGAGE SKA aims to foster radioastronomy to a new level, increasing its footprint in society and promoting advanced formation of a new breed of scientists and engineers with strong ICT skills, closely aligned with European and national programs like Horizon 2020, INCode2030, Atlantic Interactions Agenda. ENGAGE SKA has some of the Portuguese radio-telescopes for advanced training in radio-astronomy and radio-frequency engineering, computing centers and big data curation. ENGAGE SKA congregates know-how from different communities: theoreticians and observational astronomers with different backgrounds (e.g., solar physics, planetary sciences, transients, interstellar medium, galactic structure, galaxy evolution, cosmology and computational astrophysics), scientists, engineers sourcing on the industry of Information and Computing fields, and Power and Renewable Resources areas. Ultimately, our ambition is to ensure the scientific community has the resources it will need to achieve the truly transformational science potential of the SKA and contribute to the development of Portuguese industry and assets.

## ACKNOWLEDGMENTS

Enabling Green E-science for the Square Kilometre Array Research Infrastructure (ENGAGESKA), POCI-01-0145-FEDER-022217, funded by Programa Operacional Competitividade e Internacionalização (COMPETE 2020) and FCT, Portugal.

## REFERENCES

[1] Portuguese Roadmap of Research Infrastructures 2014-2020, FCT 2014, Available at: https://www.fct.pt/apoios/equipamento/roteiro/2013/docs/Portuguese_Roadmap_of_Research_Infrastructures.pdf

[2] Strategy report on research infrastructures; Available at: http://roadmap2018.esfri.eu/media/1066/esfri-roadmap-2018.pdf

[3] Barbosa, D., Paulo, C., Ribeiro, V., Loots, A., Thondikulam, V.L., Gaylard, M., van Ardenne, A., Colafrancesco,S., Bergano, J., Amador, J., Maia, R., Melo, R., "Design, Environmental and Sustainability Constraints of new African Observatories: The example of the Mozambique Radio Astronomy Observatory", Proceedings of the URSI BEJ Session 'Large Scale Science Projects: Europa-Africa Connects', IEEE Africon 2013 Conference Mauritius (9-12 Sep) 2013, IEEE Xplorer, Nov 2013, arXiv:1311.4464

[4] Bourke, T. L. et al. "Advancing Astrophysics with the Square Kilometre Array," in Proceedings of Science, 14 (2015).

[5] Barbosa, D. et al. "BIOSTIRLING-4 SKA: A cost effective and efficient approach for a new generation of solar dish-Stirling plants based on storage and hybridization; An Energy demo project for Large Scale Infrastructures," Proceedings of the 7th International Workshop on Integration of Solar Power into Power Systems, arXiv: 1712.03029 (2017).

[6] Duarte, L., **Teodoro, A., Maia, D., Barbosa, D.** "Radio Astronomy Demonstrator: Assessment of the Appropriate Sites through a GIS Open Source Application," ISPRS International Journal of Geo-Information, 5, 209 (2016).

[7] "BIG SCIENCE: What's It Worth?", Science/Business Publishing Ltd, (2015)

# DOPPLER: Development of PALOP Knowledge in Radio Astronomy

Valério A. R. M. Ribeiro[a,b], Antonio Batel Anjo[c,d], Domingos Barbosa[b], João Paulo Barraca[b,e], Diogo Gomes[b,e], Miguel Bergano[b], Dalmiro J. F. Maia[f], Maria Luísa Bastos[f], Neftalí Sillero[f], Ana Cláudia Teodoro[f,g], Mário Cunha[f], José Alberto Gonçalves[f], Sonia Anton[a], Alexandre C. M. Correia[a], Valente Cuambe[h], Dinelsa Machaieie[h], Claudio M. Paulo[h], João Fernandes[i,j,k], Nuno Peixinho[i,j,k], Teresa Barata[i,j,k], Vasco Lagarto[l]

[a] CFisUC, Department of Physics, University of Coimbra, 3004-516 Coimbra, Portugal
[b] Instituto de Telecomunicações, Campus Universitário de Santiago, 3810-193 Aveiro, Portugal
[c] OSUWELA, Avenida Patrice Lumumba N 854, Maputo, Moçambique
[d] Departamento de Matemática, Universidade de Aveiro, Campus Universitário de Santiago, 3810-193 Aveiro, Portugal
[e] Departamento de Electrónica, Telecomunicações e Informática, Universidade de Aveiro, Campus Universitário de Santiago, 3810-193 Aveiro, Portugal
[f] Faculdade de Ciências da Universidade do Porto, Rua do Campo Alegre, s/n 4169-007 Porto, Portugal
[g] Instituto Ciências da Terra (CT), polo FCUP, Rua do Campo Alegre, s/n 4169-007 Porto, Portugal
[h] Departamento de Física, Universidade Eduardo Mondlane, Avenida Julius Nyerere 3453, Maputo, Moçambique
[i] Departamento de Matemática, Universidade de Coimbra, Rua Larga Edifício Faculdade de Medicina (R/Ch. Esq.), 3004-504 Coimbra, Portugal
[j] Geophysical and Astronomical Observatory of University of Coimbra, 3040–004 Coimbra, Portugal
[k] Centre for Earth and Space Research of University of Coimbra, 3040–004 Coimbra, Portugal
[l] TICE.PT, Instituto de Telecomunicações, Campus Universitário de Santiago, 3810-193 Aveiro, Portugal

## ABSTRACT

The International Astronomical Union through its strategic plan "Astronomy for Development" actively promotes the use of astronomy as a tool for development by mobilizing the human and financial resources to connect science with economic growth and cultural change in society. This became particularly relevant for Africa because, in 2012, an international panel of astronomers awarded the bulk of the Square Kilometre Array (SKA) project to be hosted in Southern Africa. Undoubtedly, the SKA will play a major role on the African scientific renaissance. Although the participation of the African Partner countries – Botswana, Ghana, Kenya, Madagascar, Mauritius, Mozambique, Namibia and Zambia – is not expected until the mid-2020s, efforts are required - *right now* - to transfer knowledge, technology and develop the necessary skills in radio astronomy and associated fields. Hosting part of a large scientific infrastructure like the SKA, require major developments in cyberinfrastructures, big data, and renewable energies, which represents a major asset for improving Quality of Life (QoL). Furthermore, DOPPLER is a partnership between Portuguese and Mozambican institutions to foster ongoing collaborations in particular, those dealing with Earth Observations, matching similar African-Europe capacity building projects targeting Anglophone countries in the region. DOPPLER includes initiatives to further those ongoing endeavours, with advanced training on areas such as biodiversity, food security, and resource management. Of most importance, DOPPLER will foster science, industry and government linkages through training and knowledge transfer. This places DOPPLER partnership in recognition of a number of United Nation Sustainable Development Goals. DOPPLER is funded by the Aga Khan Development Network and the Fundação para a Ciência e a Tecnologia.

**Keywords:** Radio astronomy, Science cooperation agreements, Knowledge in education for all, Sustainable development, value chain, durable solutions, inclusion and education, Technological capabilities, Inclusive economic growth

## 1      INTRODUCTION

The Square Kilometre Array (SKA) will answer some of the most fundamental physics question, such as whether or not Einstein´s theory of General Relativity is correct (see, e.g., contributions in [1]). However, the SKA will also impact on a number of areas which will require developments in cyberinfrastructures, big data, and renewable energies, which represents a major asset for improving the Quality of Life (QoL). Through the ENGAGE SKA consortia, part of the Portuguese Roadmap of Research Infrastructures of Strategic Relevance, a strategic plan to integrate activities which will have societal impact, a major aim is for human capacity and a sustainability plan for Green e-Science Infrastructure which places Portuguese participation in the forefront of the SKA as an Innovation Open Living Lab. For example, ENGAGE SKA contemplates issues relating to the use of solar energy (following [2,3]) and remote sensing observations and models in order to choose sites for future radio telescopes [4]. The latter, in particular, has spin-off potential for, for instance, smart farming knowledge transfer using a network of sensors and geographical information system (GIS). This approach, where we pursue excellence in science while promoting Sustainability and Quality of Life, will be further expanded and developed in DOPPLER for the specific circumstances found in Mozambique.

Africa has embraced the endeavour of hosting a major astronomical world class large scale infrastructure with an expected lifetime of several decades. This endeavour requires a sustainable effort of building and preparing a new generation of scientists, engineers and industries deeply rooted into the development paths of the local economies. As an example, South Africa in the last decade, or so, has invested massively on student training – both in science and engineering - in preparation for a number of astronomical projects happening on the ground. First with the creation of the National Astrophysics and Space Programme (NASSP), honours and MSc, and then the SKA Bursary program, which has given opportunities to a number of students from other African countries to pursue studies in astronomy, astrophysics, cosmology and engineering. The expectation was that the students would return to their countries to develop astronomy and contribute to local development paths, either through research and teaching or through entrepreneurial engagement with local industry and services. This has largely worked, with the caveat though, that, unfortunately, a large fraction of these students returned to a heavy teaching load and any research activity soon disappeared. Perhaps, a primary reason for this is the lack of sustainability to a "*research culture*" in a large fraction of African countries. DOPPLER will take advantage of existing collaborations in order to fortify the research culture within it and align it with a socio-economic development impetus.

## 2  CONTEXTUALISING MOZAMBIQUE

Over the last two decades, Mozambique has been experiencing an accelerated economic development and social changes, which has been accompanied by a significant increase in the demand for skilled workforce. However, Mozambique has not fared well on a number of indexes from organisations such as the World Bank, the Southern African Development Community and UNESCO. In fact, one of the hardest hit sectors has been education (see, e.g., [5-7]). This undoubtedly affects all sectors and subsectors of the economy. The population of Mozambique is ~26,5 million people with 52% women and just over twice as many people living in rural areas as urban areas.

In order to metigate some of the shortfalls in education, a number of Information and Communications Technology (ICT) educational projects have been running successully in Mozambique. Particularly relevant change making these ICT projects successful was Mozambiques creation of a Science and Technology Policy in 2003. The policy aimed at brining Mozambique in line with developments in science and technology and keep pace with the fast economic growth. One such project which targeted ICT was Pensas, a joint collaboration between the University of Aveiro, the Portuguese Cooperation and the Mozambican Ministry of Education (see [8]). Pensas developed the concept of "dynamical schools" which linked a number of major schools, in the provinces capitals, with satellite schools and installed the Pensas Network Insfrastructure – servers, computers and software. Pensas developed EquaMat@Moz, Mozambique´s first ICT project for Education. Pensas run with great success and adoption from 2003 to 2013. In 2017, Osuwela took the challenge to re-ignite EquaMat@ and designed a new, simpler and more intuitive platform – developed locally.

**Astronomy in Mozambique**

Astronomy in Mozambique has enjoyed in the last decade a steady increase in interest (see, e.g., [9]). The Universidade Eduardo Mondlane (UEM) has been spearheading the capacitation of human resources in Astrophysics, particularly in radio astronomy primarily driven by the SKA. Currently there are 3 lecturers, with or about to obtain their PhDs, among a number of Mozambican students who have supported activities surrounding astronomy. In order to introduce and bring astronomy to the masses, the Department of Physics at UEM has been host to a number of projects and international initiatives including, the Joint Exchange Development Initiative (JEDI), the Radio Astronomy for Development in Africa (DARA), the Galileo Teacher Training Program (GTTP), the Portuguese Language Office of Astronomy for Development (PLOAD) and Southern African Regional Office of Astronomy for Development (SAROAD), Universe Awareness (UNAWE). All these aim to establish work partnerships, strength and capacitate the local astronomical community. More recently, a new Astrophysics Laboratory was installed, in partnership with South Africa, with 20 computers which are to be used to support astronomy courses within the Department of Physics. However, these efforts still require focused efforts to strengthen connections to industrial and infrastructure development, where most priority investment is being sought.

# 3 WORK PLAN

The SKA is expected to contribute significantly to the development of alternative energy and communication technologies as well as data processing, science visualisation and imaging, pioneering technological innovation. DOPPLER aims at providing the collaboration needed to make the best use of the opportunity provided by the SKA and make use of the fact that the research units involved in DOPPLER are active in other fields that are important for QoL and promote inclusiveness and impact across Mozambique, without being confined to the urban areas like Maputo. In particular, DOPPLER will collaborate with Mozambican teams of scientists and engineers to promote the importance of science and technology education and to foster and/or complement collaborative Europe-Africa (PALOP-PT) networks in science and engineering, with industry support whenever possible, on the African continent. The organisation of workshops, training sessions and staff mobility – both MZ-PT and PT-MZ, is mandatory for a lasting development of long term research partnerships and foster industry-specific training. It is well known that a major impact of large scale projects lay with the Human Capital Development. The highly skilled people trained through these projects, only a smaller fraction enter research [10—12], while the vast majority are highly attractive to industry and services, therefore enriching society and contributing to innovation and socio-economic development.

A project such as the SKA can create a long lasting societal development path by inspiring individuals, the research communities, industrial partners and governments to be part of a global enterprise which will last for generations to come. Furthermore, the profit and benefits to all those involved will be realised over a long timescale and in the broadest sense, promote capacity and pride for those engaged. It is important to stress that large scale science projects are the most productive on realising major indirect benefits of science, industry and government linkages. These benefits include:

- Production and dissemination of new scientific information, including synergies with commercial applications.
- Training skilled graduates.
- Supporting new scientific networks and stimulating interaction.
- Expanding the capacity for problem-solving.
- Producing new instrumentation and methodologies/techniques.
- Creating new firms and jobs.
- Providing social knowledge.
- Access to unique facilities.
- Pride and self-confidence: Being an active part of the world stage of Science.

In summary, DOPPLER is also in recognition of the United Nations Sustainable Development Goals (see Table 1) and the skills for the 21st century economies [13].

| | | How? |
|---|---|---|
| 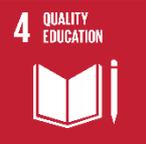 | ENSURE INCLUSIVE AND QUALITY EDUCATION FOR ALL AND PROMOTE LIFELONG LEARNING | DOPPLER will work in close collaboration with UEM and Oswela to utilise the existing relationships to further develop long-term educational programs. Furthermore, Osuwela will provide logistical support and the training platform in rural areas. We aim for local partnership in order to target specific local issues while also promoting education. |
| 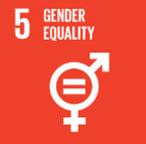 | ACHIEVE GENDER EQUALITY AND EMPOWER ALL WOMEN AND GIRLS | A number of projects and activities which focus on women and girls. The advanced training will be crucial to foster the formation of women and retain women in sciences. Where possible emphasises on the contributions of women will be made. |
| 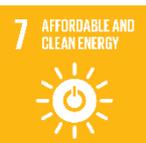 | ENSURE ACCESS TO AFFORDABLE, RELIABLE, SUSTAINABLE AND MODERN ENERGY FOR ALL | DOPPLER, through ENGAGE SKA, has competencies in delivering solar energy projects. World Pilot initiatives like Energy for SKA (E4S) through which Projects such as BIOSTIRLING-4SKA (B4S) [3] are a first demonstration, can show small-scale demo pilots can produce 9kW power – enough to power a small village. Through our industry-specific training we may be able to investigate the logistics behind introducing such a project. UEM has also a strong renewable energy group which collaboration should be explored. |
| 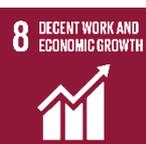 | PROMOTE INCLUSIVE AND SUSTAINABLE ECONOMIC GROWTH, EMPLOYMENT AND DECENT WORK FOR ALL | The SKA as a mega project, perhaps the largest basic science endeavour in Africa, will have a very long-life spam. Therefore, the training of specialised groups of people will ensure that not only will there be capacity to run the SKA in the long-term but also provide a platform for high-level training for those who wish to pursue a career in other sectors of the economy. More immediately, funding for MSc students and mobility of Mozambican scientists will aid in furthering the skill base and personal development [11—13]. |
| 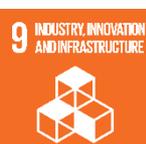 | BUILD RESILIENT INFRASTRUCTURE, PROMOTE SUSTAINABLE INDUSTRIALIZATION AND FOSTER INNOVATION | Through the industrial-specific training, DOPPLER will foster relationship with key players who will gain the necessary skills and competencies to deliver a huge a project such as the SKA. Delivery of the SKA will also entail a component of maintenance, therefore the skills and competencies will have a long lasting effect beyond the initial training [11]. Through DOPPLER, and the mobility plan, technical skills will be gained. |
| 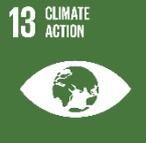 | TAKE URGENT ACTION TO COMBAT CLIMATE CHANGE AND ITS IMPACTS | One key component of DOPPLER is strengthening our Earth Observation group. Specific training will be done on using satellite imagery for specific areas of biodiversity, food security, and resource management. Furthermore, with our industry-specific training we will focus on renewable energy sources and engage in dissemination action with Oswela. |
| 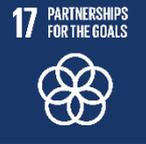 | REVITALIZE THE GLOBAL PARTNERSHIP FOR SUSTAINABLE DEVELOPMENT | Ultimately, DOPPLER will aim to bridge the local communities, universities, industry, non-governmental organisations and government in order to deliver a shared goal, vision and ensure that large science projects contribute to sustainable development. This will be achieved through the coordinated capacitation at universities, also engaging the local communities where the SKA may be hosted. |

**Table 1.** Targeted United Nations Sustainable Development Goals

# 4      SYNERGIES

Within the African context, a hand full of projects exist outside, but in collaboration with, South Africa lead by European institutions, the Development in Africa with Radio Astronomy (DARA)[1], and Joining up Users for Maximizing the Profile, the Innovation and Necessary Globalization of JIVE (JUMPING JIVE)[2]. These projects focus on introductory astrophysics lectures and hands-on training in radio astronomy observations, data reduction and analysis. However, one particular hurdle that was encountered with these projects, and particularly in Mozambique, was a language barrier. Here DOPPLER can aid and initial contacts have already been established in order to have member of the DOPPLER team help with some of the activities. Where DOPPLER differentiates itself is on staff mobility in order to create a long lasting research culture teaming with industry. This will also allow for joint supervision of students. Furthermore, thorough engagement with ONGs such as Oswela, an organization experienced in STEM education and oriented for the spread of knowledge across the country, DOPPLER will promote educational inclusiveness and utilise knowledge gained from platforms as that being developed for EquaMat@Moz. Furthermore, DOPPLER has a strong component to foster a hub for industry-specific training and development, to be aligned with major infrastructure and industrial investments. Of course, these requires a long-term and sustained plan. Last, but not least, we note DOPLLER may strengthen regional linkages between Mozambique and SKA member countries like India, South Africa and its neighbouring African-Indian Ocean partners, contributing to the regional pool of talent and critical mass.

## ACKNOWLEDGMENTS

DOPPLER is funded by the Aga Khan Development Network and the Fundação para a Ciência e a Tecnologia (FCT). VARMR acknowledges financial support from FCT in the form of an exploratory project of reference IF/00498/2015. VARMR, SA and AC acknowledges financial support from Center for Research & Development in Mathematics and Applications (CIDMA) strategic project UID/MAT/04106/2013. VARMR, SA, DB, JPB, AC, DM and DG acknowledges financial support from Enabling Green E-science for the Square Kilometre Array Research Infrastructure (ENGAGESKA), POCI-01-0145-FEDER-022217, funded by Programa Operacional Competitividade e Internacionalização (COMPETE 2020) and FCT, Portugal.## REFERENCES

[1]    Bourke, T. L. et al. "Advancing Astrophysics with the Square Kilometre Array," in Proceedings of Science, 14 (2015).

[2]    **Barbosa, D**., Lobo Márquez, G., Ruiz, V., Silva, M., Verdes-Montenegro, L., Santander-Vela, J., **Maia, D., Antón, S.** et al. "Power Challenges of Large Scale Research Infrastructures: the Square Kilometer Array and Solar Energy Integration; Towards a zero-carbon footprint next generation telescope," in Proceedings of the 2nd International Workshop on Integration on Solar Power into Power Systems, arXiv: 1210.3972 (2012).

[3]    **Barbosa, D.** et al. "BIOSTIRLING-4 SKA: A cost effective and efficient approach for a new generation of solar dish-Stirling plants based on storage and hybridization; An Energy demo project for Large Scale Infrastructures," Proceedings of the 7th International Workshop on Integration of Solar Power into Power Systems, arXiv: 1712.03029 (2017).

[4]    Duarte, L., **Teodoro, A., Maia, D., Barbosa, D.** "Radio Astronomy Demonstrator: Assessment of the Appropriate Sites through a GIS Open Source Application," ISPRS International Journal of Geo-Information, 5, 209 (2016).

[5]    Bilale, F. J. C. "Educational Performance in Mozambique: An Economic Perspective," MSc Thesis, University of Stellenbosch (2007).

[6]    Passos, A. "A Comparative Analysis of Teacher Competence And Its Effect On Pupil Performance In Upper Primary Schools In Mozambique And Other Sacmeq Countries," PhD Thesis, University of Pretoria (2009).

---

[1] https://www.dara-project.org/
[2] http://www.jive.eu/jumping-jive

[7] IBE-UNESCO "World Data on Education," International Bureau of Education UNESCO, Seventh edition (2012).

[8] Anjo, A. B., Amaro, S., Manganlal, K. "ICT in Education in Mozambique – the example of EquaMat@moz," in IST-Africa 2018 Conference Proceedgins, (2018)

[9] **Ribeiro, V. A. R. M., Paulo, C. M.** et al. "Introduction Astronomy into Mozambican Society," in Proceedings of the International Astronomical Union Symposium, 260, 522 (2011).

[10] "BIG SCIENCE: What's It Worth?", Science/Business Publishing Ltd, (2015).

[11] Hertzfeld, H. R. "Measuring the Returns to NASA Life Sciences Research and Development," (1998).

[12] The Royal Society "The Scientific Century; securing our future prosperity," The Royal Society Policy document 02/10, Issued: March 2010 DES1768, ISBN: 978-0-85403-818-3, (2010).

[13] Jayaram, S., Burnett, N., Engmann, M. "Innovative Secondary Education for Skills Enhancement (ISESE)," (2013). http://www.r4d.org/resources/innovative-secondary-education-skills-enhancement-isese-phase-ii-research/ (17/05/2018)

# Applications and Industry

# BEACON: in the next generation ground radars and radio telescopes infrastructures – the SKA project opportunity

Rogério N. Nogueira*[a,c], Vanessa C. Duarte[a,b], João G. Prata[a], Georg Winzer[b], Lars Zimmermann[b], Rob Walker[d], Stephen Clements[d], Marta Filipowicz[e], Marek Napierała[e], Tomasz Nasiłowski[e], Jonathan Crabb[f], Leontios Stampoulidis[f], Javad Anzalchi[g] and Miguel V. Drummond[a]

[a] Instituto de Telecomunicações, 3810-193, Aveiro, Portugal
[b] IHP – Innovations for High Performance Microelectronics, 15236 Frankfurt (Oder), Germany
[c] Watgrid Lda., 3810-193, Aveiro, Portugal
[d] aXenic Ltd., Sedgefield, UK
[e] InPhoTech Sp. z o.o., Warsaw, Poland
[f] Gooch & Housego, Torquay, UK
[g] Airbus Defence & Space, Stevenage, UK
* E-mail: mogueira@av.it.pt

**ABSTRACT**

The Radio astronomy community is entering a new era with the Square Kilometre Array (SKA), a telescope with 100 times improved sensitivity and a 10.000 fold survey speed compared to existing radio telescopes. For this instrument, new technologies are required for telescope design, data transport, computing and software. In particular, new technologies are being investigated to replace the classic parabolic dish with electronically steered antennas, the so-called phased arrays. Beam forming of these very large arrays, with up to 100 million elements, is an important component of the telescope. Photonic beamforming is of particular interest, due to the potential for miniaturization, high capacity and lower power consumption. The recently closed European funded project BEACON made important state of the art advances regarding photonic beamforming, which are reported on this paper.

**Keywords:** photonics, beamforming, integrated optics

## 1 INTRODUCTION

The introduction of photonics in the development of new generation ground radar systems and new generation radiotelescopes is now very close consideration. The application of photonics in such systems is bound to create a new very broad market and unique commercialization opportunity since these systems are extremely volume hungry and will require technologies that combine costoptimized mass manufacturing and compliance with harsh environment operation. The most pronounced example is the SKA (Square Kilometre Array) project, which aims to construct the world's largest radio telescope. The SKA is planned in two construction Phases, with the deployment of different sensor technologies sourcing on new generation phased antenna arrays with thousands of elements and beamforming technologies to demonstrate an aperture of up to a million square meters, built to further the understanding of the most important phenomena in the Universe. The SKA Phase 2 will require performance and technology evaluation through Advanced Instrumentation Programs aiming to guide deployments of cutting edge Aperture Array stations. Photonics is at the heart of AAs and will be critical for a planned cost-effective performance with a low power consumption for such a significant number of elements, either through beamforming or via the optical data circuitry system. This paper reports the recent state of the art advances on photonic beamforming made within the European funded project BEACON.

## 2 PROJECT BEACON

The BEACON project aimed to disrupt the introduction of photonics into terabit per second payload systems by squeezing current discrete bulk photonic components into compact array modules and generate practical photonics multi-beam systems in a scalable and power efficient way.

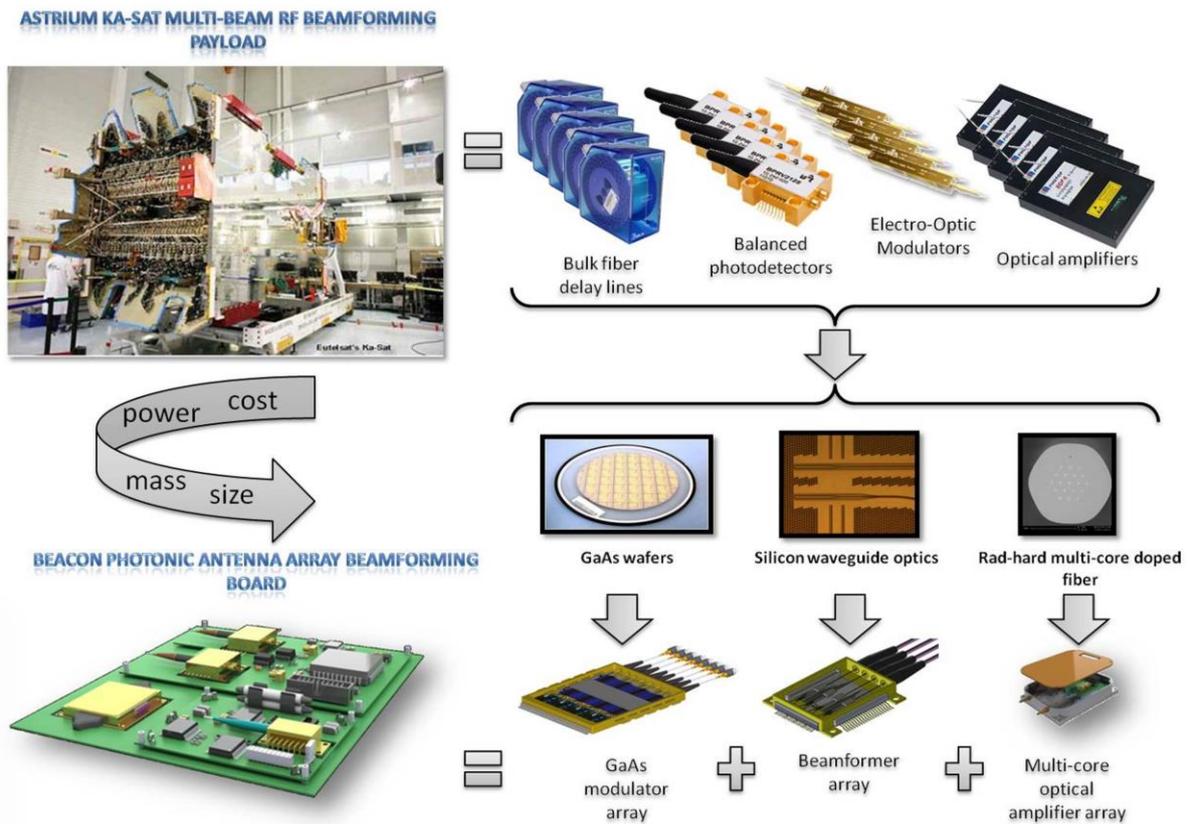

**Figure 1.** BEACON Project Concept of compact array modules to provide scalable and power efficient photonics for satellite beam array systems

The consortium consisted of:

- Component manufacturers
    - Axenic Limited, United Kingdom,
    - Constelex Technology Enablers, Greece
    - Gooch and Housego Limited (G&H), United Kingdom
    - InPhoTech, Poland
    - Watgrid, Portugal
- End users
    - Airbus Defence and Space (ADS), United Kingdom
- Research Institutions
    - IHP GmbH – Innovations for High Performance Microelectronics, Germany
    - Instituto de Telecomunicações (IT), Portugal

The partners provided a complementary set of skills and experience to match the goals and objectives of the project. The end user in the consortium is Airbus Defence and Space (ADS) who were able to direct and guide the system architecture aspects and set the specification requirements.

The core research activities on the array modules were developed by component partners as follows:

- InPhoTech developed multi-core fiber for optical amplifiers based on their experience from spinning out of leading fiber technology groups in University of Marie Curie-Sklodowska in Lublin.

- Constelex were an innovative technology company and originally the project coordinators in BEACON. Their experience was in fiber amplifier arrays and fiber systems and also in space systems.

- IHP were responsible for the design, development and fabrication of the beamformer elements based on silicon photonic interferometers based on production grade tool-sets for 0.25 and 0.13 micron technologies.

- aXenic developed compact optical modulators and their combination in arrays for efficient low weight, size and power implementations. Their background goes back through to the origins of Gallium Arsenide modulators, high quality fabrication in commercial foundries and assembly technologies.

The construction of integrated modules and their demonstration was mainly focused in the three partners: Gooch & Housego (G&H) who were responsible for the manufacture of the optical amplifiers from the multi-core fiber. Their expertise is based in precision optical components and sub-systems, including fiber components which goes back to 1985. These skills are applied particularly in demanding applications which include space missions with NASA. Watgrid is a technology based company with skills in innovative products and technology transfer services. Their business lies across photonic communication and sensor activities. They were responsible for interfacing on the beamformer and the antenna development for the demonstrator. Finally, Instituto de Telecomunicacões (IT), which is a private, not-for-profit institute co-located at the University of Aveiro with focus on research and education activities. Its expertise in optical communications and networks was deployed in the design test and characterization of the array components and their assembly into a system which included prime responsibility for accomplishing the final demonstrator.

The developments in BEACON sought to address the photonic architecture by addressing technical hurdles in the following:

- **Microwave mixing** using compact integrated electro-optics modulators which are scalable for large arrays. BEACON solution: Gallium arsenide array modulators

- **Optical amplification** which can also manage multiple-optical paths and manage to overcome the sensitivity to radiation of current off-the-shelf amplifiers. BEACON solution: multi-core fiber amplifiers

- **Optical beam-forming** using an integration platform that enables flexible functional and cost-effective integration with fast tuneable beam-forming. BEACON solution: True-time delay Silicon photonics beamformers

- **Photodetection** using co-integration of photo-diodes with the beam formers. BEACON solution: Integration of the photodiodes with the beamformers in silicon photonics

- **Cost and space** reduction using small components, integration for parallel functionality and scalable device technologies. BEACON solution: integrated compact components evaluated for space qualification.

In BEACON, the formidable challenge was not just to achieve these developments but to demonstrate all these advances in a final demonstrator combining the devices in a full working system at Ka band.

These technical activities were supported by several disseminations through exploitation activities, publications and conference presentations.

## 3 MAIN RESULTS

The BEACON project has achieved the following highlights:

The chief achievement of the project has been the first ever demonstration of a real-time photonic beamformer for processing 4 input Ka band signals (1Gbit/s QPSK at 28GHz carrier), including an array of modulators, a multi-core optical fiber amplifier and a silicon photonic integrated beamformer. The beamformer is based on a self-heterodyne architecture which transforms RF phase shifters into optical phase shifters resulting in a size reduction by a factor of 5000.

The architecture of the demonstration was based on the separate development on the project of the following components:

- The development of a 7-core booster amplifier using radiation resistant multi-core fiber amplifiers. Evaluation for baseline environmental testing for space (Fig. 3).

- The delivery of highly compact modulator arrays which are half the length of conventional modulators, halving the fiber handling space through the use of folded optics and using an expandable array architecture. The devices have performed for Ka band and are capable of much higher frequency operation (Fig. 4).

- The design and fabrication of a set of beamformers fully integrated on CMOS silicon photonics (Fig. 5).

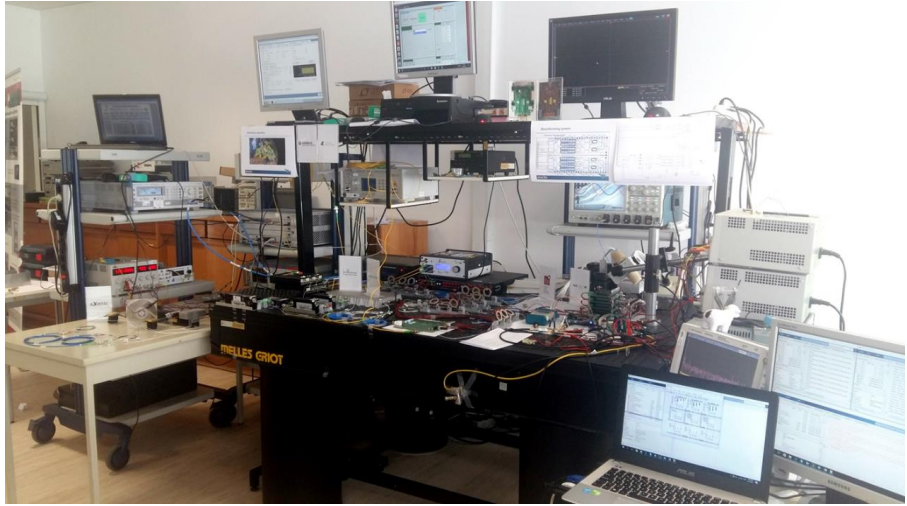

**Figure 2**. Photograph of the final demonstrator set-up at IT

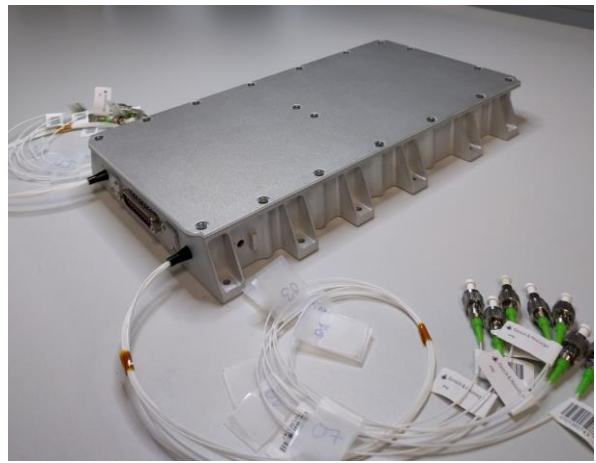

**Figure 3**. Fully assembled multi-core array fiber amplifier

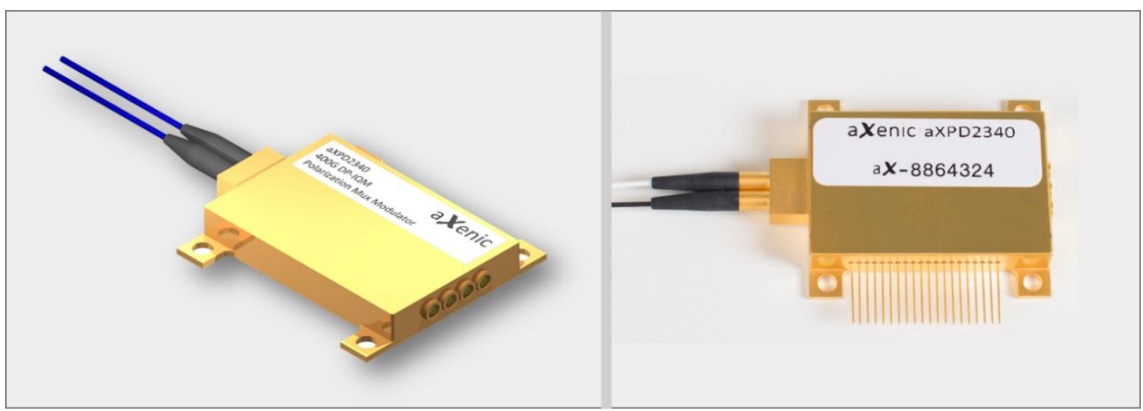

**Figure 4**. GaAs modulator array package: CAD (left) and photograph (right)

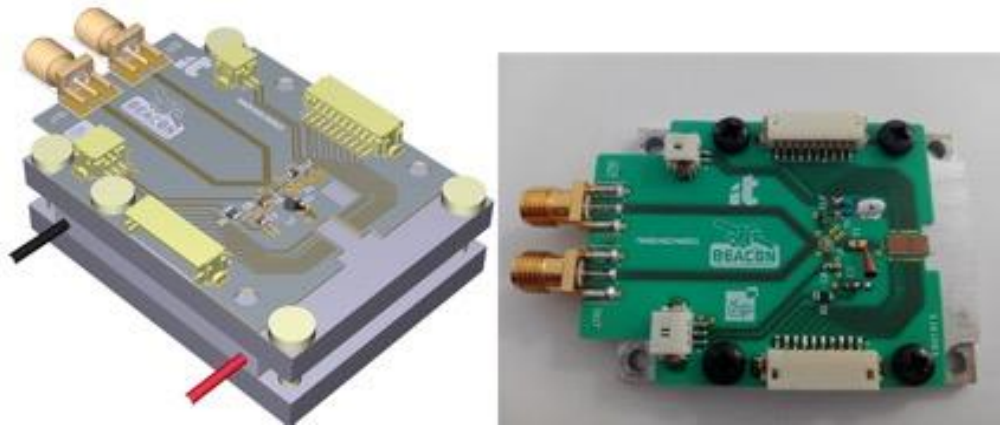

**Figure 5**. Beamformer (left) PCB design and (right) with all parts placed

## 4 CONCLUSIONS

Photonic beamforming is considered a key technology for many applications, such as space communications or ultra high capacity radio telescopes. Within the scope of project BEACON, the first ever demonstration of a real-time photonic beamformer for processing 4 input Ka band signals was made. This important milestone was possible due to the development of new architectures and components. These advances pave the way for other applications of photonics beamforming, such as the ones related to aperture arrays on the SKA telescope.

## ACKNOWLEDGEMENTS

The authors gratefully acknowledge financial supported by the European Commission through the project BEACON (FP7-SPACE-2013-1-607401).

## REFERENCES

[1] V. C. Duarte, J. Prata, C. Ribeiro, R.N. Nogueira, G WINZER, L. Zimmermann Zimmermann, R. Walker, S CLEMENTS, M. Filipowicz, M. Napierala, T. Nasilowski, J. Crabb, M. Kechagias, L. Stampoulidis Stampoulidis, J ANZALCHI, M. V. Drummond, *Modular coherent photonic-aided payload receiver for communications satellites,* **Nature Communications**, Vol. 10, No. 1, pp. 1 - 9, April, 2019,

[2] M. V. Drummond, V. C. Duarte, A. Albuquerque, R.N. Nogueira, L. Stampoulidis Stampoulidis, L. Zimmermann Zimmermann, J ANZALCHI, *Dimensioning of a multibeam coherent photonic beamformer fed by a phased array antenna,* Optics Express, Vol. 26, No. 5, pp. 6158 - 6158, February, 2018

[3] C. Duarte, M. V. Drummond, R.N. Nogueira, *Photonic True-Time-Delay Beamformer for a Phased Array Antenna Receiver based on Self-Heterodyne Detection*, IEEE/OSA Journal of Lightwave Technology, Vol. 34, No. 23, pp. 5566 - 5575, October, 2016

[4] **Patent:** M. V. Drummond, R.N. Nogueira, V. C. Duarte, *Photonic beamforming system for a phased array antenna receiver*, WO 2016/170466A1, April, 2015

# A Portuguese SKA "test site" as a case study for an Integrated Environmental Management System: synergies with Space and Earth Observation Sciences

L. Duarte[a,b], A.C. Teodoro[a,b], D. Maia[c], D. Barbosa[d*], N. Sillero[c], J.A. Gonçalves[a,e], J. Fonte[c], L. Gonçalves-Seco[c,f], L. M. Pinheiro da Luz[g], N. Santos Beja[g]

[a] Departamento de Geociências, Ambiente e Ordenamento do Território (DGAOT), Faculdade de Ciências, Universidade do Porto, Portugal
[b] Instituto de Ciências da Terra (ICT), Faculdade de Ciências, Universidade do Porto, Rua Campo Alegre, 4169-007, Porto, Portugal
[c] Centro de Investigação em Ciências Geo-Espaciais (CICGE), Faculdade de Ciências, Universidade do Porto, Observatório Astronómico Prof. Manuel de Barros, Alameda do Monte da Virgem, 4430-146 Vila Nova de Gaia, Portugal
[d] ENGAGE SKA, Instituto de Telecomunicações, Campus Universitário de Santiago, 3810-193-Aveiro, Portugal
[e] Interdisciplinary Centre of Marine and Environmental Research (CIIMAR), Universidade do Porto, Portugal
[f] ISMAI – University Institute of Maia, Av. Carlos Oliveira Campos, 4475-690 Avioso S. Pedro, Portugal
[g] Escola Superior Agrária, Instituto Politécnico de Beja, Rua Pedro Soares, 7800-295 Beja, Portugal

**ABSTRACT**

Radioastronomy Observatories usually require for its infrastructure development a combination of long term analysis of geophysical, environmental, meteorological and by definition favorable radio frequency conditions. Pilot testing for any large scale projects like the SKA implies an integrated environmental management analyses must be careful assessed. Herdade da Contenda (HC) in Alentejo, Portugal, was selected for radio astronomical testing purposes of SKA related Aperture Array prototype technologies (and to further develop a radio astronomical infrastructure). To conduct the first surveys, a Geographic Information System (GIS) open source application was created, the HC Environmental Integrated Management System (HCIEMS) which combines several functionalities and menus with different methods allowing to create multiple maps regarding the HC characteristics, such as Digital Elevation Model (DEM), Land Use Land Cover (LULC), Normalized Difference Vegetation Index (NDVI), groundwater vulnerability, erosion risk, flood risk and forest fire risk. Moreover, a decision making support tool, GeoDecision, was also developed and incorporates an algorithm through assigned weights and eliminatory factors to find the locations best suited for infrastructure development. Regarding the SKA context, the SKA will be built in phases, expanding its antenna locations through several countries that certainly may benefit from such approach tested in a suitable territory emulator. In particular, the SKA African Partner countries may benefit from an integrated, centralized application that promotes and sources on synergies between radio astronomy, GIS and Space and Earth sciences as first decision making step.

**Keywords:** GIS, decision making, radioastronomy, SKA, NDVI, DEM, LULC.

*dbarbosa@av.it.pt; phone +351 234 377 900; fax +351 234 700 901

## 1 INTRODUCTION

Large scale infrastructures like the Square Kilometre Array (SKA) require extensive testing of its chosen locations to certify spectral condition are optimal while the environmental impact of its construction den deployment is minimized and sustainable. SKA will be the largest radio telescope in the world, a sensor network dedicated to radio astronomy, spreading between two continents [1]. It will be spatially distributed with thousands of antennas to be placed in Africa and Australia. The most important factor for a construction of a radio telescope is the Radio Frequency Interference (RFI) environment [2]. Other factors such as the characteristics of the ionosphere and the troposphere, the physical characteristics of the site including climate and subsurface temperatures, the connectivity across the vast extent of the telescope itself as well as to communications networks for worldwide distribution of data produced by the SKA, infrastructure costs, including power supply and distribution, operation and maintenance costs and the long term sustainability of the site as a radio quiet zone [3] are also important for the selection of the construction of a radio telescope site.

In the framework of Portuguese participation in the SKA project, [4], Herdade da Contenda (HC) in Alentejo, Portugal, was selected for radio astronomical testing purposes and to develop a radio astronomical infrastructure. This site is a protected zone, National Hunt zone and Forestry Perimeter. Given the protected status of HC, environmental concerns and other aspects must be taken into account, so an application integrating all this information can be very useful to study the site [5].

The objective of this work was the development of a GIS open source application composed by several menus and tools allowing to create multiple maps regarding the HC characteristics and combine the resulting maps in a decision making tool, the GeoDecision, in order to identify the possible sites that are more adequate to the installation of the radio-astronomy infrastructure in suitable areas in Portugal, in the scope of the Portuguese participation in the SKA MFAA Cosnortium [6]. The application was developed in a GIS open source software, QGIS, and through Python language. The GeoDecision tool will help to support the decision of the sites. Several maps, such as forestry risk map, erosion map, groundwater vulnerability map and provides the visualization of some pre-processed information, Land Use Land Cover (LULC) map, Normalized Difference Vegetation Index (NDVI) map, calculation and representation of bioclimatic indexes, visualization techniques to identification of archaeological features, ortophoto, fauna and flora can be obtained with the HCIEMS application. This information can be combined in the GeoDecision tool and a final map can be created with the more adequate sites to the radio astronomy potential locations.

## 2 CASE STUDY AND DATASET

**Case study**

HC (Alentejo, Portugal) was chosen to be the ideal site with the conditions necessary to SKA, climatic conditions and excellent radio conditions with levels of interference comparable to SKA sites. HC is located in Serra Morena in Santo Aleixo da Restauração, Moura municipality, Beja district (Fig. 1).

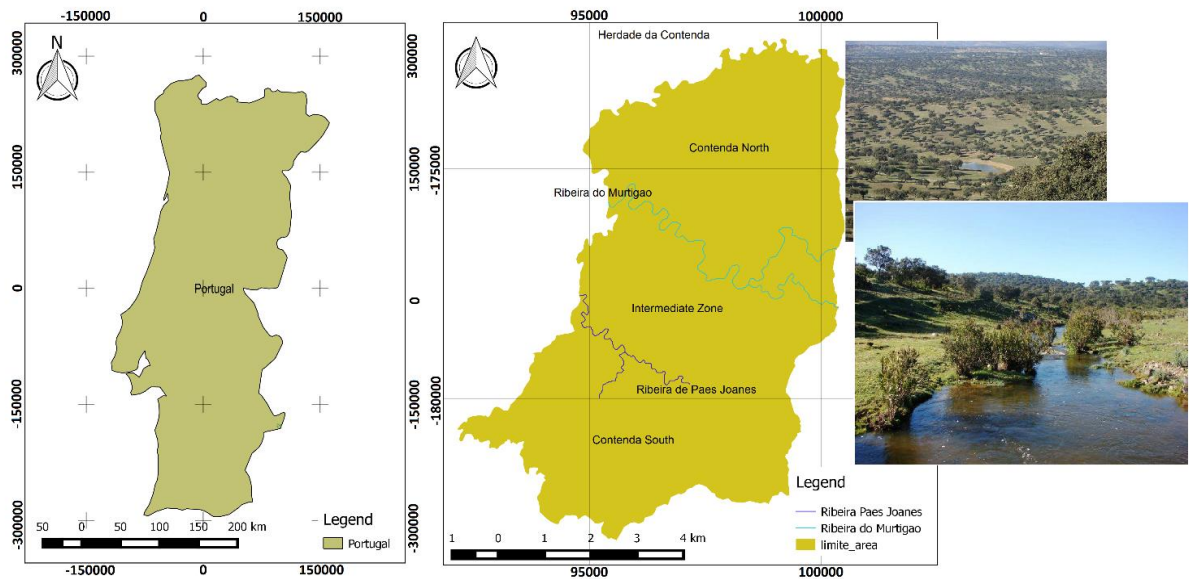

**Figure 1.** Location of HC (adapted from Duarte et al., 2016).

**Dataset**

In the study performed, different spatial data were used in order to create the required maps such as: the DEM which was generated from the flight performed in the area [5]; the geologic map (1/200 000 and different geologic information) [7]; the Land Cover Map 2007 shapefile data [8]; the economic values and vulnerability values for forest fire cartography which were assigned based on Municipal Defense Plan Against Forest Fire of Barrancos [9]; the slope map was derived from the DEM; a set of shapefiles with the burnt information of 24 years (1990/2013) from the National Institute of Forest Conservation [10] were used; meteorological observations from a period of 27 years (1981-2008) were collected in the Alentejo weather station (38°38'59,28"N, 7°32'52,82"W) and the soil map (1/35 000) was based in PGFCFP soil map [10]. The DEM used has a spatial resolution of 10 m and all the data are in ETRS89 PTTM06 (EPSG: 3763) coordinate system [6].

## 3 METHODOLOGY

The HCIEMS application was developed under a GIS open source software named QGIS using Python programming language [11-13]. To develop the application, several libraries and Application Programming Interfaces (APIs) were used: GDAL/OGR library, PyQt4 API, QGIS API, Numpy library and Python libraries [14-18]. The graphic interface was composed by several windows with parameters to define. It was composed by fourteen menus: File, DRASTIC, Forest Fire Risk, Revised Universal Soil Loss Equation (RUSLE), Flood Risk, Bioclimatic Index, Cultural Heritage, Fauna and Flora, Ortofoto, NDVI, DEM, LULC, GeoDecision and Help [5]. It was designed as a single window composed by the menus, an area where the maps will be presented, a table of contents which allows the visualization and manipulation of the layers which allows some interactions such as Show Extent, Remove Layer and Zoom to Layer. The DRASTIC, Forest Fire Risk and RUSLE menus were based on the GIS applications already developed by the authors [11, 13, 19]. The Flood Risk menu was incorporated with a method to simulate the flood risk, using *r.lake.coords*, an algorithm from GRASS [6, 20]. The Bioclimatic Index menu considers several bioclimatic indexes from PGFCFP [9] in a pdf file. The Cultural Heritage menu incorporated several algorithms from SAGA library [21] composing different relief visualization techniques [6]. The Fauna and Flora menu incorporate the data of carnivore presence and location of latrines of rabbits for summer and winter [6]. Some information was incorporated inside the HCIEMS application in order to support the maps and decisions of the user: ortophoto, DEM, LULC and NDVI maps are incorporated in the respective menus and can be used in the creation of the other maps. The GeoDecision tool was the greater advance in this application because allows to create scenarios (maps) based on several criteria and with the possibility of assigning weights given the importance degree. Considering different levels of relevance, a weight value is assigned to each map. In the study performed the most important criteria taking into account given the infrastructure installation were: fire risk, soil erosion level, flood risk and the slope values. These criteria were considered because HC is a protected zone and a forest zone, so fire risk and erosion risk are extremely important. The flood risk was considered given the fact that it is closer to the river. The higher slope values were excluded so the installation must be in flat terrain [6].

## 4 RESULTS

Several maps were created in order to obtain a final map through the GeoDecision tool. Figure 8 presents all the maps involved in the decision making tool [6].

Several scenarios were created in order to test the GeoDecision tool. Figure 9 presents the scenarios considering different weights where the results obtained ranged from 0 to 100%, where zones with values closer than 100% are the most adequate to the radio astronomy demonstrator installation. The classes of the values obtained were considered as: inadequate (0-378 10%), not adequate (10-20%), moderate adequacy (20-30%), adequacy (30-40%) and very adequacy 379 (>40%) [6].

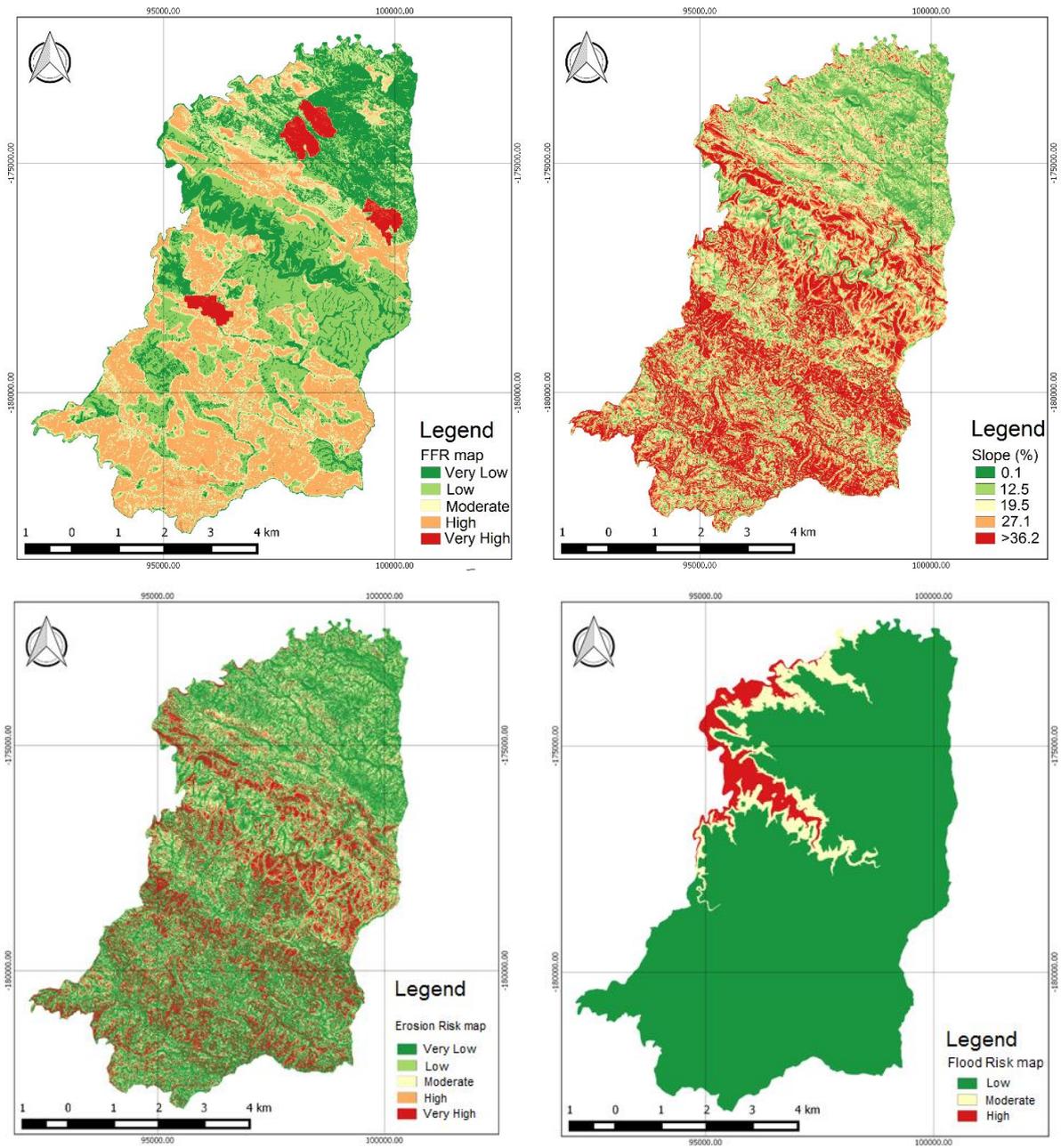

**Figure 2.** Maps required to GeoDecision tool (adapted from Duarte et al., 2016).

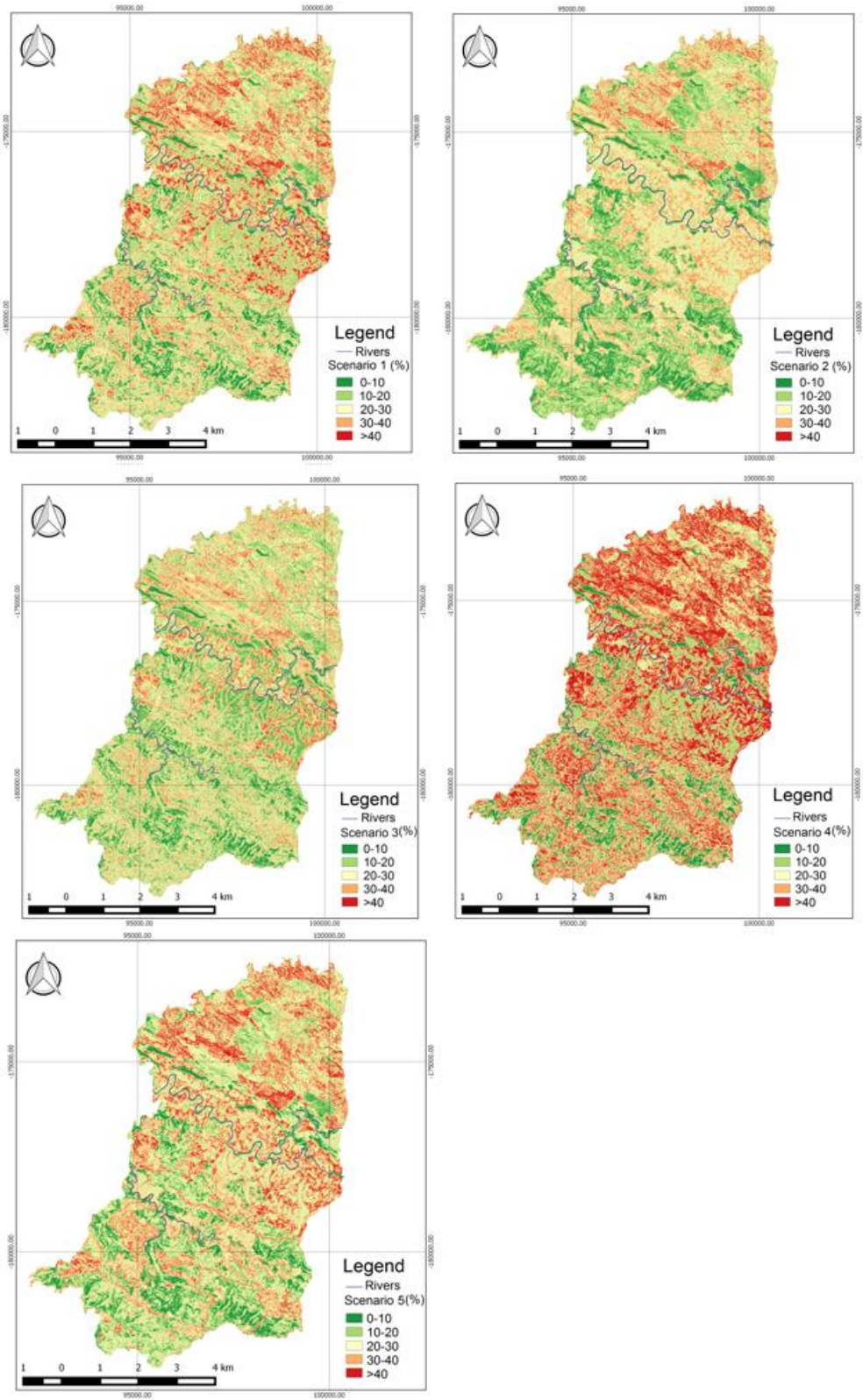

**Figure 3.** Scenarios tested (adapted from Duarte et al., 2016).

## 5 DISCUSSION AND CONCLUSIONS

Considering the maps created (Fig. 3), 38.4% of HC area was classified with very low or low values especially in the north of HC. The higher slope values were located in the south with 19.2%. These areas can be related with the high and very high forest fire risk zones, located in the south, which covers 2.2% and 32.6%, respectively (Fig. 8). So, zones with high slope values were related with high forest fire risk. 56% of HC area was classified with low and very low forest fire risk. The results obtained are coherent with the data provided in PGFCFP [9] where the south of HC presents a higher forest fire risk than the north. Regarding the Erosion Risk map (Fig. 8), the majority of high erosion risk (13.9%) is located in the south. The north presented low erosion level perhaps due to the Quercus ilex plantation which covers the total extension of the north area. Approximately 10% of the HC area was classified with high and very high erosion risk combined with high slope values, according to RUSLE method. Most of these areas were located in the south. Therefore, high slope values were related with high erosion risk level. The scenarios were tested with different weights assigned: in scenario 1 the weights were equally distributed and 404 9.88 km2 of the area corresponded to the most adequate zones to the installation, which covered 405 approximately 20% of HC. In scenario 2, the higher weight was assigned to the forest fire risk and 4.89 km2 of the area were related to zones with more adequacy and such as the previous scenario, the majority of these zones were located in the north. In scenario 3 the erosion risk was considered the most important, decreasing for slope and forest fire risk factors resulting in 5.35 km2 of the area considered as the more adequate zones and covered the north and the east part of HC. The scenario 4 considered the slope as the most important factor and 20.88 km2 of HC had the perfect conditions, which was 40% of the total area. With the maps obtained in the scenarios, it was concluded that the more adequate sites to install the demonstrator infrastructure are in the north, where the low percentage of risk was identified [6].

The HCIEMS allows to the creation of several maps, regarding different methods and the evaluation of factors such as bioclimatic indexes. The GeoDecision tool helped to identify the most adequate sites to install the radio astronomy demonstrators considering several criteria with different weights. The application is free and open source so it can adapted to other parameters and methods. Regarding the SKA context, the SKA will be built in phases, expanding its antenna locations through several countries that certainly may benefit from such approach tested in a suitable territory emulator. In particular, the SKA African Partner countries may benefit from an integrated, centralized application that promotes and sources on synergies between radio astronomy, GIS and Space and Earth sciences as first decision making step [6].

## ACKNOWLEDGMENTS

DB acknowledges support from FCT/MEC through national funds and when applicable co-funded by FEDER – PT2020 partnership agreement under the project UID/EEA/50008/2013. The authors would like to thank to José Alberto Gonçalves, Neftallí Sillero, Luís Gonçalves Seco, João Fonte, Luís Pinheiro da Luz, Nuno dos Santos Beja and Rute Almeida for the help provided in this work. We acknowledge support from the Enabling Green E-science for the Square Kilometre Array Research Infrastructure (ENGAGE SKA), POCI-01-0145-FEDER-022217, funded by Programa Operacional Competitividade e Internacionalização (COMPETE 2020) and FCT, Portugal.

## REFERENCES

[5] Schilizzi, R.T., Dewdney, P.E.F., and Lazio, T.J. "The Square Kilometre Array. In: Ground-based and Airborne Telescopes II: 70121I," Proc SPIE 7012, Marseille, France, (23 June 2008).

[6] Peng, B., Sun, J.M., Zhang, H.I., Piao, T.Y., Li, J.Q., Lei, L., Luo, T., Li, D.H., Zheng, Y.J., and Nan, R. "RFI test observations at a candidate SKA site in China." Experimental Astronomy, 2004, 17, 423-430.

[7] SKA. "SKA TELESCOPE SQUARE KILOMETRE ARRAY." Exploring the Universe with the world's largest radio telescope. 2016. (20 April 2016). https://www.skatelescope.org/

[8] Barbosa, D., Aguiar, R., Barraca, J.P., van Ardenne, A., Boonstra, A.J., Verdes-Montenegro, L., and Santander- Vela, J. "A Sustainable approach to large ICT Science based infrastructures; the case for Radio Astronomy." In IEEE International Energy Conf. – ENERGYCON. Dubrovnik, Croatia, 2014.

[9] Teodoro, A., Duarte, L., Sillero, N., Gonçalves, J.A., Fonte, J., Gonçalves-Seco, L., Pinheiro da Luz, L.M., and dos Santos Beja, L.M.R. "An integrated and open source GIS environmental management system

for a protected area in the south of Portugal." In Proc. SPIE 9644, Earth Resources and Environmental Remote Sensing/GIS Applications VI, 96440U. 2015.

[10] Duarte, L., Teodoro, A.C., Maia, D., and Barbosa, D. "Radio Astronomy demonstrator: assessment of the appropriate sites through a GIS open source application." International Journal of Geoinformation, 2016, 5(11), 209; https://doi.org/10.3390/ijgi5110209.

[11] Oliveira, J.T. Carta Geológica de Portugal Escala 1/200 000 Notícia explicativa da Folha 8 Direcção Geral de Geologia e Minas. Serviços Geológicos de Portugal, Lisboa: 1-91, 1992.

[12] dgTerritório. Direção-Geral do Território. 2015. (22 April 2016). http://www.dgterritorio.pt/cartografia_e_geodesia/cartografia/cartografia_tematica/carta_de_ocupacao_do_solo__cos_/cos__2007.

[13] PMDFCI. Associação de Produtores da Floresta Alentejana. Plano Municipal da Defesa da Floresta Contra Incêndios de Barrancos. 2010. (22 April 2016). http://www.cm-barrancos.pt/smpc/PMDFCI.pdf.

[14] ICNF. Instituto da Conservação e da Defesa das Florestas. 2009. (22 April 2016). http://www.icnf.pt/portal.

[15] Teodoro, A.C., and Duarte, L. "Forest Fire risk maps: a GIS open source application – a case study in Norwest of Portugal." International Journal of Geographic Information Science, 2013, 27(4), 699-720.

[16] Duarte, L., and Teodoro, A.C. "An easy, accurate and efficient procedure to create Forest Fire Risk Maps using Modeler (SEXTANTE plugin)." Journal of Forestry Research, 2016, pp. 1-12.

[17] Duarte, L., Teodoro, A.C., Gonçalves, J.A., Guerner Dias, A.J., and Espinha Marques, J. "A dynamic map application for the assessment of groundwater vulnerability to pollution." Environmental Earth Sciences, 2015, 74 (3), 2315-2327.

[18] GDAL. Geospatial Data Abstraction Library. 2015. (30 March 2016). http://www.gdal.org/.

[19] PyQt4 API. PyQt Class Reference. 2015. (22 March 2016). http://pyqt.sourceforge.net/Docs/PyQt4/classes.html.

[20] QGIS API. QGIS API Documentation. 2013. (22 March 2016). http://www.qgis.org/api/.

[21] Numpy API. Numpy Reference. 2015. (22 March 2016). http://docs.scipy.org/doc/numpy/reference/.

[22] Python. Python Programming Language. 2015. (22 March 2016). http://python.org/.

[23] Duarte, L., Teodoro, A.C., Gonçalves, J.A., Soares, D., and Cunha, M. "Assessing soil erosion risk using RUSLE through a GIS open source desktop and web application" Environment Monitoring Assessment, 2016, 188:351.

[24] GRASS GIS. The world's leading Free GIS software. 2013. 22 March 2016). http://grass.osgeo.org/.

[25] SAGA. SAGA-GIS Module Library Documentation. 2016. (22 March 2016). http://www.saga-549 gis.org/saga_module_doc/2.1.3/ta_lighting_3.html.

# How a Portuguese SME intends to leverage on its experience of handling TB of Data, and SKA, for the Smart Cities of the Future

Ricardo Vitorino, João Garcia, Ricardo Preto, Tiago Batista

Ubiwhere Lda, Travessa Senhor das Barrocas, 38, 3800-075 Aveiro, Portugal

## ABSTRACT

Ubiwhere is a ten-year-old software development company focused on two major domains: Telecom & Future Internet and Smart Cities. These sectors tend to generate large amounts of data, which raises a series of challenges of different natures, from operational to algorithmic. The present document reports the experience Ubiwhere has gathered over the years when dealing with Big Data in the Telecom domain and how this experience shall help the SME with future smart city deployments and large datasets from other domains or sources. Ubiwhere had started with a Research & Development project, developed for ANACOM, the Portuguese Communications Regulator, which consists of a solution capable of monitoring the nationwide Digital Video Broadcasting - Terrestrial (DVB-T) system, having been successfully running for more than four years and having generated around 4 TB of data. More recently, the company has gained more experience by working on several solutions for Smart Cities, which culminated with the creation of the Smart Lamppost, a modular urban infrastructure which provides different services for Telecom Operators and Smart Cities, such as a smart lighting system or network connectivity (4G/5G and Wi-Fi). With the ultimate purpose of fulfilling the needs of municipalities and government entities to extract valuable insights to improve the quality of life of their citizens, both these innovative services and urban infrastructure investments are required to make this secure and efficient data collection a reality.

**Keywords:** Smart Cities, Telecommunications, Future Internet, Big Data

## 1  INTRODUCTION

Ubiwhere is a Research and Innovation SME, based in Portugal, developing innovative and user-centred software solutions. Since 2007, Ubiwhere has fostered a culture of innovation and creativity by delivering the solutions that their clients need to succeed. The main objective of the company is to research and develop bleeding-edge technologies, design state-of-the-art solutions and create valuable intellectual property (by means of rich intangible assets) internally and to its clients – in order to achieve its vision of becoming an international reference in Smart Cities and Future Internet. Ubiwhere researches and develops technologies across several markets, including: Telecom and Future Internet; Transportation, Travel and Tourism; Bioeconomy; Sustainable and Efficient Resource Management; Knowledge, Collaboration and Education.

Ubiwhere's two main focus areas are Telecom & Future Internet and Smart Cities. Solutions for problems in these areas tend to generate a lot of data, either from sensors, probes or even communicated by citizens. For systems to successfully operate and be truly useful in these domains, they must be able to deal with large amounts of data and to scale accordingly [1].

Within this document, the experience Ubiwhere had working on a solution to monitor Portugal's nationwide DVB-T system will be reported along with the description of what the company is currently doing on the Smart Cities domain.

## 2  ANACOM DVB-T MONITORING SYSTEM

At the time of the transition from analogue to digital broadcasting for the Portuguese national television (also referred to as **digital switchover**), Portugal's national regulator (ANACOM) was the target of a considerable amount of criticism by the general population, as well as by consumer protection organisations such as DECO. This criticism had had its origin in service failures and lack of quality of experience perceived by end-users and associated with the new digital television service. To be able to provide the necessary regulation of this service, ANACOM launched a national project aiming to implement a nationwide Digital Video Broadcasting - Terrestrial (DVB-T) monitoring system. A consortium formed by Ubiwhere and Wavecom (two Portuguese SMEs) backed up by INESC-TEC (a Research & Development Institute) applied, having ultimately won the application and, thus, being assigned for its implementation. This system has now been in production for more than four years, with the current section describing the overall work done, concretely focusing on the components whose responsibility belonged to Ubiwhere.

On behalf of Ubiwhere, the team was composed by a pre-sales engineer, a project manager, three software developers (respectively for backend, frontend and mobile development), a Quality Assurance Engineer, a System Administrator and a UX/UI designer, who were required to follow some base guidelines for the design of this system/deployment, presented below:

- The system should be supplied as an appliance, with no operation effort to be required from the client.
- The data should be stored in-house. No data should be stored off the premises of ANACOM.
- The data should be transmitted using a secure channel, given that the Internet Service Provider (ISP) is potentially also responsible for DVB-T operations and it must be ensured that there is no integrity tampering or snooping.
- The dimension of the system should be such that it supports Big Data processing, i.e. it allows the storage of data from 400 monitoring probes, each measuring 7 distinct variables, one time per second for a duration period of two years.

It is also worth to mention that, based on the results of this project, the consortium (Ubiwhere and Wavecom) decided to launch a new product named rProbe (www.rprobe.com) providing an evolution of this solution, designed for spectrum sensing applications. It allows operators, broadcasters and regulators to remotely monitor the DVB signal in real-time, keeping a record of the network status at any time.

**Field Deployment Tool**

The early design stages of the system focused mainly on the design and implementation of the field deployment tool, as well as the architecture of the backend.

The field deployment tool was developed as an Android application, at a first stage only available on a custom image of the Android Operating System, running on a tablet device. The fact that the developed application supported only a custom image of an Android OS is related to some connectivity constraints that the DVB-T probe **had in its configuration phase.**

The end-user of this application was the field technician. The field technician would start the deployment by answering a simple questionnaire, then would deploy the hardware on the field and connect the probe to the tablet. Once connected to the probe, the deployment tool would ensure that a data connection was correctly established by the probe and would guide the technician in the final installation steps, such as adjusting the antenna and establishing communication with the central system over a secure VPN.

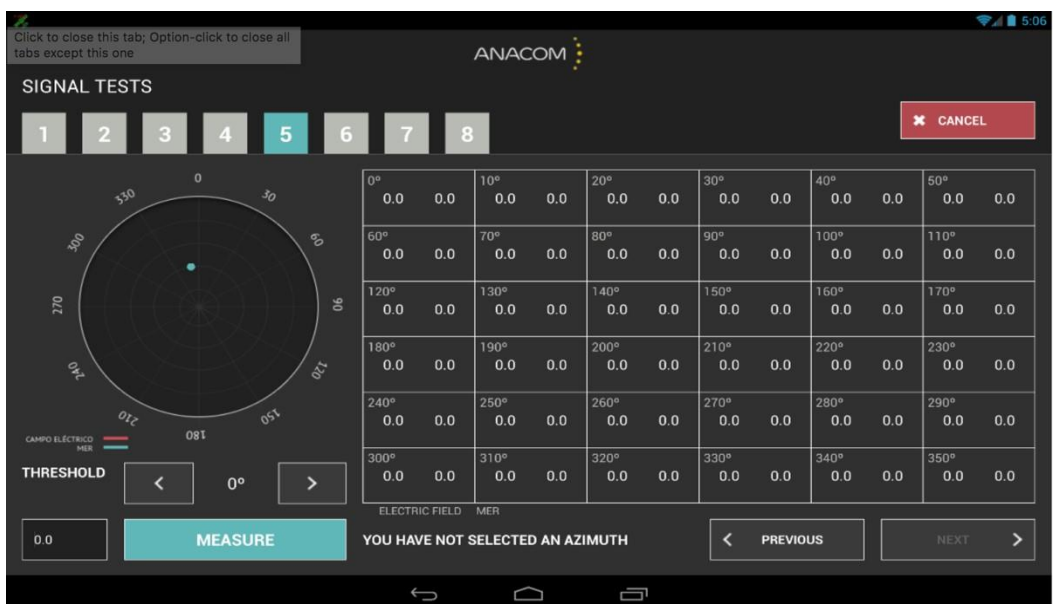

**Figure 1**. Sample screenshot of the mobile application running on an Android tablet device.

The usage of this application was critical to ensure that the probe had a stable and secure connectivity channel with the central system as well as to assert that it was correctly configured to ensure that all the collected data has auditable quality. Critical steps such as the correct alignment of the probe's antenna are covered by this application and all this information is collected and sent to the central system which allows the creation of an installation report for each probe, which can be accessed at any time to perform auditing actions.

### Central System Infrastructure

The central system infrastructure is composed of seven 1U systems supplied by Asus[1]. Two of those systems serve only as an IP frontend and VPN gateway, one is a NAS used to store a few non-critical shared filesystems and periodic on-site backups, while the remaining four systems are all identical.

Each has an instance of Gunicorn, Glassfish, PostgreSQL, Pgpool and Cassandra. All the systems' networking was configured in a redundant way, so that if for some reason one has to disconnect a cable for a maintenance task, there would be no impact on the running system. Also, other than the NAS, every system is redundant - a failure or shutdown of a single node will not result in unavailability for the service.

### Software Architecture

Other than the deployment tools, the system consists of a central API implemented in Java EE running on a Glassfish server, with a frontend server implemented in Django running on a Gunicorn process. The central API uses two databases, the first is a relational database that stores all the user information and the probe metadata, the second is a Cassandra database that stores probe measurements. Hazelcast is also used as a non-persistent data store for runtime locking mechanisms and job information.

At the time, the team considered the possibility of basing the architecture in micro-services, however some of the architecture patterns and tools that are nowadays stable were on their infancy at the time, so the choice was to go for a traditional Glassfish cluster instead, also taking into account the skillset of Ubiwhere's team at the time.

### Polling the probes for data

The polling system periodically contacts each of the probes asking for a set of data for a specified period. The task should not be isochronous, it should start after a configurable amount of time after the previous instance of the task finishes. This pause between instances of the same task allows the system to run other tasks and avoid the possibility of overlapping requests to the same probe in case a previous request is still running after the interval.

In a non-distributed system this issue would be a trivial problem to solve, just by using one of many scheduling systems and create one recurring job for each probe. However, on a distributed system the problem becomes a lot more complicated. One option would be to use a gossip-based solution to figure out which system is responsible for polling each probe. Unfortunately, Java EE 7 does not have a standard way (that the team was aware) to get the instance id in a clustered environment, so a lot of complexity would have to be added to achieve this. Another option would be to elect a master for the cluster, but that would introduce a lot of recovery issues.

The elected solution ultimately was a bespoke system tailored to the needs of the system, taking advantage of its characteristics. For instance, if the whole cluster is always restarted when performing maintenance, why should it be needed to figure out a way to redistribute probes among nodes at runtime?

The developed system periodically lists all the probes is the database in a random order and tries to schedule a pooling task for each one of them. The scheduling task is purposely performed at a slow rate allowing the other members of the cluster that are starting up at the same time to take for themselves some of the probes as well (taking advantage of a controlled race condition). When a probe is scheduled, a scheduling record is placed on the Hazelcast instance, hence when another node tries to schedule a given probe, it can skip it if a recent enough record exists. Given that the task runs periodically, it will automatically re-schedule the polling for any probe that is left orphaned as the scheduling record ages past a set amount of time. The polling task itself checks what is the latest measurement polled from the probe, and requests up to 6 hours of measurements from the probe internal storage (the probe can hold about two weeks of measurements). It also performs sanity checks on the retrieved data and stores it on the Cassandra database, and finishes with the re- scheduling of a polling task for that probe, overwriting the record on Hazelcast with a fresh one.

---

[1] Technical specifications of a sample ASUS 1U Commercial Server Workstations - https://www.asus.com/CommercialServers-Workstations/RS300E7PS4/

There is no hard-locking mechanism that avoids dual scheduling. If for some reason the probe gets scheduled on two nodes at the same time, it is likely that both will perform the scheduling task. However, it is highly unlikely that the tasks will finish at the same time, in part due to the design of the probe. This means that after the first scheduling collision, when trying to re-schedule a task, the last node to finish will read a very recent record from Hazelcast and skip the rescheduling step. The fact that the probe was asked twice for the same measurements is also not a problem as this is very uncommon, and the returned data will be the same, so the only problem would be the additional data consumed if this was a common situation. The same race condition is allowed during the initial bootstrap process, where each node on the cluster takes over a random set of probes.

### Storing measurement data

Measurement data is stored on a four-node Cassandra cluster. This cluster was originally installed on Cassandra 2.0, and is now running Cassandra 3.0. Being always upgraded while running in production and throughout the whole life of the project, the node only needed to be wiped once due to an unspecified problem that would severely degrade the performance of that single node. The cluster is repaired every night during the off hours, and from time to time a rolling restart is performed without interruption to the service.

However, the operation of the Cassandra cluster has not always been so smooth, given that early versions of Cassandra would sometimes hang during the repair process, leading to a less than controlled restart of the node or even of the cluster. While upgrading our SSTable binary format, a process that takes a very long time when each node contains more than one terabyte of data, some issues were faced.

An early decision performed on this system was to use an ORM on top of Cassandra, specifically one that implemented the JPA specification. The original rationale for this decision was that this would allow a new developer to easily understand the data model and how to operate on it. On the more recent rProbe implementation, this ORM was abandoned in favour of a more stable driver.

### Post processing of the collected data

In order to allow the client to visualise data over long periods of time, in an effort to detect long time trends, the received data is post processed after being stored. The post processing process is executed at 15-minute interval, and it gets information from all probes that have at least one complete hour of measurements. For each variable, the system calculates the median and average as well as standard deviation (linearising the data when the data is logarithmic) and calculates a 100-bucket histogram of that data. This pre-calculation allows the client to visualise long term trends once they become significant. The results are also stored on the Cassandra datastore for later retrieval.

### Operation

After being collected from the probes, the data is stored on a Cassandra datastore. This allows the client to visualise the data and replay it on a complex event processing system that enables advanced analysis about it. However, disk space at private facilities is not infinite nor is it easy to upgrade. For the first year of the system operation, the development team kept a close eye on the evolution of the disk space usage and started a plan to remove data while keeping the largest possible dataset on the system datastore for online analysis.

The plan is simple, let the data grow until it uses about 40% of the available disk space, then use those remaining 10% (avoid going over 50% as a scrub may require that much free space) to write tombstones that will eventually delete 10% of the existing data when the SSTables are compacted. However, the Cassandra datastore way of operating is sometimes unpredictable, with tombstones not being removed until the SSTable that contains the data is compacted with the SSTable that contains the tombstone.

A backup tool was designed when the version of Cassandra was still 2.2, and range deletes were not supported, hence a single delete statement must have been created for each deleted record. As the system is updated to the latest 3.0 release, the development team may review this strategy after performing unit and integration testing. The backup tool creates an export directory with the exported data and the respective delete statements. Those delete statements are moved to a second directory, and the exported data is compacted and moved to an external cold storage system. In order to allow the data to be easily handled should the need arise, the data is chunked into one-month periods and compacted on disk.

### A brief history of the system

The system design allows for the failure of every single component but the top of rack switch or the Internet connection (those are the only points of failure that we are aware of to this day). The initial system setup was based on Ubuntu server 12.04, Java 7, Cassandra 2.1 and PostgreSQL 9.1. During its production run, all of these were updated at least once, in some cases with extensive work involved.

Early during the production run the dataset grew very quickly, and a large quantity of bugs appeared while trying to run some routine repairs on the cluster. During the first months into the production run, it was also figured out that a choice elsewhere was preventing the performing of a rolling restart of the cluster without causing application downtime. At this point the operations team was spending a lot of time trying to figure out how to increase the system stability, applying every single update to the datastore and hoping that the show stopper bug (not always the same) that was keeping the cluster from operating at full potential would go away. Attempts at running the repair, scrub or clean-up tasks after each update were made, and always for one reason or another the task would fail. Here, praise must be given to Datastax, as their bug tracker and community support was always stellar. Usually when a snag is hit, a quick bug tracker search would quickly reveal the fix was already committed and due for release soon.

After the update to Cassandra 2.2, (and the migration to JDK 8 from oracle), the stability of the system began to increase, and for a long time, regular operation run scripted from cron without human intervention. The update to Cassandra 3 was performed during a major system revamp that allowed us to migrate from Kundera's thrift driver to Datastax's native driver, which gave us a significant performance increase as well as the ability to perform rolling restarts of the cluster.

While operating Cassandra 2.2 (an early release of 2.2), and holding about 1TB of data per node, the system became quite sluggish. After a series of tests, it was determined that one of the nodes was systematically timing out queries. The first thing that was tried was to repair the node partition range. The repair failed due to one of the bugs referenced earlier. Then a node scrub was tried, however scrubbing that amount of data takes quite a long time and eventually after about a week this attempt was dismissed and assumed there was something wrong with the data. Amazingly, this node was not writing any error messages to the log files that would help determine why it was misbehaving.

After long consideration, it was decided that the best course of action would be a full node rebuild. Doing so exposed our client to a situation where another node's failure would cause data loss as the replication factor is 2 and the cluster is composed of four nodes. The node was shut down, its data directory was removed and it was then restarted. After some struggle, the node joined the cluster, and a repair process rebuilt its SSTables from the data stored on the remaining nodes. This whole process took about a week, the rate was kept slow to avoid impact to the running system. After repairing the whole cluster, a scrub and cleanup was run on each node, ensuring that data that was added while the cluster had only three nodes was moved to the proper nodes and no garbage remained on each node.

After this episode, up to the time of writing this document, this was the single big issue we faced on the whole cluster that caused some concern both to the team and to the client. As Ubuntu has a very short support cycle for an enterprise Linux, the planning of a migration from 12.04 to 16.04 was started about one year before the EOL of 12.04. Plenty of tests were performed, and the team tried to ensure that it knew exactly what could go wrong and how could this process fail. The team simulated the upgrade of every single software package it was running, from the parts of the service, to support software such as keepalived and Pgpool. As such, it seemed like the update could go ahead, and about one month before the EOL of 12.04, a week of downtime was scheduled and the team started working on the update.

Unfortunately, right away the update to the first node failed, the only thing that had never happened during the system testing. After a lot of debugging the cause of the failure was found. The team needed to uninstall anything related to Pgpool2, to Java 7, and the monitoring system (Icinga) from each node before performing the update. The update was performed in two steps. The first step was to update from 12.04 to 14.04 and migrate the Postgres cluster from 9.1 to 9.3, the second step was updating to 16.04, migrating the Postgres cluster from 9.3 to 9.5, and writing a few systemd units to start the servicers on boot. Repeat the process six times (4 application nodes and 2 frontend systems) and everything should be fine. Unfortunately, that was not the case. The team removed the Pgpool system and needed to rebuild it. The team took the opportunity to rebuild the Pgpool with streaming replication instead of query replication by Pgpool, as this methodology is a lot more reliable and since setting it up never had any failure.

When the service was first deployed it ran on a Glassfish 4.0 application server. With time, it was updated to 4.1, but the reality is that the Payara release is a lot more stable, and it packages a managed instance of Hazelcast, a service that is used within Ubiwhere to maintain state across the cluster where needed. The update of the system to run on Payara required a bit of application development, the team stopped launching the embedded Hazelcast instance, and started injecting the managed instance where needed. Overall, this decreased the deployment time of the war and reduced the applicational code, something that always helps with maintainability.

**Conclusions from this experience**

This whole process was (and keeps on being) a learning experience for Ubiwhere, from system design to architecture, to operational procedures and client expectations management. This system proved that a small team can build reliable software that handles large amounts of data and has a huge impact on the quality of a public service on a country. Ubiwhere is proud to be a part of it and of our contribution to increase the quality of the DVB-T signal in Portugal. By using this solution, ANACOM is now able to monitor 24/7 the quality of the digital television service from the end user's point of view. With this tool, the regulator can detect potential problems even before a complaint is reported by a citizen. Furthermore, for each complaint, ANACOM is able to consult this monitoring system and assess if this problem is on the user's own TV installation (and if confirmed, help the user to correct its installation) or, in fact, it's a problem of coverage to be solved by the operator. In what regards the initial considerable amount of problems reported by users, ANACOM was able to collect a set of monitoring data and provide detailed reports to the service operator exposing that there were real problems in the digital television signal grid. In fact, the granularity time stamping, and contextualised features of the monitoring system allowed ANACOM to detect signal failures associated with weather conditions, tides as well as signal collisions between multiple signal transmitters. These problems would be ultimately impossible to detect without such a monitoring system since they are only perceived by the population and therefore sensed at certain (and sometimes short) time of the day. With this powerful tool, ANACOM already asked for a considerable set of interventions in the television digital broadcasting system by the service provider thus considerably increasing the quality of this service in Portugal.

As mentioned in the beginning of the document, a new product was created based on the building blocks of this solution. rProbe (www.rprobe.com) is an evolution of the previously described system being able to monitor a wider spectrum of radio communications. Both the monitoring probe (rNode) as well as the central system (rCenter) benefited from the experience gathered in this project and were enhanced with extra functionality in order to be deployed in other ecosystems that do not use DVB-T. rCenter was also evolved to a multi-tenant solution and can now accommodate more than one client in a cloud environment. However, due to some constrains of potential clients, rProbe solution can still be deployed in a private client environment (such as what was done for ANACOM) without losing any of its functional and nonfunctional features.

# 3     SMART CITIES & FUTURE INTERNET

One of Ubiwhere's main focus is on the Smart Cities domain. This area has seen a lot of evolution in the last few years, growing from a focus on R&I to large and consistent deployments in several municipalities. With this development, cities have been able to gather data regarding its various services and resources, which allows the development of solutions that can solve various issues, from optimisation of cities' resources to improving the quality of life of its citizens [2].

**Mobility Backend-as-a-Service**

With this in mind, Ubiwhere has developed a product called Mobility Backend-as-a-Service (MBaaS). This is a web platform focused on Urban Mobility capable of collecting, harmonising, processing, storing and provisioning mobility data and services to third-party applications or information systems. These capabilities enable the platform to act as a data management middleware between city service providers (parking operators, public transportation agencies, traffic managers, bike-sharing operators) and municipalities/citizens/developers (providing open data, open APIs and intelligent web services such as analytics, routing and alerts). Looking at the wide domains that such a platform can operate, it can be understood that it needs to have the capacity of handling large amounts of real-time events data, as well as the ability to process large historical datasets. When you have large amounts of sensors regularly providing data, as well as events communicated by users (e.g. reporting incidents on the roads), the quantity of data the system has to process can easily run out of control.

As such, the design of the system, apart from providing the tools to obtain values from the data (analytics, predictions, services), needs to have the capacity to scale horizontally. To reach this goal, Ubiwhere follows the approach of using open-source tools that have a proven record of functioning well in such a Big Data environment. With the purpose of defining the best architecture for these requirements, there have been several tools successfully used in this domain. Apache Airflow has been used to programmatically schedule and monitor ETL (extract, transform, load) jobs, allowing the team to better control the vast number of tasks designed to prepare the data to whatever is needed, from analytics, compliance to standards or predictions. Apache Spark has been used for large-scaling data processing, allowing the system to efficiently handle the incoming data, be it in real-time or through batch processing. Druid has been deployed to as a high performance, column-oriented distributed data store, allowing for the ability to store event data and efficiently provide analytics. The tools Superset and Metabase have also been used in different occasions to provide interactive visualisations over the data. MBaaS is being designed as an Urban Platform that can lead with multiple Smart Cities domains, taking into account the dimension that such a system can reach. As such, the tools being used in its architecture are carefully selected and experimented in order to guarantee the capability of the system to handle Big Data.

**Smart Lamppost**

While the previous section focuses on the Smart Cities domain, Ubiwhere has decided to use its vast knowledge and experience in the two areas (together with Telecom & Future Internet) to create a product that can provide a wide array of services to cities: The Smart Lamppost.

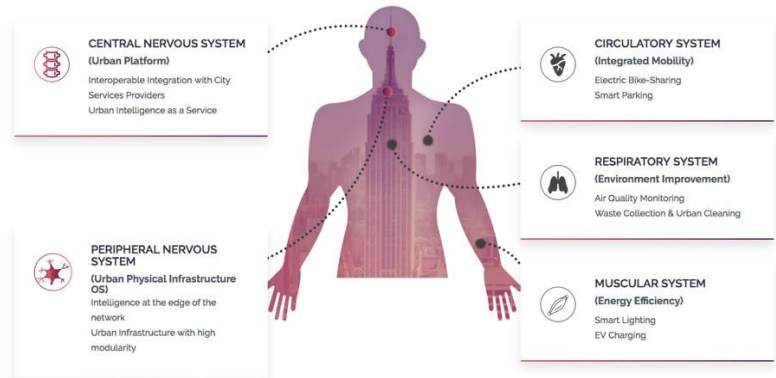

**Figure 2**. The Smart Lamppost System.

Smart Lamppost is a solution which provides not only efficient LED lighting with a companion smart lighting system but also telecommunication (4G/5G and Wi-Fi) and EV charging capabilities. Being modular by design, it targets different applications, as it can adjust to the specific use-case needs, combining only the necessary modules. Targeting smart cities, municipalities can leverage the built-in smart lighting system to manage its infrastructure and program their operational mode, based on a set of pre-programmed options. LED technology allows for a running cost reduction, given its efficiency, but the software component brings this reduction to another level when it comes to maintenance and programmability.

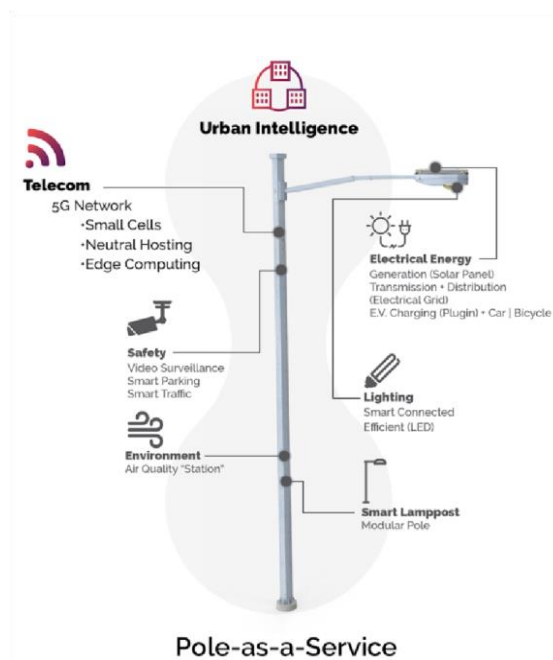

**Figure 3**. Smart Lamppost Modular System.

With its clean and minimalistic design, the optional Small Cell module can also be incorporated, giving municipalities the chance to monetise their original investment, by allowing Mobile Network Operators (MNOs) to rent this infrastructure's networking capabilities. As 5G technology is emerging, MNOs are investing largely in new infrastructure to meet their clients' demands. With the built-in Small Cell module and our set of NFV and SDN-based cloud services, MNOs can drastically reduce their initial investment, by renting already deployed and available infrastructure, effectively increasing their current capacity and coverage in a matter of minutes. For this, we are developing an innovative and powerful database of deployed infrastructure and its coverage area, where MNOs can simulate how much coverage, capacity and bandwidth they can achieve by renting the already deployed and available Small Cells.

In what relates to Big Data challenges, the Smart Lamppost opts for a different approach from the traditional cloud-based systems: by integrating computational modules in each pole, it is possible to move the logic of the system (or part of it) to the edge of the network. Thus, communication costs (which can become very high) can be avoided by computing what is possible at the edge of the network, avoiding unnecessary communication with the cloud and permitting fast response to certain incidents. One concrete example where this scenario makes sense is a camera that, through computer vision, is counting the number of cars waiting on a traffic light. Instead of communicating the video to the cloud, it can process it locally and only send the number of cars present at the moment.

There are several use cases where the Smart Lamppost provides value to the city, either by optimising communication costs, or by providing services that need fast response and cannot wait for data to be transmitted to the cloud. Such a locally fast response system can be very useful in vehicular communications. Nowadays, a car is capable of constructing a 3D map of its surroundings to become able to detect objects on the road and avoid collisions. But this is only a very small view of the local area. Through the lamppost, it will be possible for cars to communicate their views with others, and thus retrieve a bigger picture of the scene, which will help cars navigate through blind curves or have a better understanding of the overall situation. Another use case will be the understanding of the situation at each road junction, and thus providing a better picture of the whole city's traffic status, all through the communication of the cars with the lamppost. This shall also help guiding ambulances through least congested roads, by becoming able to change traffic lights dynamically to provide a faster route to such vehicles.

As already stated before, the lamppost could be very helpful in computer vision situations, as the ones promoted in SKA. Apart from the given example, the pole will also be useful in high-processing scenarios, such as detecting traffic accidents, reporting crimes, among others. Given that videos can easily scale in size, Ubiwhere's product avoids transmitting such large amounts of information, easing the load of the network and reducing communication costs. A final example on how the lamppost can be useful in a Smart City is through content distribution. If a popular show is on television, instead of sending the video feed to every household, the operators could instead distribute the video feed through the network of Smart Lampposts, which could cache the video and forward it to the various local households. This will reduce the large quantities of data being transmitted, once again reducing costs.

The Smart Lamppost takes a different approach in dealing with Big Data scenarios, by shifting computational logic to the edge of the network, optimising the communication costs of transmitting the data to the cloud. It is clear that if cities invest in this sort of infrastructure, they could optimise their services, and offer the possibility of creating new services and applications through the capabilities of the network.

# 4 UBIWHERE

As stated before, Ubiwhere focuses its commercial activities and research & development efforts on two main areas of software development: Telecom & Future Internet and Smart Cities. These activities involve multiple tasks of data management, from database setup and design, where data is stored up to the interfaces that collect the information (typically from hardware equipment) and that provide friendly and usable interaction with end-users.

Concretely for the Telecom area, Ubiwhere has developed tools to monitor the Quality of Service (QoS) and Quality of Experience (QoE) of different types of networks (such as LTE, Long-Term Evolution and DTT, Digital Terrestrial Television), having also been involved in several Research & Development projects for 5G (fifth generation of Mobile Networks), based on the implementation of innovative solutions like Network Functions Virtualization (NFVs) and Software-Defined Networks (SDNs).

For Smart Cities, on the other hand, Ubiwhere's strategy passes by offering an open and interoperable software platform that can support the interoperability of different "vertical" solutions on the domains of Mobility and Environment. With the purpose of supporting this vision, Ubiwhere is involved on several standardisation initiatives, as a full member of ETSI and a gold member of the FIWARE Foundation. On this field, Ubiwhere supports the usage and works with Open Data, allowing municipalities and city service providers to obtain more value from their operational resources. On the data front, Ubiwhere has experience in the usage of Big Data tools to support the platform, for it is expected for an urban environment to generate large amounts of data, through thousands of sensors providing readings regularly, devices used for service provisioning or citizens contributing with updates about their surroundings. Ubiwhere has also highly invested in data analytics, so that it can provide cities with true value from their data, by understanding the key metrics and providing in understandable interfaces, the information decision makers need.

Whenever possible, Ubiwhere opts for the integration of open-source tools, for they tend to be more flexible, secure and more regularly updated. Moreover, such pieces of software can be enriched with new features or error corrections. As already stated, Ubiwhere has experience in using Cassandra[1] at scale, but the company also has experience with other data storage tools, such as Elasticsearch[2] and Druid[3]. Airflow[4] is also used within the company to programmatically author, schedule and monitor workflows and Apache Spark[5] has been used for large-scaling data processing. In order to provide diverse and efficient means of data visualisation, Ubiwhere is proficient in achieving this functionality with tools like Metabase[6] and Superset[7].

As described in the use cases provided in this document, Ubiwhere has on the market solutions successfully managing multiple Terabytes of data. As such, the company is confident about its expertise to work on projects of this nature, regardless of the domain. Therefore, given the large amount of data expected to be obtained from the telescopes of the SKA project, Ubiwhere is ready to bring its experience and provide the support needed for this issue.

## 5      CONCLUSIONS

Throughout this document, we have described our experience leading with Big Data situations, both on the Telecom and the Smart Cities domains. Big Data provides a diverse set of challenges, either from an operational point of view, or when applying algorithms to obtain value from data. Ubiwhere will continue to invest in this area, which is ever more relevant as the amount of data grows exponentially. However, much of the work done within the company is in the service of municipalities, government entities or public companies and, as such, it is necessary that these entities invest in the infrastructure of the cities to obtain and handle large quantities of data so that companies can provide innovative services over these data and improve the quality of life of their citizens.

## REFERENCES

[1]    Gavalas, D., Nicopolitidis, P., Kameas, A., Goumopoulos, C., Bellavista, P., Lambrinos, L., & Guo, B. (2017). Smart Cities: Recent Trends, Methodologies, and Applications. Wireless Communications and Mobile Computing, 2017.

[2]    Han, G., Guizani, M., Lloret, J., Chan, S., Wan, L., & Guibene, W. (2018). Emerging Trends, Issues, and Challenges in Big Data and Its Implementation toward Future Smart Cities: Part 2. IEEE Communications Magazine, 56(2), 76-77.

---

[1] http://cassandra.apache.org

[2] https://www.elastic.co

[3] http://druid.io/

[4] https://airflow.apache.org

[5] https://spark.apache.org/

[6] https://www.metabase.com

[7] https://superset.apache.org

# Power Monitoring and Control for Large Scale projects: SKA, a case study[¥]

Domingos Barbosa*[a], João Paulo Barraca[a,b], Miguel Bergano [a], Diogo Gomes[a,b], Dalmiro Maia[c], Luis Seca[d]

[a]Instituto de Telecomunicações, Campus Universitário de Santiago, 3810-193 Aveiro, Portugal;
[b]Universidade de Aveiro, Campus Universitário de Santiago, 3810-193 Aveiro, Portugal;
[c]Faculdade de Ciências da Universidade do Porto, Rua do Campo Alegre, 4169-007 Porto, Portugal;
[d]Center for Power and Energy Systems (CPES), INESC TEC - INESC Technology and Science Porto, Portugal;

**ABSTRACT**

Large sensor-based science infrastructures for radio astronomy like the SKA will be among the most intensive data-driven projects in the world, facing very high demanding computation, storage, management, and above all power demands. The geographically wide distribution of the SKA and and its associated processing requirements in the form of tailored High Performance Computing (HPC) facilities, require a Greener approach towards the Information and Communications Technologies (ICT) adopted for the data processing to enable operational compliance to potentially strict power budgets. Addressing the reduction of electricity costs, improve system power monitoring and the generation and management of electricity at system level is paramount to avoid future inefficiencies and higher costs and enable fulfillments of Key Science Cases. Here we outline major characteristics and innovation approaches to address power efficiency and long-term power sustainability for radio astronomy projects, focusing on Green ICT for science and Smart power monitoring and control.

**Keywords:** Radioastronomy, SKA, infrastructure, Power, Green Computing, Smartgrid

## 1    INTRODUCTION

The Square Kilometre Array (SKA) is an international multipurpose next-generation radio interferometer, an Information and Communication Technology machine with thousands of antennas linked together to provide a collecting area of one square kilometer [1,5]. Activities toward Large science projects toward a next generation of Radio Astronomy telescopes, enable innovative approaches to a reduced power footprint. This is done through pushing innovative approaches to renewable energies as well as to low power computing with improved algorithms, in a domain were high density big data processing and imaging computing is mandatory. The Energy Sustainability of large-scale scientific infrastructures led to consider the impact of their carbon footprint and Power costs into the respective development path and lifetimes SKA presents the opportunity for a combination of low power computing, efficient data storage, local data services, inclusion of newer Smart Grid power management, and inclusion of local energy sources, including potential Renewable Energies.

*dbarbosa@av.it.pt; phone +351 234 377 900; fax +351 234 700 901

---

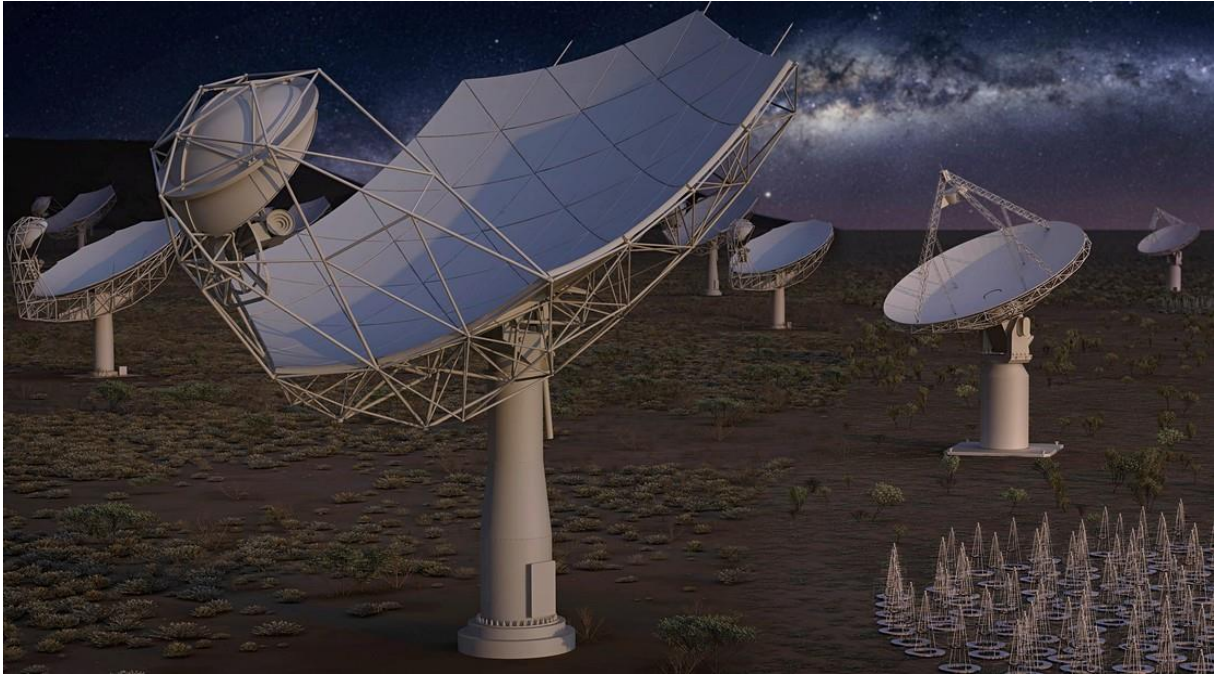

**Figure 1**. An artist vision of the SKA Core site, with some of the projected 3000 15-meter parabolic dishes and low frequency dipoles. From http://www.skatelescope.org.

The SKA is broken down in to various elements that will form the final SKA Observatory managed by an international consortium comprising several world leading experts in their respective fields. We should note that SKA will be in fact constituted by two Telescopes, SKA1-Low (Australia) and SKA1-MID (South Africa), each one providing their own Elements subsystems. Table 1 provides the characteristics of each SKA Phase 1 telescope (hence SKA1). A distinct Element product tree will be designed for each Telescope during the Pre-Construction Phase. These products will be based on a common architecture and design as far as possible, except for the Dish products, to be deployed in South Africa and the Low Frequency Arrays products to be deployed in Australia.

|  | **SKA1_LOW (Australia)** | **SKA1_MID (South Africa)** |
|---|---|---|
| **Sensors type** | 130 000 dipoles | 197 Dishes (including 64 MeerKAT) |
| **Frequency range** | 50-350 MHz | 0.45-15 GHz |
| **Collecting Area** | 0.4 Km$^2$ | 32 000m$^2$ |
| **Max baseline** | 65 Km (between stations) | 150 Km |
| **Raw Data Output** | 0.49 Zettabyte/year | 122 Exabyte/year |
| **Science Archive** | 128 Petabyte/year | 1.1 Exabyte/year |

**Table 1**. SKA Phase1 Telescopes Broad Characteristics

Like any major large-scale astronomy projects, installed usually in remote locations, the associated (power hungry) data processing location is conditioned by the experiment, and not by the computational facilities, resulting in far from optimal efficiency, higher capital expenditure (CAPEX) and higher operational expenditure (OPEX). Addressing both the reduction of electricity costs and the generation and management of electricity at system level is paramount to avoid future inefficiencies and higher costs. Remote locations also imply development of customized supply grid that may be built in phases, preceding the deployment of the experiments. This means usually power caps may be imposed with consequences on Key Science prioritization. Phasing of projects also alleviates concerns with Infrastructure and power budgets: it is easier to aggregate sensors (antennas) and upgrade processing facilities once Infrastructure including power is expanded to cope with the planning of Science Operations. However, fulfillment of certain Science Cases including observations of Transient or other Virtual Observatory (VO) triggered observations may produce sudden peak power loads.

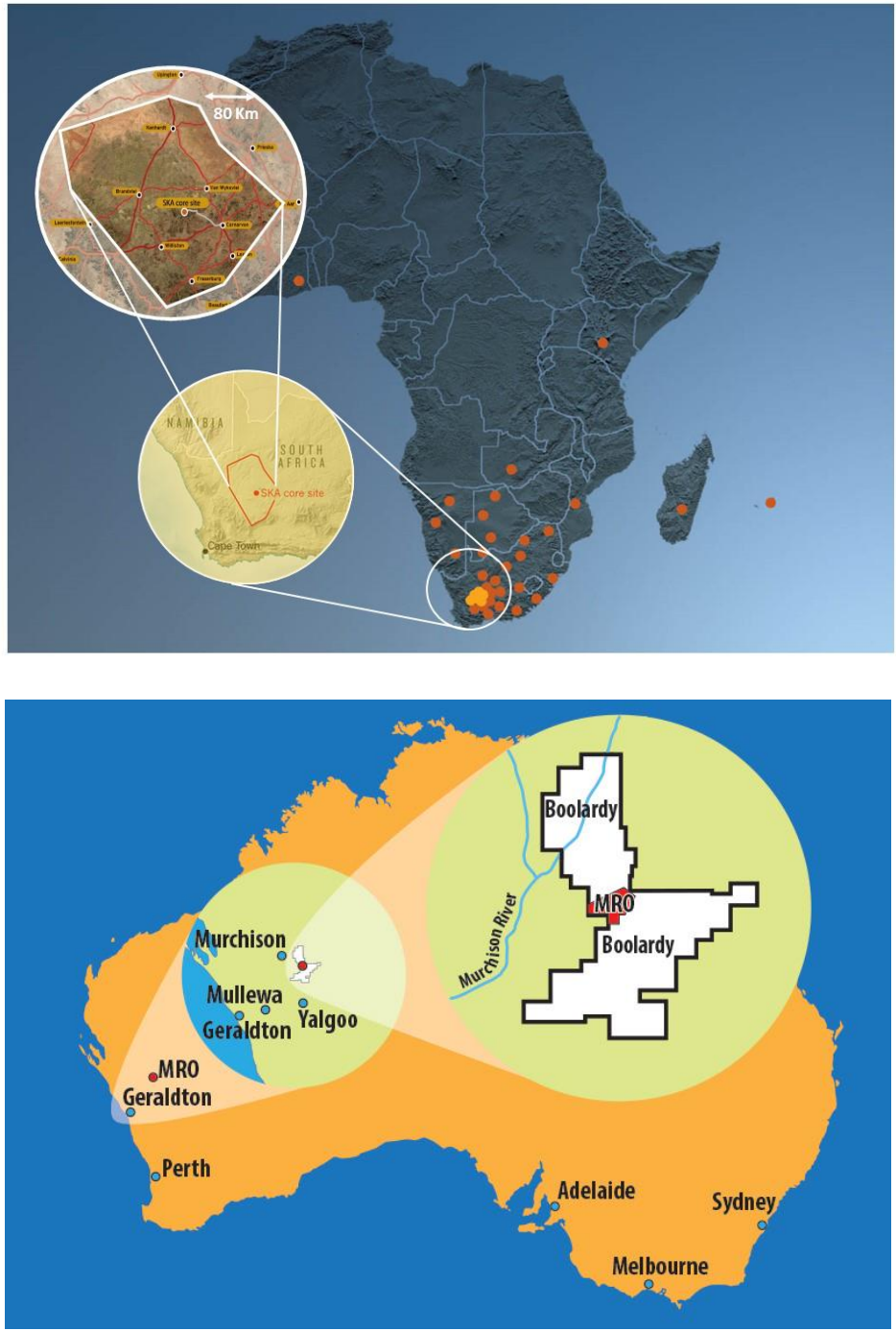

**Figure 2**. Top: the SKA1 Radio Quiet Zone, in the Karoo, South Africa. Also shown are the planned SKA2 locations in the SKA Africa Partner Countries. Bottom: The SKA1 Radio Quiet Zone of the Murchison Shire of Western Australia (@ SKA Australia).

As an example, these unexpected, yet extremely relevant astronomical events will not only require precise power metering, but also the capacity to manage overall system dispatch, considering technical and economic constraints. So it becomes clear that this would imply that to deal with these operational modes, it is necessary to forecast load and generation but also to remotely control some of the telescope components to not compromise the capacity to register these phenomena. Hence, to be able to support an adequate operation for this system, we should plan in detail the dimension of the local renewable based generation, the storage needs and at the same time some flexibility from demand, by curtailing unnecessary services/appliances whenever possible. The local electrical network should also have the capacity to adapt to the different operation scenarios, namely by reconfiguration actions and also by multi-temporal management of distributed resources.

## 2      POWERING LARGE SCALE PROJECTS

Energy Sustainability of large-scale scientific infrastructures and the related control of OPEX means inclusion of the impact of power provision costs and associated their carbon footprint into the respective development path and lifetimes [2]. Additionally, the ESFRI has indicated that it is paramount that a multitude of test facilities and Research Infrastructures should lead the world in the efficient use of energy, promote new renewable forms of energy, and develop low carbon emission technologies, to be adopted as part of a future Strategic Energy Technology Plan [2]. Radio astronomy projects will be among the most data-intense and power hungry projects. Recent experiences with SKA precursors and pathfinders like ASKAP, MeerKAT and LOFAR, MWA reveal that an important part of the life cycle cost of these large-scale radio astronomy projects will be power consumption [6],[7]. As an example, a 30-meter radio telescope requires approximately during operation 1GWh for a typical 6h VLBI observation experiment, enough to power a small village, while new infrastructures based on Aperture Arrays, promising huge sky survey speeds, may require even more, based on estimated digital processing needs [7]. Among the large scale well-known science infrastructures facilities, SKA will set the highest constrains on power consumption and availability, surpassing current figure considerably as can be perceived by Table 2. This is due to geographical spread of the SKA sensor network, ancillary facilities deployed in remote locations in the Karoo savanna in South Africa and the Western Australia desert

| SKA Phase 1&2 | South Africa | Australia |
|---|---|---|
| Sparse Arrays | | 3.36 MW |
| Mid Dishes | 2.5MW | |
| Survey Dishes | | 1.2MW |
| On-site Computing | 4.7MW | 1.32MW |
| Totals/site | 5.7MW | 4.8MW |
| SKA Phase2 incl. Dense Arrays | >40MW (SKA Phase 2 configuration not known yet) | |
| Off-site Computing | ~20-40MW (SKA Phase 2 configuration not known yet) | |

**Table 2**. SKA Energy Budgets

Addressing both the reduction of electricity costs and the generation and management of electricity is paramount to avoid future inefficiencies and higher costs. For instance, the Atacama Large Millimeter Array (ALMA) interferometer and the Very Large Telescope (VLT) in the Chilean Andes are powered from diesel generators, leading the European Southern Observatory (ESO) to consider potential inclusion of local greener energy sources to its Very Large Telescope (VLT) facilities in Paranal [8]. For VLT, electrical power is produced in off-grid mode using a combination of efficient multi-fuel turbine generators (2.6MWe at the site) that can use fuel sources with lower carbon footprint like natural gas, or Liquefied Petroleum Gas (LPG), combined with diesel generators connected to a 10kV power grid. However, fossil based fuels experience strong market fluctuations, with the overall long term trend showing a steep price increase. In a recent study, it was pinpointed the impacts of the increasing costs in electricity provision in Chile: between 2003 and 2010 the price rose by 7% per year according to statistics from the Organization for Economic Cooperation and Development (OECD) [6, 8]. Therefore, the ALMA permanent power system plant, capable of providing up to 7MW peak in "island -mode" is already prepared to connect to a renewable power plant, and the European Extremely Large Telescope (E-ELT) might include options for local renewable provision when market options in Chile make these technologies economically accessible [8]. Hence, from the pure power provision point of view and associated control of operational costs, fossil fuel price fluctuations and longer term availability and associated price rises represent a challenge in terms of planning a suitable energy mix supply, in particular for remotely located infrastructures. Overall, the main characteristics concerning the SKA power system can summarized as:

- Many Antennas nodes are far away from civilization centers and power grid in climates with high thermal amplitudes.

- Exquisite control of Radio Frequency Interference and EMI from Power systems is needed, since RFI would impair the radio telescope sensitivity.

- Different Power requirements over large distances.

- Continuous operation (meaning 24/7 availability) for sky surveying points out that some storage capabilities may be required, and power supply for night operations must be carefully considered.

- Power balance and control: control of large power peaks, for operation, cooling, computing and telescope management and monitoring, as a mean to maximize the integration of renewables based energy sources.

- Scalability: the power infrastructure should scale from SKA Phase 1 to the later, more extended, and more power demanding Phase2 (see Table ).

# 3 THE IMPACT OF GREENER ICT

The biggest computing challenge within radio astronomy lies within the architecture of the correlator of big synthesis radio telescopes and the second tier processing and storage infrastructures. The correlator processes the data streams arising from the large number of antenna elements of say, with N>1000 antennas. The optimum architecture is planned to minimize power consumption as much as possible by following several approaches: minimizing I/O (storage media, and network interconnects) and memory operations, implying preference for a matrix structure over a pipeline structure and avoiding the use of memory banks. For instance, the ALMA correlator selected for its core design the StratixII 90nm technology based on considerations on power dissipation and logic resources while much lower power technologies are available now. The SKA, under the Central Signal Processor Element Consortium, is currently developing design concepts in a power efficient way for design for N>2000 and over 1 GHz frequency bandwidth, based on Application-specific integrated circuits (ASICs) fabricated in a ~20nm CMOS process, still better than 20nm for FPGAS with low power considerations. Also, the integrated circuit (IC) design that performs digital cross-correlations for arbitrarily many antennas in a power-efficient way using intrinsically low-power architecture in which the movement of data between devices is minimized. Excluding antenna data pre-processing, the SKA correlator is estimated to consume less than 100 kW [3].

Hence it is expected a great advance on the low-power processing capabilities of SKA correlators. After data is integrated by the correlator and further processed to create calibrated data, it must be stored in a permanent media, such as the case of massive Storage Area Networks (SANs), relying in rotational technologies such as hard disks.

| Telescope | Status   | Technology  | Design Freeze | N of elements | Power efficiency |
|-----------|----------|-------------|---------------|---------------|------------------|
| VLA       | Obsolete | ASIC        | 1975          | 27            | 171k             |
| JVLA      | Existing | ASIC 130nm  | 2005          | 32            | 4270             |
| ALMA      | Existing | ASIC 250nm  | 2002          | 64            | 992              |
| LEDA      | Existing | GPU 28nm    | 2011          | 256           | 977              |
| CHIME     | Existing | GPU 28nm    | 2013          | 128           | 769              |
| SKA1-Low  | Proposed | FPGA 16nm   | 2017          | 512           | 74               |
| SKA1-Low  |          | ASIC 32nm   | 2015          | 512           | 4.8              |
| SKA1-MID  | Proposed | FPGA 16nm   | 2017          | 197           | 103              |
| SKA1-MID  |          | ASIC 32nm   | 2015          | 197           | 12.7             |
| SKA2      | Planned  | TBD         | 2021          | >2000         | TBD              |

**Table 3**. Power efficiency of Radiotelescope Correlators. From D'Addario and Wang (2016)

ALMA can output several TeraBytes of data per project that must be stored, and the future SKA infrastructure is expected to produce closer to an Exabyte/day of raw information, prior to further processing and data reduction (see Table **1**). All these data must be made available in large facilities for further reduction by researchers (eg, using CASA [13] or other parallelized data handling software pipeline requiring most probably a high degree of automatisms). Due to the amount of information, and the costs of transmitting data through long distance optical links, it is vital the use of computation facilities located in close proximity to the source of information, but also close to researchers, in order to reduce latency and cost of the post-analysis process. Hence at SKA, most of the compute power will be located in Central Processing Facilities, properly shielded, in the vicinity of the Radio Quiet Zones in the Karoo and Western Australia.

The typical approach is to create computational behemoths capable of handling the entire operation of the instruments, storage, and frequently further processing of the data produced. However, lessons learned from similar large infrastructures show firsts years' operations to have frequent interruptions caused by detection of erroneous or unexpected behavior, requiring further tuning, or even due to integration of components arising from phasing of project deployment. Even after entering into its normal operational status, instruments are, among other factors, affected by maintenance downtime, and also by weather conditions limiting observations. As an example, according to the ALMA cycle 0 report, over the course of 9 months (total of ~6500 hours), the instrument was allocated for 2724 hours of observation time, and this resulted in 38% (1034 hours) of successful observation [9]. This results in a considerable efficiency loss, considering all the processing infrastructure that must be available, independent of the observation status. Although we believe the initial processing must be done close to the location of the sensors, data processing should be shared or co-located as much as possible to other already existing infrastructures, exploiting time multiplexing as a way of increasing power efficiency. Moreover, further offline reduction methods can be improved as they currently typically use dedicated hardware and facilities, which are only used after a successful observation is obtained, further increasing the OPEX and the carbon footprint of science.

From this, it is clear SKA requires signal and data processing capacities exceeding current state-of-the art technologies. The DOME project [18,19] is a 5-year collaboration between ASTRON, South Africa and IBM, aimed at developing emerging technologies for large-scale and efficient (green) exascale computing, data transport, storage, and streaming processing. Based on experience gained with a retrospective analysis of LOFAR, the DOME team analyzed the compute and power requirements of the telescope concepts for the first phase of the SKA [12]. These initial estimates indicate that the power requirements are challenging (up to order ten peta operations per second (OPS) in the station processing and correlation), but especially the post correlation processing (order 100 peta OPS to exa OPS) is dominating the power consumption [12, 18]. The study also poses mitigation strategies, such as developing more efficient algorithms, fine-tuning the calibration and imaging processing parameters, and phased-implementation of novel accelerator technologies.

From the perspective of Green Computation, there are several aspects that have been tackled in order to increase the efficiency and decrease operational costs (OPEX) of current infrastructures such as: location, infrastructure reuse, equipment selection (servers, racks, networking), and cooling parameters. Recently, the reuse of devices reaching their end-of-life has also been addressed as a way to reduce the ecological footprint of a given system. Ultimately, if considering the operational stage of a datacenter, the most common metrics for evaluating the efficiency of a computational infrastructure are FLOPS per Watt (F/W) and Power User Efficiency (PUE), where PUE= (Total Facility Energy/ Information Equipment Energy). An ideal PUE value would be 1, whereas state-of-art is already PUE ≤ 1.2 for some greener large datacenters. Most of these metrics can also be applied to large Science Infrastructures. In addition, some tasks may be off-loaded to public clouds having lower PUE values.

Location is a major aspect driving the development of a large computational cloud facility. Ideally, a data center should be placed next to a power source so that the price is minimum, and losses in the power grid are minimized (est. 17% is lost in the power grid [25]). For projects with grid supply difficulties and with ecological aspects, as it is common to observe, presence of water dams, wind turbines and solar panels may be considered, provided strict compliance the Radio Quiet Zone maximum allowed interference requirements. Moreover, if possible, to decrease OPEX there must be interconnectivity to the global Internet through multiple providers although in remote locations like the Karoo and Western Australia, this job may be provided only by the local National Research and Education Networks (NRENs) like the TENET/SANReN (South Africa) and AARNet (Australia). Climate and geography also play an important role with great impact in temperature control, and overall security of the infrastructure, although at the desertic SKA locations, free cooling can certainly be a problem due to the absence of water planes (rivers and oceans) or wind.

However, for most large science facilities location is conditioned by the experiment, and not by the computational facilities, which results in far from optimal efficiency, higher capital expenditure (CAPEX) and higher OPEX. As an example, ALMA, with its correlator located in the middle of the Atacama Desert, at an altitude of 5km, far from power sources, and with a thin atmosphere, presents serious engineering challenges even for keeping basic operation, and just without addressing efficiency concerns. Infrastructure reuse is another important aspect that is always considered, and at multiple levels.

In the area of computing and Internet service provisioning, it is possible to increase the usage rate of computational resources (servers), by exploring virtualization and service-oriented technologies, mostly due to the intermittent resource consumption pattern shown by almost any application or service. By combining multiple, unrelated services in the same hardware resources, processing cycles can be multiplexed, ensuring that overcapacity is reduced to a minimum. Using this technique, servers are optimized and redesigned to become highly power efficient. As a practical example, some commercial cloud providers exploit these properties by providing spot pricing for their resources, according to the laws of demand and supply. In this aspect, Cloud Computing technologies have emerged as a promising Green ICT solution, which can be exploited by Big Data Centers and Science Organizations [6]-[8], thus addressing also the management and power concerns of large scale science infrastructures. Hence, the concepts of Infrastructure as a Service (IaaS), Platform as a Service (PaaS) or even the lately developed Software as a Service (SaaS), can provide abstraction from physical compute infrastructure and potentiate data center operators to trim energy costs and reduce carbon emission. Furthermore, software development sourcing on the emergent DevOps ideas from the telco/IT sectors promote agile resource management, automating the process of software delivery and infrastructure through microservice architectural delivery with high modularity. DevOps practice ensures a set of Architecturally Significant Requirements (ASRs) such as deployability, modifiability, testability, and monitorability. These ASRs require a high priority, allowing the architecture of an individual service to emerge through continuous refactoring, hence reducing the need for a big upfront design, reconfiguration of physical infrastructure underneath and reducing the time to market introduction of well-developed software services via frequent software releases early and continuously.

## 4      THE MONITORING TECHNOLOGIES: SMART GRIDS

The specificity of demand, namely by a significant amount of scenarios with high levels of power requirements, support the implementation of the smart grid paradigm for the local distribution network. A truly smart grid will rely on adequate monitoring, communication and control over existing network, including flexibility coming form generation, storage and demand. This flexibility bears in mind that load and generation forecast are also included, being each of the SKA sites run by a Supervisory Control and Data Acquisition (SCADA) Distribution Management System (SCADA-DMS) that will support a more efficient, reliable and sustainable operation of each of the sites. Power monitoring of antennas and ancillary systems, Correlators, HPC facilities or related data center tiered systems must include advanced remote metering technologies, efficient distribution automation and power Network Operation Centers (NOC).

SCADA protocols are designed to be very compact, and we do expect SCADA-DMS system filtered information to be provided by INFRA element to the Telescope Manager (The Operational, Monitoring and Control Element of the SKA). SCADA also improve reliability, increase resource utilization and contribute to OPEX reduction.

In a typical configuration, power substations are controlled and monitored in real time by a Programmable Logic Controller (PLC) and by power-specialized devices like circuit breakers and power monitors. PLCs and the associated devices communicate data to SCADA node located at the substation. The links between the substation PCs and the central station PCs are generally Ethernet-based and may be implemented via the SKA Non-Science Data Network intranet. In some cases, the information could even use private versions of cloud computing. SCADA systems feature built-in redundancy and backup systems to provide adequate reliability, and can be deliver much faster-acting automated control that can greatly benefits utilities and consumers, in this case can benefit Large Scale Infrastructures like SKA. Capabilities include systemic problem detection with alarm handling and trigger adjustments and corrections, often preventing an outage when more serious problems may arise. These SCADA-DMS capabilities largely benefit extended sensor networks since they enable maintenance teams to identify the exact location of outage or any other major critical problems that may affect the Telescopes performance, and thus significantly increase the power stability and the speed of power restoration in the case of an outage via fast rerouting o power for unaffected regions without the need for maintenance visual inspections.

Besides allowing system operators to use powerful trending capabilities to forecast future problems, SCADA system allow storage of data for profiling the quality of power supply properties (voltage levels, power factors, other system parameters) across grid and hence across any SKA component subsystem. If economically viable, inclusion of any local power source generation (like renewable sources) can make power quality more difficult to achieve, thus requiring more automated responses since power supplied to the distribution system would come from multiple sources in addition to the large base-load power stations.

# 5 DEMONSTRATORS

To enable a good understanding of power reduction options and respect good Radio Frequency Interference, the weighting different solutions emerging from innovative approaches to renewable energies as well as to low power computing with improved algorithms are important. In this context, the BIOSTIRLING - 4SKA (B4S) was a EU demonstration project dealing with the implementation of a cost-effective and efficient new generation of solar dish-Stirling plants based on hybridization and efficient storage at the industrial scale. The main goal of the B4S demonstration project was the generation of electric power using simultaneously solar power and gas to supply an isolated system and act as a scalable example of potential power supply for many infrastructures, including future sustainable large scientific infrastructures. B4S build an interdisciplinary approach to address reliability, maintainability and costs of this technology. In April 2017, B4S successfully tested in Portugal the first world Stirling hybrid system providing about 4kW of power to a phased array of antennas, overcoming challenges in Stirling and hybridization and smartgrid technologies. B4SKA Consortium, with fourteen companies from six European countries, has performed the engineering, construction, assembly and experimental exploitation, under contract signed with the European to develop on off-grid demonstrator in Contenda (Moura) Portugal.

As an example connected to B4S, astronomical infrastructures are usually built in remote locations, turning power supply a potential sizeable fraction of capital and investment costs. Distributed facilities over hundreds of Kms like the Square Kilometre Array (SKA), to be installed in Africa and Australia deserts present serious and very stringent energetic demands. SKA, is a project to become the largest science infrastructure of XXI century, has been considered an ideal example to have as a reference for B4S, to guide some of the specifications of a demo plant. Additionally, a demonstrator of one of SKA Advanced Instrumentation technologies installed at Moura (Portugal), has been used as a reference isolated system to be fed by the B4S demo plant.

The expected average power usage of the whole SKA will be between 50-100 MW, but over an extended location (up to 3000 Km diameter), with many different nodes, and sparse occupation of that terrain beyond the central core. Since SKA will scan the sky continuously, it will not present strong power peaks and power fluctuations, requires a smooth consumption profile. Energy generation at a continental scale for this facility, with different load profiles at different locations, means that modular power generators are needed, presenting an ideal scenario for development of innovative solutions with its own degree of customization and grid connectivity. Another consideration is that SKA, by definition, requires 24/7 observation operations. Another consideration is that SKA, as a radiotelescope, can observe the sky 24/7, so power consumption should also be maintained night and day, and free of radiointerferences, given the extremely faint signals to be observed. Because of these technical requirements, the power supply considerations for SKA present an opportunity for inclusion of smart, more efficient and low-carbon technologies. It is at this point where B4S and SKA are connected.

**New Concepts**

Apparatus for the radio measurements included 4 tiles of an MFAA prototype based on Electronic Multi Beam Radio Astronomy ConcEpt (EMBRACE) [4,5]. EMBRACE demonstrates the design readiness of the phased array technology for the Square Kilometre Array (SKA). There are two major EMBRACE stations, one in Nançay, France, and the other one at the Westerbork Synthesis Radio Telescope (WSRT) in the Netherlands [8]. However, for deployment in South Africa EMBRACE requires qualifying for environmental factors and compatibility studies with the inclusion of sustainable energy sources. The former is also addressed with installation of EMBRACE modules similar to those in Contenda in the SKA South Africa Karoo site. Renewable's inclusion in a system with new SKA technologies is therefore performed as a world's first in Contenda. On the concept level the reliance on a "software telescope" such as these phased array concepts, reduces the number of stations as the collecting area per station is larger. Interesting enough this reduces the requirements of the central processing units and hence of the corresponding power needs. As now, the station requires a probably higher power as a result of the pre-processing and aggregate needs of the receivers, this invites the needs for a station level power solution. At the same time, this may reduce the costly copper requirement to the central processing when using fiber optic interconnect-and network solutions, which simultaneously increase radio interference and lightning (if any) effects.

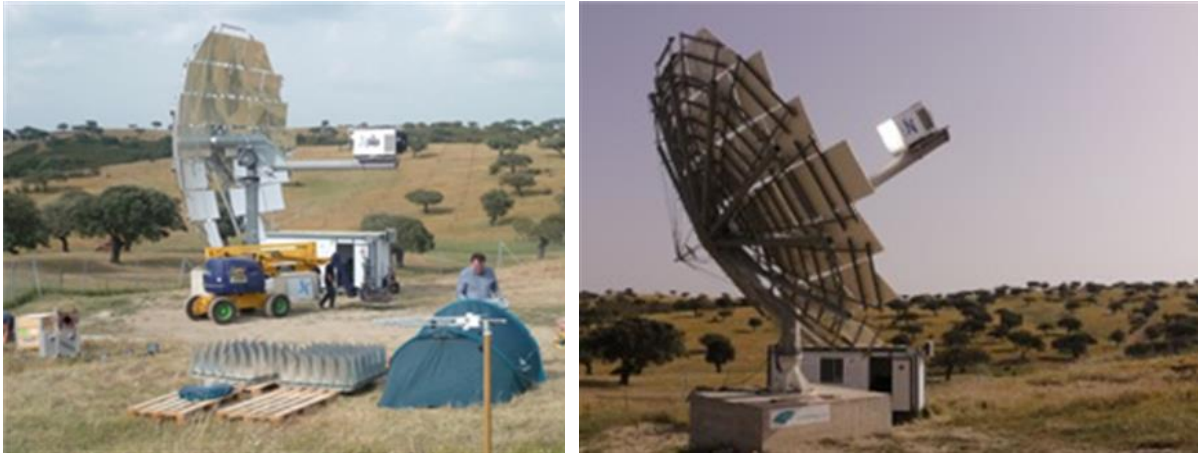

**Figure 3.** : left: EMBRACE Aperture Array tiles in RFI testing; right: The Dish-Stirling unit in routine Sun operations.

The B4S project is a demo project, with a high Technology Readiness Level (TRL) and aimed to push forward the solar dish–Stirling technology towards full dispatchability in the smart grid field or power islands, reduce cost and increase efficiency and life-time of these concentrating solar power systems. The B4S Power Plant was connected to the SKA Mid Frequency Aperture Array (MFAA) demonstrator and started working in simulated real conditions as the only electric supplier of the radio-astronomical system as a whole. MFAA tiles require a stringent steady state power supply, besides a good control of Radio Frequency Interference (RFI) levels, in agreement with SKA RFI standards. For Assembly, Integration and Verification operations (AIV), ASTRON provided the demonstrator including 4 antennas tiles, 1 antenna beam former, protection huts and processing unit, electrical circuitry and control. ASTRON was also responsible for its assembly and integration, while IT and CSIC were responsible for the verification tasksThe Bio-Stirling-4SKA (B4SKA) Consortium, with fourteen companies from six European countries, led by Gonvarry (Spain) has performed the engineering, construction, assembly and experimental exploitation, under contract with the European Commission for the Seventh Framework Program (7FP) of a solar concentration system powering a set of demonstration astronomical EMBRACE - MFAA antennas installed in Contenda Forest (Moura, Portugal).

This new plant achieved "first light" on 25th of April 2017, Freedom day in Portugal, by providing about 4kw of power to a set of EMBRACE – MFAA antenna tiles (Figures 7,8). Astronomical institutes were responsible for requirements, RFI measurements and deployment planning of the system. Besides the testing of dispatchability and compatibility with a radioastronomical system, B4S actually implemented the first world example of an hybrid concentrator engine, opening new avenues for further innovations of green autonomous radioastronomical systems with greater economic impact.

# 6    CONCLUSIONS

Here we outline the major characteristics and innovation approaches to address power efficiency, life cycle impact, and long-term power sustainability for radio astronomy projects like the SKA. The current trends in the area of Green ICT, embodied in the Cloud Computing technologies and DevOps software ideas are influencing the design of compute infrastructures like Data centers and will be key in improving the power operational cost of SKA. The inclusion of Smart Grid technologies will raise greater efficiency and will provide capabilities including detailed power forecasts, improved service reliability, more efficient power asset management, and better operational planning. Power monitoring of antennas and ancillary systems, Correlators, HPC facilities or related data centre tiered systems must include advanced metering technologies, efficient distribution automation and Network Operation Centres (NOC).

Powering with green concepts new radioastronomical Aperture Array stations in the hugely beneficial conditions at the "SKA" sites, appears therefore an obvious choice advantageously positioning the SKA as a "green" telescope while reducing the operational costs and its carbon footprint. Such were key arguments to advance innovative solutions like explored in the project Biostirling for SKA ("B4S"). This also constitute some lessons learned for greening sizeable parts of the SKA with more reliable and affordable photovoltaic technology.

# ACKNOWLEDGMENTS

DB and JPB acknowledge support from FCT through national funds and when applicable co-funded by FEDER – PT2020 partnership agreement under the project UID/EEA/50008/2013. DM acknowledge support from FCUP. The Portuguese team contributed through the ENGAGE SKA RI, grant POCI-01-0145-FEDER-022217, funded by COMPETE 2020 and FCT, Portugal and from grant UID/EEA/50008/2019 funded by FCT. We warmly thank Arnold van Ardenne for invaluable comments and for the inspirational collaboration towards the "greening" of large scale science and the enhancement of its societal impact.

# REFERENCES

[1] R.T. Schilizzi, P. E. F. Dewdney, T. J. Lazio, "The Square Kilometre Array", Proc. of SPIE, Vol. 7012, 70121I, (2008)

[2] European Cooperation in Science and Technology (COST), "Benefits of Research Infrastructures Beyond Science (The Example of the SKA)," COST Strategic Workshop Final Report, Rome, Italy, (2009)

[3] D'Addario, L.R., Wang, D, "An Integrated Circuit for Radio Astronomy Correlators", Journal of Astronomical Instrumentation, Vol. 5, No. 2 (2016)

[4] Helix Nebula, "Strategic Plan for a Scientific Cloud Computing infrastructure for Europe", CERN-OPEN-2011-036, 08/08/2011, (2011)

[5] P.J. Hall, "Power considerations for the Square Kilometre Array (SKA) radio telescope", Invited Paper, In proceedings of: General Assembly and Scientific Symposium, 2011 XXXth URSI, 09/2011; DOI:10.1109/URSIGASS.2011.605120, (2011)

[6] OECD: http://www.oecd-ilibrary.org/economics/country-statistical-profile-chile_20752288-table-chl

[7] A.W. Gunst,, "Cost and power usage evaluations for the various processing architectures: A hardware and operational cost comparison of two architectures for large scale focal plane array beamforming and correlation.", SKA Memo 110, (2014)

[8] U. Weilenmann1, "Renewable Energy for the Paranal Observatory", The Messenger, ESO, 148, 39 June (2012)

[9] R. E. Hills, R. J. Kurza, A. B. Pecka, "ALMA: Status Report on Construction and Early Results from Commissioning", in Ground-based and Airborne Telescopes III, SPIE Conference 7733-38, (2010).

[10] Richard Martin , "Green Data Centers", report by Pike Research, (2012)

[11] P. Chris Broekema, et al, "DOME: Towards the ASTRON & IBM Center for ExaScale Technology", Proc. of the 2012 workshop on HPC for Astronomy (AstroHPC12), Delft, the Netherlands, June, 2012

[12] M. Schmatz, et al, "Scalable, Efficient ASICs for the Square Kilometre Array: from A/D Conversion to Central Correlation", ICASSP 2014, Florence, Italy, May (2014)

[13] CASA Synthesis & Single Dish Reduction, "REFERENCE MANUAL & COOKBOOK", Version May 29, 2013, ©2013 National Radio Astronomy Observatory, Editor: Jürgen Ott – Project Scientist, Jeff Kern – CASA Project Manager, see http://casa.nrao.edu/, (2013)

[14] The Climate Group for the Global e-Sustainability Initiative, "Smart 2020: Enabling the low carbon economy in the information age", (2008)

[15] U.S. Environmental Protection Agency ENERGY STAR Program , "Report to Congress on Server and Data Center Energy Efficiency Public Law", 109-431, (2007)

[16] Chen, Lianping, "Continuous Delivery: Huge Benefits, but Challenges Too", IEEE Software 32 (2): 50. doi:10.1109/MS.2015.27, (2015)

[17] Chen, Lianping; Ali Babar, Muhammad, "Towards an Evidence-Based Understanding of Emergence of Architecture through Continuous Refactoring in Agile Software Development", The 11th Working IEEE/IFIP Conference on Software Architecture (WICSA 2014). IEEE, (2014)

[18] R. Jongerius, S. Wijnholds, R. Nijboer, and H. Corporaal, "End-to-end compute model of the Square Kilometre Array", 2014, IEEE Computer, (2014)

[19] Green Data Centers, Richard Martin, report by Pike Research, (2012)

[20] Mattman, C.A., Hart, A., Cnquini, L., Lazio, J., Khudikyan, S., Jones, D., Preston,R., Bennett, T., Butler, B., Harland, D., Glendenning, B., Kern, J., Robnett, J., Scalable Data Mining, Archiving, and Big Data Management for the Next Generation Astronomical Telscopes, on Big Data Management, Technologies, and Applications, IGI Global book series in Advances in Data Mining and Database Management (ADMDM), ISBN 978-1-4666-4699-5, 196-221, (2014)

# Annexes

# Annex A – Multi-messenger astronomy Synergies

The Square Kilometre Array (SKA) will be entering a new era in radio astronomy and in parallel we are entering an even more exciting period, the multi-messenger astronomy era. This is where we apply the information from across the whole of the electromagnetic emission, neutrinos, cosmic-rays and gravitational waves in order to understand the physical phenomena across the Universe. We highlight below a number of synergies which will support a Key Science cases.

Table 1.

| Electromagnetic radiation | |
|---|---|
| | *SKA &* |
| Radio & Microwave | **VLBI –** <ul><li>Planetary Missions: high precision angular tracking via VLBI phase referencing using in-beam calibrators;</li><li>The SKA1 baseline design covers VHF/UHF frequencies appropriate for some planetary atmospheric probes (band 1) as well as the standard 2.3 GHz deep space downlink frequency allocation (band 3).</li><li>SKA leading role and Discovery of multiple/binary Super-Massive Black Holes (SMBH)- In the next decade, large-scale surveys with the SKA will make significant contributions, driven by superior angular resolution and sensitivity, negligible dust and gas attenuation at GHz frequencies, and the enhanced nuclear accretion that appears to take place in kpc-scale dual and triple AGN.</li><li>SKA2 (particularly when combined with existing VLBI arrays) will completely revolutionize this field once again (following SKA1-MID/SUR) in the low separation parameter space and bridge pulsar timing array results with what is gleaned from mas-scale continuum surveys. This naturally leads to the question: will SKA2-VLBI resolve SMBH binaries that can be detected by pulsar timing arrays? The answer is : YES.</li></ul> **CMB –** <ul><li>better understanding of the Galactic foreground; key astrophysical information for the separation of CMB and the cosmological HI 21 cm emission</li><li>Primordial magnetic fields versus seed magnetic field</li><li>Non-Gaussianity from joint analyses of CMB and radiosources</li><li>Integrated Sachs-Wolfe effect and constraints on dark energy through combination of SKA1 and Planck Surveyor maps.</li><li>Measurement of radio source counts at GHz frequencies to improve spectral knowledge of future space CMB spectrum experiments</li></ul> |
| Submm band | **ALMA –** <ul><li>dust content vs synchrotron emission in AGN/SF systems. Novae are prolific dust formers. Exploring the detailed Radio-Submm-Optical morphologies of novae will allow us to understand how dust is expelled into the interstellar medium.</li><li>Observations conducted with ALMA and the SKA will reveal the process of mass assembly and accretion onto young stars and will be revolutionary for studies of star formation.</li><li>The SKA location in the southern hemisphere makes it particularly suitable to complement ALMA, which is already giving exciting results both on the local and the more distant Universe. By the time the SKA will start observing, ALMA will have already imaged many nearby galaxies in the southern hemisphere, for which no low frequency data at comparably high spatial resolution will be available. The SKA will fill this gap, and have a profound impact on the studies of nearby galaxies, making valuable contributions to our understanding of star formation processes, and of the role of magnetic fields and cosmic rays in them</li></ul> |

| | | |
|---|---|---|
| Infrared band | | **EUCLID** –<br>• SKA-MID matched in resolution, further insight in AGN/SF<br>• Baryonic Acoustic Oscillations<br>• Weak gravitational lensing and galaxy clustering: Euclid + SKA MID photometry + emission line galaxy analyses; Euclid + SKA redshifts,<br>• Discovery of the QSO in the EoR<br>**JWST** –<br>• SKA-MID matched resolution will allow us to compare the structures of novae and reveal unprecedented detail<br>• Galaxy formation and evolution, physics and dynamics Surveys of galaxies at high and intermediate redshifts Mass assembly and star formation, mergers, cold accretion<br>• Early galaxies and black holes z=10-6 |
| Optical band | | |
| | Small telescopes (< 1m) | **MeerLICHT: a simultaneous optical-radio observatory** – These provide great potential as discovery machines. Their wide field-of-view allow us to explore the sky quickly searching for new transient events.<br><br>• MeerLICHT, twinned with MeerKAT, the SKA precursor, provides a new test case for truly simultaneous optical-radio telescopes with great potential for scalability in SKA phases 1 and 2. MeerKAT LSPs will have deep stares of faraway galaxies which will allow us gain insight into how many novae occur in external galaxies.<br>• **Where do Fast Radio Bursts occur?** – With simultaneous optical-radio observations it will be possible to pin point the location of fast radio burst.<br>• **Detecting gravitation waves electromagnetic counterparts** – In order to localize the electromagnetic counterpart of gravitational wave mergers, large field-of-views are required. These can be achieved with optical small telescopes. |
| | Large Telescopes ( > 8m) | **ELT**<br><br>• Detection of the HII region around QSO, at high redshift, discovered by SKA<br>• E-ELT HIRES instrument and SKA to probe the chemical enrichment by the first stars (Signatures of Pop III SNe in Second Generation (SG) Stars/ in gas clouds at high redshifts)<br>• Follow-up joint polarimetric observations between the SKA and the E-ELT for magnetic field estimations (Detailed field structure in the ISM); Stellar jets (with Emission lines)<br><br>**LSST -**<br><br>**The SKA and LSST are the two major ground-based survey telescopes of the next decade.**<br><br>• Weak gravitational lensing and galaxy clustering: The SKA has the potential to provide a large amount of redshift information through observations of HI emission, which can then be used to calibrate LSST objects through crosscorrelation (Newman 2008; McQuinn & White 2013).<br>• The SKA and LSST are well-suited to each other to obtain the necessary observational cosmology data.<br>• Galaxy evolution with LSST and SKA<br>• The evolution of hydrogen; High-redshift galaxies and reionization |
| X-ray & Gamma-ray | | **e-Rosita**<br><br>• **The X-ray satellite eROSITA is expected to be launched in 2016 and** will observe about 105 clusters including 103 high-redshift (z > 1) clusters. With a weak lensing survey by the SKA, the estimation of halo mass will become |

possible for a drastically large number of clusters and we will be able to calibrate the scaling relation much more precisely Colafrancesco et al. (2015).

**Athena**

- **The Advanced Telescope for High Energy Astrophysics (Athena)** is the X-ray observatory large mission selected by the European Space Agency (ESA), within its Cosmic Vision 2015-2025 programme, to address the "Hot and Energetic Universe" scientific theme (Nandra et al. 2013), and it is provisionally due for launch in the early 2030s.
- Understanding the early Universe, including the sources responsible for the reionization of the Universe at $z > 7$ and the formation of the first generation of stars.
- Unveiling the growth of supermassive black holes over cosmic time, and determining its relationship to star formation and the evolution of galaxies.
- Investigating the role of black-hole feedback in shaping galaxy clusters, via the determination of the physical properties of the gas in cluster cores from X-ray observations, and radio studies of the non-thermal cavity contents.
- Determining the nature of non-thermal phenomena in galaxy clusters, including the relationship between cluster radio halos and turbulence, and the connections between X-ray shock structures and radio relics.
- Revealing and illuminating the Cosmic Web of baryons, with the exciting possibility of detecting both thermal and non-thermal emission of cosmic filaments to constrain the plasma conditions at strong accretion shocks, in a hitherto poorly known environment.
- Combining multiple probes of accretion and outflow physics in X-ray Binaries (XRBs), transients, active galactic nuclei (AGN) and Tidal Disruption Events (TDEs).
- Pushing forward our understanding of the life cycles of stars in our Galaxy, including young stellar objects (YSOs) and ultra-cool dwarfs, star-planet magnetic interaction, massive stars, pulsars and supernova remnants (SNR).

| Gravitational waves |
|---|
| *SKA &* LISA |
| Merger of black holes |

| Cosmic rays |
|---|
| *SKA &* Auger |
| Are relativistic radio jets the origin for very high energy cosmic rays? |

| Neutrinos |
|---|
| *SKA &* IceCub*e* |
| Are AGNs the origin of cosmic neutrinos ? |

# Synergies with AIR CENTRE

| | SKA & |
|---|---|
| Radio & Microwave | **RAEGE St. Maria /Flores–**<br><br>• Astrometry, International Reference Systems.<br>• Space debris monitoring, LEO orbits (with radar installed).<br><br>**S.Miguel SATCOM-32 m – (potential station for EVN)**<br>• Planetary Missions: Deep Space network Support; high precision angular tracking via VLBI<br>• Very high resolution astronomy with VLBI enabled (Black Holes, SNs and transients).<br>• QSO monitoring<br>• OH Masers;<br>• Spectroscopy<br>• Space debris monitoring experiments (MEO-GEO orbits)<br><br>**SVOM satellite**<br><br>• SVOM (Space-based multiband astronomical Variable Objects Monitor) is a joint mission of the China National Space Administration (CNSA) and CNES that is set to send aloft a satellite to observe gamma-ray bursts (GRBs) from a 600-km Earth orbit. GRBs are some of the highest-energy phenomena known in the Universe, generated from the explosion of massive stars more than 20 times the mass of our Sun, and from the merger of compact objects like neutron stars or black holes.<br>• Gamma-ray bursts (GRBs) are sudden and intense bursts of X- and gamma-ray light.<br>• cataclysmic formation of black holes, either by the merger of two compact stars (neutron stars or black holes)<br>• Sudden explosion of a massive star, 20 to 100 times larger than our Sun. |
| | |

# Annex B – List of contributors

**Jason Adams**
IT and University of Aveiro,
ASML, The Netherlands

**João Afonso**
Instituto de Astrofísica e Ciências do Espaço, Universidade de Lisboa
Departamento de Física, Faculdade de Ciências, Universidade de Lisboa

**Rui Aguiar**
DETI/IT, University of Aveiro

**Alan Alves**
CIDMA and Physics Department, University of Aveiro

**Stergios Amarantidis**
Instituto de Astrofísica e Ciências do Espaço, Universidade de Lisboa
Departamento de Física, Faculdade de Ciências, Universidade de Lisboa

**António Batel Anjo**
OSUWELA, Maputo
Mathematics Department, University of Aveiro

**Sonia Antón**
CIDMA and Physics Department, University of Aveiro

**Miguel Avillez**
Mathematics Department, University of Évora

**Teresa Barata**
Universidade de Coimbra, Coimbra

**Maria Luísa Bastos**
Faculty of Sciences, University of Porto

**Domingos Barbosa**
IT, University of Aveiro

**João Paulo Barraca**
IT, University of Aveiro

**Tiago Batista**
Ubiwhere Lda, Aveiro

**D. Bartashevich**

IT, University of Aveiro

**Miguel Bergano**
IT, University of Aveiro

**Orfeu Bertolami**
Physics and Astrophysics Department, University of Porto

**Tjarda Boekholt**
CIDMA and Physics Department, University of Aveiro

**José Jasnau Caeiro**
Department of Engineering, Politécnico de Beja

**Bruno Coelho**
Instituto de Teelcomunicações, Aveiro

**Alexandre Correia**
CFisUC, Department of Physics, University of Coimbra, 3004-516 Coimbra, Portugal

**Pedro Costa**
CFisUC, Department of Physics, University of Coimbra

**Valente Cuambe**
Departamento de Física, Universidade Eduardo Mondlane, Maputo

**Mário Cunha**
Faculty of Sciences, University of Porto

**Miguel V. Drummond**
IT, University of Aveiro

**Vanessa Duarte**
IT, IHP – Innovations for High Performance Microelectronics, Germany

**João Fernandes**
Universidade de Coimbra

**Márcio Ferreira**
CFisUC, Department of Physics, University of Coimbra

**Mercedes Filho**
CENTRA/SIM, IST and Departamento de Engenharia Física, FEUP

**João Garcia**
Ubiwhere Lda, Aveiro

**Cláudio Gomes**
Faculty of Sciences, University of Porto

**Diogo Gomes**
IT / University of Aveiro

**José Alberto Gonçalves**
Faculty of Sciences, University of Porto

**Andrew Humphrey**
Instituto de Astrofísica e Ciências do Espaço, CAUP, University of Porto

**Vasco Lagarto**
TICE.PT, Universidade de Aveiro

**Luis Lucas**
Critical Software

**Dinelsa Machaieie**
Departamento de Física, Universidade Eduardo Mondlane, Maputo

**Dalmiro Maia**
Faculty of Sciences, University of Porto

**João Carlos Martins**
Department of Engineering, Politecnico Beja

**António Morais**
CIDMA and Physics Department, University of Aveiro

**P. Moniz**
Departamento de Física e Centro de Matemática e Aplicações, Universidade da Beira Interio

**Bruno Morgado**
Faculty of Sciences, University of Porto

**Israel Matute**
Instituto de Astrofísica e Ciências do Espaço, Universidade de Lisboa
Departamento de Física, Faculdade de Ciências, Universidade de Lisboa

**Rogério N. Nogueira**
IT, University of Aveiro
Watgrid Lda, 3810-193, Aveiro

**Helena Pais**
CFisUC, Department of Physics, University of Coimbra

**Ciro Pappalardo**

Instituto de Astrofísica e Ciências do Espaço, Universidade de Lisboa
Departamento de Física, Faculdade de Ciências, Universidade de Lisboa

**Claudio M. Paulo**
Departamento de Física, Universidade Eduardo Mondlane, Maputo.

**João G. Prata**
Instituto de Telecomunicações

**Renan Pereira**
CFisUC, Department of Physics, University of Coimbra

**Nuno Peixinho**
Universidade de Coimbra, Coimbra

**João G. Prata**
IT, University of Aveiro

**Ricardo Preto**
Ubiwhere Lda, Aveiro

**Constança Providência**
CFisUC, Department of Physics, University of Coimbra

**João G. Rosa**
CIDMA and Physics Department, University of Aveiro

**Tom Scott**
Instituto de Astrofísica e Ciências do Espaço, Universidade do Porto, CAUP

**Neftalí Sillero**
Faculty of Sciences, University of Porto

**Nuno Silva**
Critical Software, Coimbra

**Ana Cláudia Teodoro**
Faculty of Sciences, University of Porto
Instituto Ciências da Terra (CT), polo FCUP

**Ema Valente**
CIDMA and Physics Department, University of Aveiro

**Ricardo Vitorino**
Ubiwhere Lda, Aveiro